\documentclass{aa}
\usepackage{graphicx,rotating}
\usepackage{times}
\usepackage{latexsym}
\usepackage{epsfig}
\usepackage{txfonts}
\usepackage{natbib}
\usepackage{longtable,lscape,multirow,supertabular}
\usepackage{amssymb}
\usepackage{color}

\def\kms{km\,s$^{-1}$}

\def\ie{\emph{i.e.}}
\def\eg{\emph{e.g.}}
\def\Mdot{$\dot{M}$}

\bibpunct{(}{)}{;}{a}{}{,} 
\begin{document}

\title{A far-infrared survey of bow shocks and detached shells around AGB stars and red supergiants~\thanks{
Herschel is an ESA space observatory with science instruments provided by European-led Principal Investigator consortia 
and with important participation from NASA.}}

\author{N.\,L.\,J.~Cox \inst{1} \and
	F.~Kerschbaum \inst{2} \and
	A.-J.~van Marle \inst{3} \and
	L.~Decin \inst{1,4} \and
	D.~Ladjal \inst{1}\thanks{presently at: Department of Physics and Astronomy, University of Denver} \and
	A.~Mayer \inst{2} \and
	M.\,A.\,T.~Groenewegen \inst{5} \and
	S.~van Eck \inst{6} \and
	P.~Royer \inst{1} \and
        R.~Ottensamer \inst{2} \and
	T.~Ueta \inst{7} \and
	A.~Jorissen \inst{6} \and
	M.~Mecina \inst{2} \and
	Z.~Meliani \inst{3,8} \and
        A.~Luntzer \inst{2} \and
	J.\,A.\,D.\,L.~Blommaert \inst{1} \and
	Th.~Posch \inst{2} \and
	B.~Vandenbussche \inst{1} \and
	C.~Waelkens \inst{1}
	}

\institute{Instituut voor Sterrenkunde, K.U.Leuven, Celestijnenlaan 200D, B-3001 Leuven, Belgium  \and
	   University of Vienna, Department of Astronomy, T{\"u}rken\-schanz\-stra\ss{}e 17, A-1180 Wien, Austria \and
	   Centre for Plasma-astrophysics,  K.U.Leuven, Celestijnenlaan 200B, B-3001 Leuven, Belgium \and
	   Sterrenkundig Instituut Anton Pannekoek, Universiteit van Amsterdam, Science Park 904, NL-1098 Amsterdam, The Netherlands \and
	   Royal Observatory of Belgium, Ringlaan 3, B-1180 Brussels, Belgium \and
	   Institut d'Astronomie et d'Astrophysique, Universit\'e Libre de Bruxelles, CP.226, Boulevard du Triomphe, B-1050, Bruxelles, Belgium \and
	   Department of Physics and Astronomy, University of Denver, 2112 E. Wesley Ave., Denver, CO 80208, United States \and
	   Laboratoire Univers et Th\'eories, Observatoire de Paris, UMR 8102 du CNRS, Universit\'e Paris Diderot, F-92190 Meudon, France
     	   }

\offprints{N.L.J.~Cox, \email{nick.cox@ster.kuleuven.be}} 
\date{Received 18 August 2011}

\abstract{}
{%
Our goal is to study the different morphologies associated to the interaction of stellar winds of AGB stars and red 
supergiants with the interstellar medium (ISM) to follow the fate of the circumstellar matter injected into the interstellar medium.
}{
Far-infrared Herschel/PACS images at 70 and 160~$\mu$m of a sample of 78 Galactic evolved stars are used to study 
the (dust) emission structures, originating from stellar wind-ISM interaction. 
In addition, two-fluid hydrodynamical simulations of the coupled gas and dust in wind-ISM interactions are used 
to compare with the observations.
}{%
Four distinct classes of wind-ISM interaction (\ie\ \emph{``fermata''}, \emph{``eyes''}, \emph{``irregular''}, 
and \emph{``rings''}) are identified and basic parameters affecting the morphology are discussed.
We detect bow shocks for $\sim$40\% of the sample and detached rings for $\sim$20\%. 
The total dust and gas mass inferred from the observed infrared emission is similar to the stellar mass loss over a period of 
a few thousand years, while in most cases it is less than the total ISM mass potentialy swept-up by the wind-ISM interaction.
De-projected stand-off distances ($R_0$) -- defined as the distance between the central star and the nearest point of the interaction 
region -- of the detected bow shocks (\emph{``fermata''} and \emph{``eyes''}) are derived from the PACS images and compared to 
previous results, model predictions and the simulations. All observed bow shocks have stand-off distances smaller than 1~pc. 
Observed and theoretical stand-off distances are used together to independently derive the local ISM density.
}{%
Both theoretical (analytical) models and hydrodynamical simulations give stand-off distances for adopted stellar 
properties that are in good agreement with the measured de-projected stand-off distance of wind-ISM bow shocks. 
The possible detection of a bow shock -- for the distance limited sample -- appears to be governed by its physical 
size as set roughly by the stand-off distance. In particular the star's peculiar space velocity and the density of the ISM appear
decisive in detecting emission from bow shocks or detached rings.
In most cases the derived ISM densities concur with those typical of the warm neutral and ionised gas in the Galaxy, though some cases point
towards the presence of cold diffuse clouds.
Tentatively, the \emph{``eyes''} class objects are associated to (visual) binaries, while the \emph{``rings''} generally appear not to 
occur for M-type stars, only for C or S-type objects that have experienced a thermal pulse.
}

\keywords{Stars: AGB and post-AGB - Stars: supergiants - Stars: mass-loss - Stars: winds, outflows 
	  - Stars: circumstellar matter - Infrared: ISM - Hydrodynamics}

\titlerunning{Bow shocks and detached shells in the far-infrared}
\authorrunning{Cox et al.}

\maketitle

\section{Introduction}\label{sec:intro}

The chemical enrichment of the Universe plays an important role in explaining stellar and galaxy evolution. Apart from
supernovae, asymptotic giant branch (AGB) stars and supergiants play a crucial role in supplying heavy elements into the
interstellar medium (ISM). It has been well established that evolved stars lose material during their AGB and supergiant phase
via a dusty wind, and this wind material will eventually be dissipated into the ISM (\eg\ \citealt{2005ASPC..341..605T}). The
presence of extended envelopes around AGB stars and supergiants is a direct result of their stellar mass loss. The properties
of these (detached) shells are directly affected by the mass-loss history of low- to intermediate-mass stars and thus contain
information on this late stage of stellar evolution.

Both the chemical composition and physical conditions of the stellar-wind material returned to the ISM can be altered by
several processes, such as mixing of nucleosynthesis products into the envelope, dust formation
(\citealt{2006A&A...447..553F}), shock-induced chemistry in the inner wind envelope (\citealt{1999A&A...341L..47D}), as well
as photo-dissociation due to interstellar cosmic rays and UV photons penetrating the outer envelope
(\citealt{1997A&A...324..237W}; \citealt{2010Natur.467...64D}). There is a marked difference between the dust and gas formed
in circumstellar environments and that present in the ISM (\eg\ \citealt{2001RSPTA.359.1961J}). 
To reconcile these differences in the composition of circumstellar
and interstellar dust additional processes are expected  to play an important role. One such mechanism involves processing and
alteration of dust in shocks that can occur when the stellar wind material interacts with the local ISM. Shocks are not only
important in that they can alter the composition of the dusty wind but they also generate turbulence and acoustic noise, which
affect the evolution of the ISM (\citealt{1974ApJ...189L.105C}; \citealt{1977ApJ...218..148M}) and are believed to be of
importance to the initial phases of star formation (\citealt{1975ApJ...195..715M}; \citealt{1982ApJ...262..315S}).

IRAS observations already hinted at the existence of low temperature circumstellar dust (\citealt{1988A&A...194..125V}). Clear
observational evidence for detached shells around evolved stars was found by \eg\ \citet{1988A&A...196L...1O,
1988AJ.....95..141S,1990A&A...229L...5H}. Following these discoveries, infrared observations with IRAS and ISO revealed a
growing number of AGB stars with extended and sometimes clearly detached thermal dust emission
\citep[e.g.][]{1993ApJS...86..517Y,1993ApJ...409..725Y,1994A&A...281L...1W,1996A&A...315L.221I,1998Ap&SS.255..349H}. 
At the same time, CO radio line emission revealed a number of objects with detached gas shells 
\citep[and references therein]{1996Ap&SS.245..169O}. A number of detached-shell objects was also imaged in the optical
\citep[e.g.][]{2001A&A...372..885G,2010A&A...511A..37M,2010A&A...515A..27O}. Interferometric radio maps uncovered their
detailed structure, which turned out to be remarkable spherical thick CO-line-emitting, geometrically-thin shells
(\citealt{1999A&A...351L...1L}; \citealt{2000A&A...353..583O}). These observations showed that possibly short phases of
intense mass-loss rate are required to form such large detached (spherical) molecular shells of swept up wind and interstellar
material (\citealt{1986MNRAS.222..273R,2002ApJ...572.1006Z,2007MNRAS.380.1161L}). The enhanced mass-loss rate events could be
induced by a thermal pulse (\citealt{1988A&A...194..125V}; \citealt{1993ApJ...413..641V}) or, alternatively, the He core-flash
at the end of the RGB phase (\citealt{1984ApJS...55...27D}). 

IRAS also observed for the first time the infrared signatures of bow shocks associated to \eg\ ultra-compact \ion{H}{ii}
regions (\citealt{1990ApJ...353..570V}; \citealt{1991ApJ...369..395M}) and runaway OB stars (\citealt{1997AJ....114..837N}). 
\citet{1988AJ.....95..141S} presented the first IRAS detection of a bow shock around a red supergiant, $\alpha$\,Ori. These
data were re-processed by \citet{1997AJ....114..837N} which revealed a detached bow shock at $\sim$6\arcmin\ as well as a
linear bar at 9\arcmin. Only recently, infrared space telescopes such as Spitzer and AKARI reveal much more detail on the
infrared signatures of bow shocks around evolved stars, such as R\,Hya (\citealt{2006ApJ...648L..39U}), $\alpha$~Ori
(\citealt{2008PASJ...60S.407U}), R Cas (\citealt{1998Ap&SS.255..349H,2010A&A...514A..16U}), and U\,Hya (\citealt{2011A&A...528A..29I}). The
Herschel Space Observatory now reveals these infrared shells and bow shock regions at the best spatial resolution ever.
Detached infrared shells around TT\,Cyg, U\,Ant, and AQ\,And are discussed in \citet{2010A&A...518L.140K} and on more objects
in \citet{Kerschbaum2011}. Bow shocks are reported for CW\,Leo (\citealt{2010A&A...518L.141L}), X\,Her and TX\,Psc
(\citealt{2011A&A...532A.135J}), and $\alpha$~Ori (Decin et al. in preparation). For $o$ Cet the inner part of the stellar wind
bubble bounded and formed by the termination shock is seen (\citealt{2011A&A...531L...4M}). For some of these infrared bow
shocks, counterparts are detected in UV emission ($o$\,Cet: \citealt{2007Natur.448..780M}; CW\,Leo:
\citealt{2010ApJ...711L..53S}). These observations indicate that the wind material undergoes additional processing in wind-ISM
shocks and that such processing is perhaps more common than previously envisioned.

\begin{figure}[th!]
 \centering
 \includegraphics[angle=0,width=0.24\textwidth]{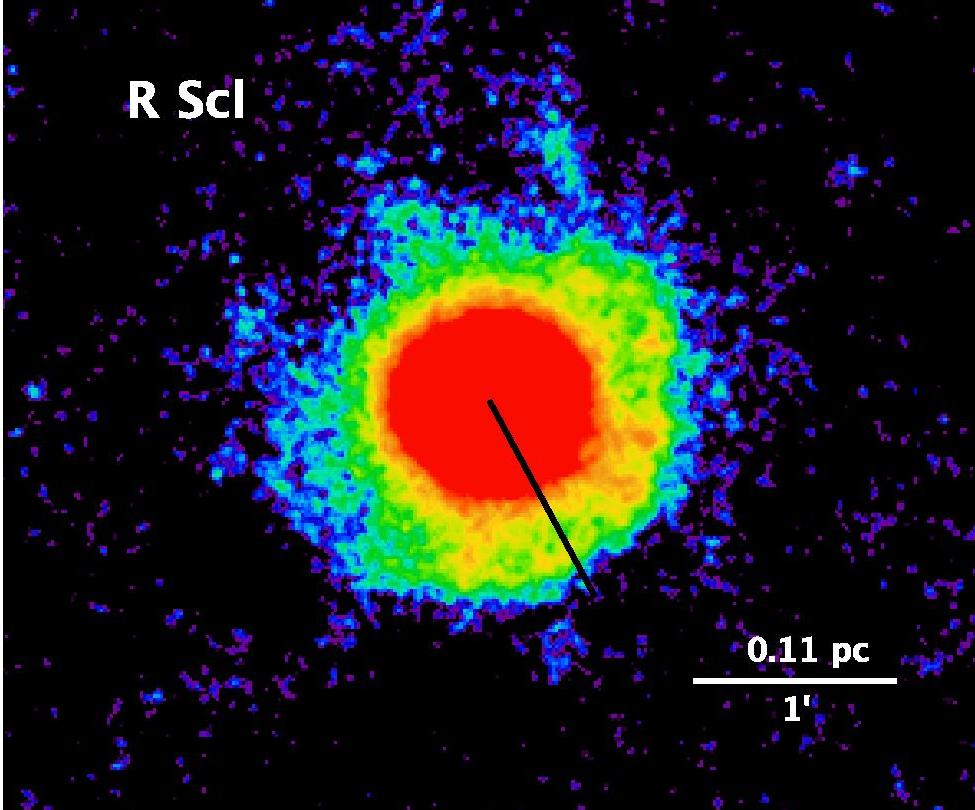}
 \includegraphics[angle=0,width=0.24\textwidth]{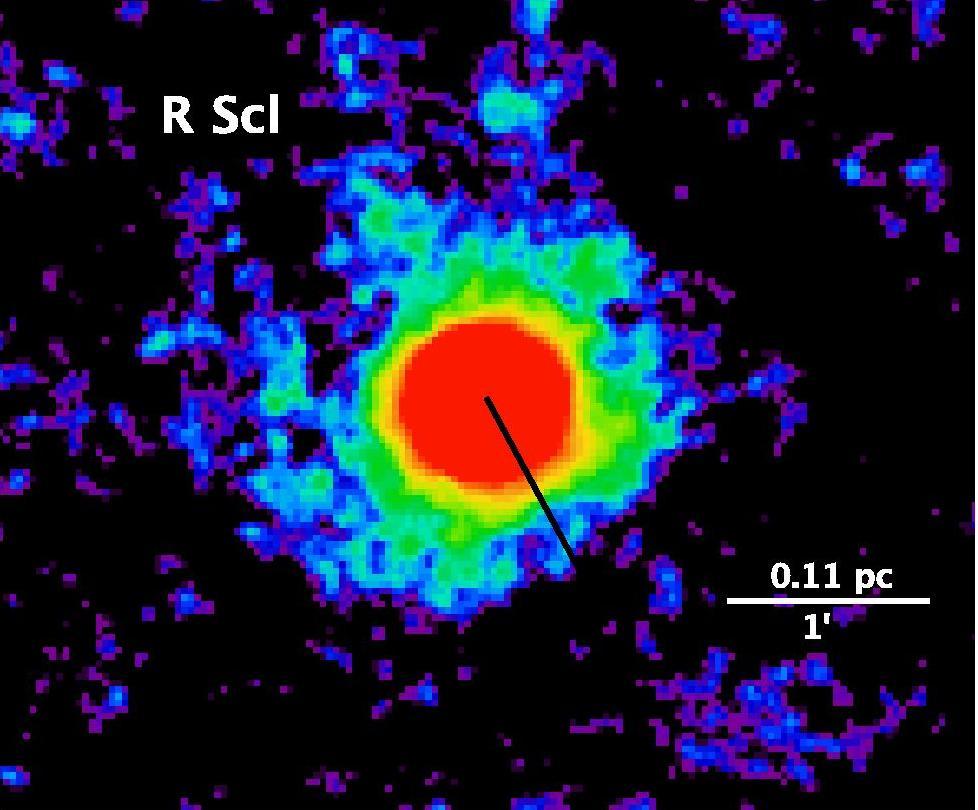}
 \medskip
 \includegraphics[angle=0,width=0.24\textwidth]{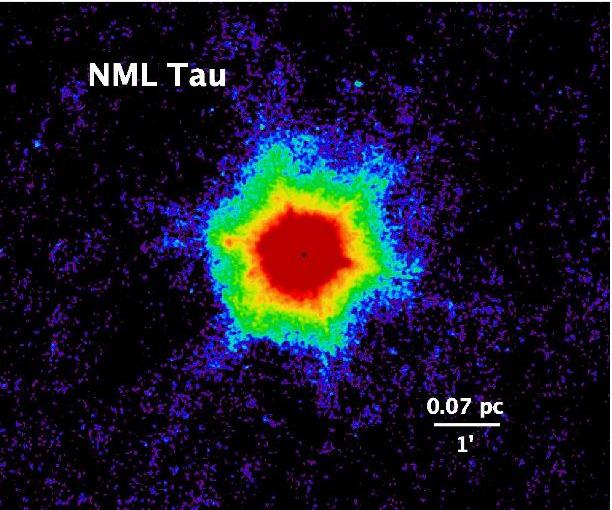}
 \includegraphics[angle=0,width=0.24\textwidth]{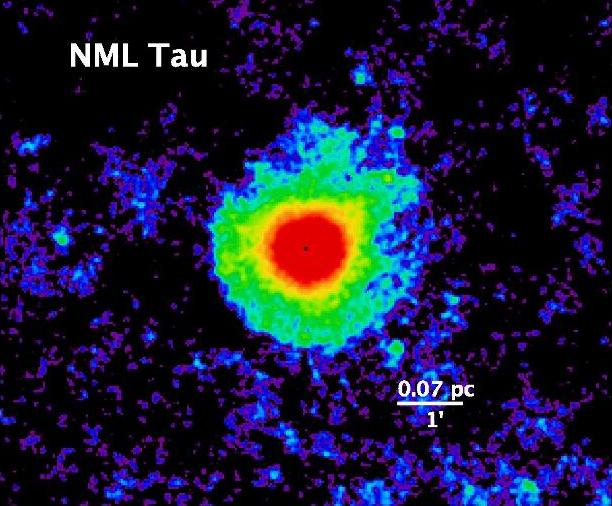}
 \medskip
 \includegraphics[angle=0,width=0.24\textwidth,clip]{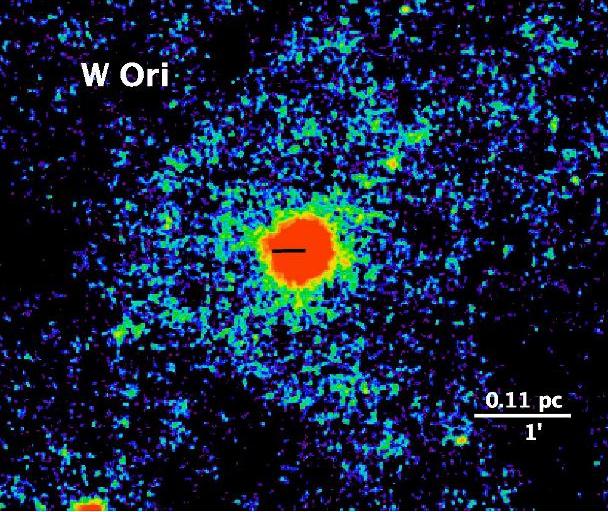}
 \includegraphics[angle=0,width=0.24\textwidth,clip]{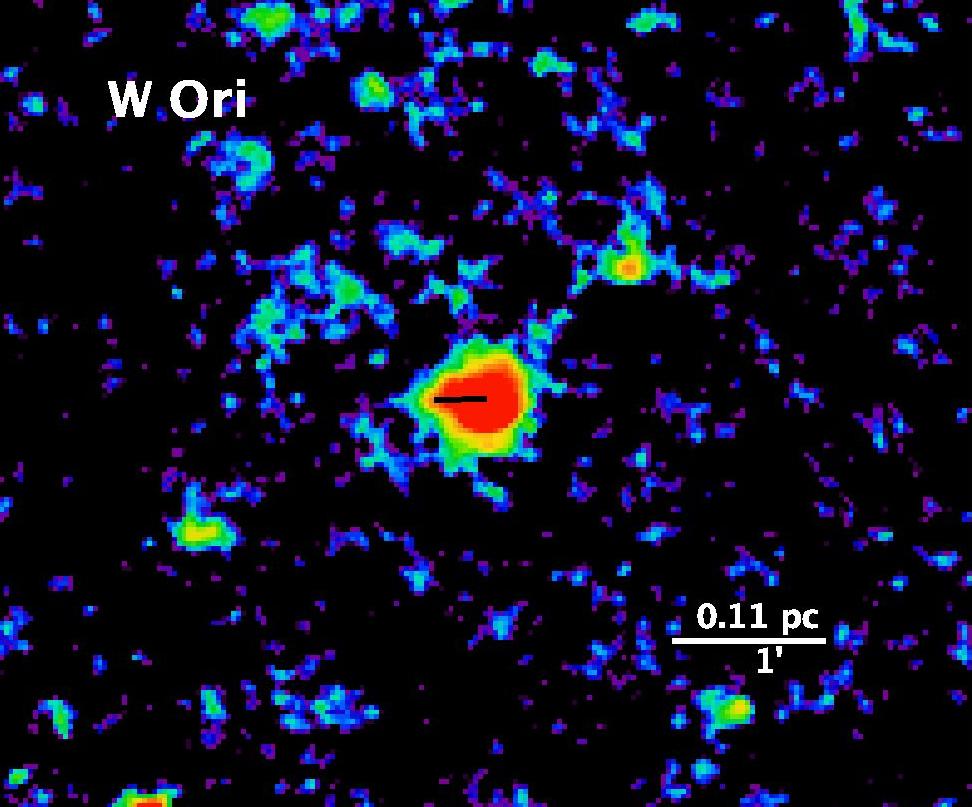}
 \medskip
 \includegraphics[angle=0,width=0.24\textwidth]{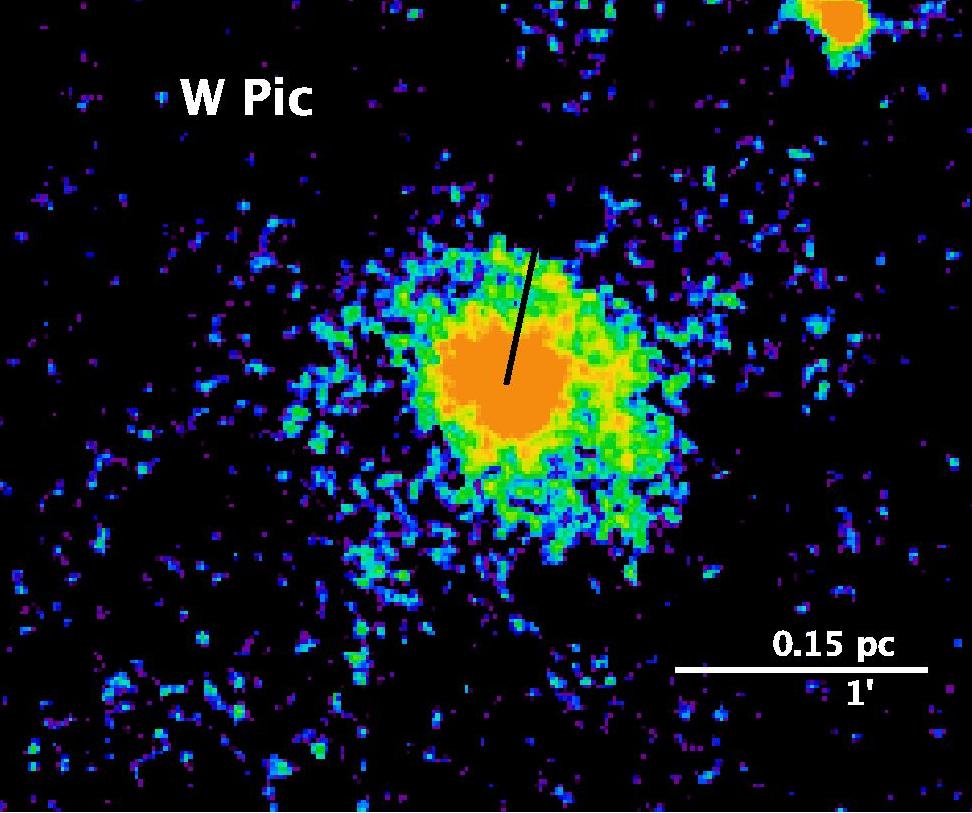}
 \includegraphics[angle=0,width=0.24\textwidth]{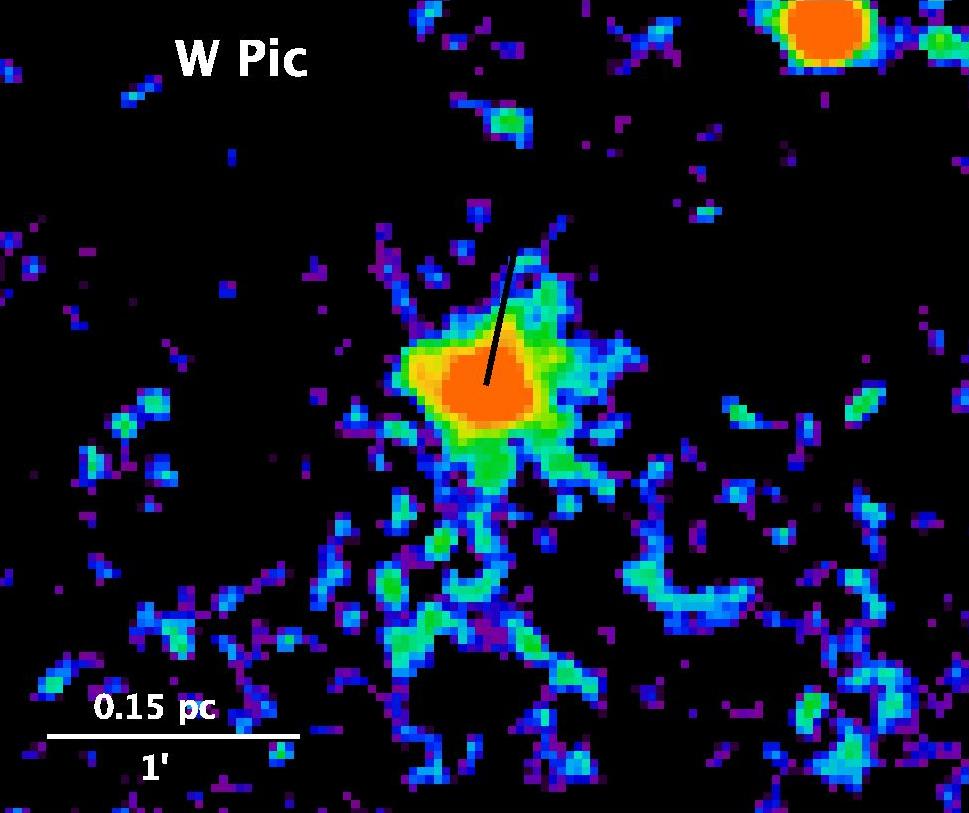}
 \medskip
 \includegraphics[angle=0,width=0.24\textwidth]{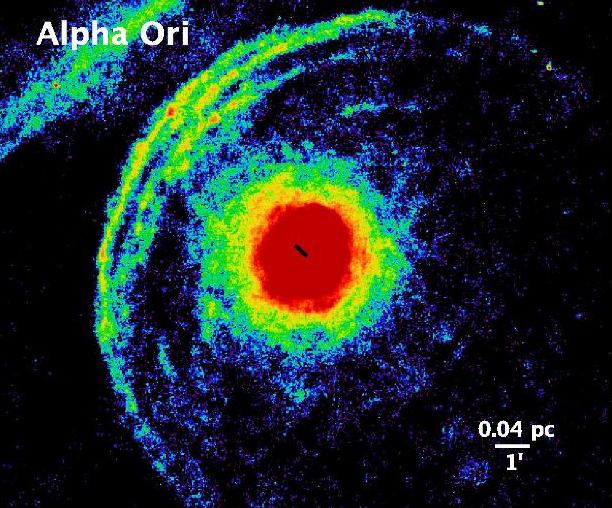}
 \includegraphics[angle=0,width=0.24\textwidth]{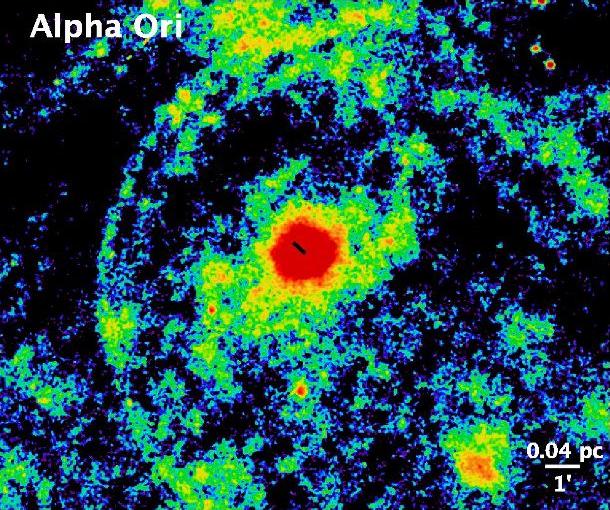}
 \caption{Interaction type \emph{``fermata''} (Class\,I). PACS 70~$\mu$m (left) and 160~$\mu$m (right).
 The white horizontal bar in each panel is 1\arcmin\ in length,
 annotated with the corresponding physical size. All panels have north up and east to the left.
 The black line indicates the direction of the space velocity of the star (adopting a scale such that 1~\kms\
 corresponds to 1\arcsec on the image).
 Note: R Scl also has an inner spherical shell (not visible here).
 One needs to multiply flux values by $4.25 \cdot 10^{4}$ to convert from Jy arcsec$^{-2}$ to MJy sr$^{-2}$.}
 \label{fig:fermata-1}\label{fig:firstmap}
\end{figure}

\addtocounter{figure}{-1}

\begin{figure}[th!]
 \centering
 \includegraphics[angle=0,width=0.24\textwidth]{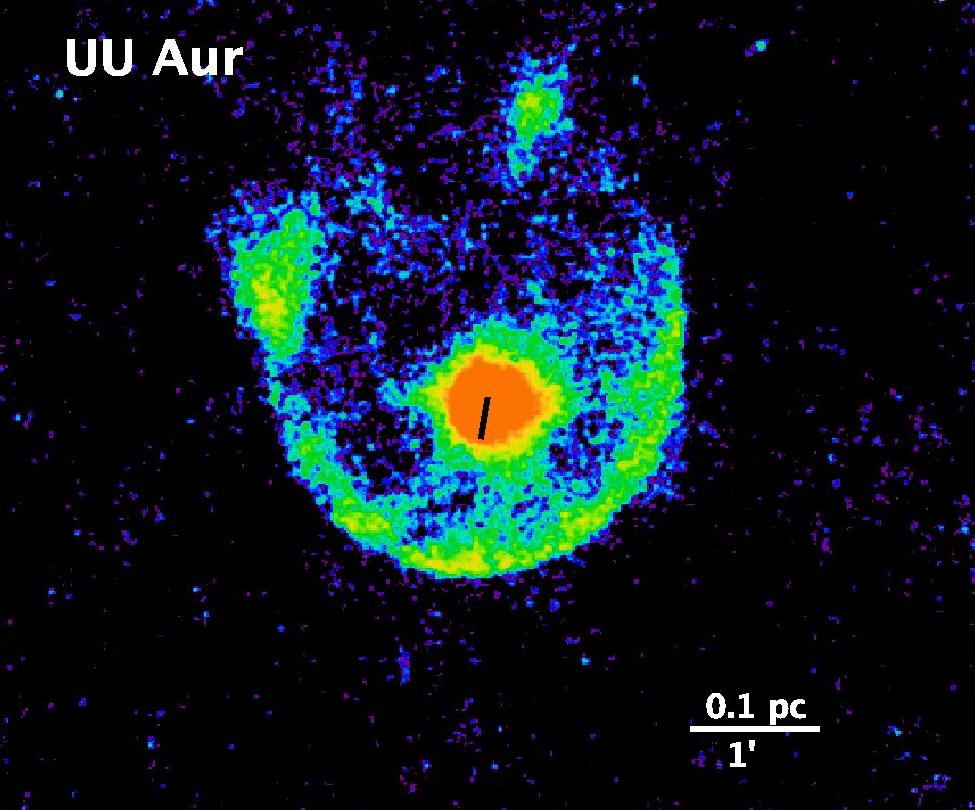}
 \includegraphics[angle=0,width=0.24\textwidth]{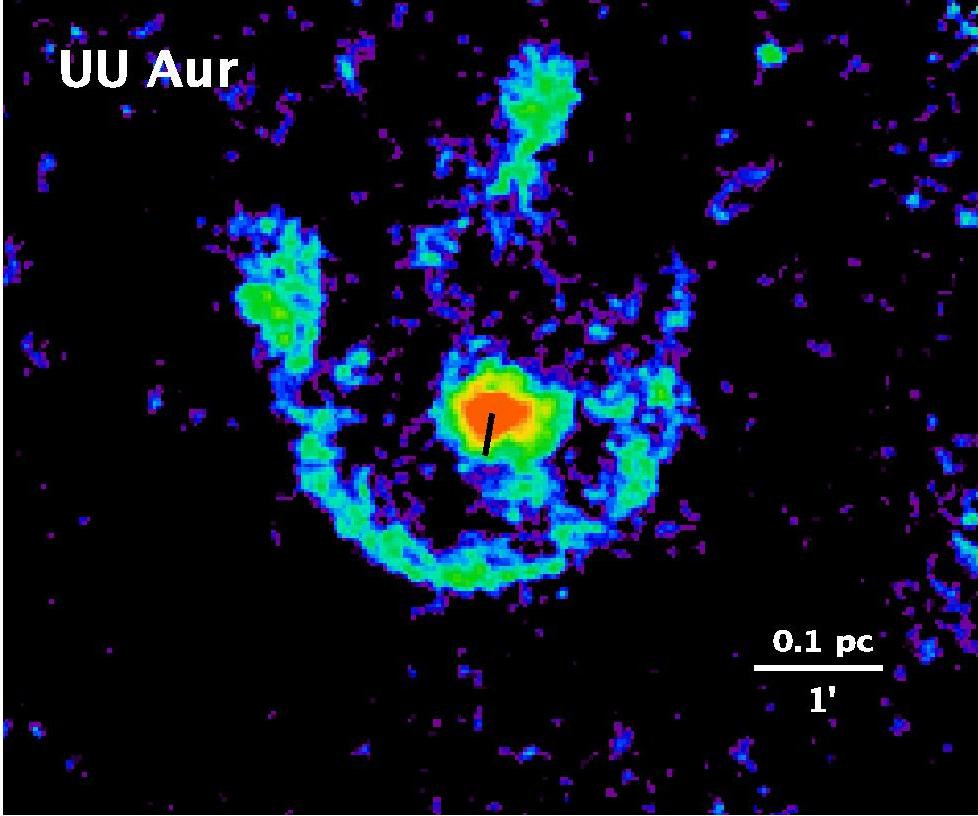}
 \medskip
 \includegraphics[angle=0,width=0.24\textwidth]{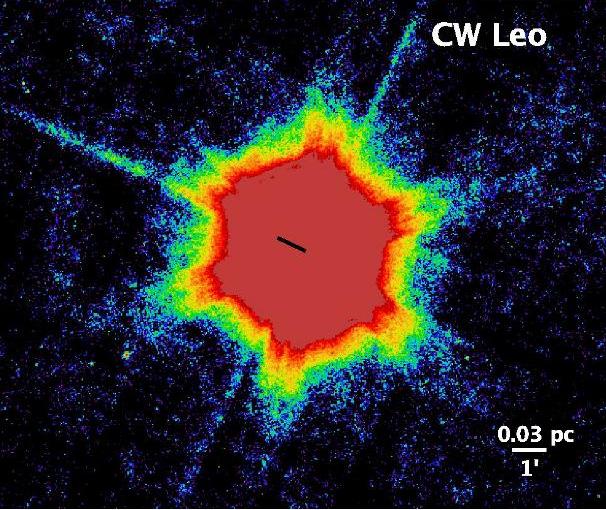}
 \includegraphics[angle=0,width=0.24\textwidth]{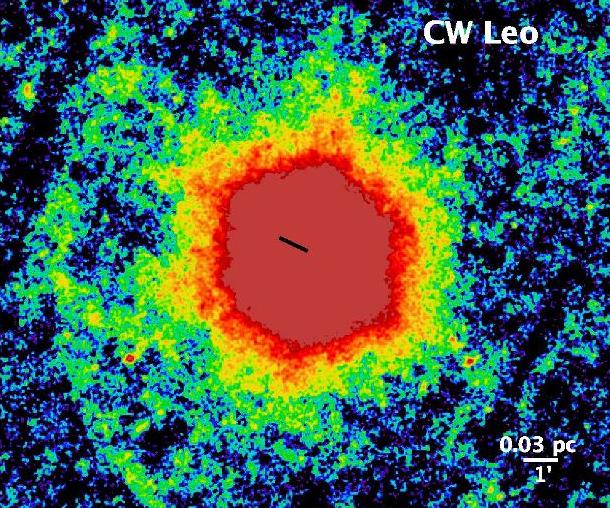}
 \medskip
 \includegraphics[angle=0,width=0.24\textwidth]{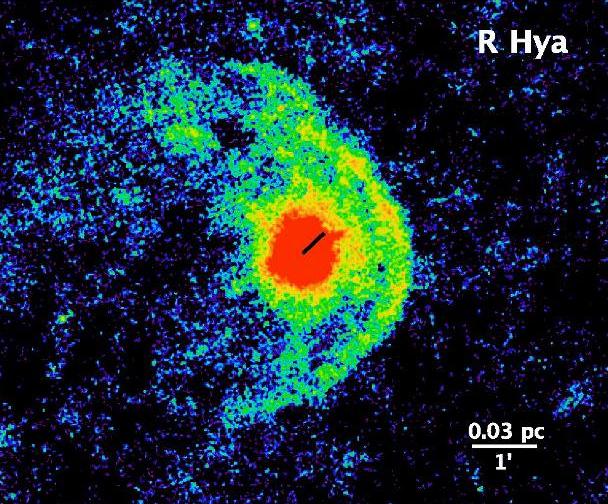}
 \includegraphics[angle=0,width=0.24\textwidth]{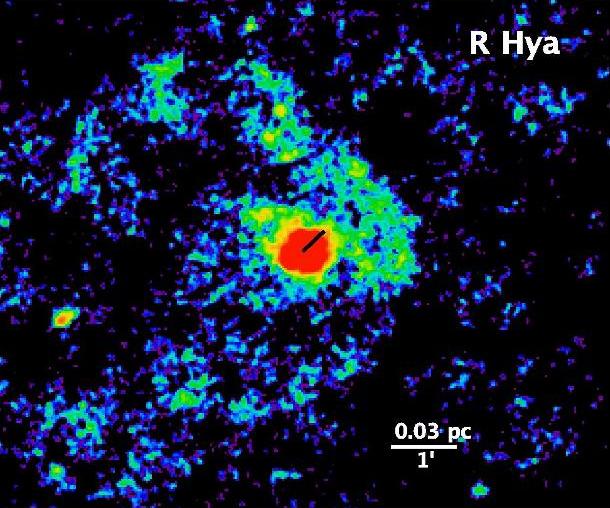}
 \medskip
 \includegraphics[angle=0,width=0.24\textwidth]{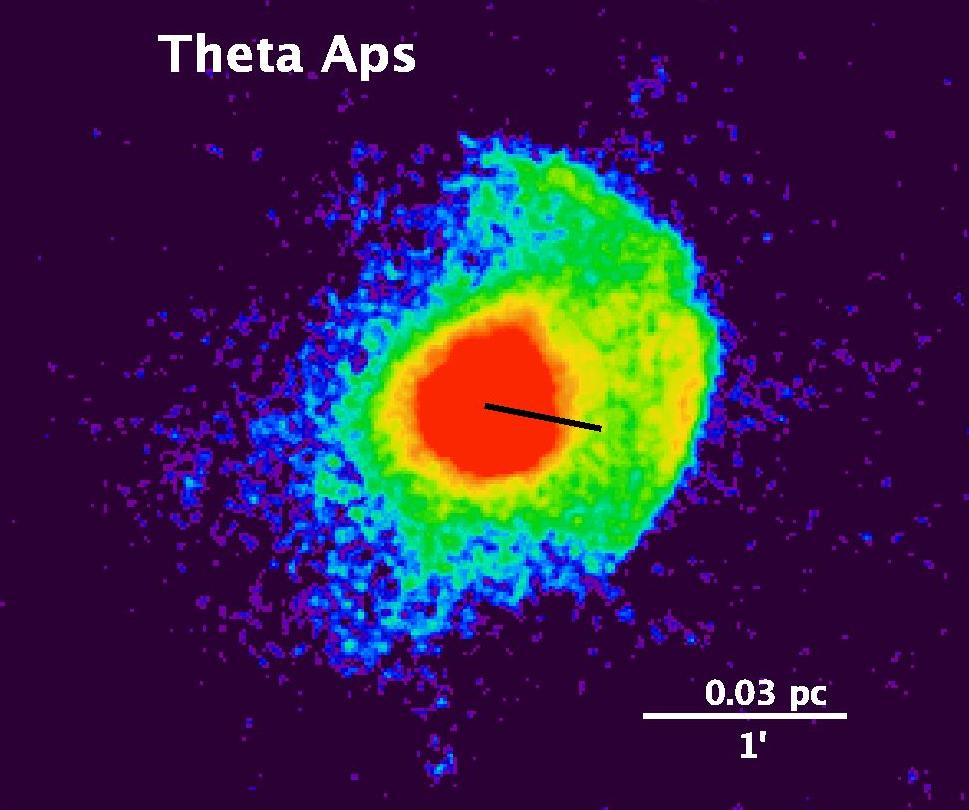}
 \includegraphics[angle=0,width=0.24\textwidth]{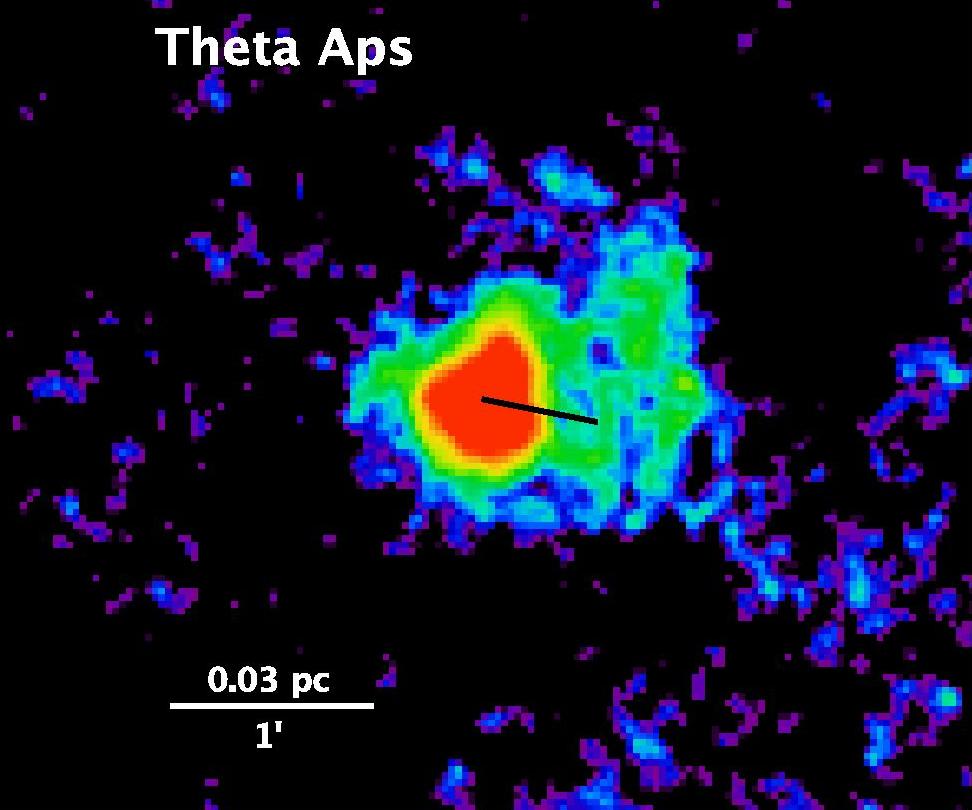}
 \medskip
 \includegraphics[angle=0,width=0.24\textwidth]{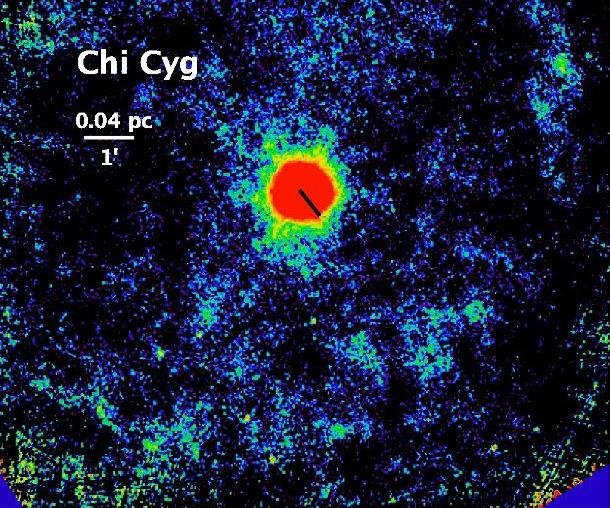}
 \includegraphics[angle=0,width=0.24\textwidth]{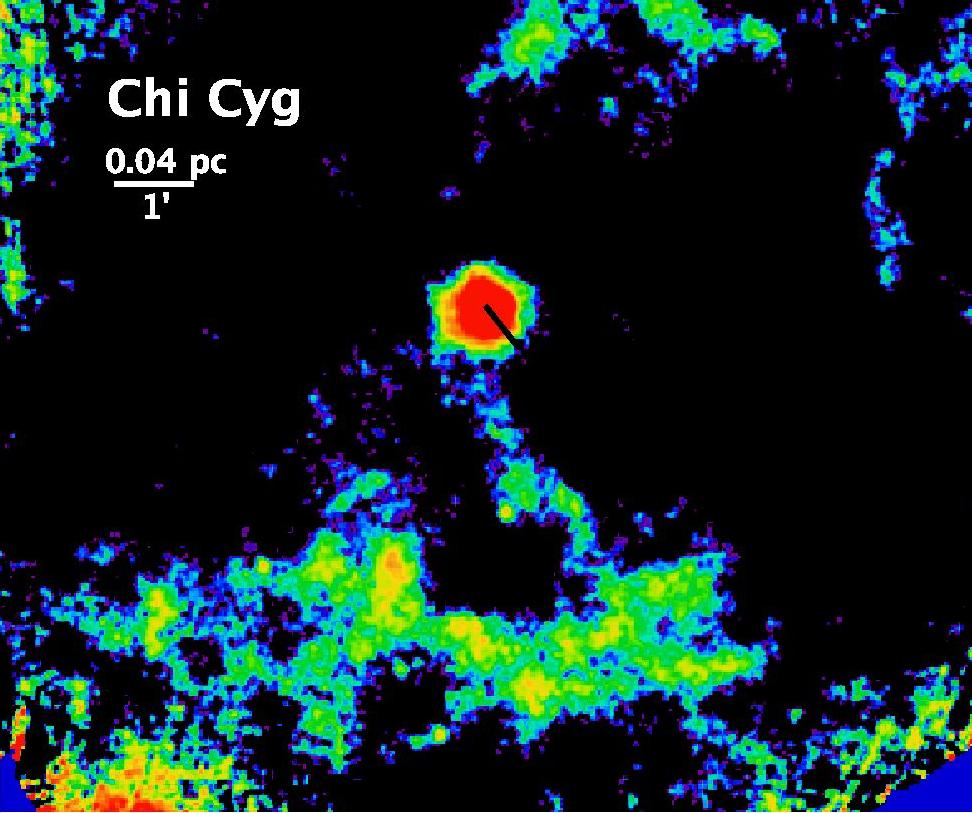}
 \caption{(continued). Interaction type \emph{``fermata''} (Class\,I). PACS 70~$\mu$m (left) and 160~$\mu$m (right).
 For CW Leo see also Ladjal et al. (2010) and for $\alpha$~Ori see Decin et al. (in preparation).
 The bow shock of $\chi$\,Cyg is at several arcminutes to the south (not readily discernible in the image).}
 \label{fig:fermata-1}\label{fig:firstmap}
\end{figure}

\addtocounter{figure}{-1}

\begin{figure}[th!]
 \centering
 \includegraphics[width=0.24\textwidth,clip]{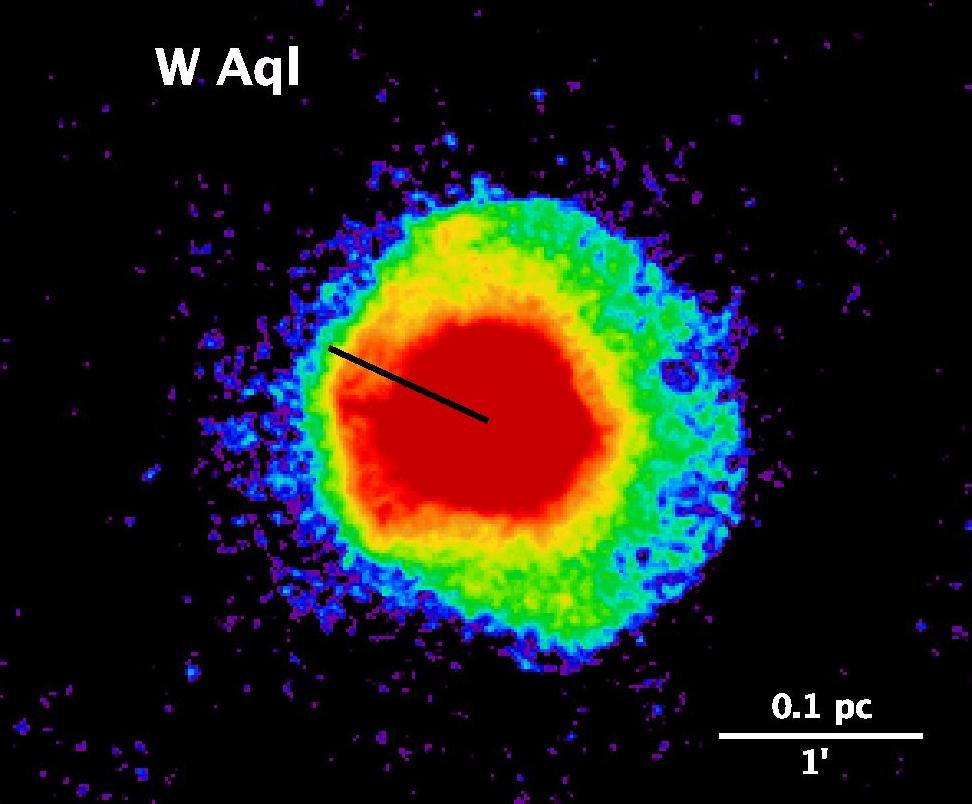}
 \includegraphics[width=0.24\textwidth,clip]{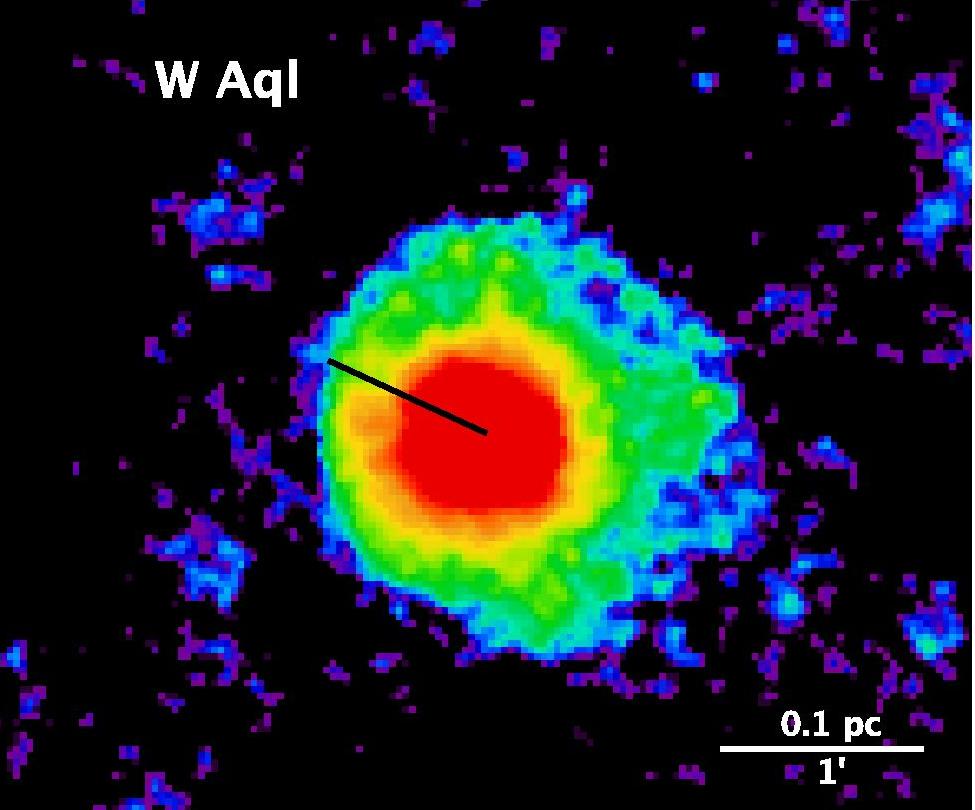}
 \includegraphics[width=0.5\columnwidth]{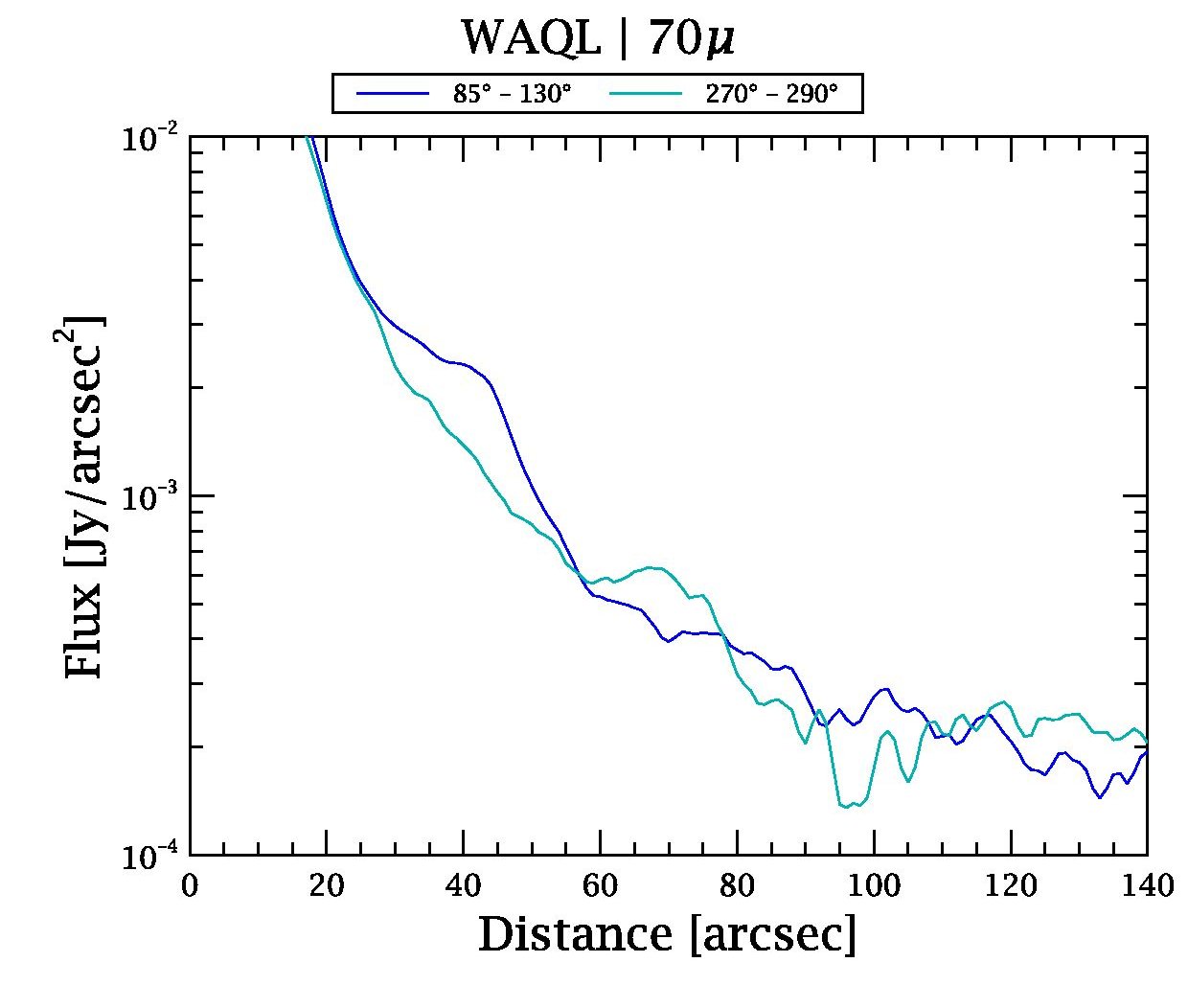}\\
 \medskip
 \includegraphics[width=0.24\textwidth,clip]{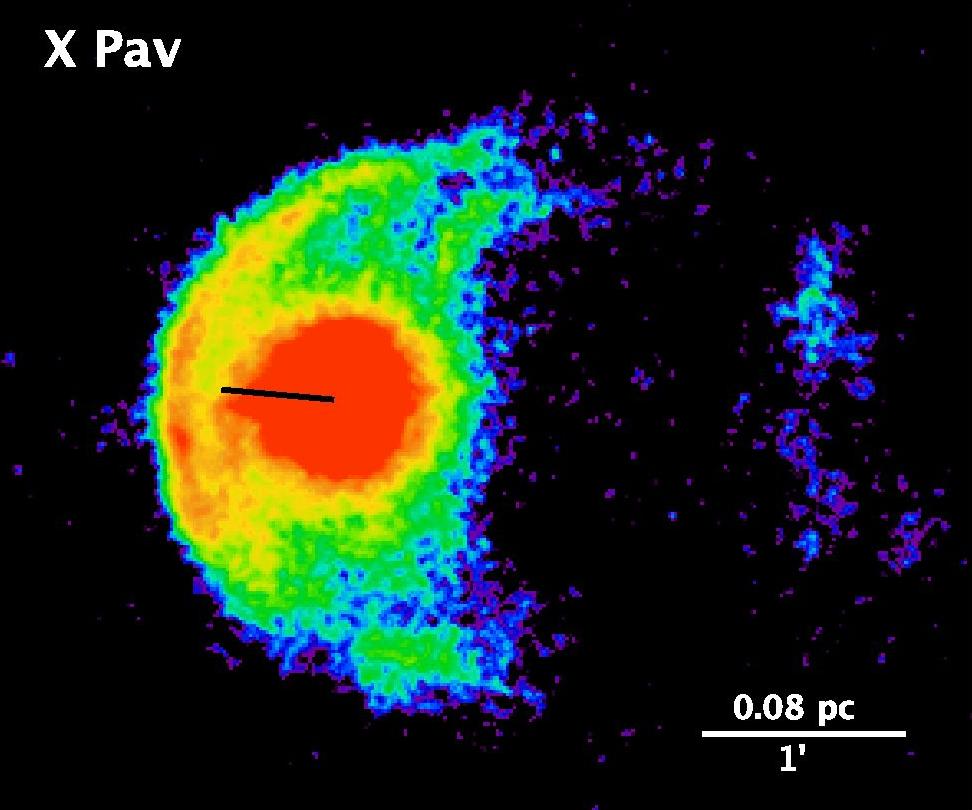}
 \includegraphics[width=0.24\textwidth,clip]{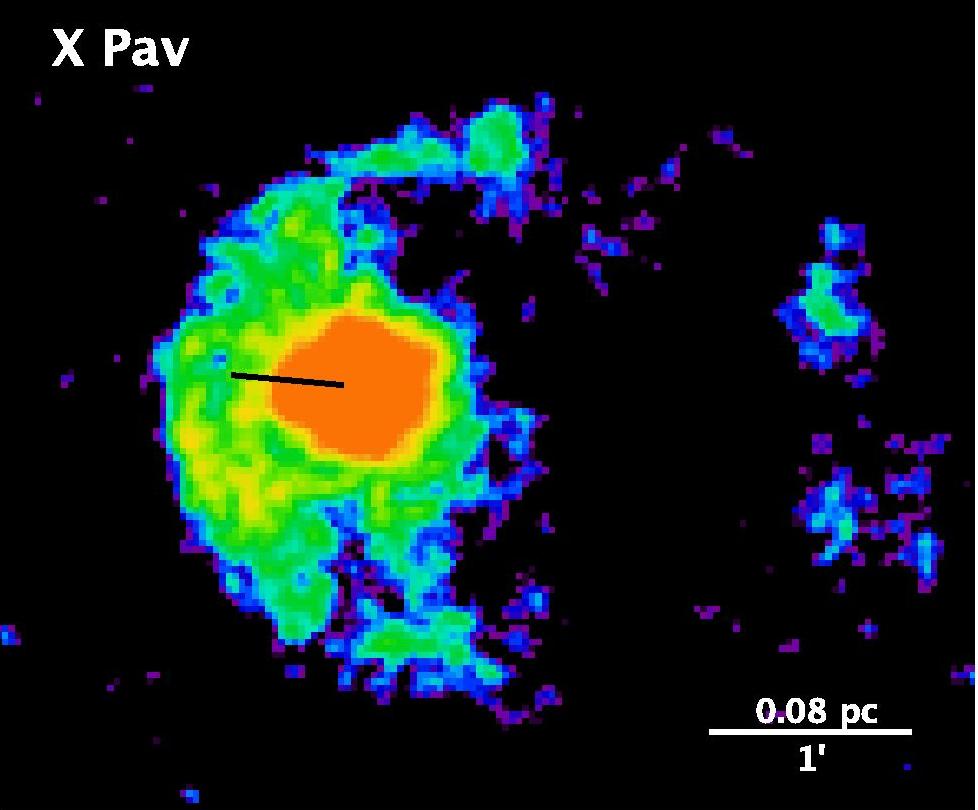}
 \medskip
 \includegraphics[width=0.24\textwidth,clip]{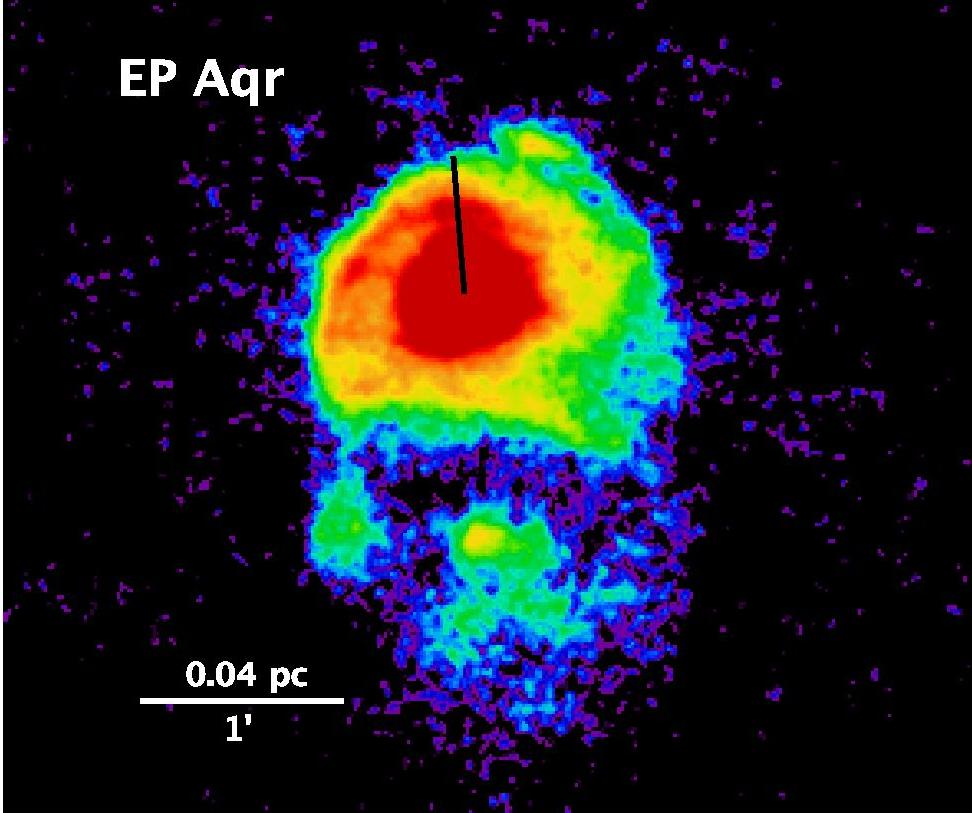}
 \includegraphics[width=0.24\textwidth,clip]{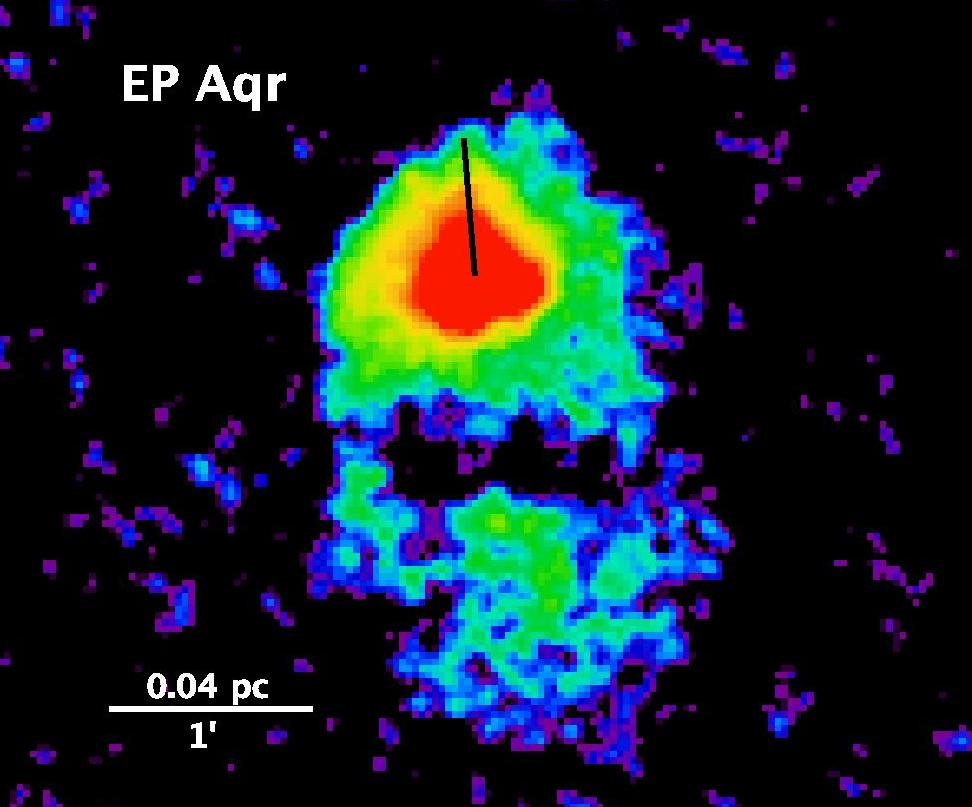}
 \medskip
 \includegraphics[width=0.24\textwidth,clip]{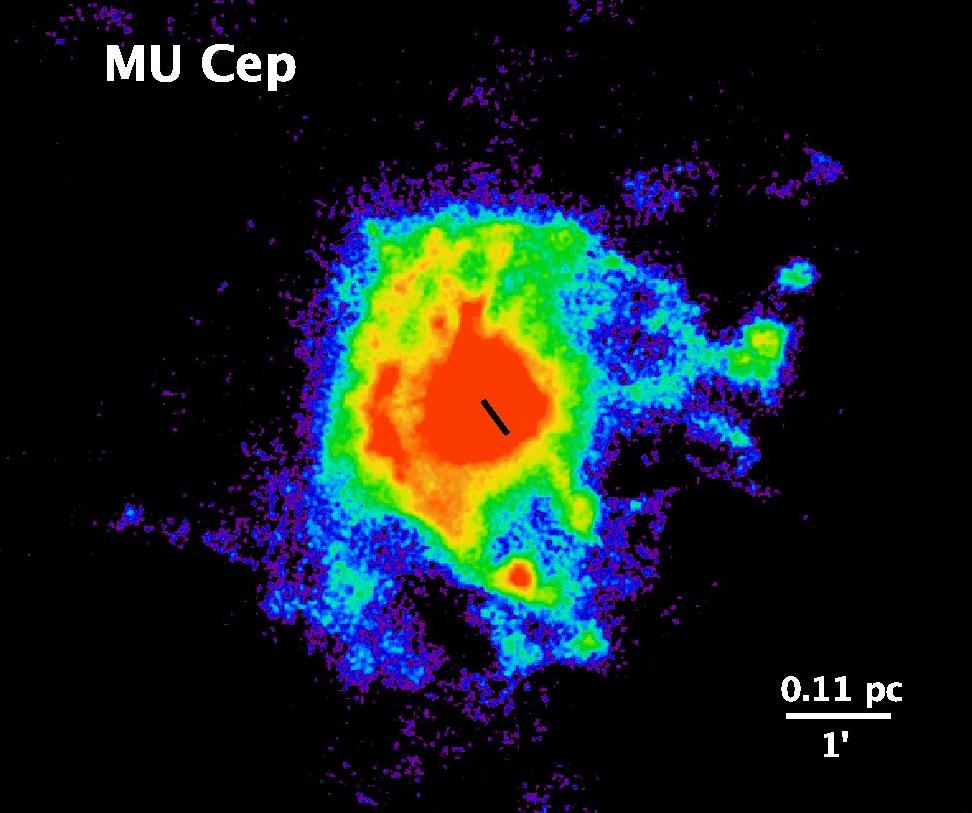}
 \includegraphics[width=0.24\textwidth,clip]{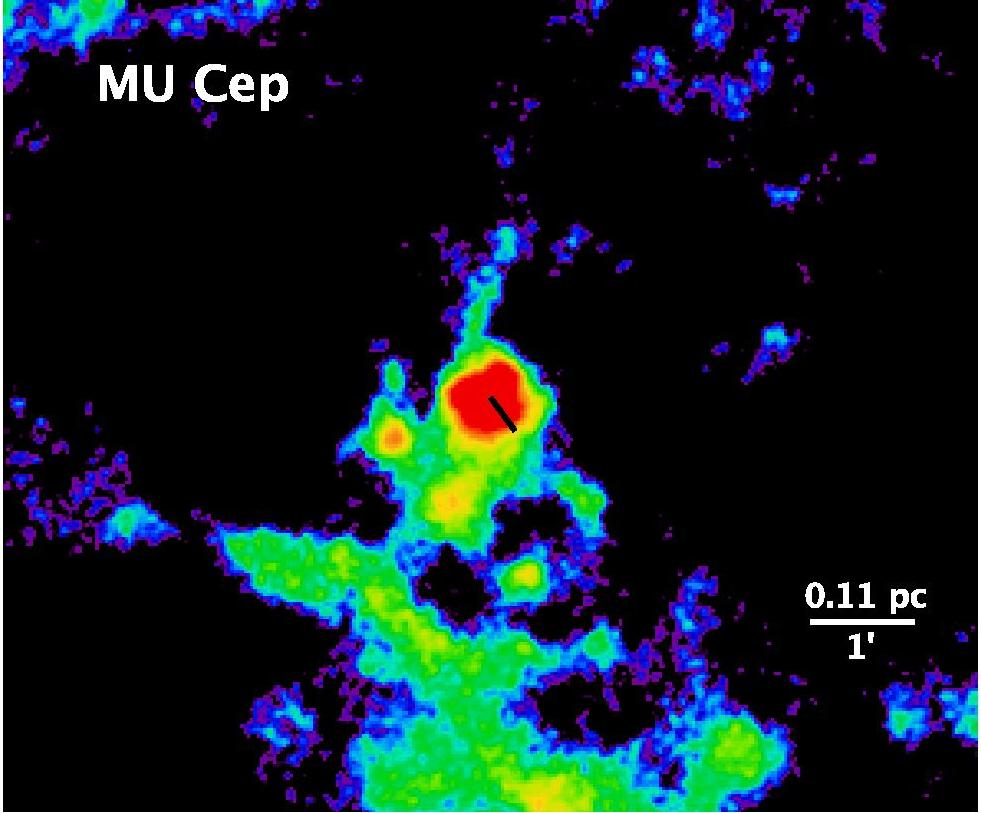}
 \caption{(continued). Interaction type \emph{``fermata''} (Class\,I). PACS 70~$\mu$m (left) and 160~$\mu$m (right).
 Radial profiles (azimuthally averaged over two opening angle ranges: 85-130\degr\ and 270-290\degr) 
 are shown for 70~$\mu$m image of W\,Aql (bottom right).
 Note the presence of turbulent instabilities in $\mu$\,Cep.}
 \label{fig:fermata-bullets-instabilities}
\end{figure}

\addtocounter{figure}{-1}

This paper presents the properties -- such as morphology, size, and brightness -- of detached or spatially extended far-infrared
emission associated to bow shock interaction regions and detached shells around a large sample of AGB stars and red
supergiants.  The paper is structured as follows. First we discuss the observations, the data processing and map-reconstruction 
scheme, and present the infrared images in Sect.~\ref{sec:obs}. In Sect.~\ref{subsec:classification} we
introduce a simple morphological classification system differentiating observed shapes, in particular we make a distinction
between bow shocks and detached rings. In Sects.~\ref{subsec:radialprofile} and~\ref{subsec:flux} we present radial profiles
as well as infrared flux measurements of the extended emission. In order to make a qualitative and quantitative comparison of
the observed bow shocks and detached rings we review first the basic physics and relevant stellar and interstellar parameters
underlying the formation of a bow shock or detached shell (Sect.~\ref{sec:bowshock}). In Sect.~\ref{sec:properties} we discuss
and collect the relevant parameters in order to predict, for example, the size of the bow shock. Next, in
Sect.~\ref{sec:hydro}, we present hydrodynamical simulations to elaborate on the origin of the observed morphologies, in
particular, the features related to (turbulent) instabilities. The observations are compared to the theoretical predictions
and the simulations in Sect.~\ref{sec:discussion}. Furthermore, the observations are interpreted in the context of stellar
mass-loss and ISM properties as well as with respect to the interaction between these two material flows. Then we discuss the
requirements to observe wind-ISM interaction, and thus address the frequent occurrence of bow shocks for our entire sample of
AGB stars and red supergiants. The main results and conclusions are summarised in Sect.~\ref{sec:conclusion}.

\begin{figure}[t!]
 \centering
 \includegraphics[width=0.24\textwidth,clip]{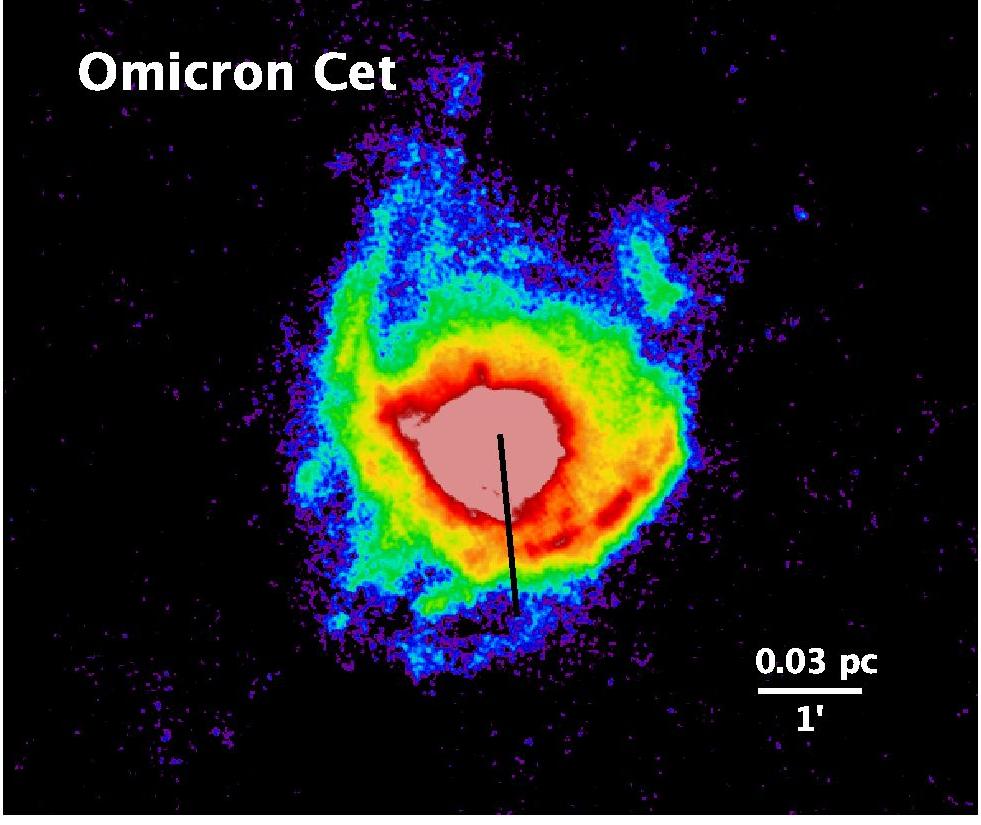}
 \includegraphics[width=0.24\textwidth,clip]{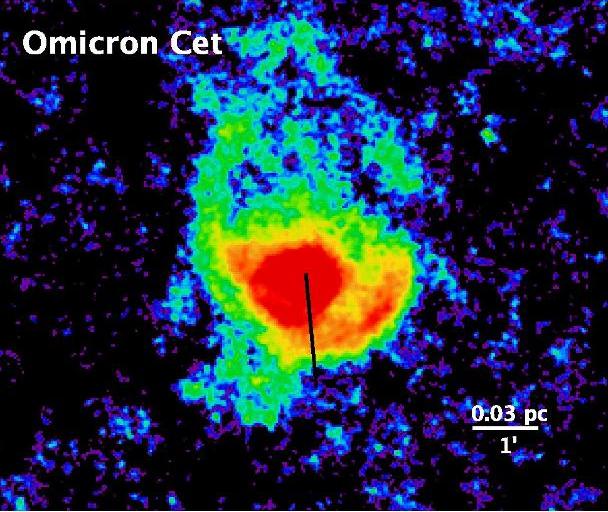}
 \medskip
 \includegraphics[width=0.24\textwidth,clip]{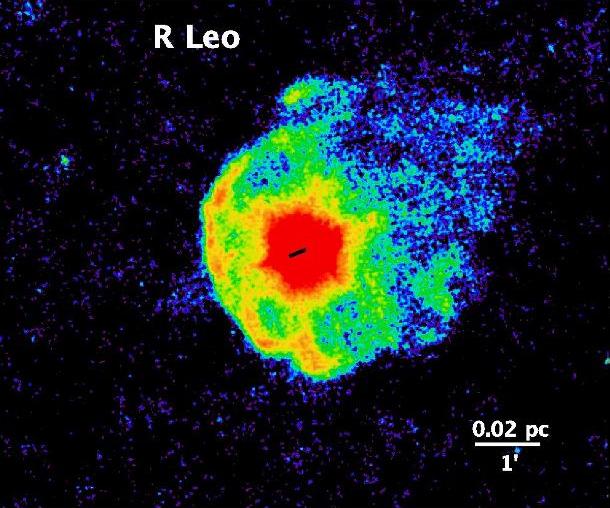}
 \includegraphics[width=0.24\textwidth,clip]{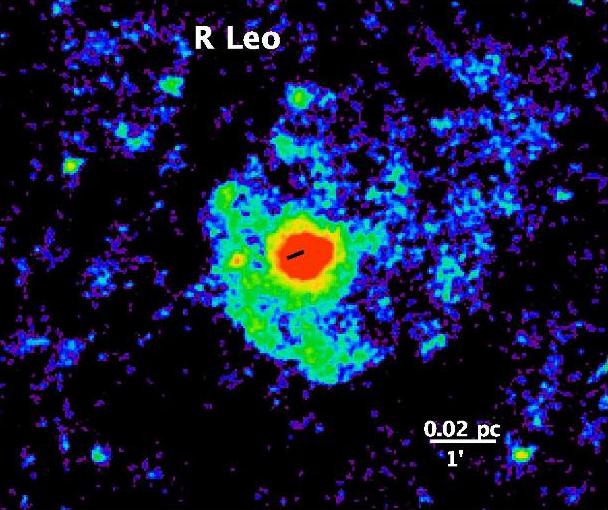}
 \medskip
 \includegraphics[width=0.24\textwidth,clip]{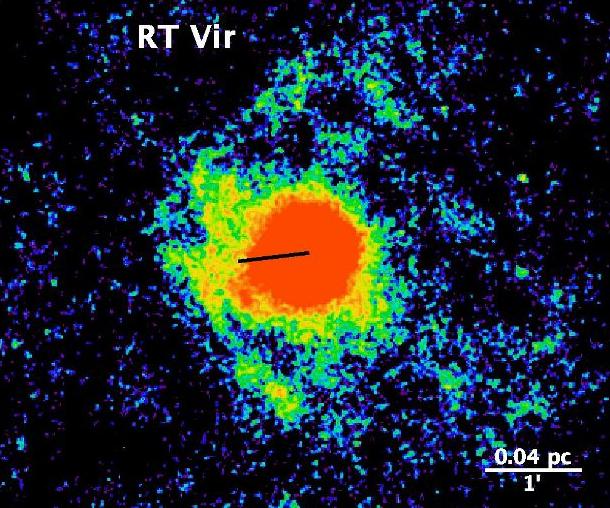}
 \includegraphics[width=0.24\textwidth,clip]{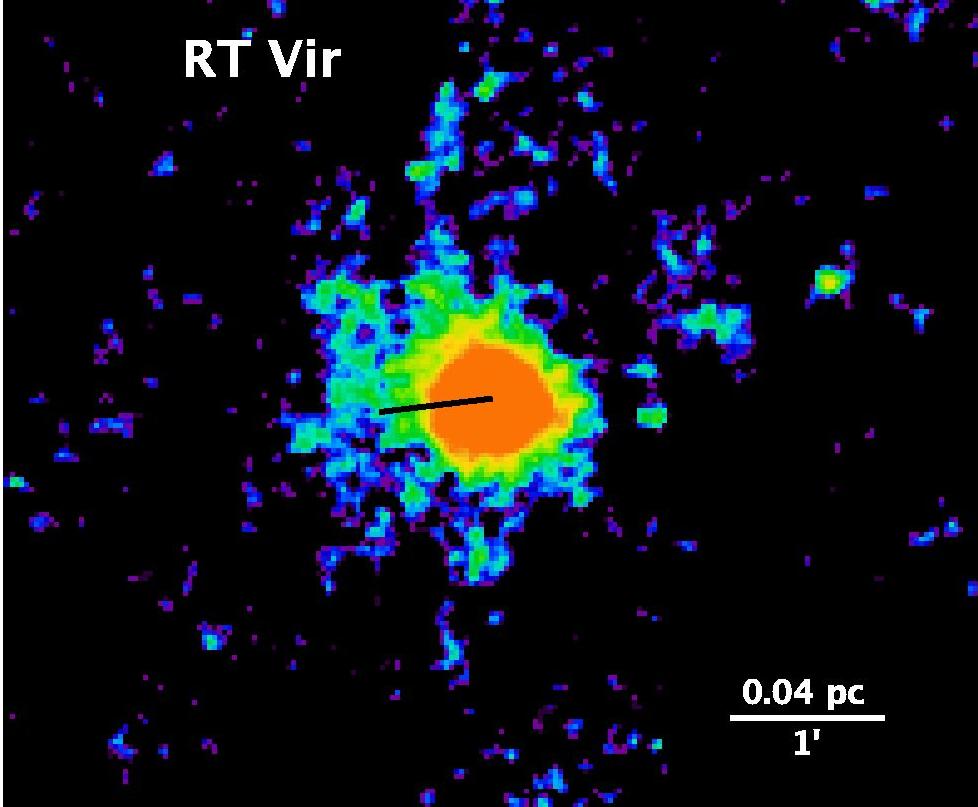}
 \medskip
 \includegraphics[width=0.24\textwidth,clip]{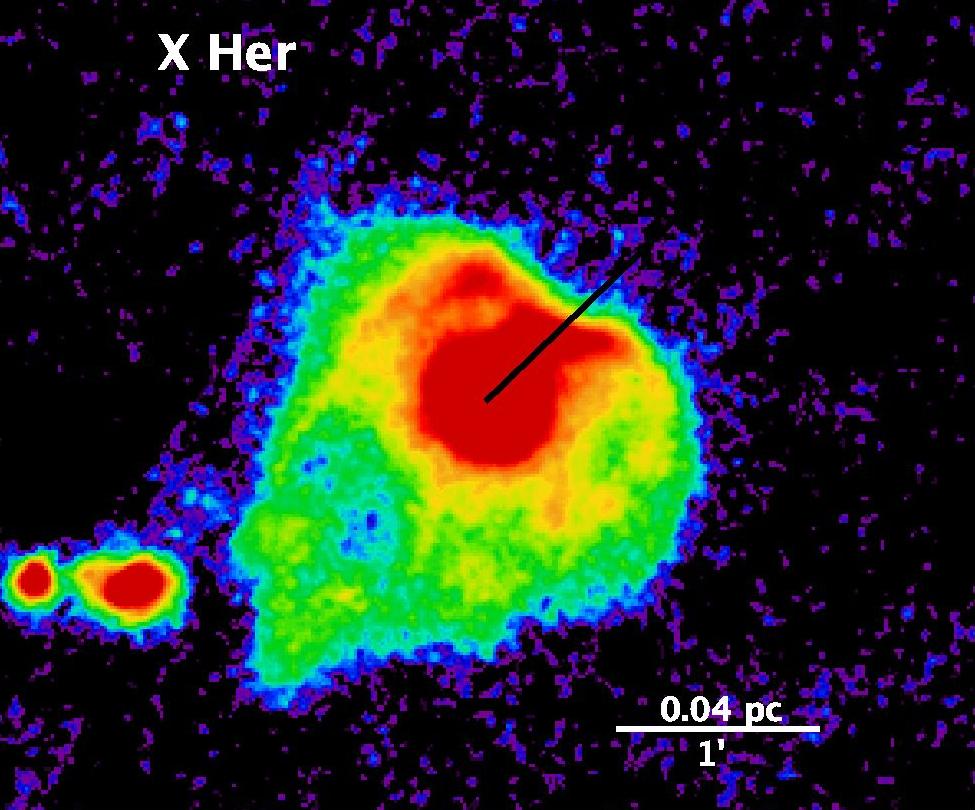}
 \includegraphics[width=0.24\textwidth,clip]{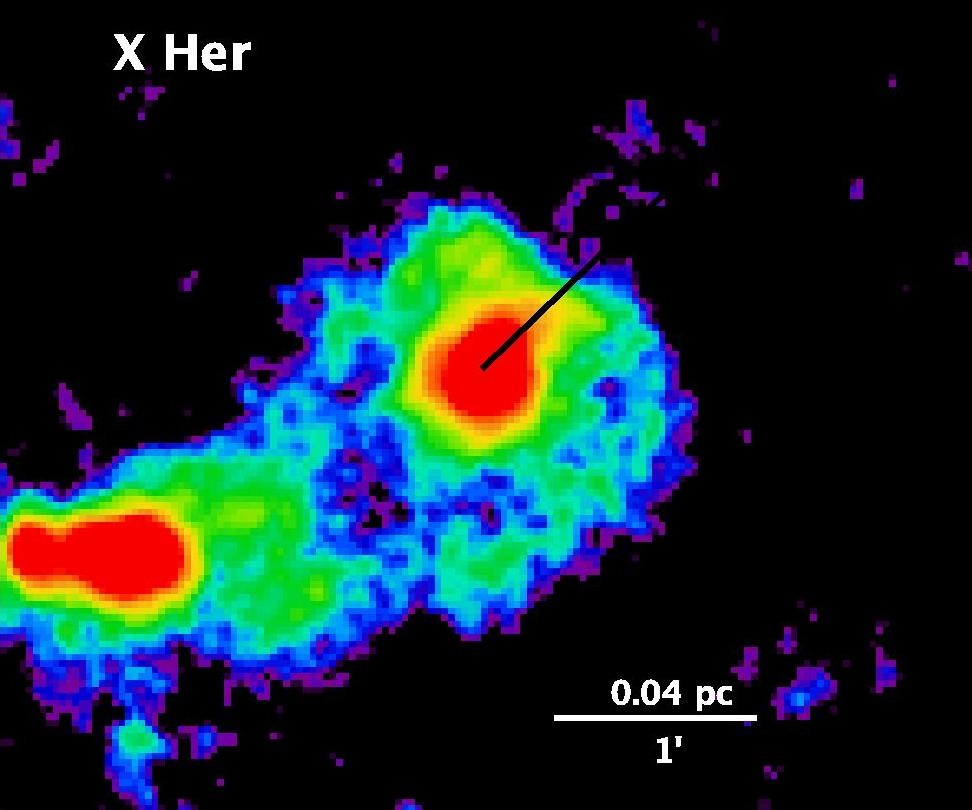}
 \caption{(continued). Interaction type \emph{``fermata''} (Class\,I). PACS 70~$\mu$m (left) and 160~$\mu$m (right).
 $o$\,Cet, R\,Leo, RT\,Vir, and X\,Her reveal RT and/or KH instabilities.
 For $o$\,Cet see also Mayer et al. (2011).
 The sources visible eastward of X\,Her are background galaxies (see also Jorissen et al. 2011).}
 \label{fig:fermata-bullets-instabilities}
\end{figure}

\addtocounter{figure}{-1}

\begin{figure}[ht!]
 \centering
 \includegraphics[width=0.24\textwidth,clip]{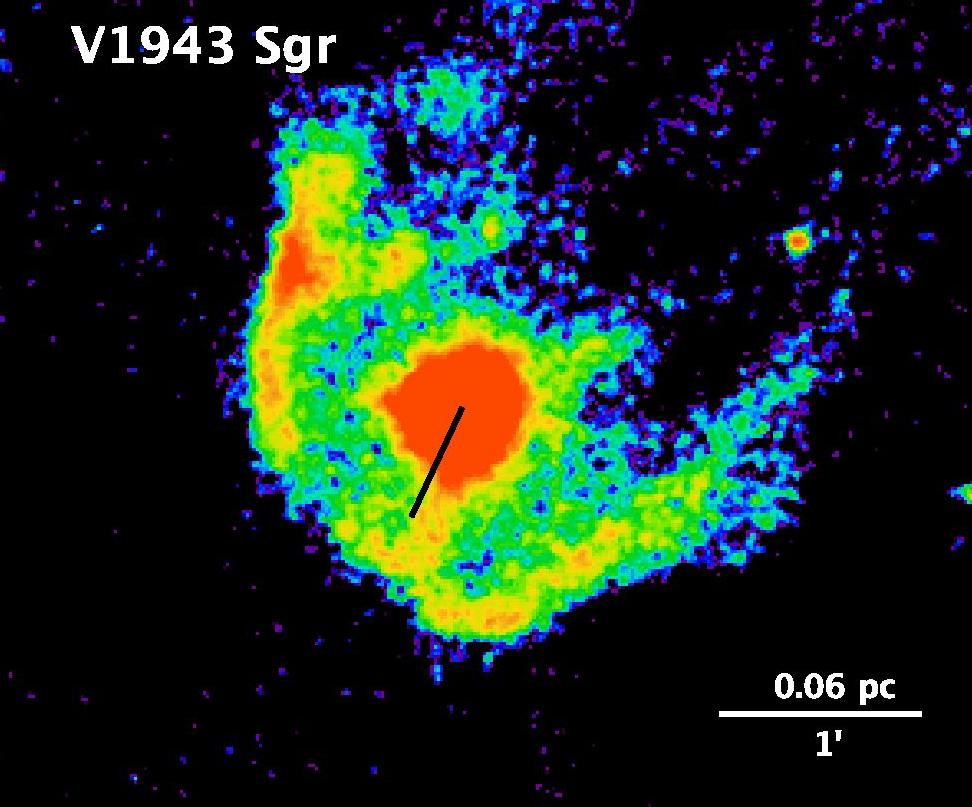}
 \includegraphics[width=0.24\textwidth,clip]{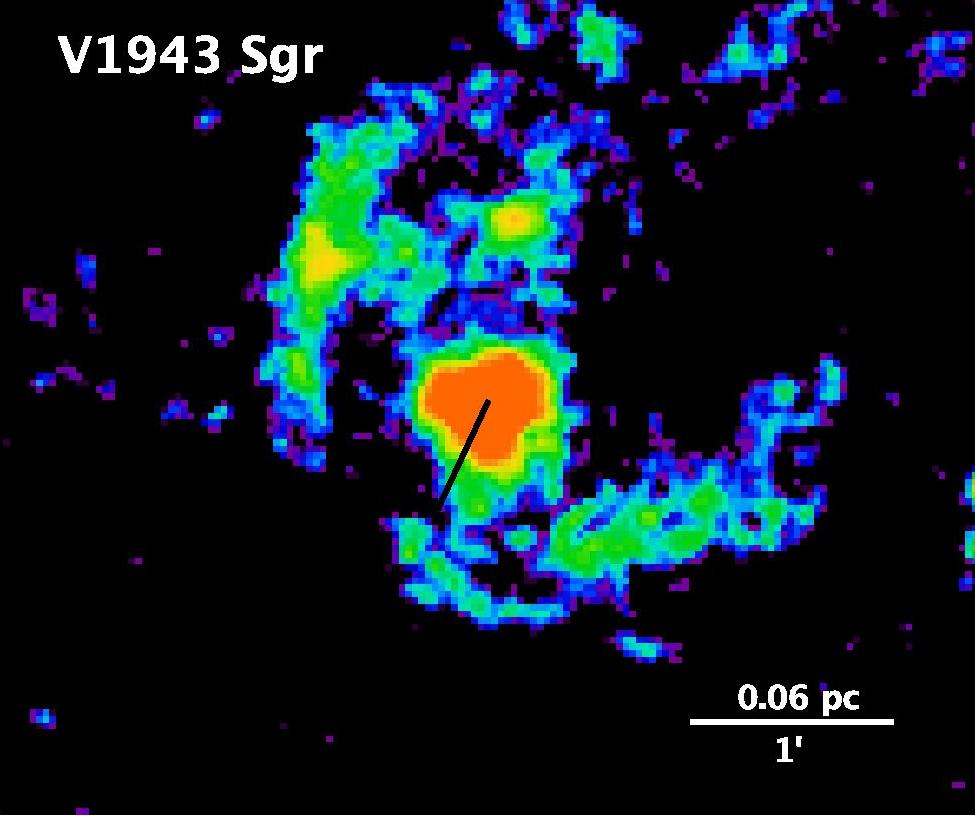}
 \medskip
 \includegraphics[width=0.24\textwidth,clip]{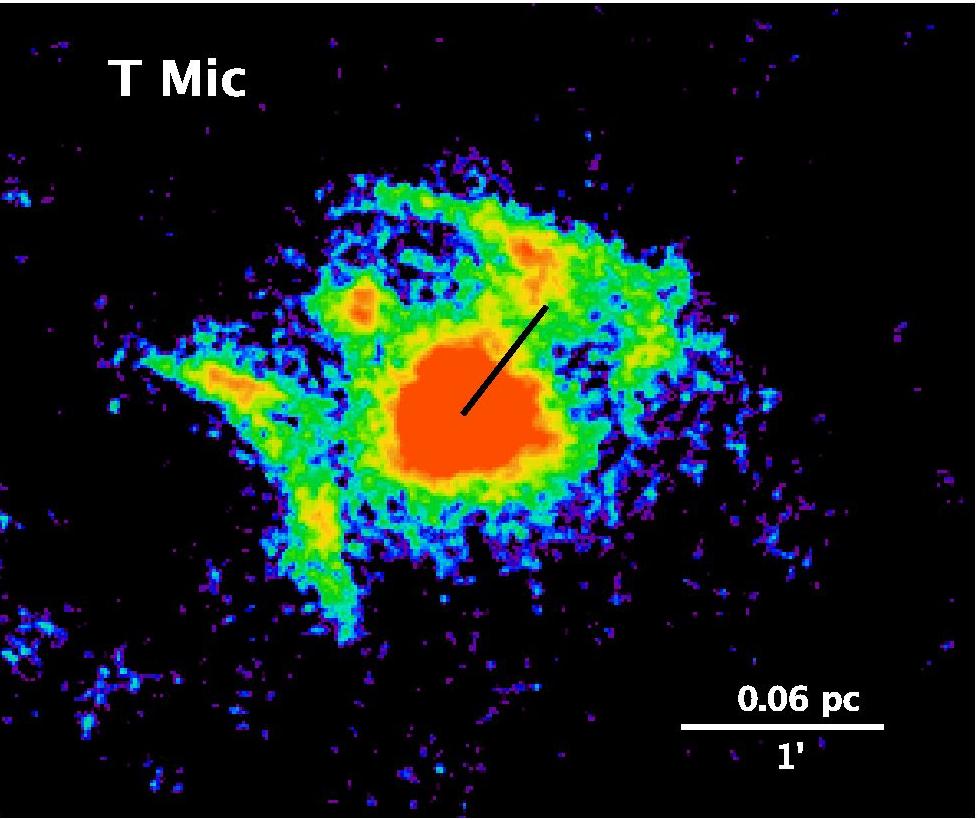}
 \includegraphics[width=0.24\textwidth,clip]{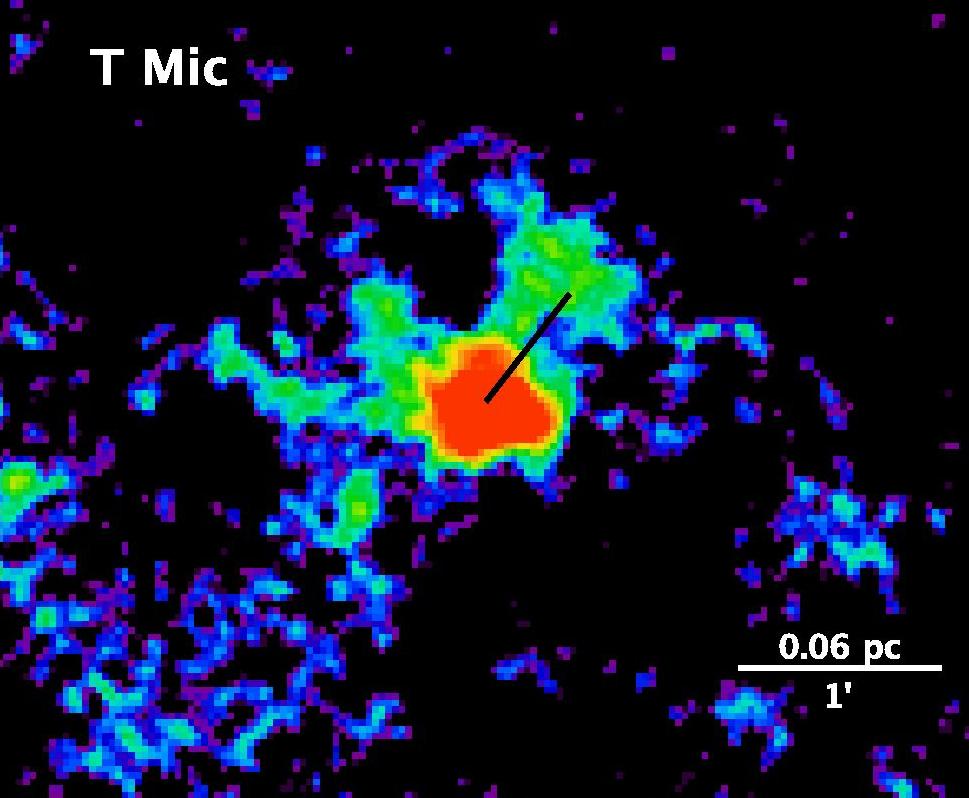}
 \medskip
 \includegraphics[width=0.24\textwidth,clip]{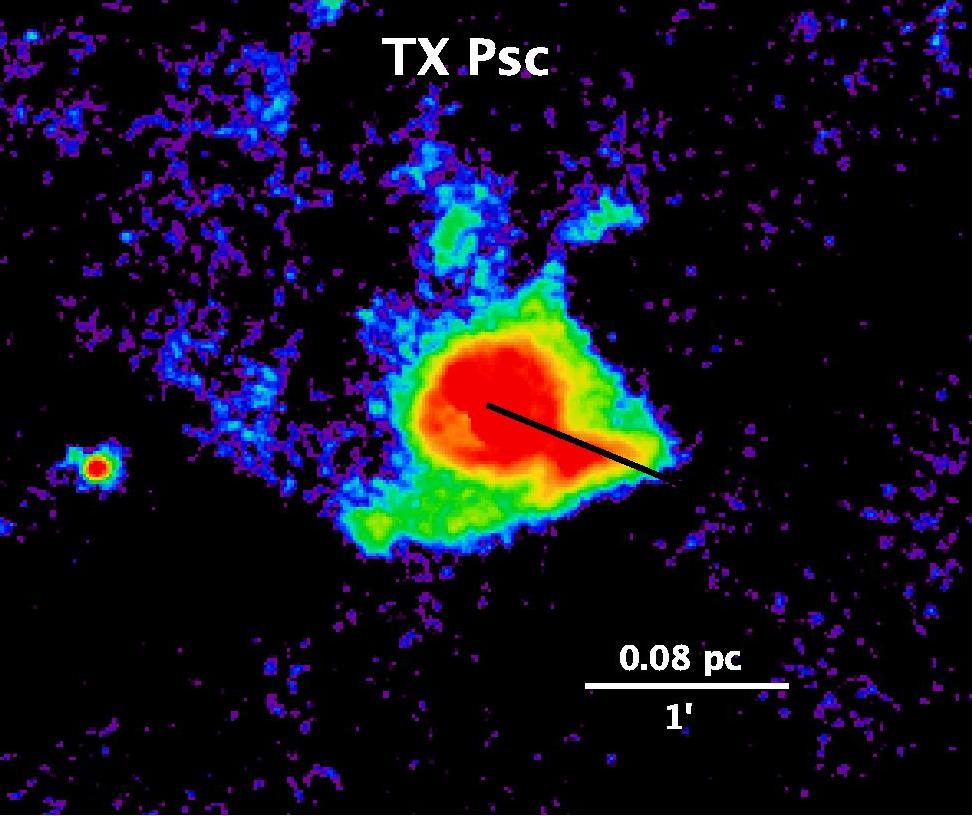}
 \includegraphics[width=0.24\textwidth,clip]{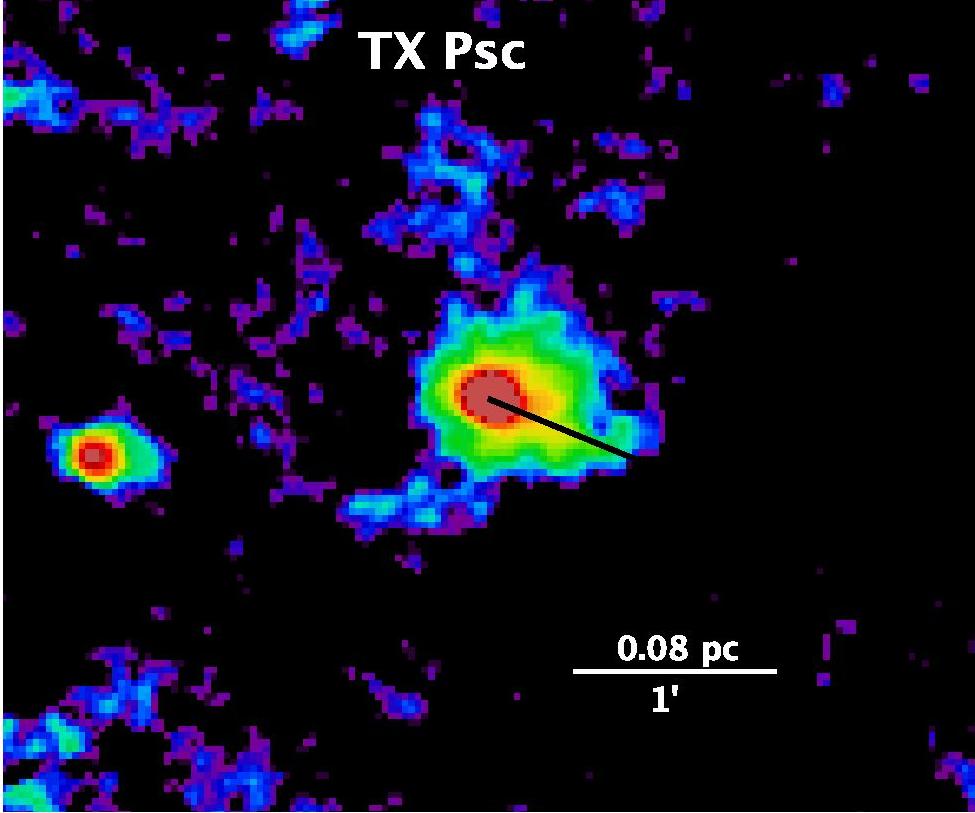}
 \medskip
 \includegraphics[width=0.24\textwidth,clip]{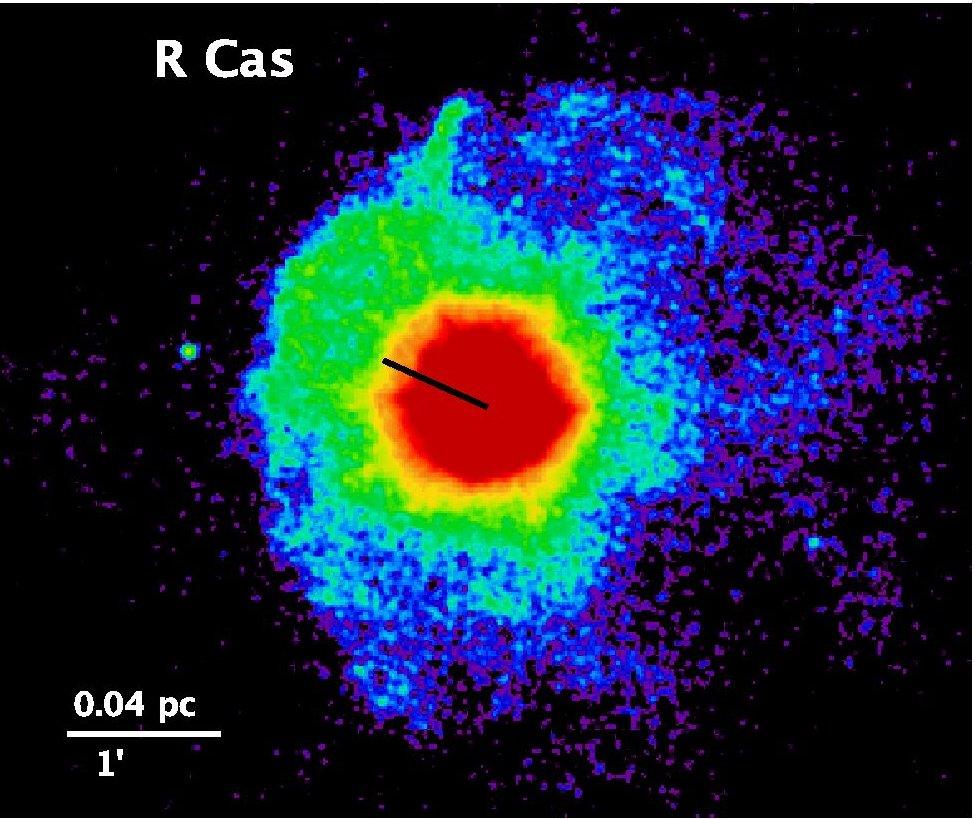}
 \includegraphics[width=0.24\textwidth,clip]{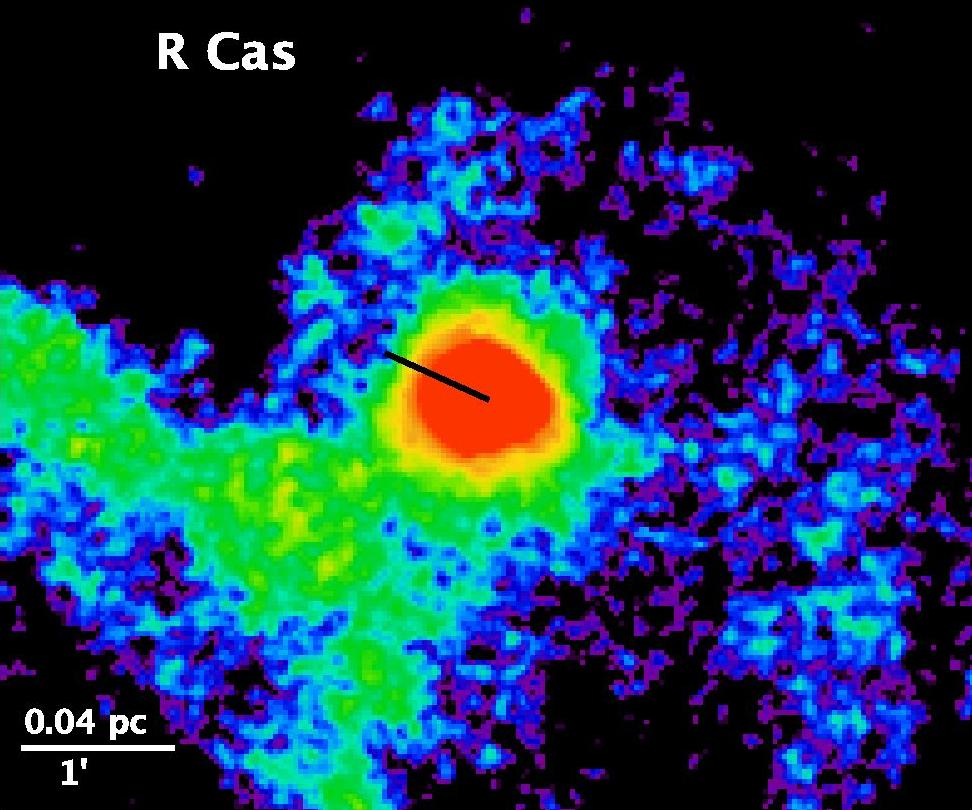}
 \caption{(continued). Interaction type \emph{``fermata''} (Class\,I). PACS 70~$\mu$m (left) and 160~$\mu$m (right).
 V1943\,Sgr, T\,MIrc, TX\,Psc, and R\,Cas reveal RT and/or KH instabilities.
 Note: TX\,Psc also has an inner spherical shell (not visible here).The sources visible eastward of TX\,Psc are
 background galaxies (see also Jorissen et al. 2011).}
 \label{fig:fermata-bullets-instabilities}
\end{figure}

\section{AGB stars and supergiants in the MESS Herschel Key Program}\label{sec:obs}

The MESS (Mass-loss of Evolved StarS) program (\citealt{2011A&A...526A.162G}) is a large far-infrared Herschel
(\citealt{2010A&A...518L...1P}) survey aimed at studying the mass-loss history of evolved stellar objects, from AGB stars to
post-AGB, planetary nebulae, luminous blue variables and supernovae. One key goal of the MESS program is to resolve the
circumstellar envelopes around a representative sample of evolved stars (from AGB stars to PNe), thereby studying the global
evolution of mass-loss and circumstellar envelope structure. The details of the program, the complete target list, data
processing approach, and first science results are summarised in \citet{2011A&A...526A.162G}.

 \begin{figure*}[th!]
 \includegraphics[width=0.24\textwidth,clip]{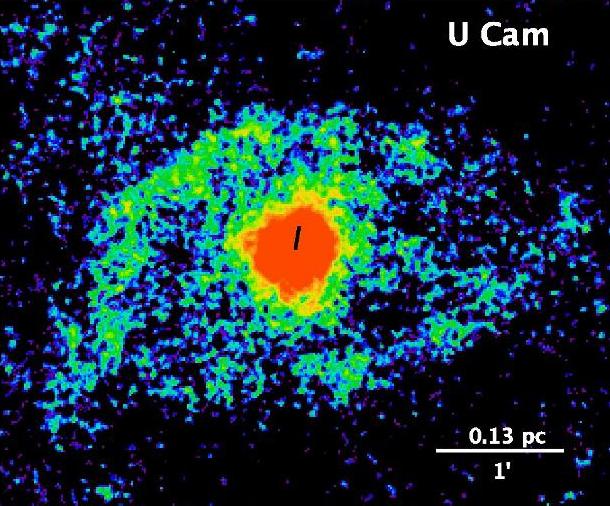}
 \includegraphics[width=0.24\textwidth,clip]{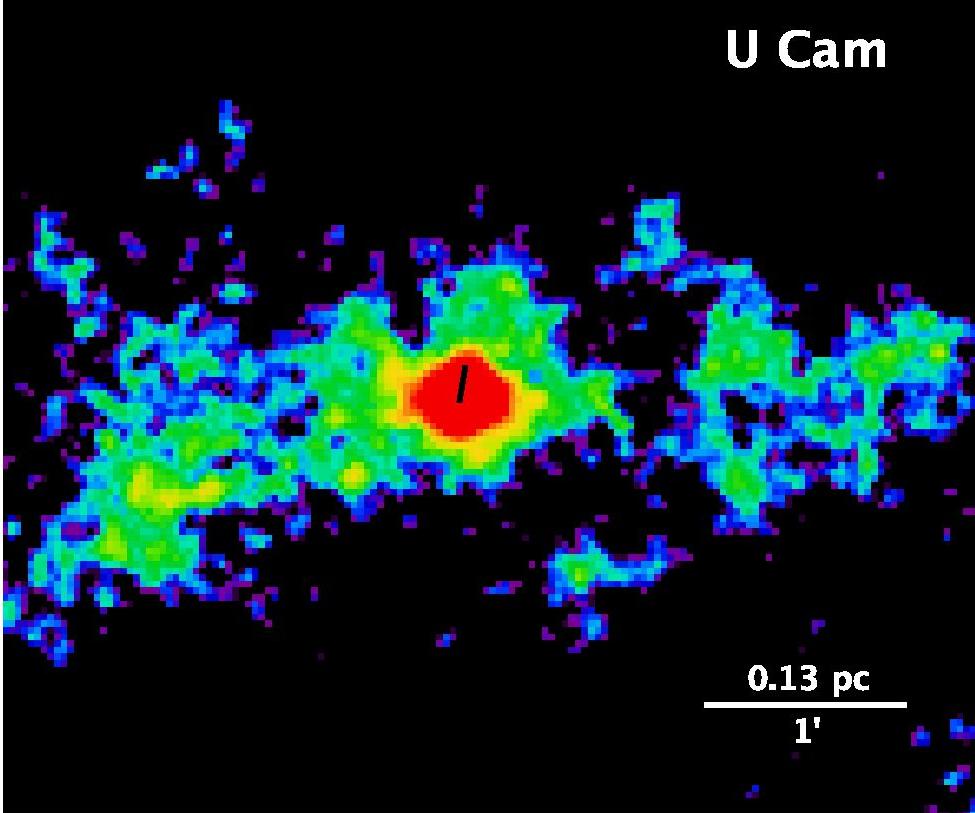}
 \includegraphics[width=0.5\columnwidth]{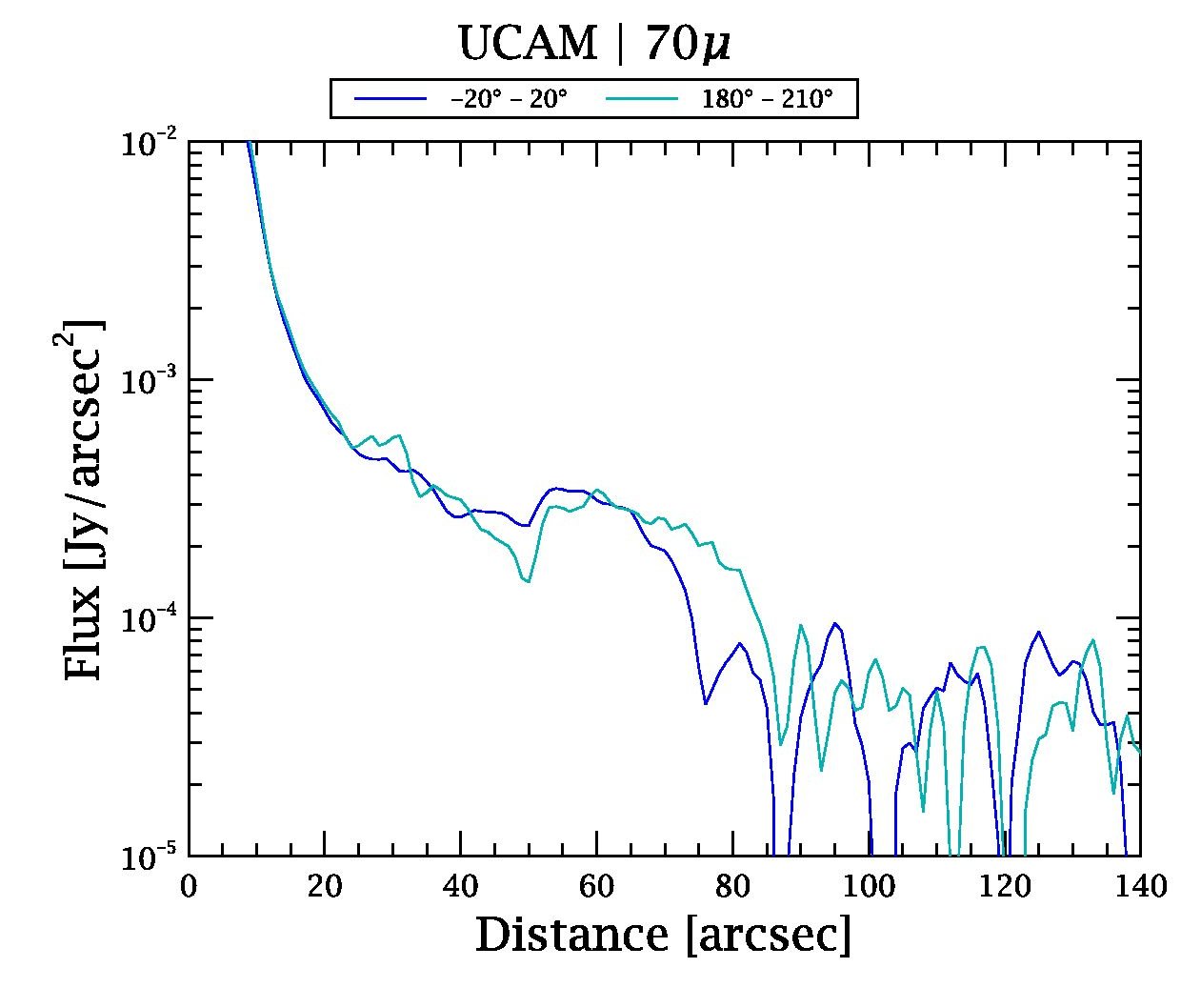}\\
 \medskip
 \includegraphics[width=0.24\textwidth,clip]{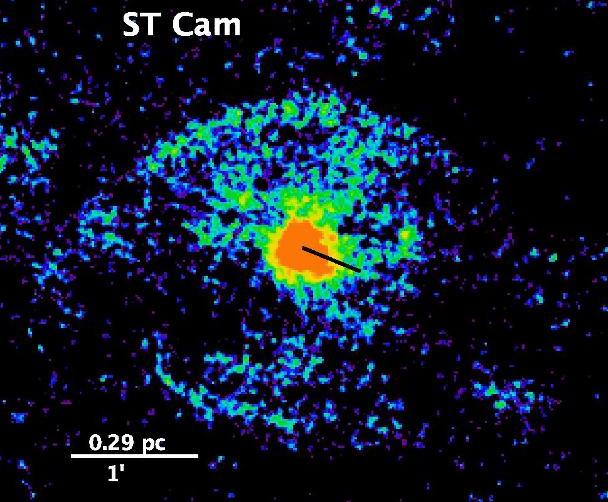}
 \includegraphics[width=0.24\textwidth,clip]{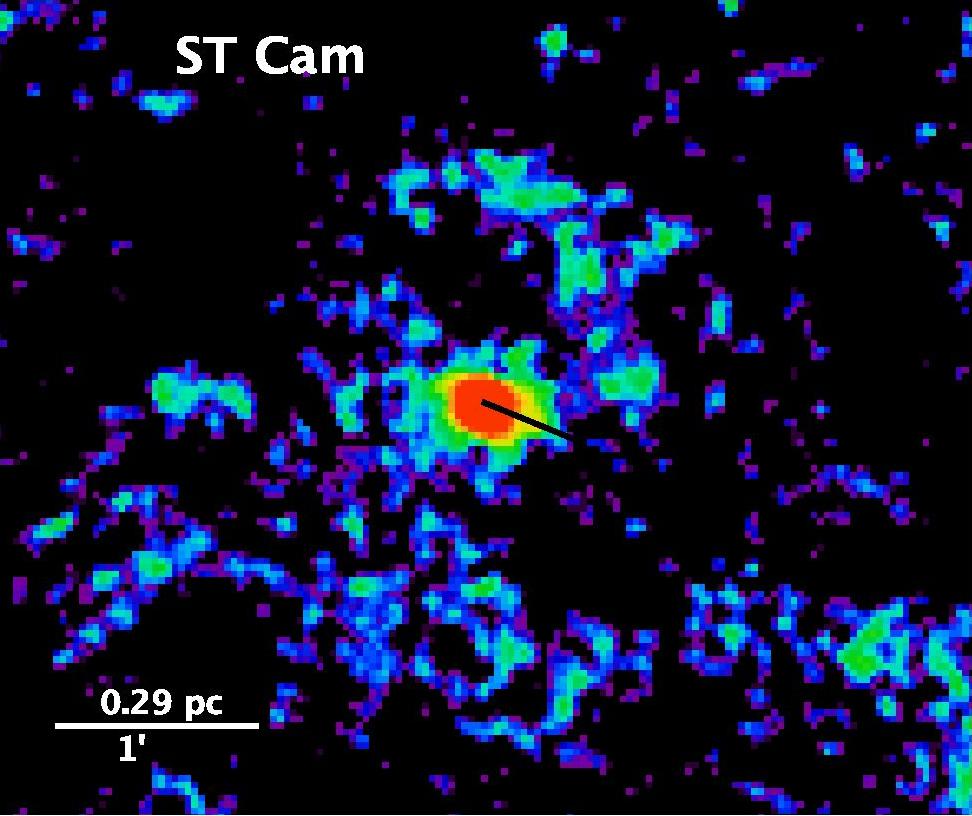}
 \includegraphics[width=0.5\columnwidth]{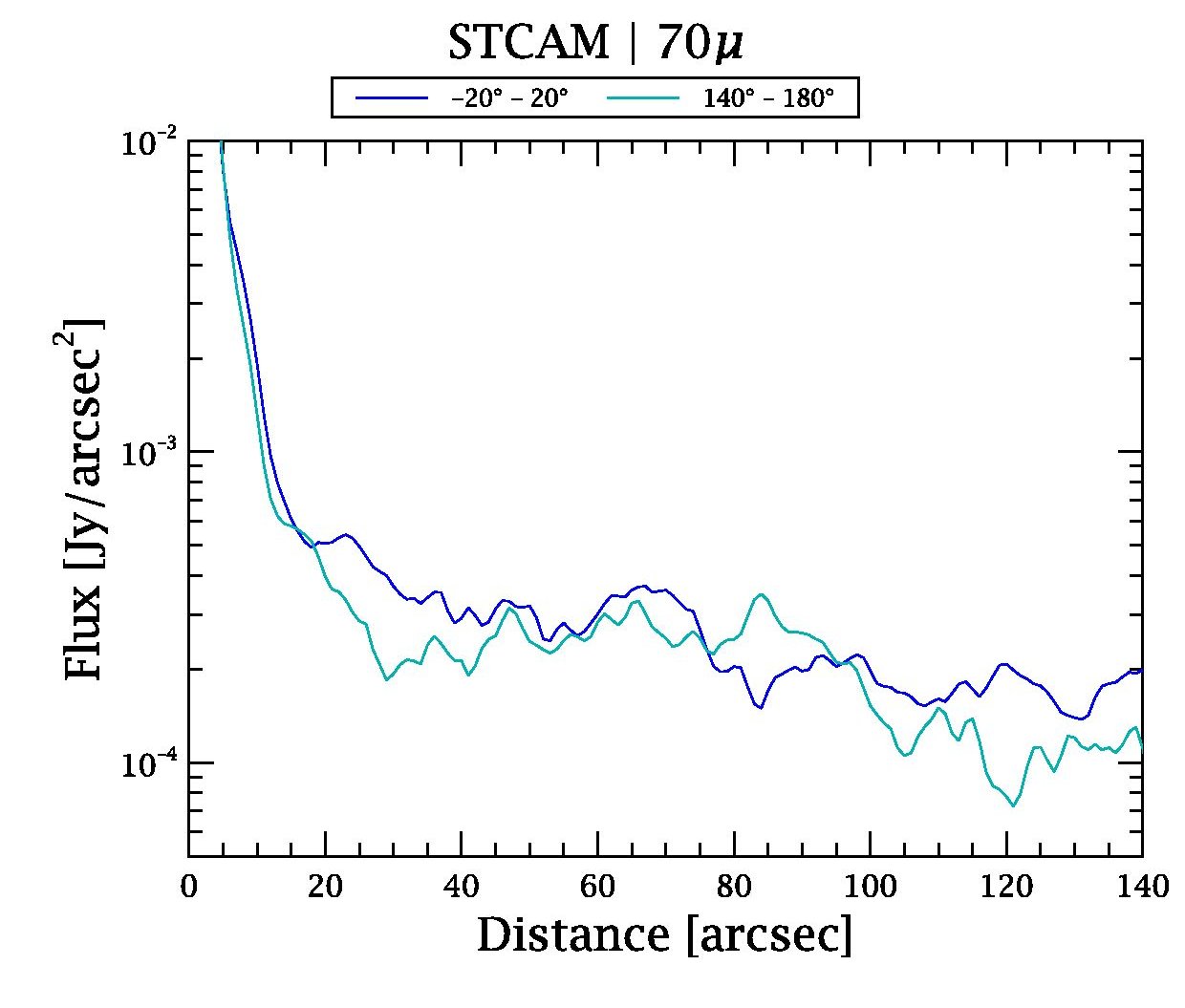}\\
 \medskip
 \includegraphics[width=0.24\textwidth,clip]{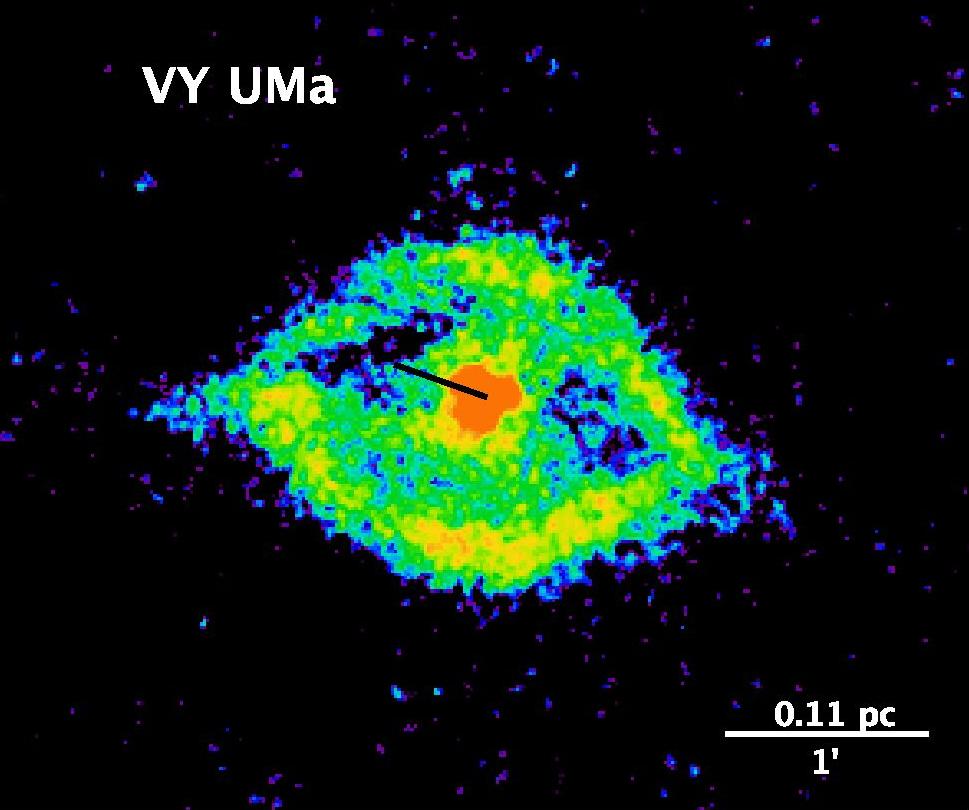}
 \includegraphics[width=0.24\textwidth,clip]{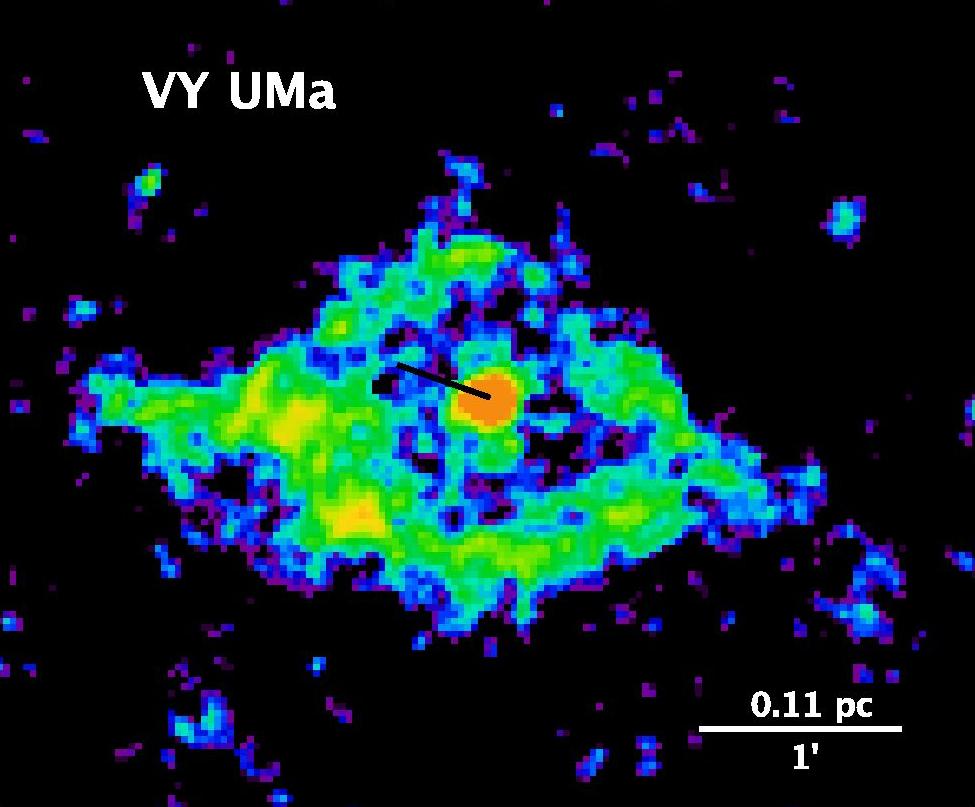}
 \includegraphics[width=0.5\columnwidth]{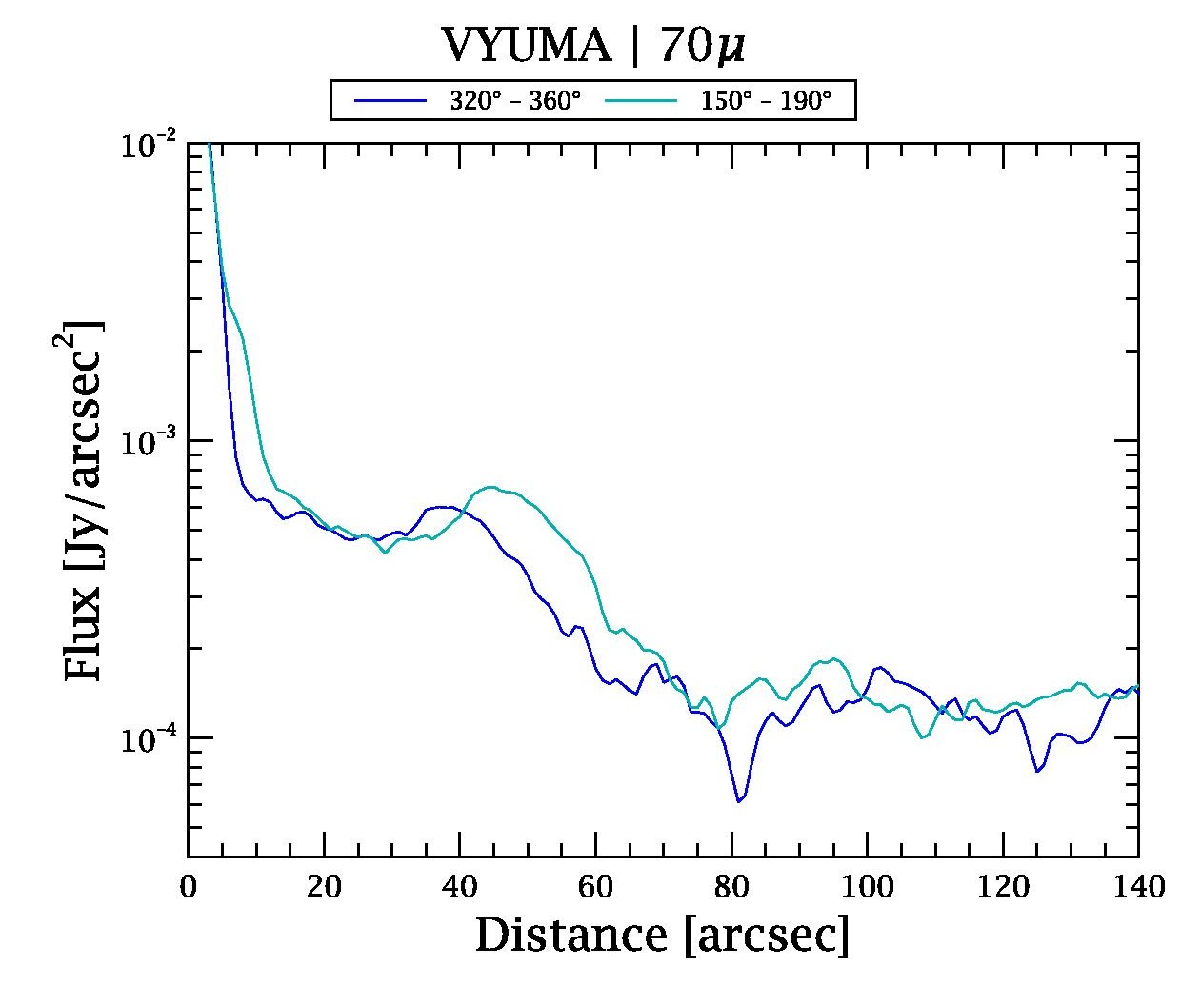}\\
 \medskip
 \includegraphics[angle=0,width=0.24\textwidth]{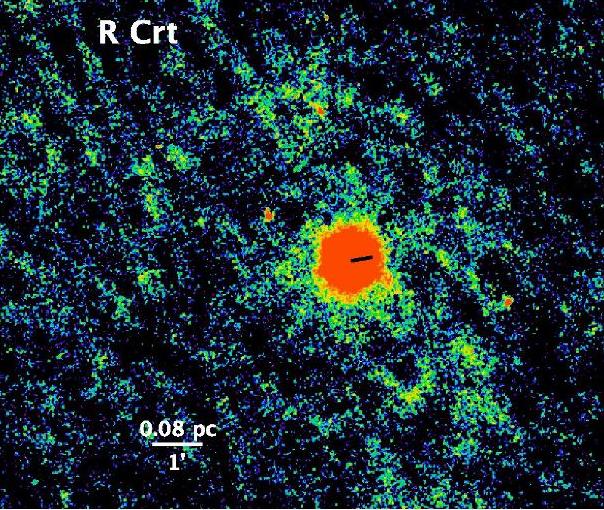}
 \includegraphics[angle=0,width=0.24\textwidth]{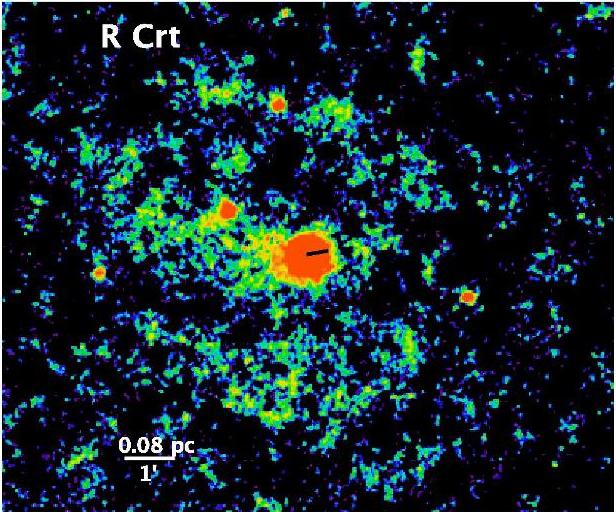}\\
 \medskip
 \includegraphics[width=0.24\textwidth,clip]{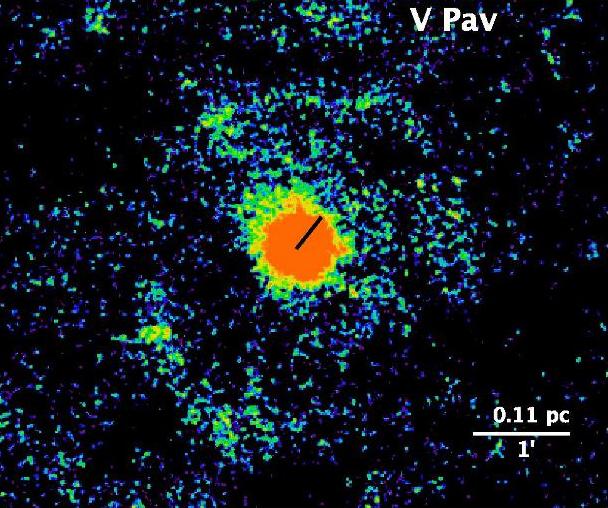}
 \includegraphics[width=0.24\textwidth,clip]{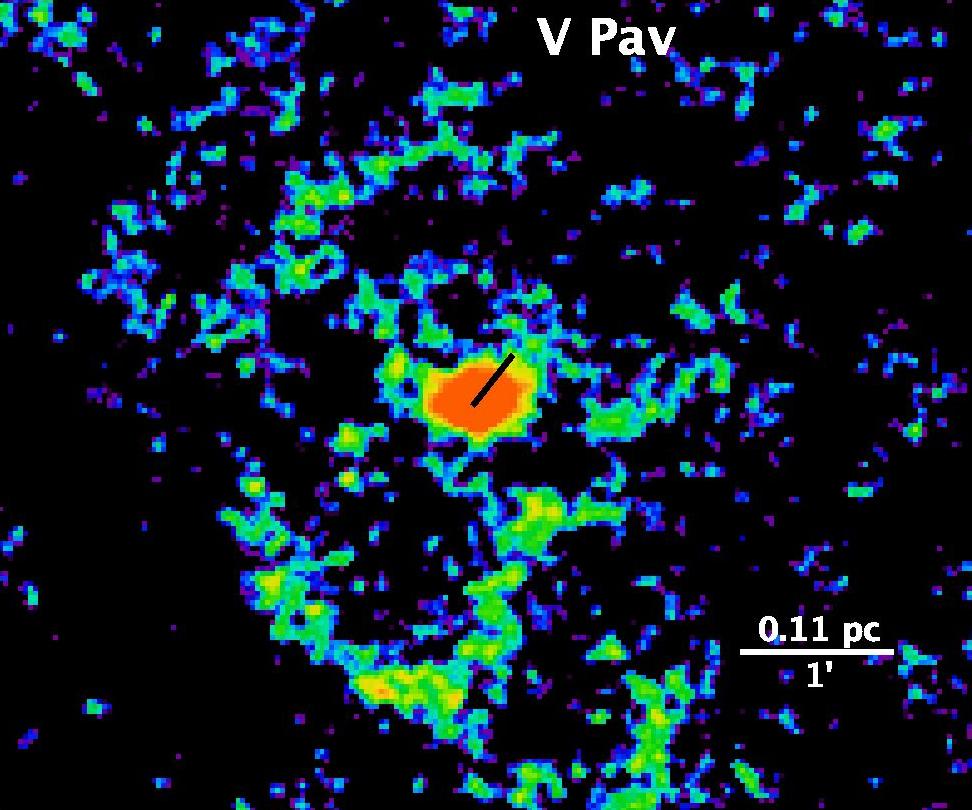}
 \includegraphics[width=0.5\columnwidth]{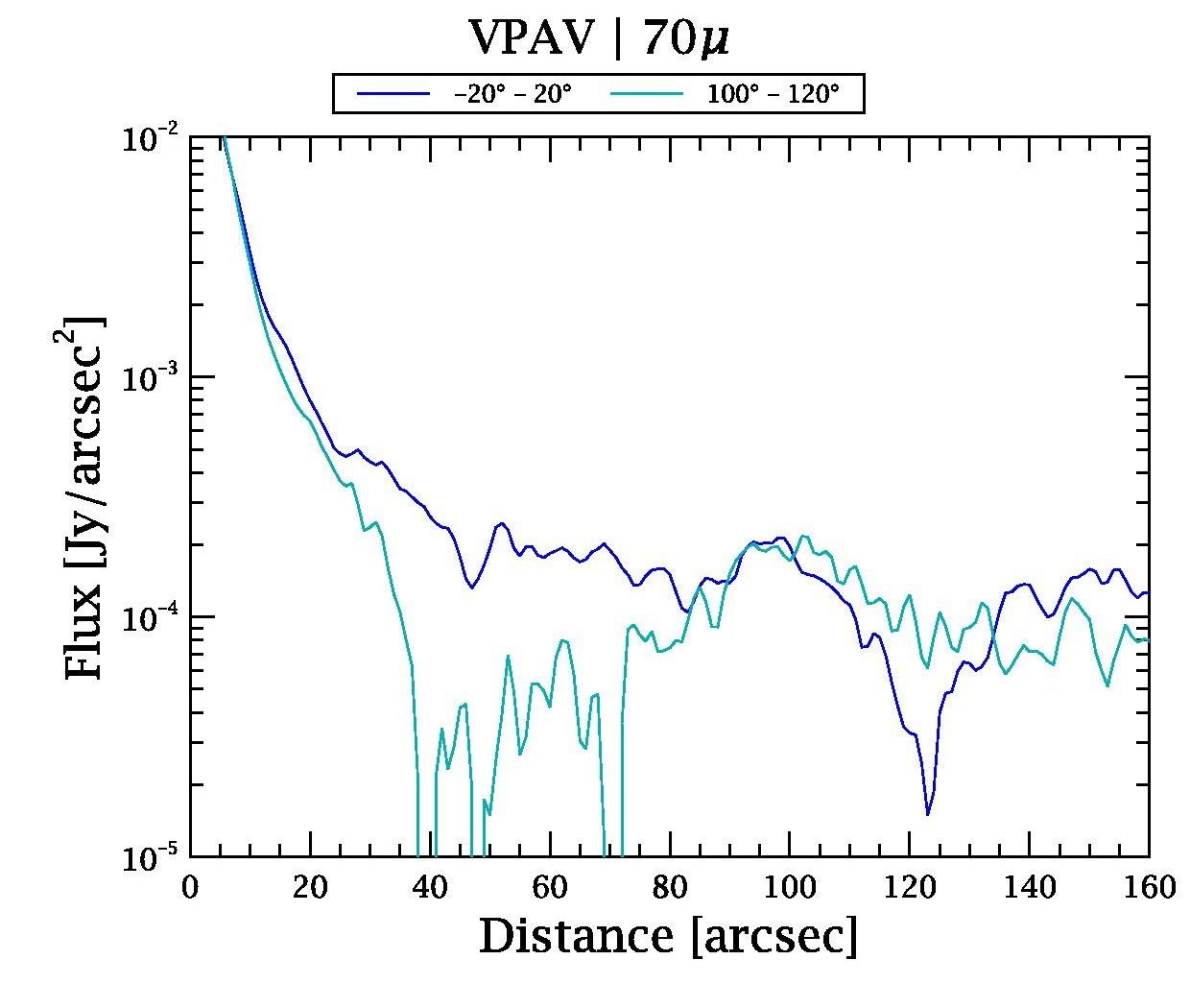}\\
 \caption{Interaction type \emph{``eyes''} (Class\,II). PACS 70~$\mu$m (left) and 160~$\mu$m (middle).
 Azimuthally averaged radial profiles are shown for 70 and 160~$\mu$m (right). Azimuth opening angles
 adopted for the radial profile are indicated at the top of each radial profile panel. Note: U Cam also has an inner spherical shell (not visible here).}
 \label{fig:eyes}
\end{figure*}

\addtocounter{figure}{-1}

 \begin{figure*}[ht!]
 \includegraphics[width=0.24\textwidth,clip]{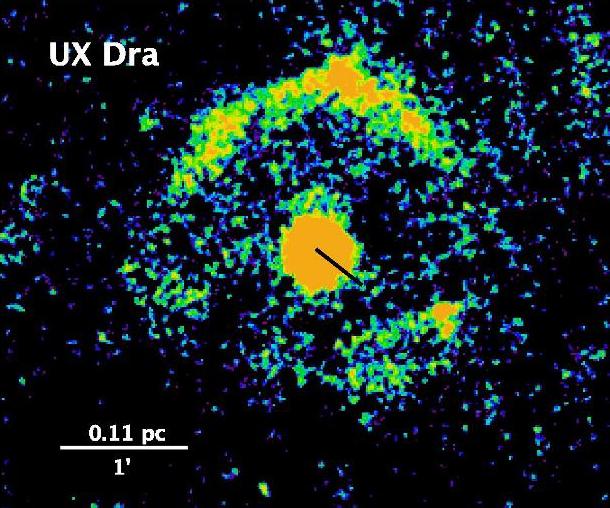}
 \includegraphics[width=0.24\textwidth,clip]{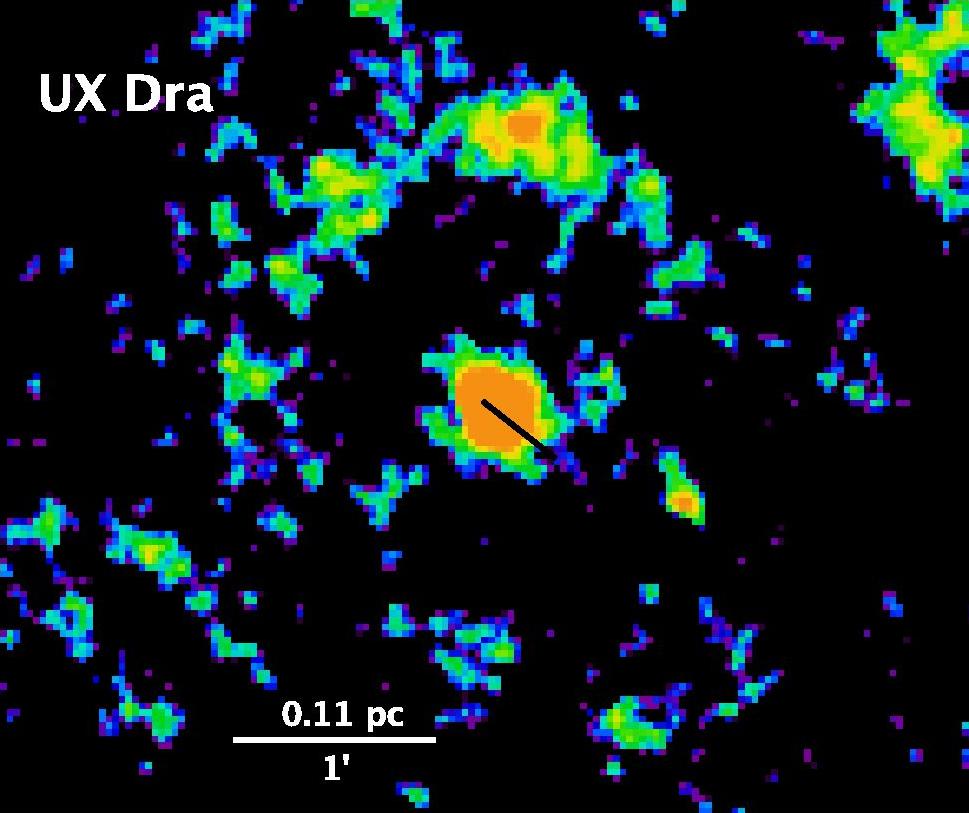}
 \includegraphics[width=0.5\columnwidth]{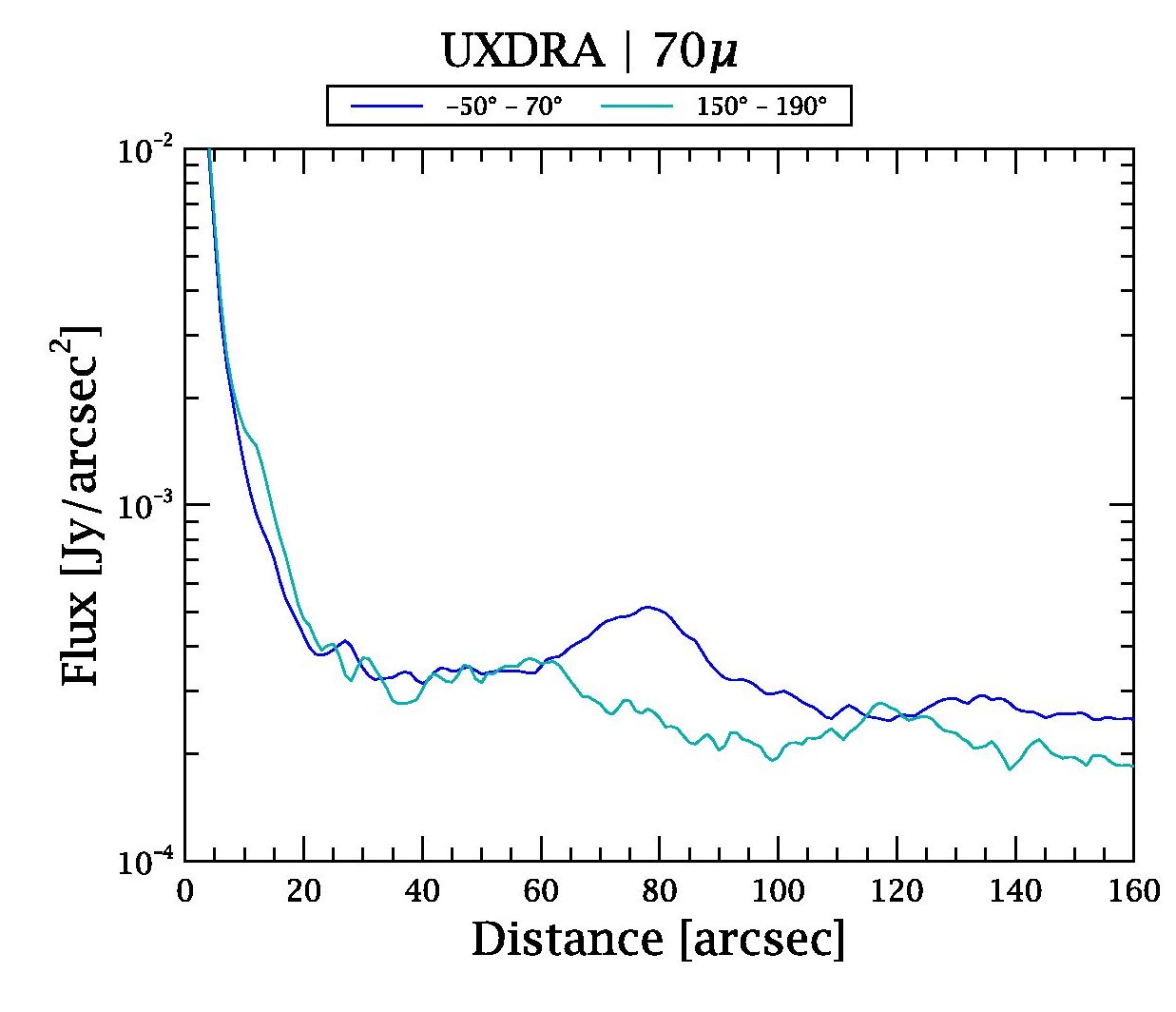}\\
 \medskip
 \includegraphics[width=0.24\textwidth,clip]{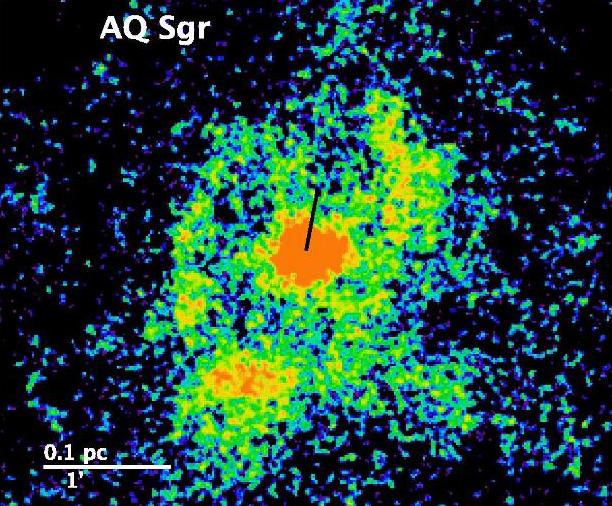}
 \includegraphics[width=0.24\textwidth,clip]{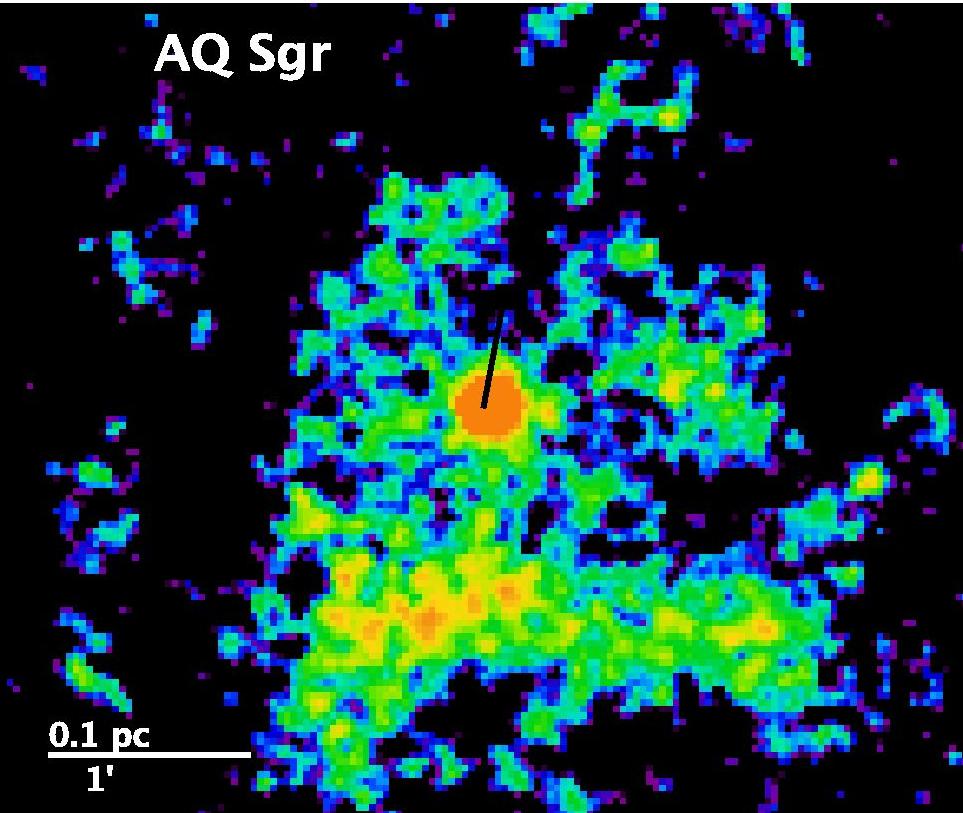}
 \includegraphics[width=0.5\columnwidth]{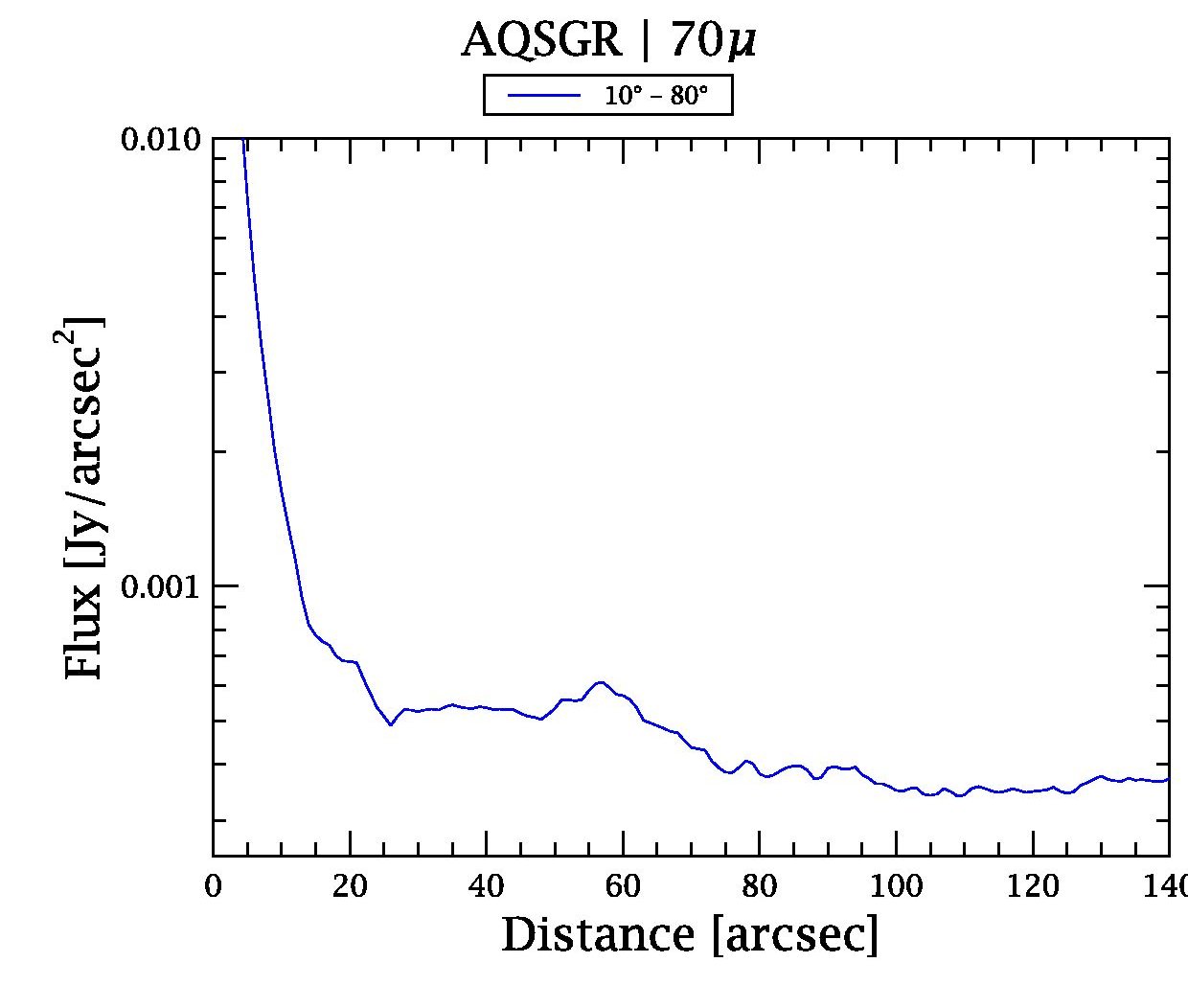}
 \caption{(continued) Interaction type \emph{``eyes''} (Class\,II). PACS 70~$\mu$m (left) and 160~$\mu$m (middle).
 Azimuthally averaged radial profiles are shown for 70 and 160~$\mu$m (right).  Azimuth opening angles adopted 
 for the radial profile are indicated at the top of each radial profile panel.}
 \label{fig:eyes}
\end{figure*}

\begin{figure*}[h!]
 \includegraphics[width=0.23\textwidth,clip]{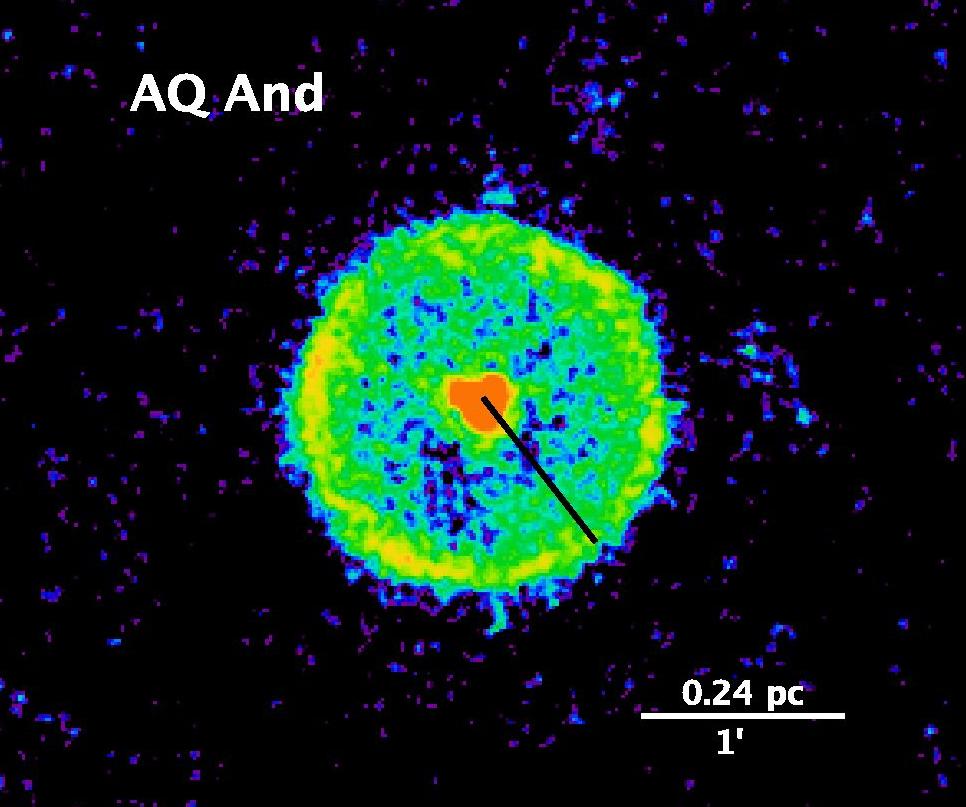}
 \includegraphics[width=0.23\textwidth,clip]{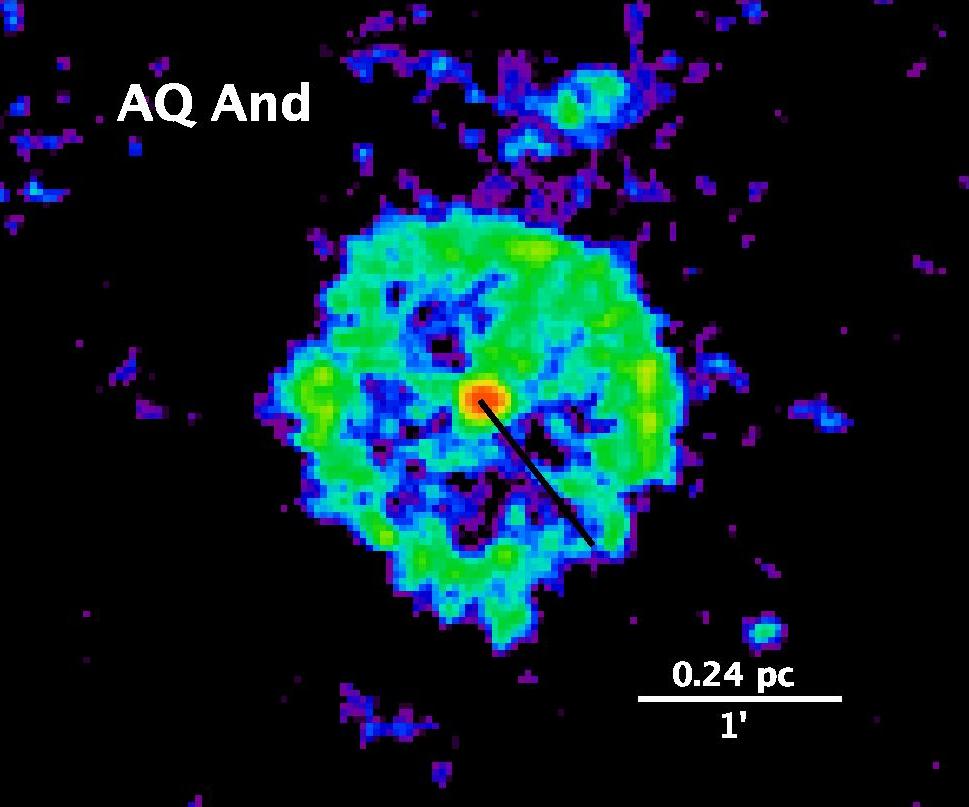}
 \includegraphics[width=0.45\columnwidth]{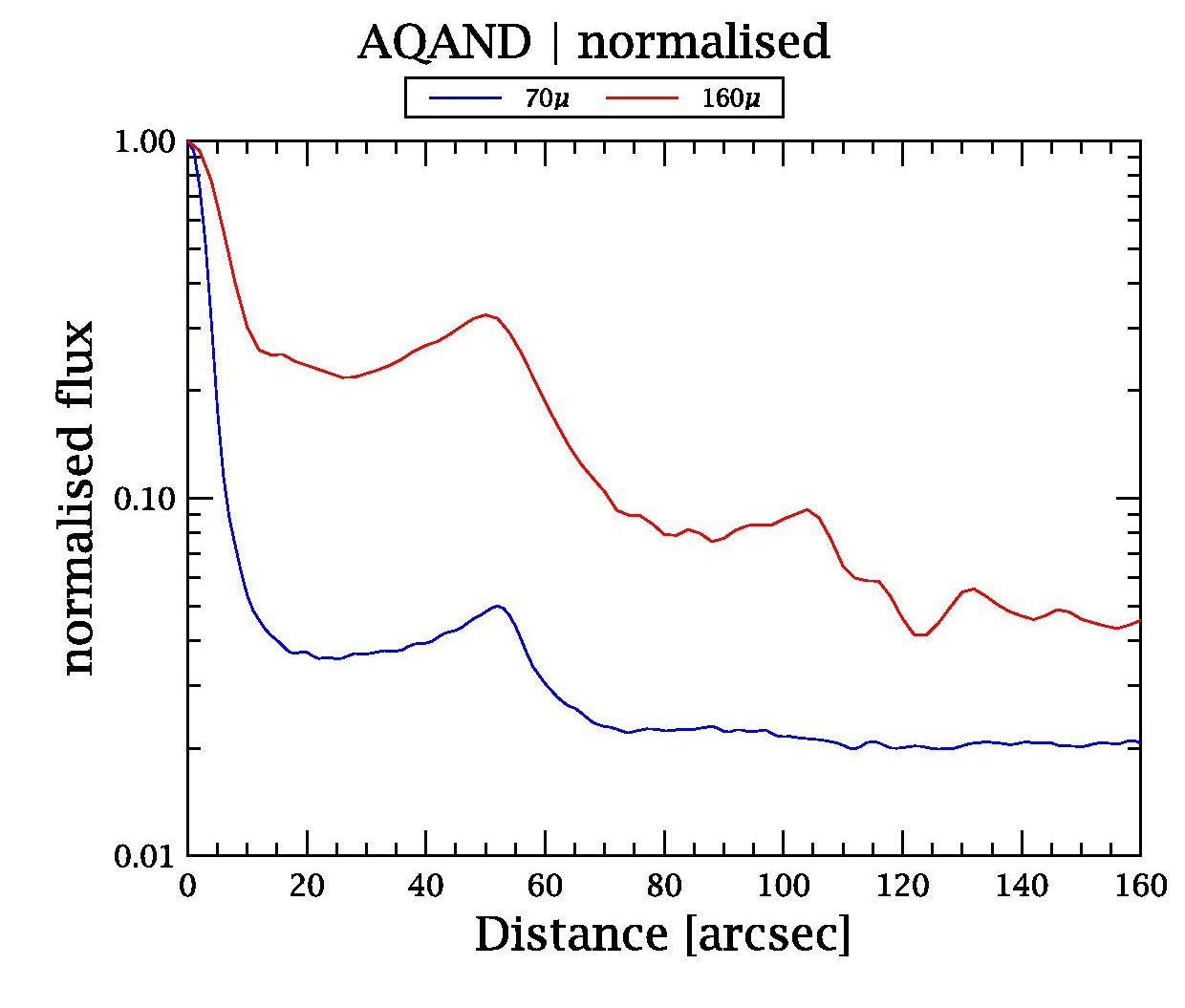}\\
 \medskip
 \includegraphics[width=0.23\textwidth,clip]{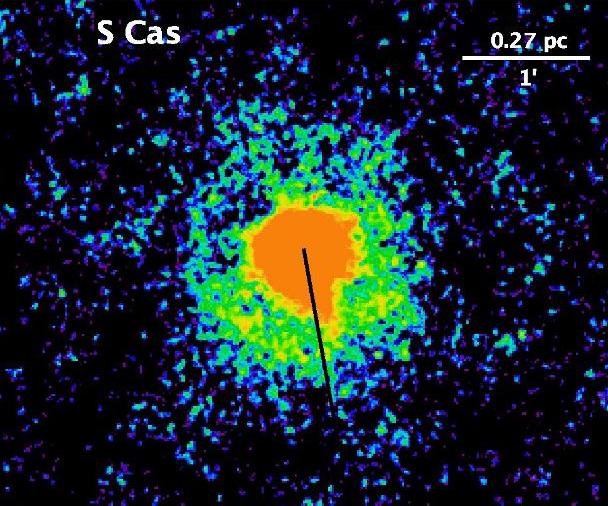}
 \includegraphics[width=0.23\textwidth,clip]{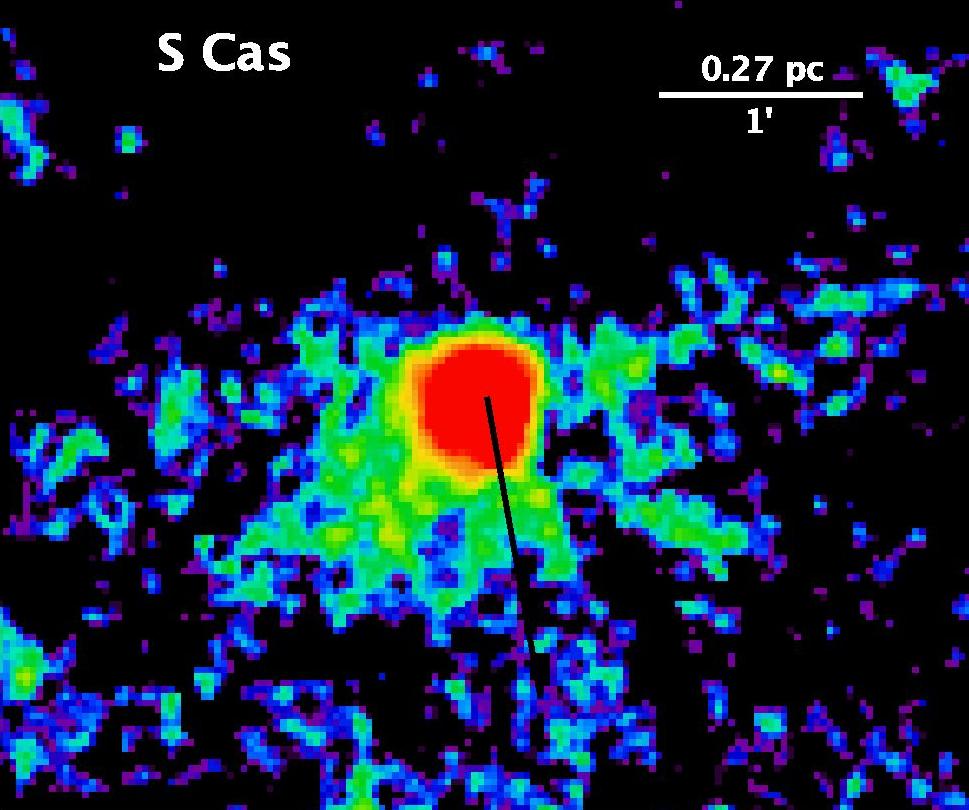}
 \includegraphics[width=0.45\columnwidth]{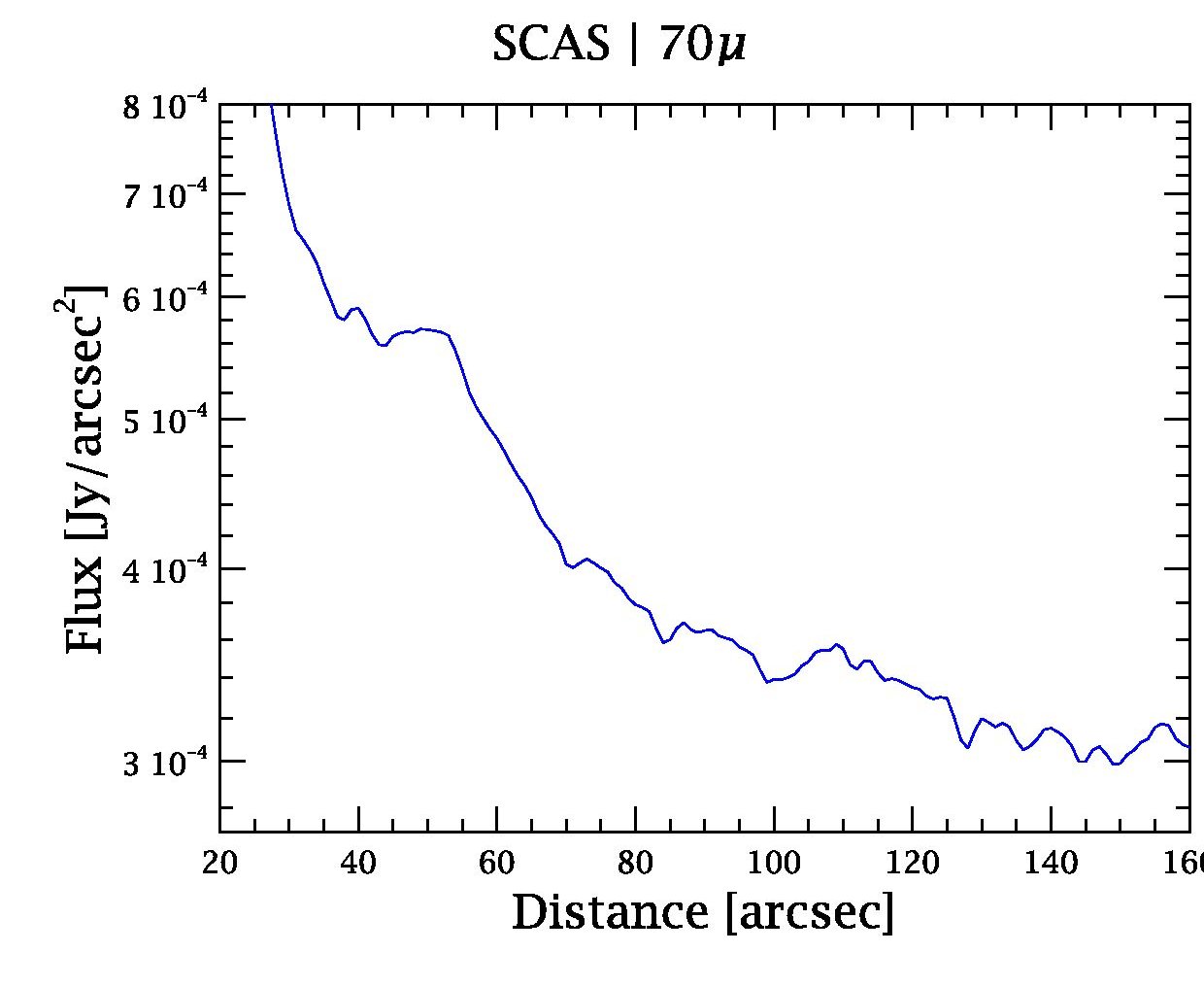}\\
 \caption{Interaction type \emph{``rings''} (Class\,III). PACS 70~$\mu$m (left) and 160~$\mu$m (middle).
 Azimuthally averaged (360\degr) radial profiles are shown for 70 and/or 160~$\mu$m (right).
 The ring towards S Cas is only faintly visible in the current display, but its presence is 
 verified from the radial profiles. For AQ\,And see also Kerschbaum et al. (2010).}
 \label{fig:rings}
\end{figure*}

\addtocounter{figure}{-1}

Briefly, infrared scan maps of a large sample of AGB stars and supergiants are obtained with the Photodetector Array Camera
and Spectrometer (PACS; \citealt{2010A&A...518L...2P}) using the ``scan map'' observing mode with the ``medium'' (20\arcsec\
s$^{-1}$) scan speed in the blue (70~$\mu$m) and red (160~$\mu$m) filter setting. To create a uniform coverage, avoid striping
artefacts, and increase redundancy, two observations at orthogonal scan directions are systematically concatenated. Scan
lengths range from 6 to 34\arcmin, scan leg steps are 77.5\arcmin\ or 155\arcmin, and repetition factors ranged from 2 to 8.

Data processing was performed applying the standard pipeline steps. In particular, a careful correction of the detector signal
drifts is critical to reveal faint extended emission structures with PACS. After deglitching, we applied two different
map-making algorithms to assess the quality of the final map, and to verify the faint emission associated to the wind-ISM
interaction; these are ``PhotProject'' with high-pass filter (HIPE v.7.0.) and ``Scanamorphos'' (\citealt{2011Roussel}). The
maps shown here are those obtained with the latter method as these are less susceptible to the filtering of true extended
emission. All frames are projected onto an image with a pixel size of 1\arcsec\ and 2\arcsec\ for the blue (70~$\mu$m) and red 
(160~$\mu$m) bands, respectively. Thus, the final image over-samples the instrumental point-spread functions, which have full-width 
at half-maxima of 5.6\arcsec\ and 11.4\arcsec\ at these wavelengths, respectively.

Aperture photometry of the central point-like sources observed with PACS at 70 and 160~$\mu$m shows that both the
``PhotProject'' and ``Scanamorphos'' maps give flux densities that agree within 5\%, well within the 15\% calibration
uncertainties. However, the processing with the updated calibration files in HIPE 7.0 results in 70 and 160~$\mu$m fluxes that
are 11\% and 15\% lower, respectively, compared to the values presented in \citet{2011A&A...526A.162G} using HIPE 4.4.

\begin{figure*}[h!]
 \includegraphics[width=0.23\textwidth,clip]{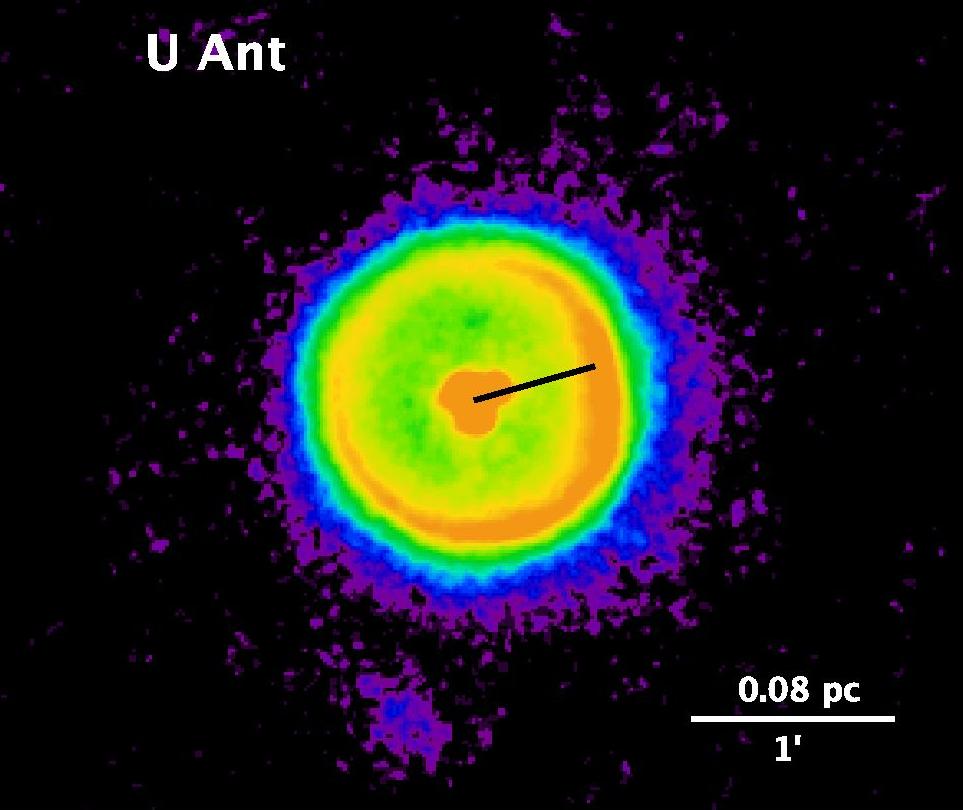}
 \includegraphics[width=0.23\textwidth,clip]{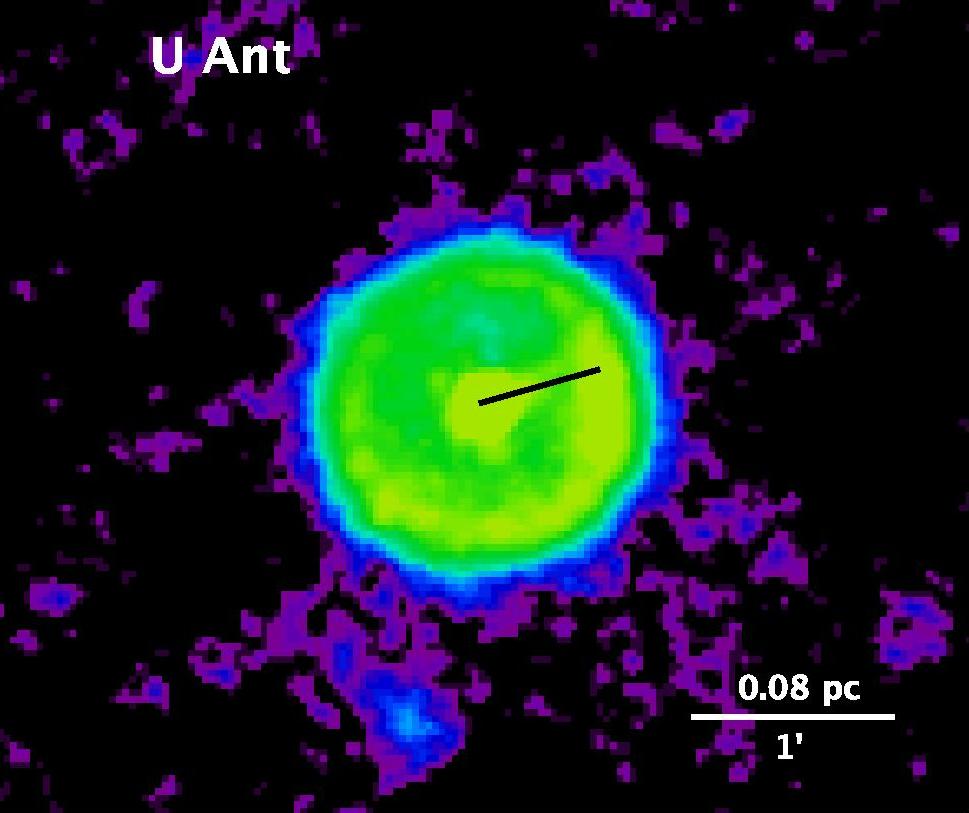}
 \includegraphics[width=0.45\columnwidth]{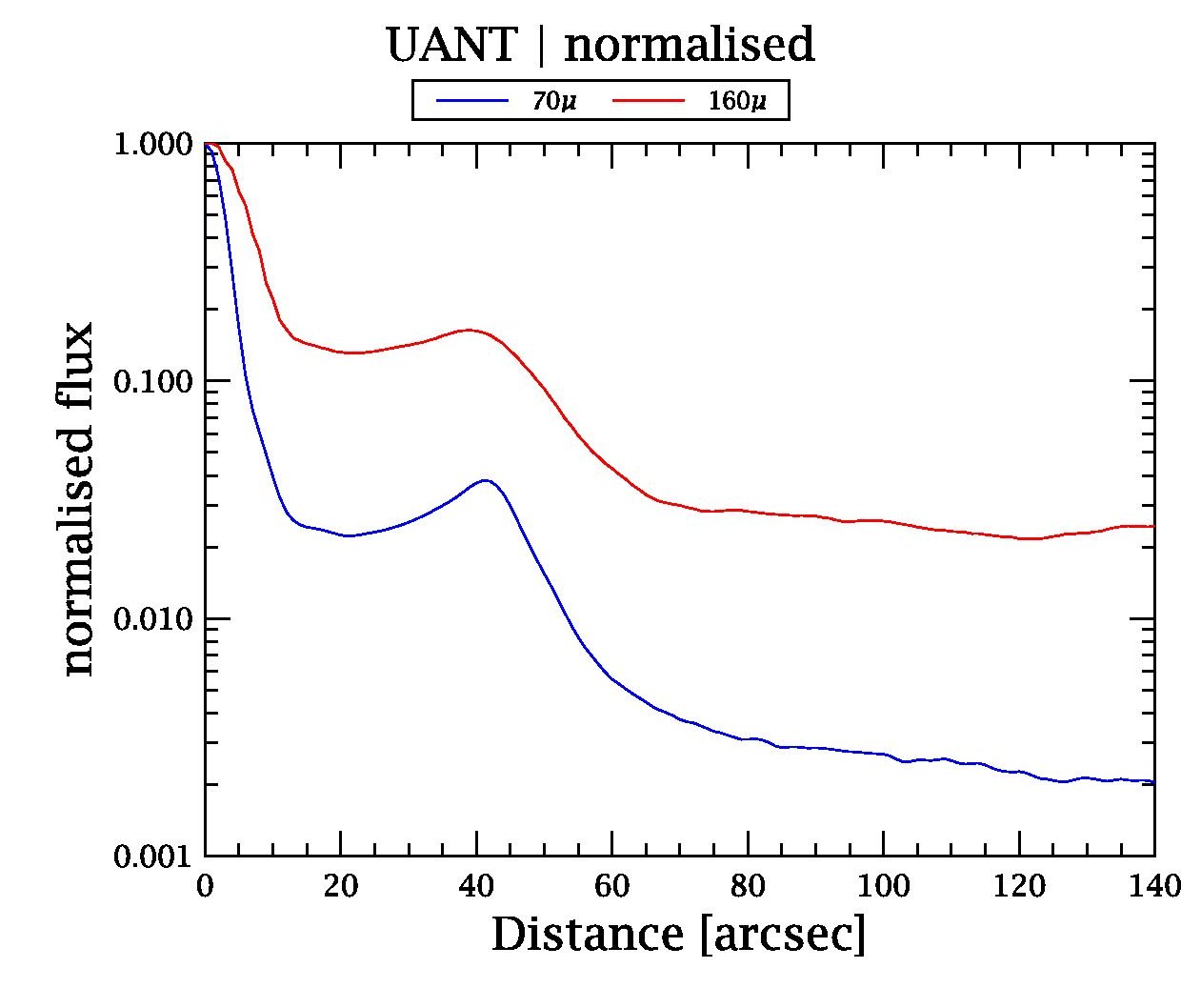}\\
 \medskip
 \includegraphics[width=0.23\textwidth,clip]{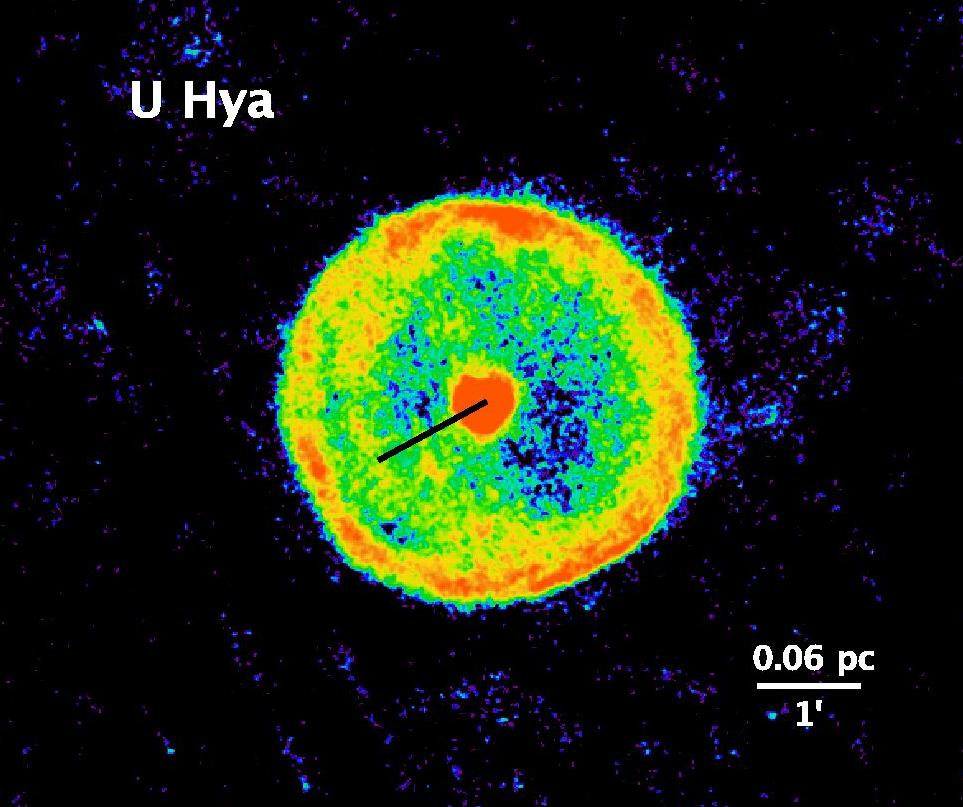}
 \includegraphics[width=0.23\textwidth,clip]{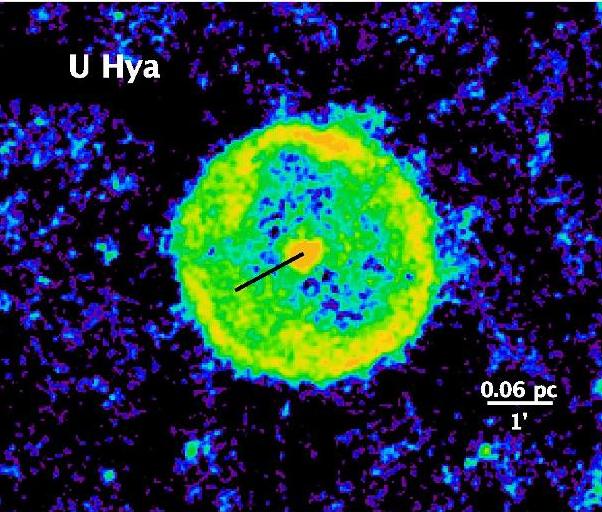}
 \includegraphics[width=0.45\columnwidth]{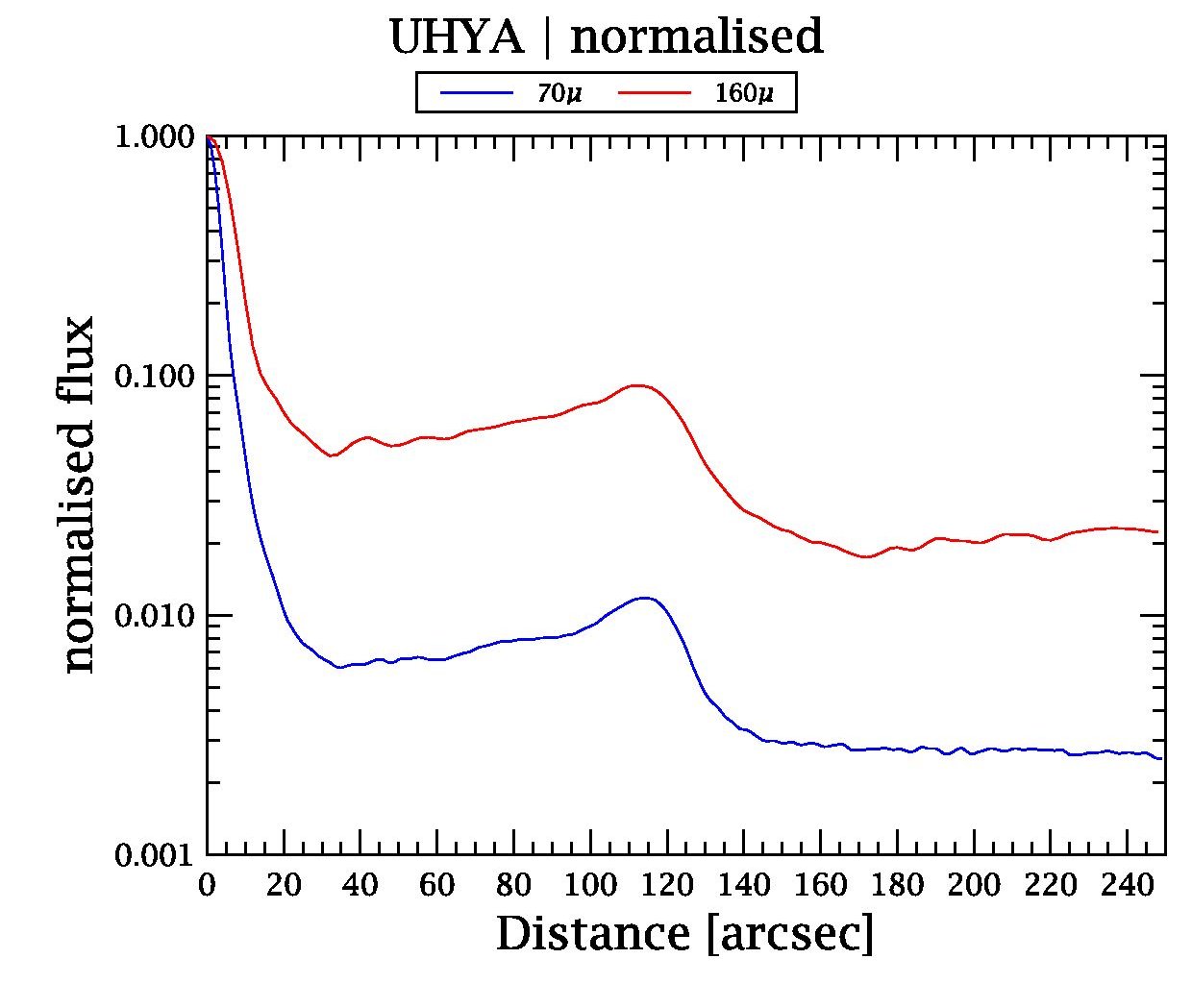}\\
 \medskip
 \includegraphics[width=0.23\textwidth,clip]{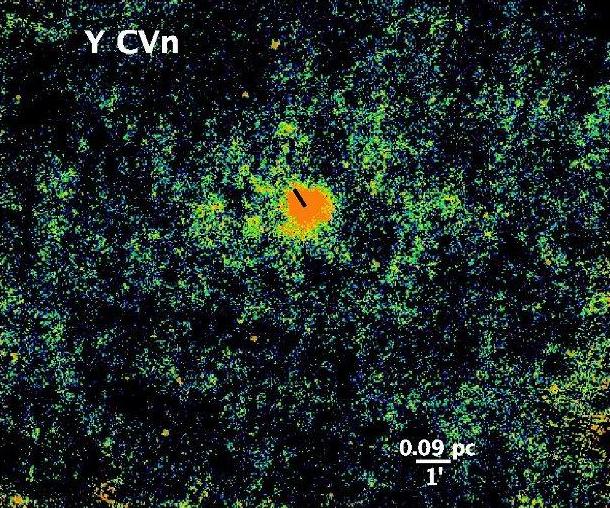}
 \includegraphics[width=0.23\textwidth,clip]{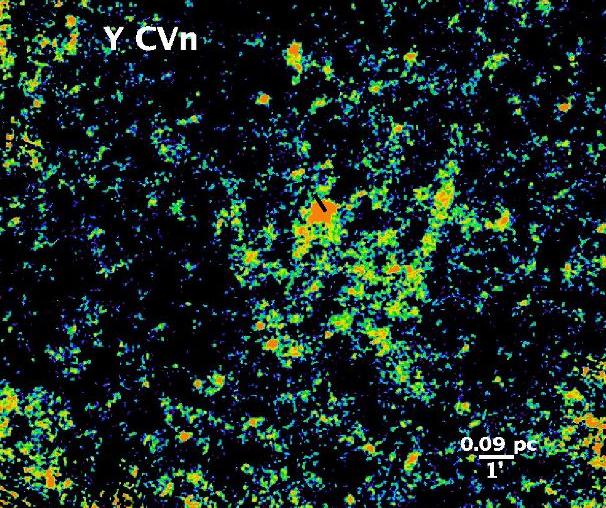}
 \includegraphics[width=0.45\columnwidth]{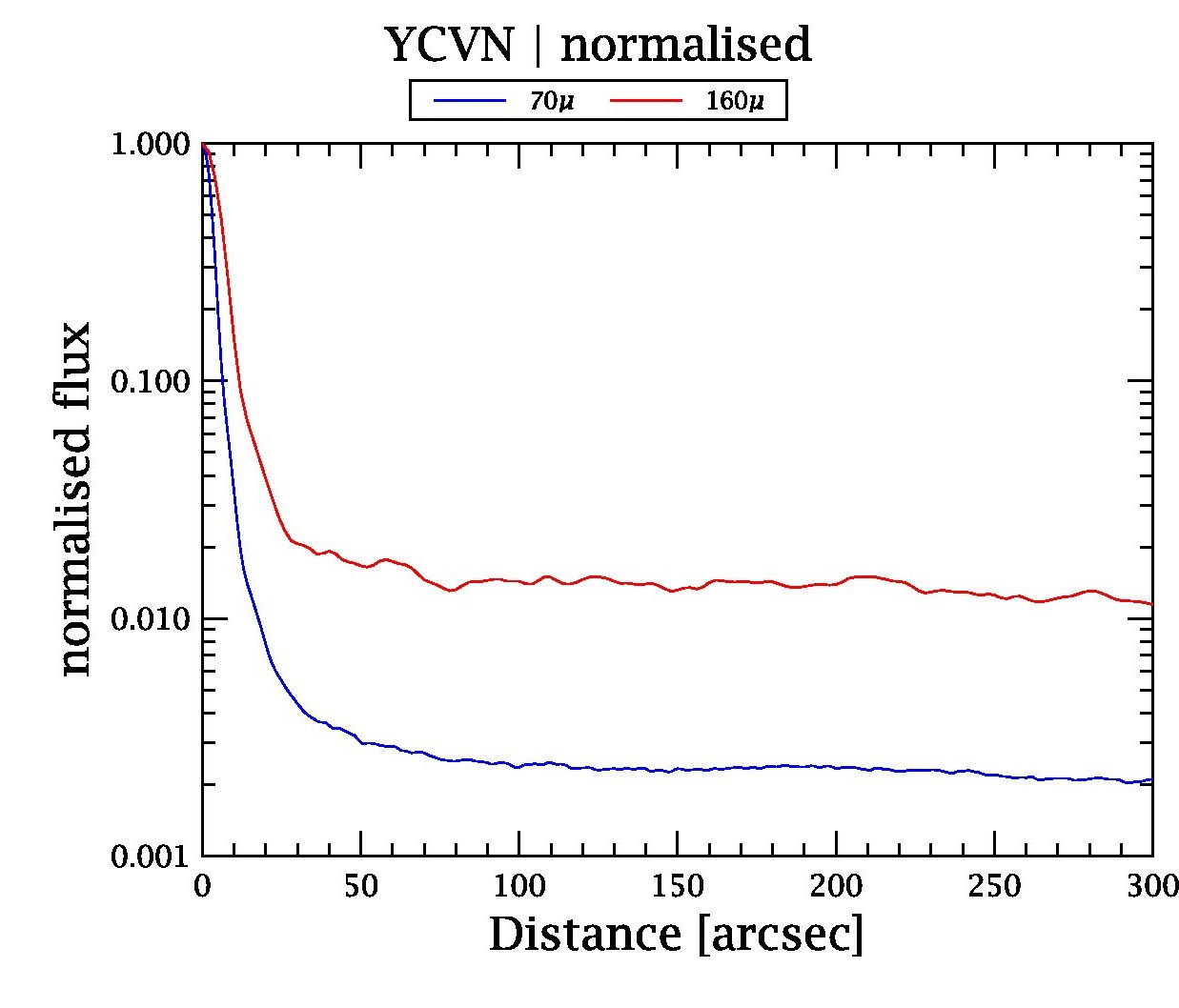}
 \caption{Interaction type \emph{``rings''} (Class\,III). PACS 70~$\mu$m (left) and 160~$\mu$m (middle).
 Azimuthally averaged (360\degr) radial profiles are shown for 70 and/or 160~$\mu$m (right).
 For U\,Ant see also Kerschbaum et al. (2010).
 The ring towards Y CVn is only faintly visible in the current display, but its presence is 
 verified from the previous detection with IRAS (\citealt{1996A&A...315L.221I}).}
 \label{fig:rings}
\end{figure*}

\addtocounter{figure}{-1}

Figures~\ref{fig:firstmap} to~\ref{fig:lastmap} presents the objects in our sample which reveal signs of extended and/or detached 
dust emission around the central star in either or both the PACS 70 and 160~$\mu$m bands. Colour versions of these figures are 
available in the electronic edition providing enhanced visibility. The different observed morphologies are divided in several classes 
(see Sect.~\ref{sec:morphology}). Out of 78 AGB stars and supergiants imaged with PACS as part of the MESS survey, 32 stars show
clear evidence of wind-ISM interaction, 15 stars show detached rings, and 6 show extended irregular emission
(Tables~\ref{tb:properties1} and~\ref{tb:properties1b}). The remaining 30 stars do not show evidence for wind-ISM interaction
(Table~\ref{tb:properties2}). 
Tables~\ref{tb:properties1} to~\ref{tb:properties2} provide basic properties of the observed AGB stars and supergiants; 
IRAS identifier (col. 1), Name (col. 2), Distance (col. 3), Mass-loss rate (col. 4), $z$ (col. 5); height above the Galactic plane, 
$n_H$ (col. 6); local ISM density, $\mu$ (col. 7); proper motion, $v_\star$ (col. 8); space velocity, P.A. (col. 9); the proper motion 
position angle measured from north to east, $i$ (col. 10); the inclination angle of the space motion with respect to the plane of the sky, 
and $v_w$ (col. 11) the terminal wind velocity. In addition, we also provide the predicted stand-off distance, $R_0$ in arcminutes (col. 12) 
and parsec (col. 13), as well as, for Tables 1 and 2, the measured stand-off distance $R_0$ in arcminutes (col. 14), parsec (col. 15), 
the position angle $\theta$ (col. 16) and the inferred local ISM density, $n_H$ (col. 17). Columns 18 and 19 give information on the 
spectral type / circumstellar chemistry and binarity, respectively.

\begin{figure*}[ht!]
 \includegraphics[width=0.23\textwidth,clip]{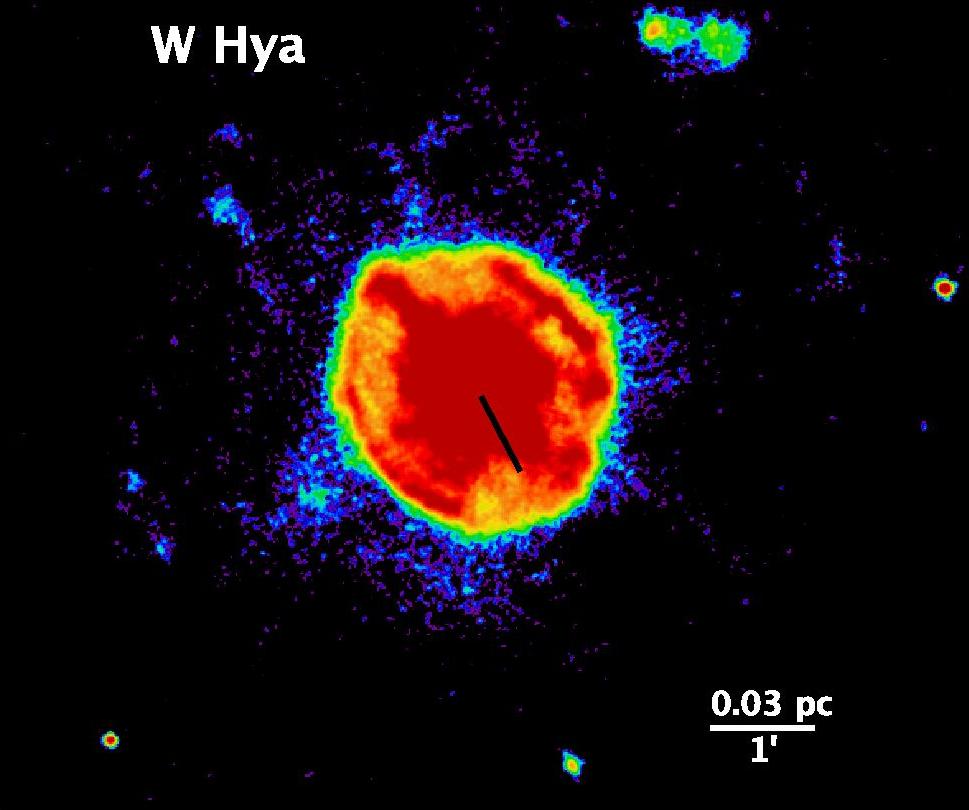}
 \includegraphics[width=0.23\textwidth,clip]{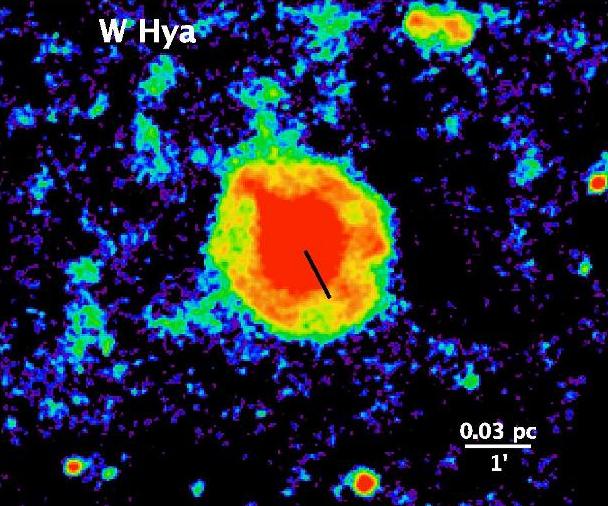}
 \includegraphics[width=0.45\columnwidth]{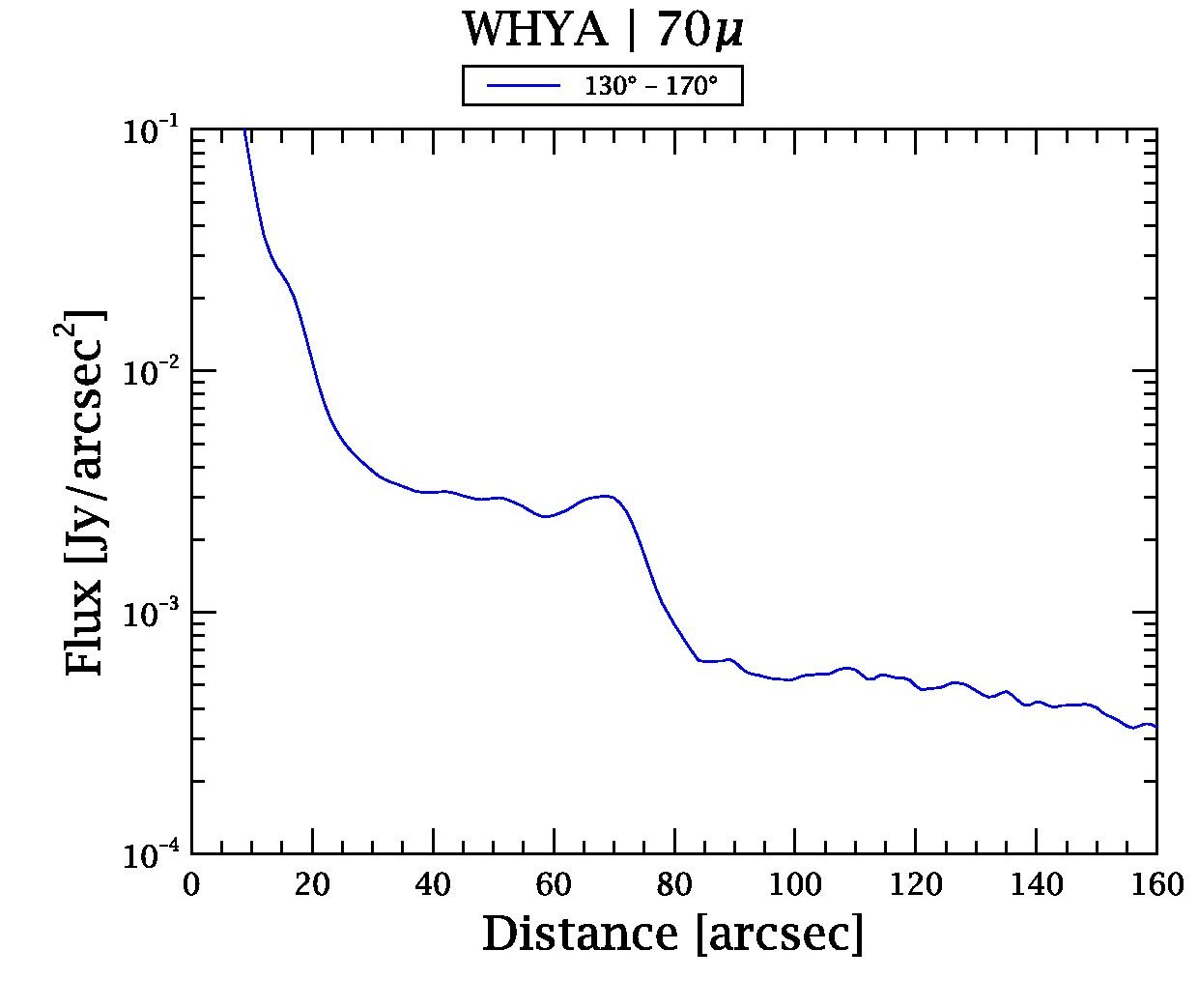}
 \includegraphics[width=0.45\columnwidth]{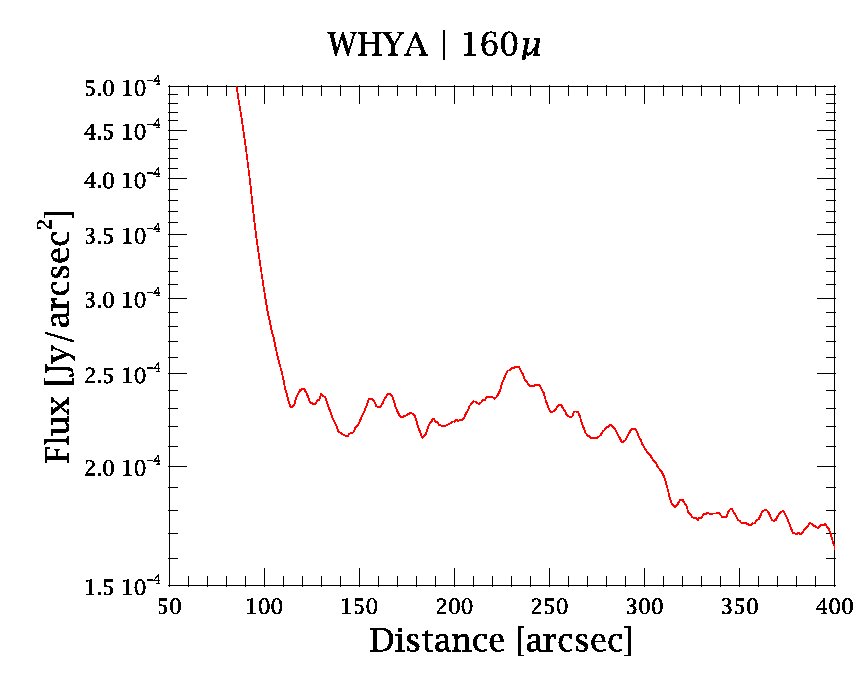}\\
 \medskip
 \includegraphics[width=0.23\textwidth,clip]{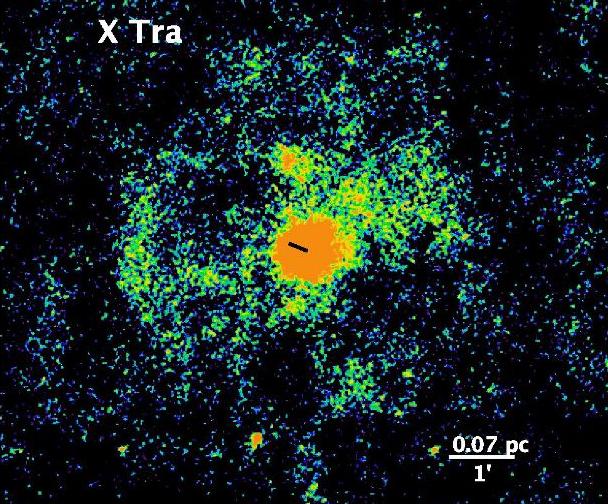}
 \includegraphics[width=0.23\textwidth,clip]{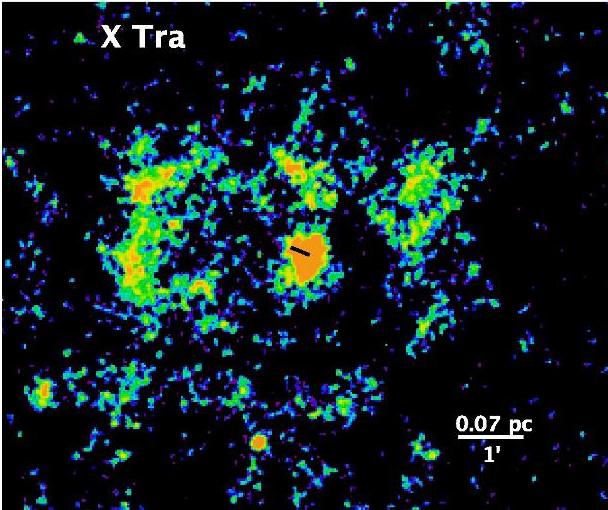}
 \includegraphics[width=0.45\columnwidth]{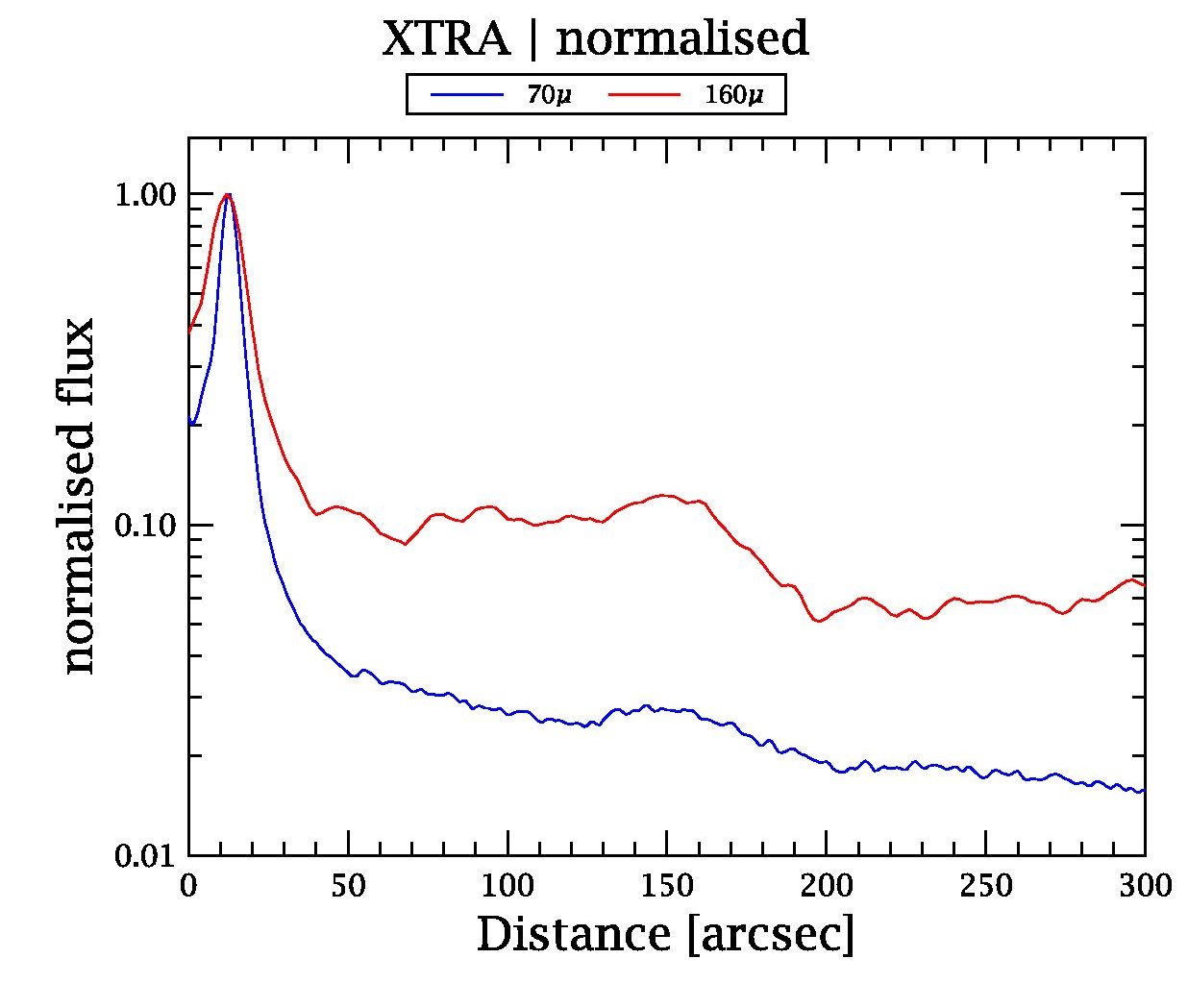}\\
 \medskip
 \includegraphics[width=0.23\textwidth,clip]{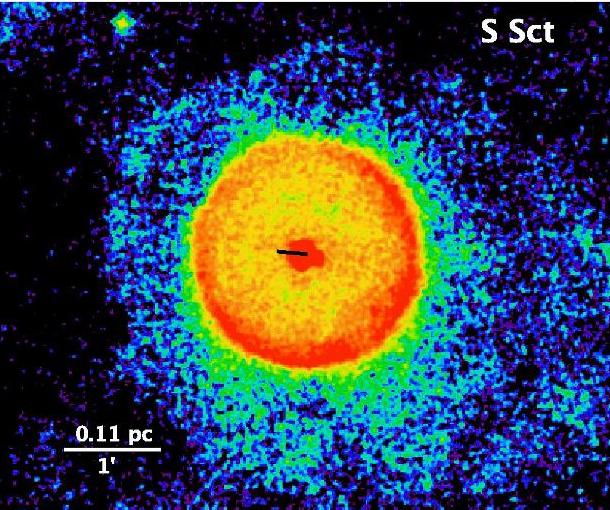}
 \includegraphics[width=0.23\textwidth,clip]{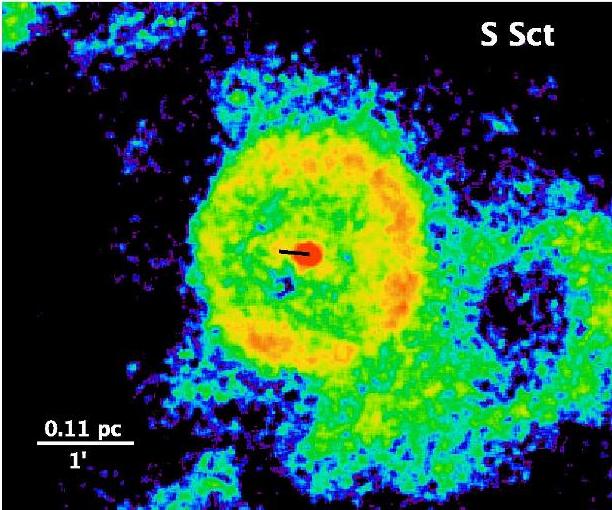}
 \includegraphics[width=0.45\columnwidth]{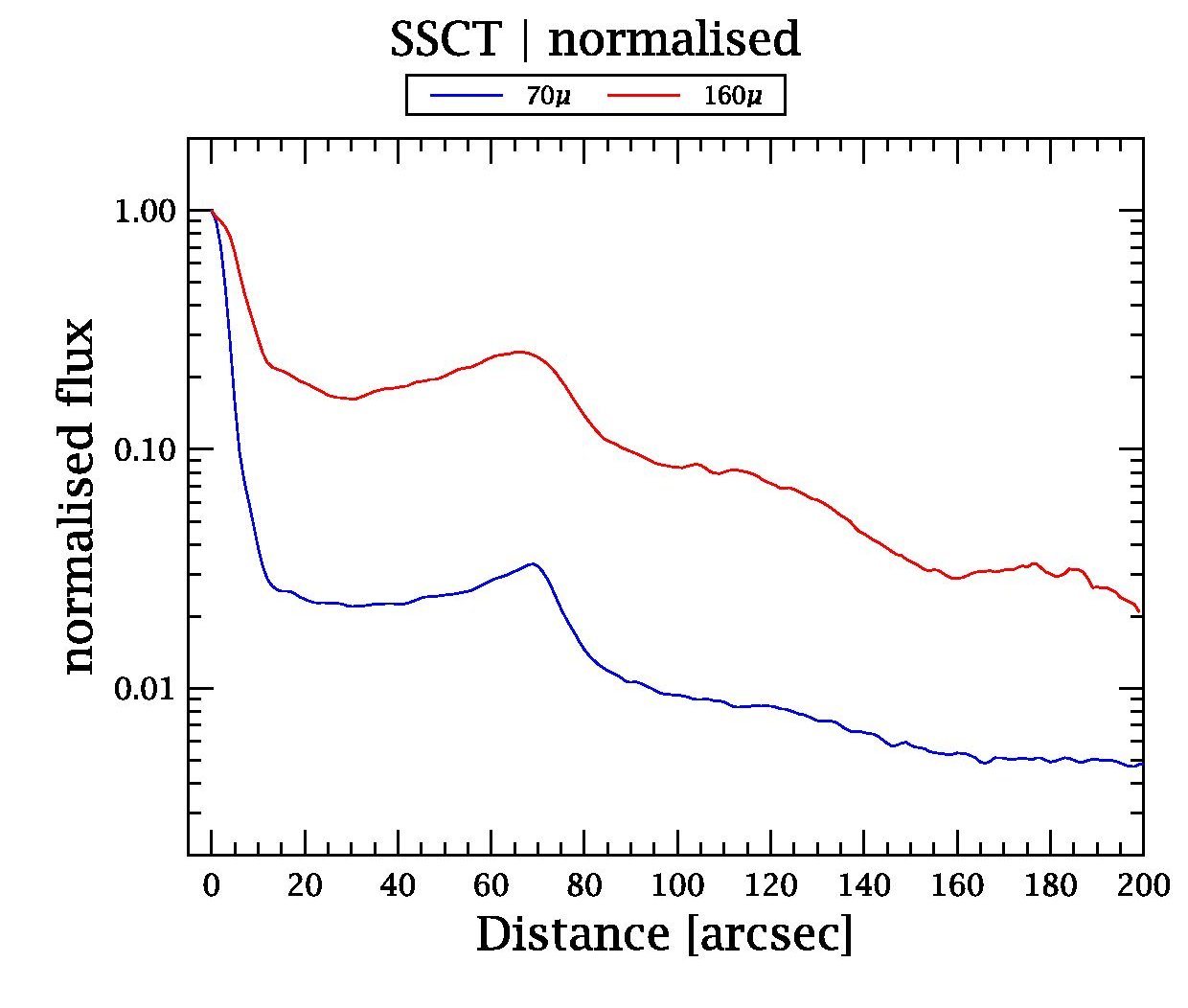}\\
 \medskip
 \includegraphics[width=0.23\textwidth,clip]{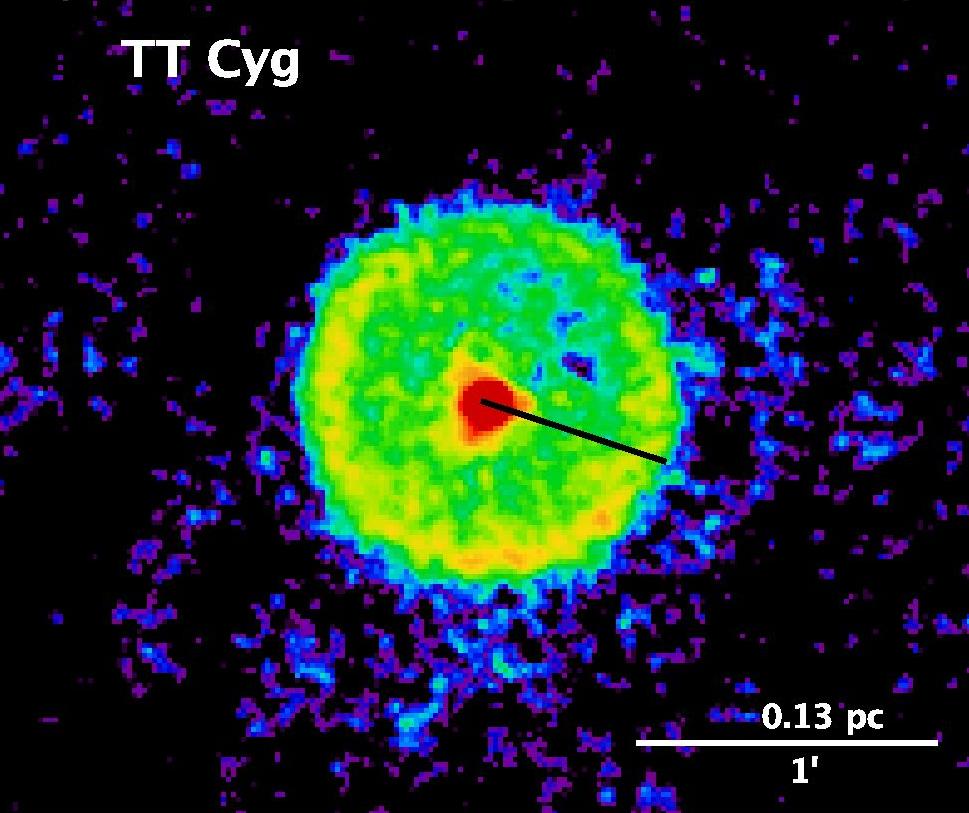}
 \includegraphics[width=0.23\textwidth,clip]{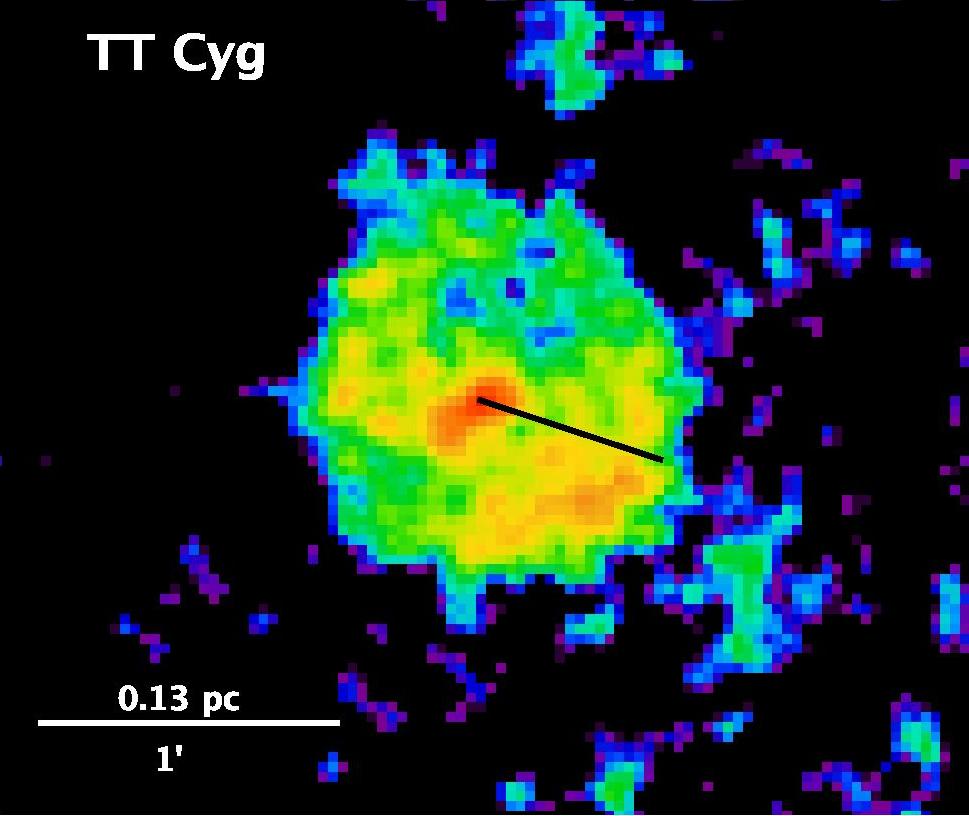}
 \includegraphics[width=0.45\columnwidth]{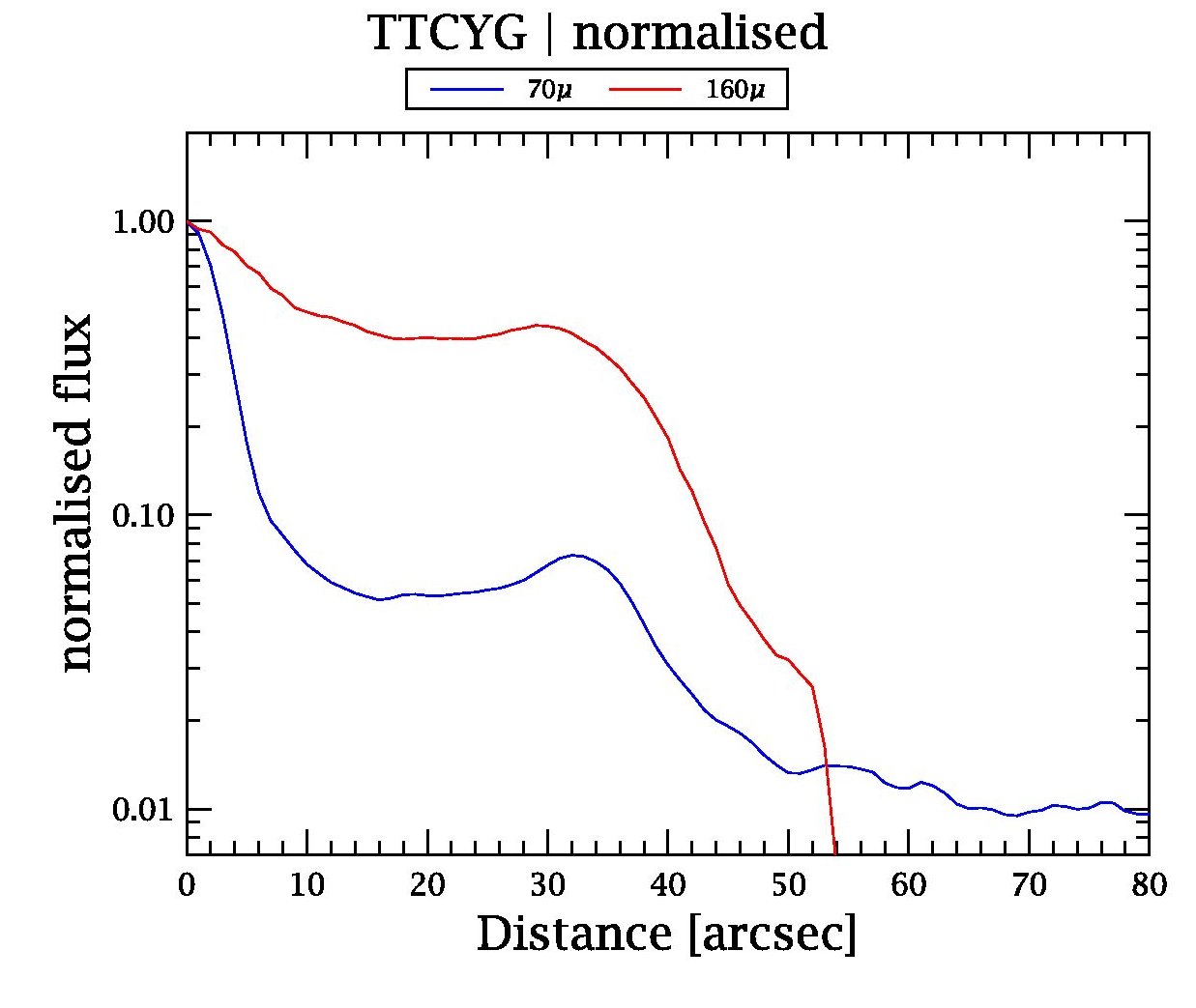}\\
 \medskip
 \includegraphics[width=0.23\textwidth,clip]{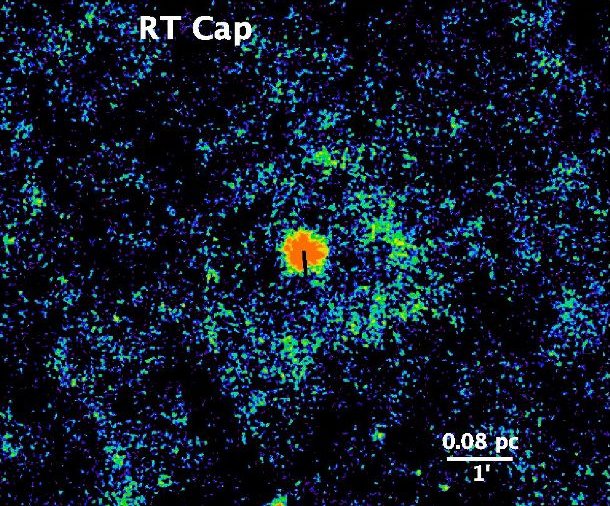}
 \includegraphics[width=0.23\textwidth,clip]{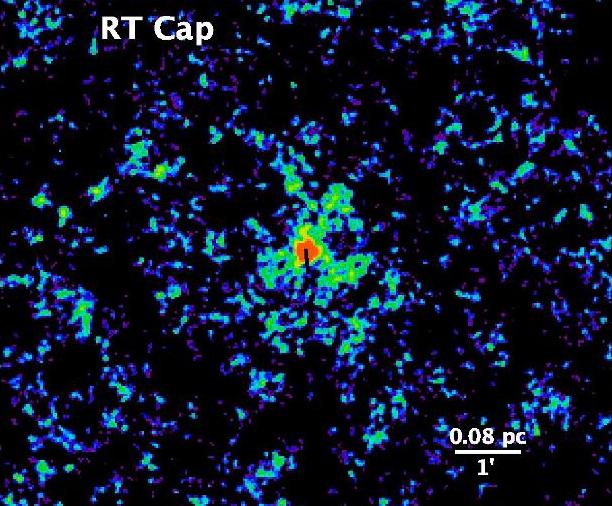}
 \includegraphics[width=0.45\columnwidth]{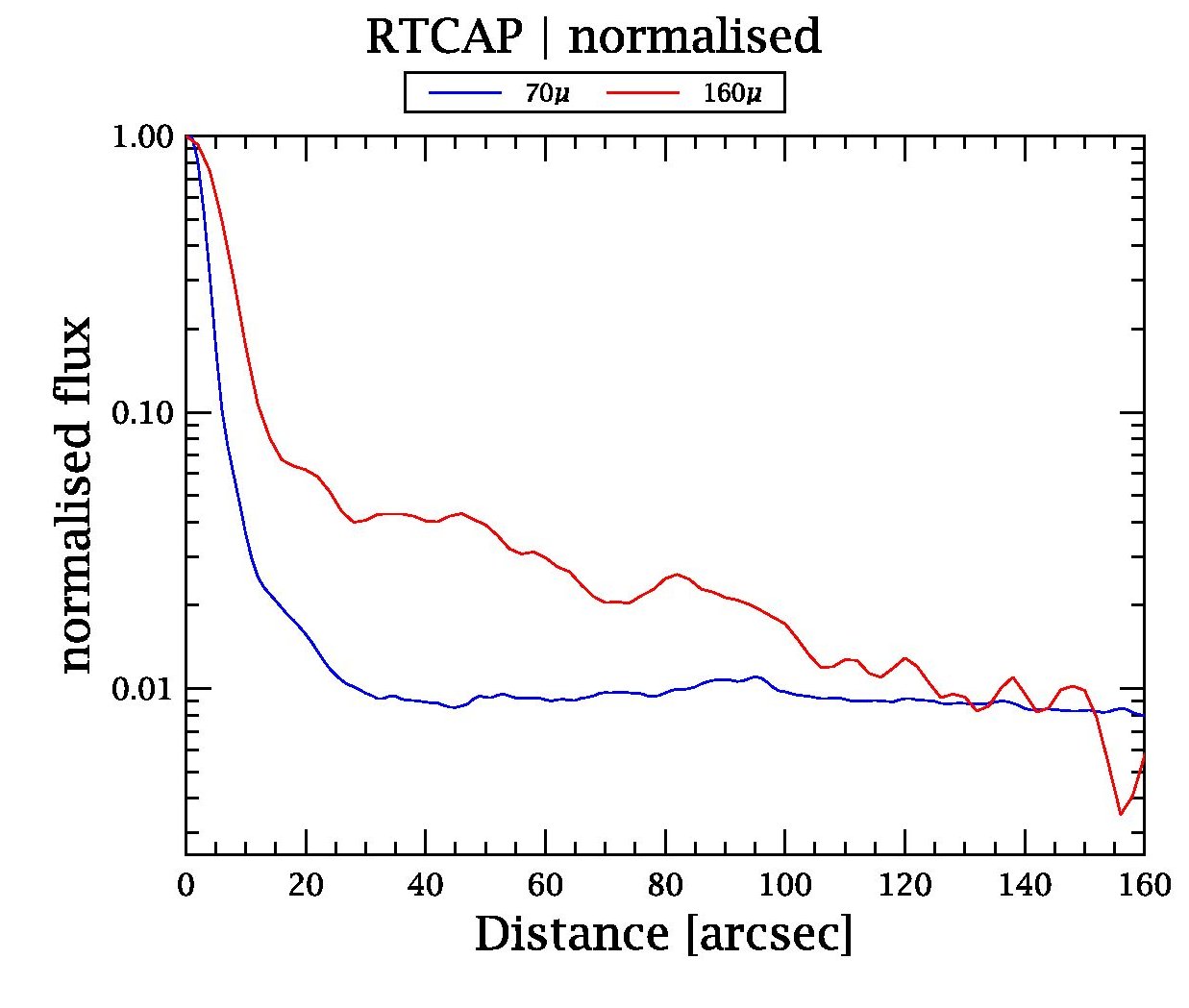}\\
 \medskip
 \includegraphics[width=0.23\textwidth,clip]{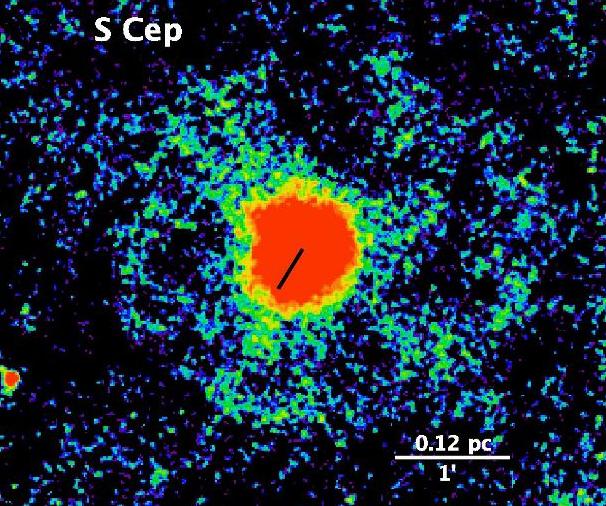}
 \includegraphics[width=0.23\textwidth,clip]{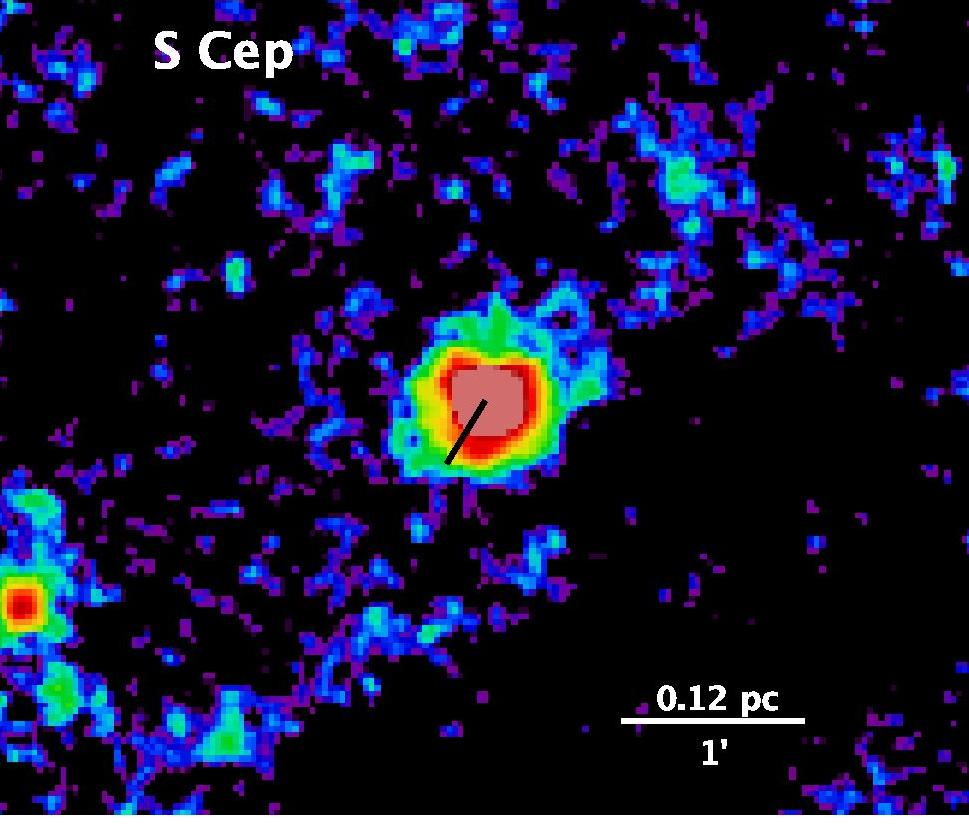}
 \includegraphics[width=0.45\columnwidth]{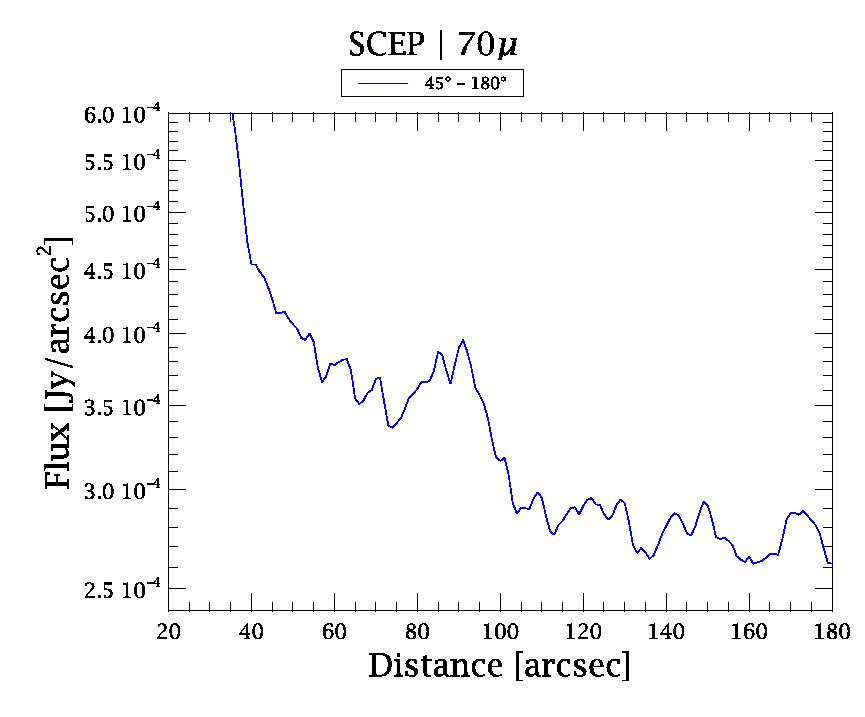}
 \caption{(continued) Interaction type \emph{``rings''} (Class\,III). PACS 70~$\mu$m (left) and 160~$\mu$m (middle).
 Azimuthally averaged (360\degr) radial profiles are shown for 70 and/or 160~$\mu$m (right).
 For TT\,Cyg see also Kerschbaum et al. (2010).
 The ring around W\,Hya is incomplete and deviates slightly from sphericity, also a jet-like structure
 is visible in the north-east direction.}
\end{figure*}

After deconvolution, three stars (R\,Scl, TX\,Psc, U\,Cam) reveal evidence for both a wind-ISM
interaction as well as a detached ring. To measure the ring of TX~Psc a de-convolved 2D-image was used. 
Detailed studies of some individual (classes of) objects included in the survey here are being presented in companion papers.
The complex multiple wind-ISM interaction shells associated to $\alpha$ Ori are presented, together with complementary
observations, in Decin et al. (in preparation). The interaction around X\,Her and TX\,Psc is described by
\citet{2011A&A...532A.135J} and that of $o$~Cet by \citet{2011A&A...531L...4M}. First results on the detached ring-like shells
around AQ\,And, TT\,Cyg and U\,Ant  were already discussed by \citet{2010A&A...518L.140K}, while CW\,Leo (IRC+10\,216) was
discussed previously by \citet{2010A&A...518L.141L} and \citet{2011A&A...534A...1D}. 
Also, the objects with both inner shells and large detached bow shock regions will be discussed in more detail in a forthcoming paper.
This notwithstanding, these sources are included here to complete the sample of observed AGB stars and red supergiants.

\begin{figure*}[ht!]
 \includegraphics[width=0.23\textwidth,clip]{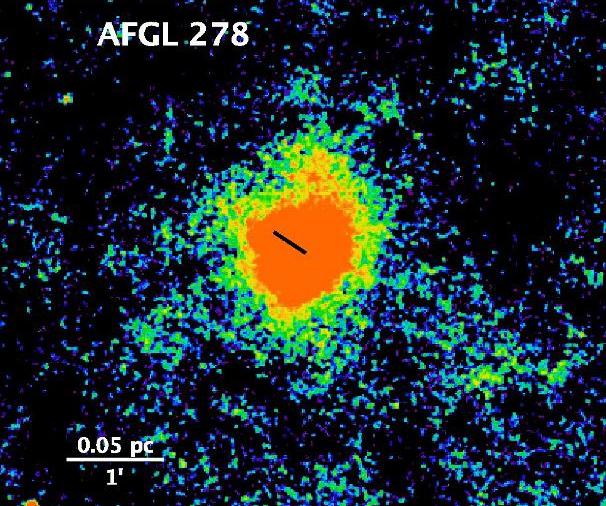}
 \includegraphics[width=0.23\textwidth,clip]{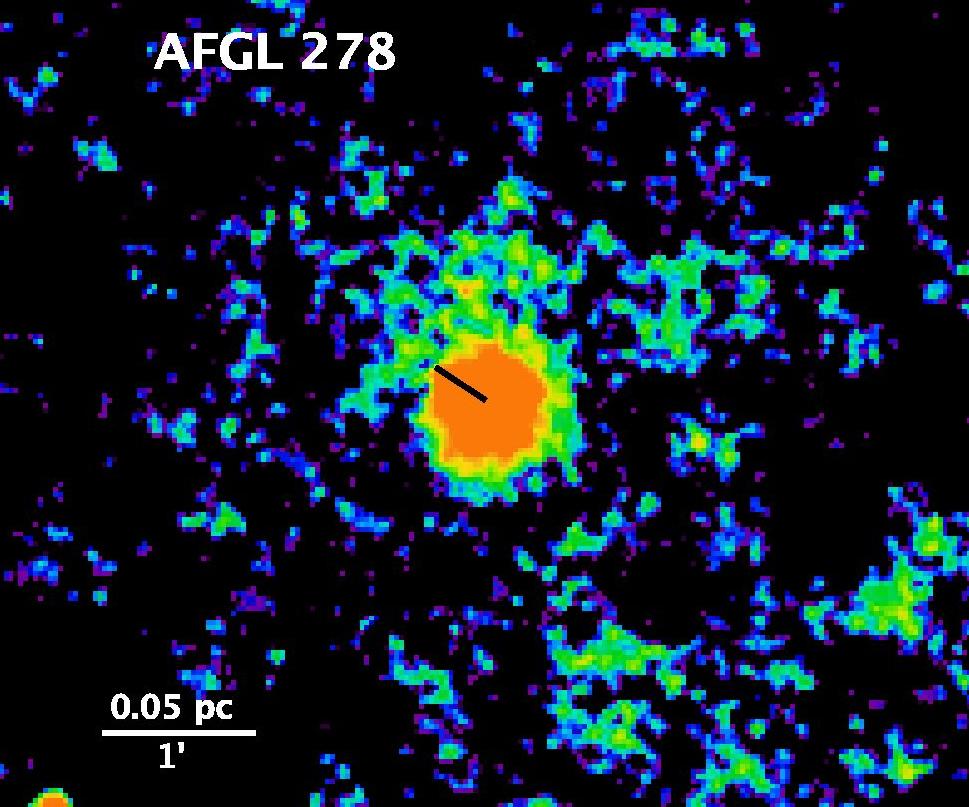}
 \hspace{5mm}
 \includegraphics[width=0.23\textwidth,clip]{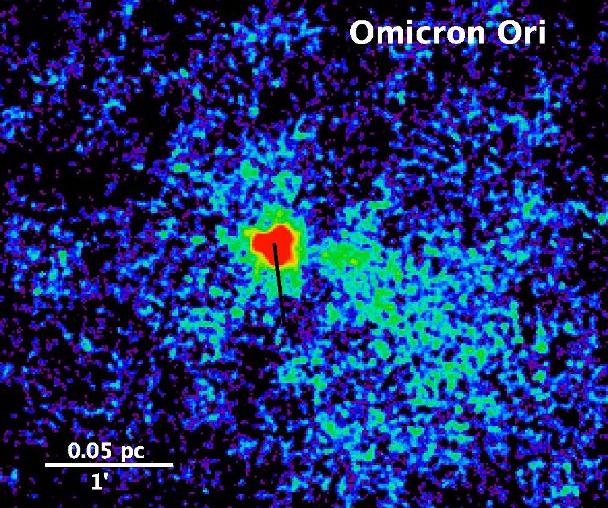}
 \includegraphics[width=0.23\textwidth,clip]{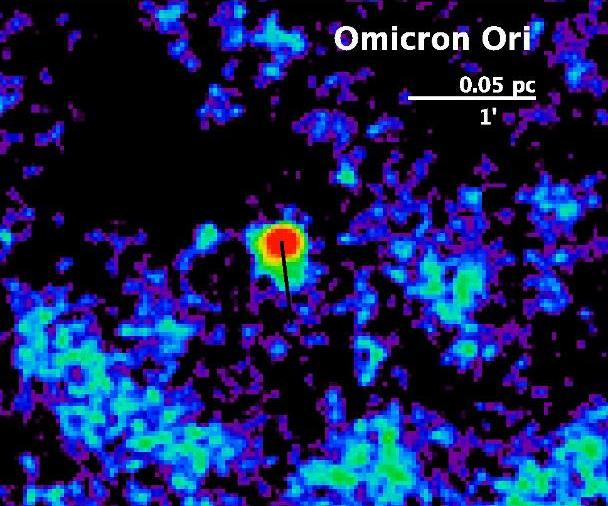}\\
 \medskip
 \includegraphics[width=0.23\textwidth,clip]{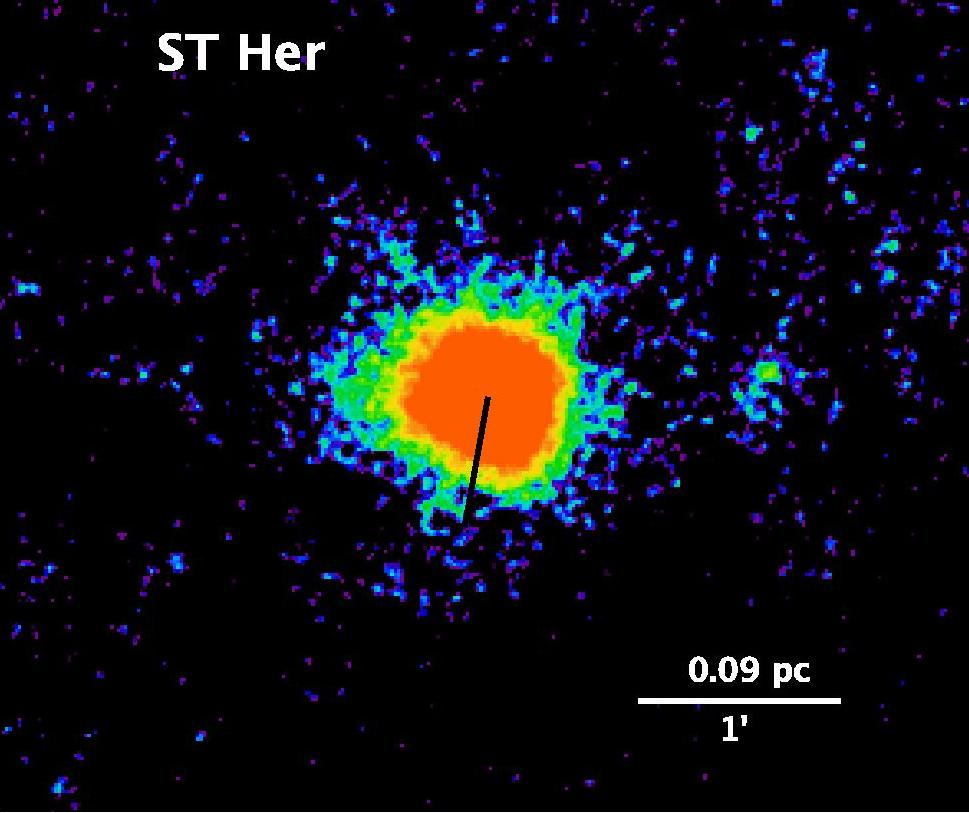}
 \includegraphics[width=0.23\textwidth,clip]{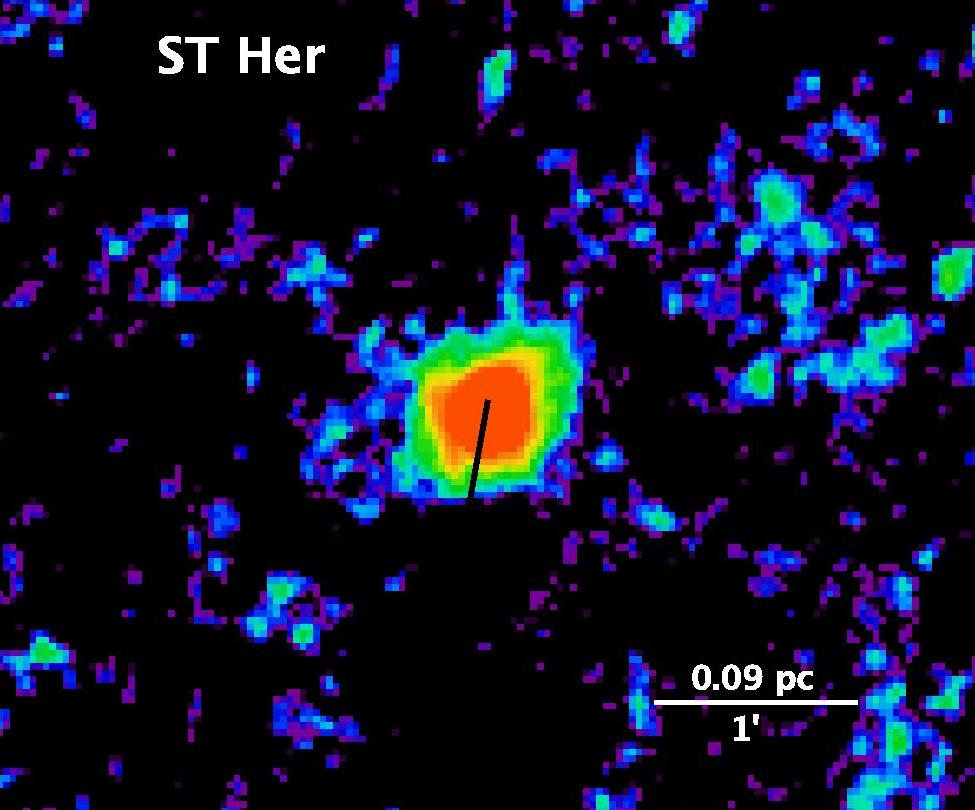}
 \hspace{5mm}
 \includegraphics[width=0.23\textwidth,clip]{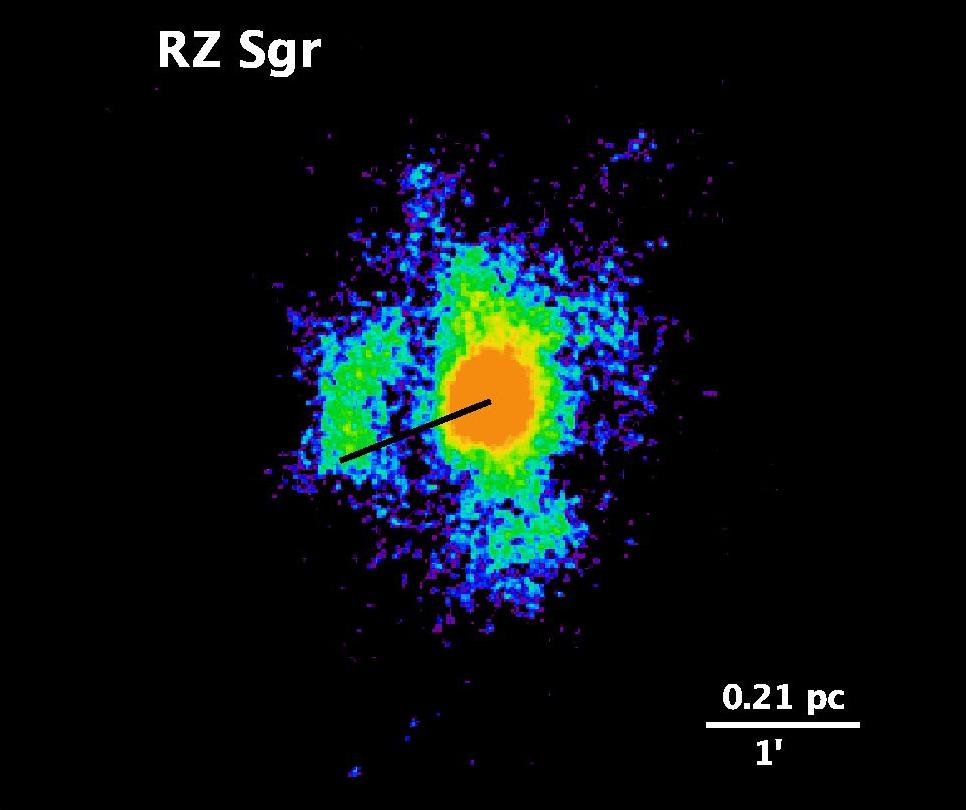}
 \includegraphics[width=0.23\textwidth,clip]{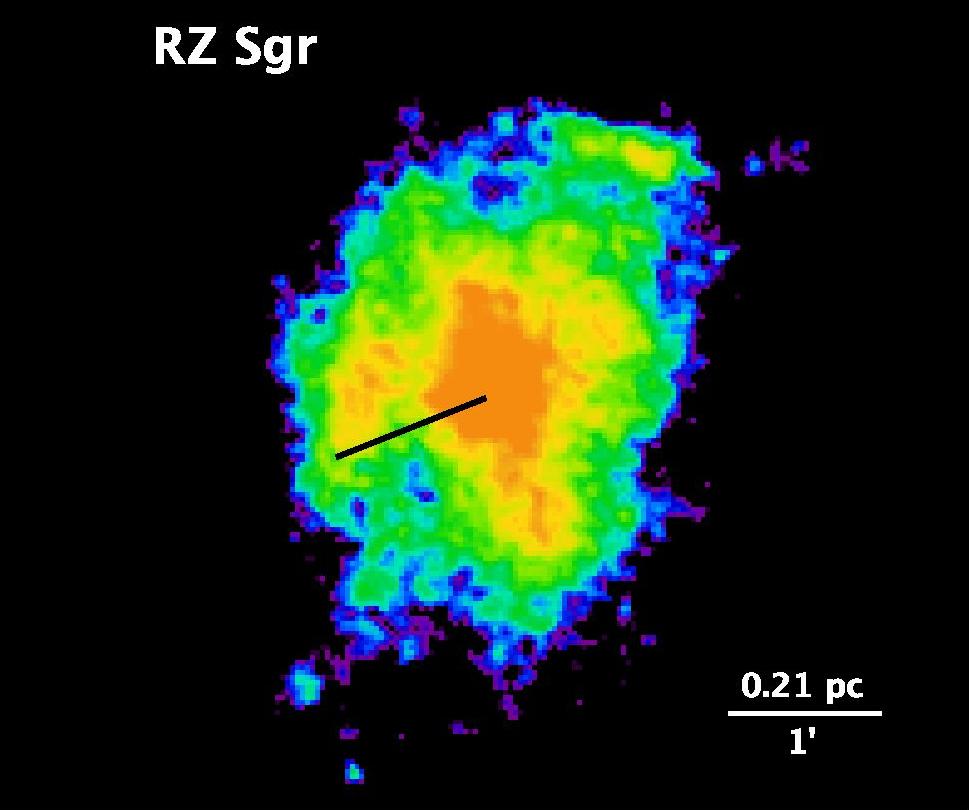}\\
 \medskip
 \includegraphics[width=0.23\textwidth,clip]{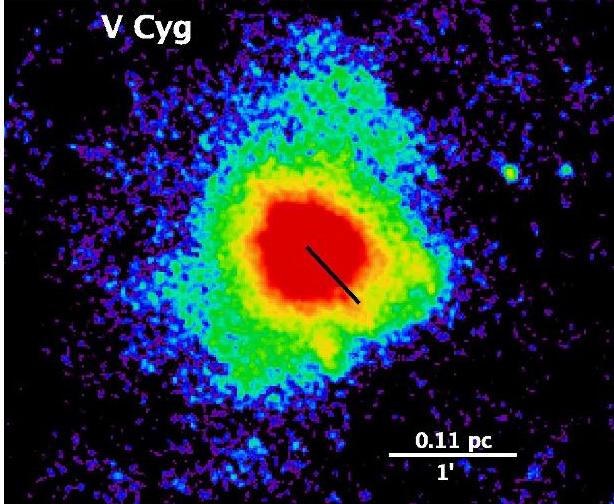}
 \includegraphics[width=0.23\textwidth,clip]{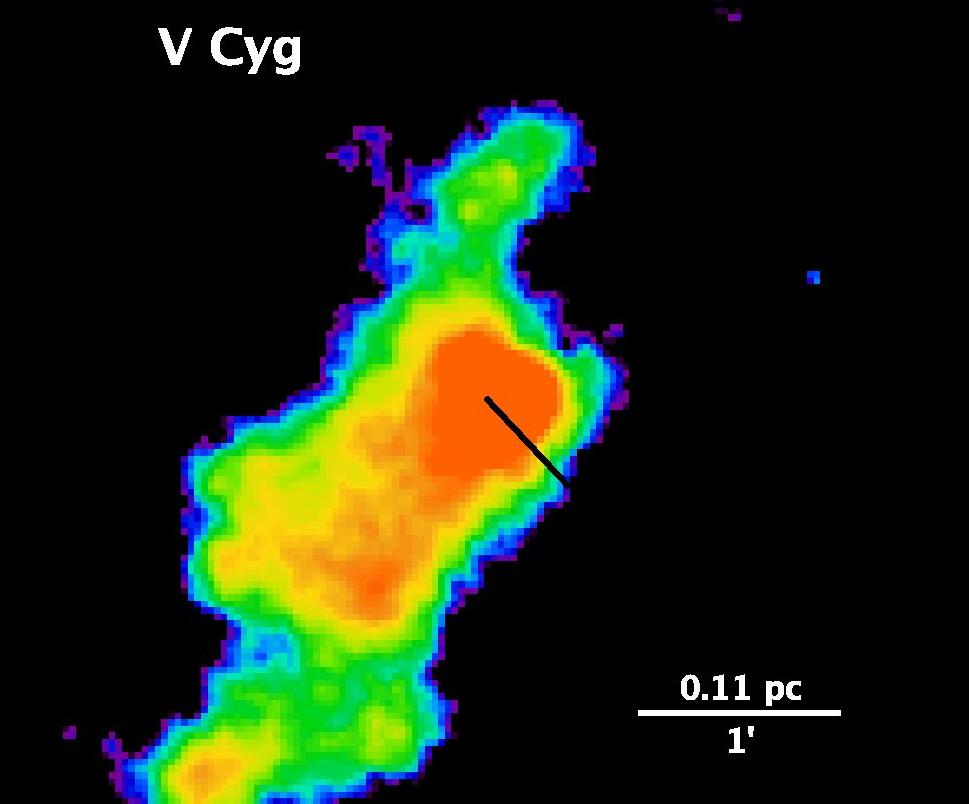}
 \hspace{5mm}
 \includegraphics[width=0.23\textwidth,clip]{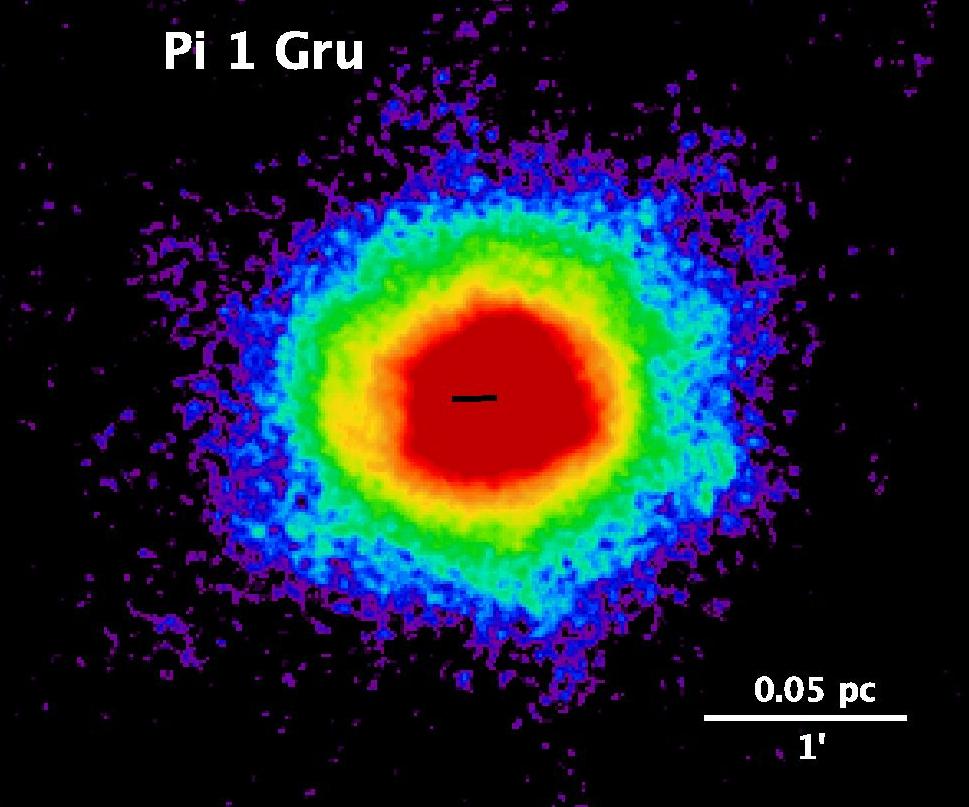}
 \includegraphics[width=0.23\textwidth,clip]{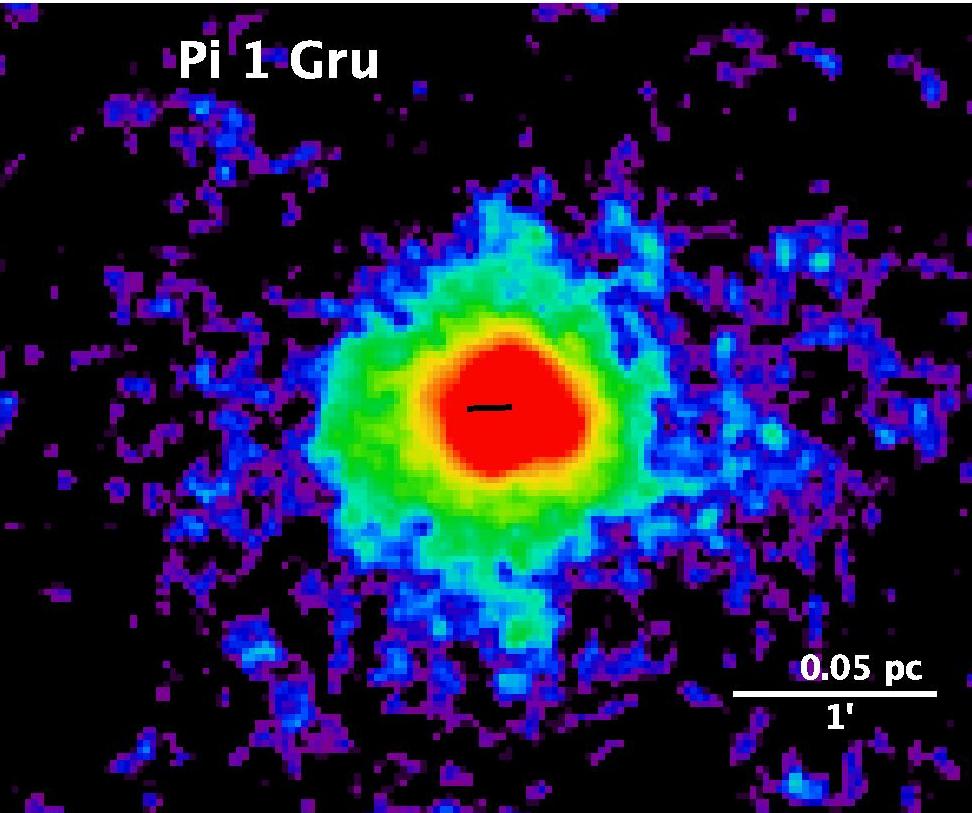}\\
 \medskip
 \includegraphics[width=0.23\textwidth,clip]{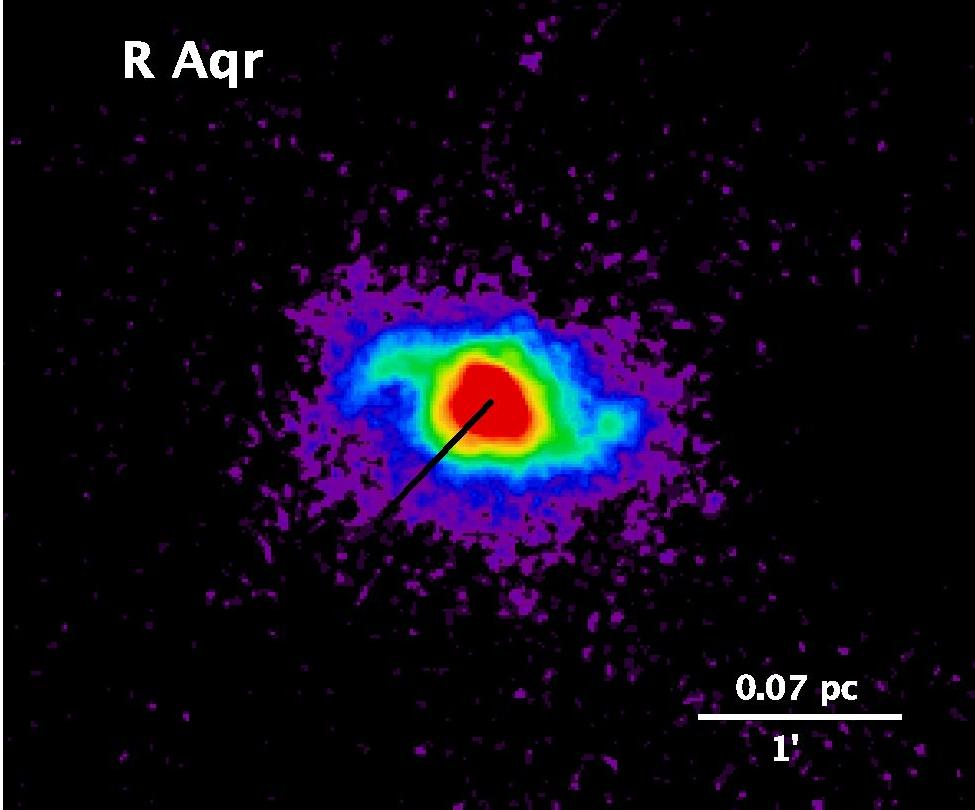}
 \includegraphics[width=0.23\textwidth,clip]{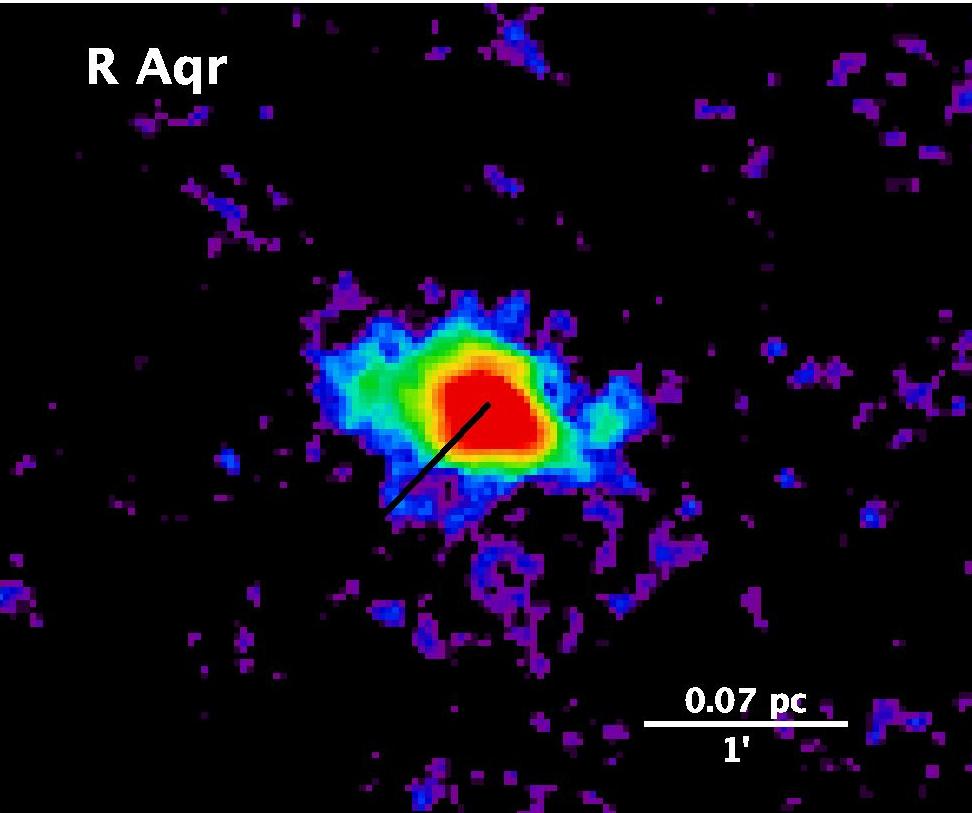}\\
 \caption{Interaction type \emph{``irregular''} (Class\, IV). PACS 70~$\mu$m (left) and 160~$\mu$m (right).
 These objects all show some evidence of irregular, diffuse extended emission. However, their
 morphology is neither a clear (double) arc nor a ring (possibly simply due to observational limitations
 in sensitivity and/or spatial resolution) and are therefore included in Class\,IV.}
 \label{fig:ClassIV}
 \label{fig:lastmap}
\end{figure*}

\clearpage

\longtabL{1}{
\scriptsize
\begin{landscape}
\begin{longtable}{llp{15mm}p{15mm}llllllllllllllll}
\caption{Key properties of AGB stars and supergiants that show wind-ISM bow shock interaction at 70~$\mu$m and/or 160~$\mu$m. 
The objects marked with $*$ show a combination of two types of interaction: an inner spherical ring together with an outer extended structure.
And, the object with $**$ is either a typical Fermata bow shock or a spherical ring.}\label{tb:properties1}
\endfirsthead
\caption{Notes and references}\\
\hline\hline							        	   				 																																				    
\endhead
\endfoot
\\
\hline\hline
IRAS identifier		 & Name		& Distance\tablefootmark{A}           & \Mdot			       & $z$\tablefootmark{E}	& $n_H$\tablefootmark{D}    & $\mu$\tablefootmark{H}	    & $v_\star$\tablefootmark{G}    & P.A.	    &  $i$	 & $v_w$\tablefootmark{F}    &\multicolumn{2}{c}{predicted $R_0$\tablefootmark{B}}  &&\multicolumn{3}{c}{observed $R_0$\tablefootmark{C}}   & $n_H$\tablefootmark{I} & Type\tablefootmark{J}	& Binary  \\ \cline{12-13}\cline{15-17}
	     		 &		&  (pc)	   			      & ($10^{-7} M_\odot$ yr$^{-1}$)	       & (pc)			& (cm$^{-3}$)		    &(mas yr$^{-1}$)	    & (\kms)			    &(\degr\ N-E)   &  (\degr)   & (\kms)		     &(\arcmin)     & (pc)				    &&(\arcmin)& (pc)	    & ($\theta$)		    &(cm$^{-3}$)    &	    		&         \\ \hline 
\hline
{\bf I ``Fermata''} \\ \cline{1-1}
 \object{00248+3518} & AQ And $^*$&  781$\pm$555$^6$,	  {\bf 825}		     & 6.5$^7$  		      & -360	  &  0.05      &  13.4    & 54.1	& 218	     &  17	  &		 & 1.5     & 0.37	 &&         &          &	    &      	 & C		&	    \\  	       
 \object{01246-3248} & R Scl $^*$ &			    {\bf 370}$^5$, 290$^6$   & 16$^2$			      & -350	  &  0.06      &  36.7    & 66.3	& 208.3      & -15	  & 17.0$^2$	 & 4.4     & 0.48	 && 0.88    & 0.095    & 225	    &  1.5 	 & C		&	    \\  	       
 \object{02168-0312} & $o$ Cet    & {\bf 92$\pm$10}$^h$,     110$^5$		     & 2.5$^2$  		      & -63	  &  1.07      &  223.3   & 107.7	& 185.3      &  26	  & 8.1$^2$	 & 0.7     & 0.02	 && 1.23    & 0.033    & 181	    &  0.4	 & O - M7IIIe	& yes$^9$   \\  		   
 \object{03507+1115} & NML Tau    &			    245$^7$		     & 32$^1$			      & -113	  &  0.65      &  --	  & $>$33.8	&	     &  -	  & 18.5$^2$	 &$\leq$5.9& $\leq$0.42  &&$\sim$1.4& 0.10     &	    &  11.6	 & O - M6me	&	    \\  	       
 \object{05028+0106} &W Ori$^{**}$&  377$\pm$135$^h$				     & 2.3$^7$  		      & -131	  &  0.54      &  10.5    & 18.8	& 91.5       &  0	  & 8.6$^{16}$   & 1.4     & 0.15	 && 0.7     & 0.077    &	    &  2.1	 & C		&	    \\  	       
 \object{05418-4628} & W Pic	  & 775$\pm$300$^h$	     670$^{26}$,  {\bf 512}  & 3.0$^{15}$		      & -246	  &  0.17      &  12.5    & 38.2	& 348	     &  33	  &7.0$^{25}$ (13.9$^{15}$)& 1.2  & 0.18 && 0.57    & 0.085    & 292	    &  0.8	 & C		&	    \\ 
 \object{05524+0723} &$\alpha$ Ori& 153$\pm$19$^h$,  {\bf 131}$^2$		     & 2.1$^2$, 31$^3$  	      & -5	  &  1.89      &  38.7    & 24.3	& 47.7       &  8	  & 14.0$^2$	 & 7.7     & 0.29	 && 4.98    & 0.19     & 54	    &  4.6	 & O - M2Iab	&	    \\ 
 \object{06331+3829} & UU Aur	  &			    341 		     & 2.7$^7$, 74$^{28}$	      & 96	  &  0.76      &  10.4    & 17.9	& 170.4      &  24	  & 11.0$^{16}$  & 3.3     & 0.33	 && 1.32    & 0.131    & 200	    &  4.9	 & C		&  yes$^1$  \\ 
 \object{09448+1139} & R Leo	  & {\bf 71$\pm$13$^h$},	100$^5$, 82$^2$)     & 0.92$^2$ 		      & 65	  &  1.05      &  40.4    & 13.6	& 112.3      &  -4	  & 9.0$^2$	 & 4.7     & 0.10	 && 1.25    & 0.026    & 117	    &  14.9	 & O - M8IIIe	&	    \\ 
 \object{09452+1330} & CW Leo	  & 150$^1$, 140$^{19}$, {\bf 120$^2$}, 110$^{14}$   & 16$^2$			      & 100	  &  0.74      &  82.6    & 53.3	& 64.5       & -29	  & 14.5$^2$	 & 4.5     & 0.16	 && 6.6     & 0.23     & 82	    &  0.3	 & C - C9.5e	&	    \\ 
 \object{10350-1307} & U Hya $^*$ &  208$\pm$10$^h$,	  161$^2$		     & (0.92$^1$, 2.0$^{14}$),0.49$^2$& 143	  &  0.48      &  66.8    & 71.9	& 118.5      & -26	  & 8.5$^2$	 & 0.3     & 0.02	 &&         &          &	    &      	 & C		&	    \\ 
 \object{13001+0527} & RT Vir	  &  136$\pm$15$^h$				     & 5$^{17}$ 		      & 141	  &  0.49      &  61.9    & 43.9	& 96.8       &  24	  & 8.9$^1$,(7.8$^{17}$)& 2.6& 0.10	 &&	    &	       &	    &		 & O - M8III	&	    \\ 
 \object{13269-2301} & R Hya	  &  124$\pm$11$^h$,	  130$^5$, {\bf 118}$^2$     & 1.6$^2$  		      & 89	  &  0.82      &  42.9    & 25.3	& 313.7      & -23	  & 12.5$^2$	 & 2.7     & 0.09	 && 1.55    & 0.053    & 284	    &  2.5	 & O - M7IIIe	&  yes$^1$  \\ 
 \object{14003-7633} &$\theta$ Aps&  113$\pm$6$^h$,				     & 1.1$^7$  		      & -13	  &  1.75      &  63.3    & 34.2	& 258.8      &   2	  & 4.5$^{16}$   & 0.7     & 0.02	 && 1.25    & 0.041    & 297	    &  0.6	 & O - M6.5III  &  yes$^6$  \\ 
 \object{16011+4722} & X Her	  &  137$\pm$8$^h$				     & 1.5$^{11}$		      & 116	  &  0.62      &  79.6    & 91.2	& 313.9      & -53	  & 6.5$^{11}$   & 0.4     & 0.01	 && 0.60    & 0.024    & 330	    &  0.5	 & O - M8III	&	    \\ 
 \object{19126-0708} & W Aql	  &			    610$^1$, {\bf 340}$^4$   & (25$^{15}$), 130$^2$	      & -35	  &  1.40      &  27.6    & 51.4	& 65.3       & -29	  & 20.0$^2$	 & 4.0     & 0.40	 && 0.8/1.3 & 0.08/0.13&	    &  35	 & S - S6.6+FV  &  yes$^7$  \\ 
 \object{19486+3247} & $\chi$ Cyg &  181$\pm$36$^h$,	  170$^5$, {\bf 149}$^2$     & 2.4$^2$  		      & 24	  &  1.58      &  52.1    & 37.6	& 218.3      &  16	  & 8.5$^2$	 & 1.0     & 0.05	 &&$\sim$6.6& 0.29     &	    &  0.04	 & S - S6+/1e	&	    \\ 
 \object{20038-2722} & V 1943 Sgr &  197$\pm$23$^h$				     & 1.3$^{17}$		      & -77	  &  0.93      &  34.7    & 35.2	& 155.0      & -23	  & 5.4$^{17}$   & 0.6     & 0.04	 && 1.1	    & 0.06     & 135	    &  0.3	 & O - M7III	&	    \\ 
 \object{20075-6005} & X Pav	  &			    270$^2$		     & 5.2$^2$  		      & -133	  &  0.53      &  21.9    & 32.6	& 84.9       & -35	  & 11.0$^2$	 & 1.9     & 0.15	 && 0.78    & 0.061    & 88	    &  3.2	 & O - M7III:p  &	    \\ 
 \object{20248-2825} & T Mic	  &  211 $\pm$ 45$^h$				     & 0.8$^{17}$		      & -98	  &  0.75      &  30.9    & 39.9	& 322.2      &  39	  & 4.8$^{17}$   & 0.4     & 0.03	 &&	    &	       &	    &		 & O - M7III	&	    \\ 
 \object{21419+5832} & $\mu$ Cep  &  390$^8$					     & 20$^2$			      & 17	  &  1.69      &   5.1    & 21.9	& 216.3      &  50	  & 35.0$^2$	 & 2.8     & 0.3	 && 0.83    & 0.094    & 85	    &  19.3	 & O - M2Iae	&	    \\ 
 \object{21439-0226} & EP Aqr	  &  114 $\pm$ 8$^h$,	    {\bf 135}$^2$	     &3.1$^2$, 0.2$^{28}$, 10.0$^{29}$& -70	  &  0.99      &  31.2    & 40.2	& 4.6	     & -57	  & 11.5$^2$	 & 1.8     & 0.07	 && 0.45    & 0.018    & 22	    &  15.8	 & O - M8IIIvar &  yes$^8$  \\ 
 \object{23438+0312} & TX Psc $^*$&  275 $\pm$ 30$^h$				     & 0.91$^{12}$		      & -212	  &  0.24      &  50.2    & 66.4	& 247.7      &  11	  & 10.5$^{12}$  & 0.3     & 0.03	 && 0.62    & 0.050    & 238	    &  0.2	 & C - C5-7,2	&	    \\ 
 \object{23558+5106} & R Cas	  &{\bf 127$\pm$16}$^h$, 176$_{-92}^{+45}$$^w$       & 4.0$^2$, 12$^{28}$	      &  -8	  &  1.84      &  63.4    & 44.0	& 65.7       &  35	  & 13.5$^2$	 & 1.6     & 0.06	 && 1.5     & 0.055    & 65	    &  2.1	 & O - M7IIIe	&  yes$^1$  \\ 
\\ 
{\bf II ``Eyes''}\\ \cline{1-1}
 \object{03374+6229} & U Cam $^*$ &  430$^6$					     & 10$^7$			      & 60	  &  0.66      &  4.2	  & 9.8 	& 350.4      &  49	  & 20.6$^{21}$  & 5.3     & 0.66	 && 0.95/1.0& 0.12/0.13 &	    &  30	 & C		&  yes$^{1,14}$  \\ 
 \object{04459+6804} & ST Cam	  &  419$^2$, {\bf 990} 			     & 4.5$^7$  		      & 270	  &  0.14      &  4.9	  & 28.7	& 247.8      & -31	  & 10.5$^{25}$  & 1.1     & 0.31	 && 1.1/1.4 & 0.32/0.40 &	    &  0.1	 & C		&	    \\ 
 \object{10416+6740} & VY UMa	  &  380$\pm$51$^h$				     &  			      & 286	  &  0.12      & 15.6	  & 28.2	& 70.5       & -1	  &  7.9$^{25}$  & 1.5     & 0.16	 && 0.6/0.8 & 0.07/0.09 &	    &  0.4	 & C		&  yes$^2$  \\ 
 \object{10580-1803} & R Crt	  &  261$\pm$65$^h$				     & 6.9$^1$, 1.5$^{24}$, 8$^{17}$  & 173	  &  0.37      & 15.8	  & 23.8	& 279.2      &  28	  & 10.8$^{1,17}$& 3.8     & 0.29	 &&$\sim$2.3& 0.18	&	    &  1.0	 & O - M7III	&  yes$^1$  \\ 
 \object{17389-5742} & V Pav	  &  370$\pm$73$^h$				     & 3.4$^{14}$		      & -76	  &  0.89      &  5.3	  & 24.0	& 321.7      &  56	  & 16.0$^{25}$  & 1.2     & 0.13	 && 1.6/1.7 & 0.17/0.18 &	    &  0.5	 & C		&  yes$^1$  \\ 
 \object{19233+7627} & UX Dra	  &  386$\pm$43$^h$				     & 1.6$^h,r$		      & 176	  &  0.36      & 13.0	  & 26.7	& 231.9      &  34	  & 4.0$^{15}$   & 0.7     & 0.08	 && 1.3/0.9 & 0.15/0.10 &	    &  0.2	 & C		&  yes$^4$  \\ 
 \object{19314-1629} & AQ Sgr	  &  333$\pm$74$^h$				     & 2.5$^{15}$		      & -81	  &  0.85      &  8.0	  & 29.1	& 349.4      &  66	  & 10.0$^{15}$  & 0.8     & 0.08	 && 0.95    & 0.09	&	    &  0.6	 & C		&	    \\ 
\hline
\\
\multicolumn{20}{c}{
\tablefoot{
\tablefoottext{A}{The distance, $d$, is computed from the Hipparcos parallax ($h$) \mbox{\citep{2007A&A...474..653V}} or via the Period-Luminosity relation
		 (Sect.~\ref{sec:distance}). For unavailable and uncertain Hipparcos measurements we adopt the PL distance (indicated in boldface).}
\tablefoottext{B}{$R_0$ is the stand-off distance of the bow shock with respect to the star.  
     The theoretical $R_0$ is derived from Eq.~\ref{eq:standoff}, using the adopted stellar parameters for $v_\star$, $v_w$, \Mdot, $n_H$ 
     (this Table and Sect.~\ref{sec:standoff}).}
\tablefoottext{C}{For Class I objects the observed (de-projected) $R_0$ is derived from the measured angular distance $A$ and $B$ (Sect.~\ref{sec:standoff}).
     For Class II and III objects we give the (angular and physical) radial distance for each arc and the ring, respectively.}
\tablefoottext{D}{$n_H$ (cm$^{-3}$) from Eq.~\ref{eq:nH2}.}
\tablefoottext{E}{$z$ is the distance from the Galactic plane ($z$(pc) = $d$ sin $b$ + 15, with $d$ the distance and $b$ the Galactic latitude); 
		(Sect.~\ref{sec:properties}).} 
\tablefoottext{F}{$v_w$ is the terminal velocity of the CO envelope.}
\tablefoottext{G}{The peculiar space velocity (and proper motions and radial velocities) have been corrected for the solar motion [(U,V,W) = (11.10, 12.24, 7.25)~\kms] 
	          (\citealt{2010MNRAS.403.1829S}).}
\tablefoottext{H}{Proper motion, $\mu$ (mas yr$^{-1}$) from $\mu_\alpha$ and $\mu_\delta$ (\mbox{\citealt{2007A&A...474..653V}}) and corresponding 
		  sky position angle from north to east.}
\tablefoottext{I}{{\bf $n_H$ inferred from observed $R_0$ and Eq.~\ref{eq:standoff}.}}
\tablefoottext{J}{Type gives circumstellar envelope chemistry (oxygen (O) vs. carbon (C) rich) and the central star's spectral type.}
}
}\\
\multicolumn{20}{c}{
\tablebib{
(1) \citet{1998ApJS..117..209K}; (2) \mbox{\citet{2010A&A...523A..18D}}; (3) \mbox{\citet{2006ApJ...646.1179H}}; (4) \mbox{\citet{2008A&A...488..675G}}; 
(5) \citet{2008MNRAS.386..313W}; (6) \mbox{\citet{2003A&A...403..993K}}; (7) this work (Sect.~\ref{sec:properties});
(8) \citet{2008ApJ...685L.141R}; (9) \citet{2000ApJ...545..945R}; (10) \mbox{\citet{1996Ap&SS.245..169O}}; 
(11) \mbox{\citet{2003A&A...399.1021G}}; (12) \citet{1993ApJS...87..267O}; (13) \mbox{\citealt{1996MNRAS.281.1347G,1998A&A...337..797G}}); 
(14) \mbox{\citet{2002A&A...390..511G}}. Updated distances (\mbox{\citealt{2007A&A...474..653V}}) have been taken to re-scale the (gas) mass-loss rates; 
(15) \mbox{\citet{2006A&A...454L.103R}}; (16) \mbox{\citet{2001A&A...368..969S}}; (17) \mbox{\citet{2002A&A...391.1053O}}; (18) \citet{2001ApJ...551.1073H};
(19) \citet{2006MNRAS.369..783M}; (20) \mbox{\citet{2003A&A...407..213V}}; (21) Linquist et al. (1999); (22) Olofsson et al. (2000); (23) Lombaert et al. (in prep);
(24) via Period-\Mdot\ relation; (25) \mbox{\citet{1993A&AS...99..291L}}; (26) Bergeat \& Chevallier (2005); (27) Choi et al. (2008);
(28) Matthews \& Reid (2007); (29) Winters et al. (2003).\\
Binarity: (1) visual binary, \mbox{\citet{1981A&AS...44..179P}}; (2) hot companion, \citet{2008ApJ...689.1274S}; (3) symbiotic Mira, \citet{1981IBVS.1961....1W,1997ApJ...482L..85H};
(4) suspected by \citet{1984BAICz..35...65V}, but could be RV Tau variable instead \citep{1998Ap.....41..367P}; (5) white dwarf companion; \citet{1988ApJ...327..214A};
(6) \citet{2005AJ....129.2420M}; (7) composite spectrum; \citet{1965VeBam..27..164H}; (8) composite CO line; \citet{1998ApJS..117..209K};
(9) visual binary, \citet{2002ApJS..139..249P}; proper-motion binary, \citet{2005AJ....129.2420M}; \mbox{\citet{2007A&A...464..377F}}; 
(10) composite spectrum, \citet{1953MNRAS.113..510F}; orbital motion, \mbox{\citet{1992A&A...253L..33S}}; 
proper-motion binary, \citet{2005AJ....129.2420M}; \mbox{\citet{2007A&A...464..377F}};
(11) eclipsing system, \mbox{\citet{1999A&A...351...97K}}; hot companion, \citet{2008ApJ...689.1274S}; visual binary, \mbox{\citet{1981A&AS...44..179P}};
(12) spiral arm indicative of binary, \mbox{\citet{2006A&A...452..257M}};
(13) spiral arm indicative of binary, \mbox{\citet{2010A&A...523A..59C}}; 
(14) The Hipparcos Double Star solution (position angle is 79\degr, and separation of 0.17\arcsec) gives additional evidence for a close companion,
and potential a triple system (if the other visual companion is at 208\arcsec\ (Alain Jorissen; private communication);
(15) \citet{2003A&A...399.1167P} found it to be a 'variability-induced mover' in the Hipparcos catalogue which is a strong indication for a close visual companion.
}
}
\end{longtable}
\end{landscape}
}

\longtabL{2}{
\scriptsize
\begin{landscape}
\begin{longtable}{llp{15mm}p{15mm}llllllllllllllll}
\caption{Key properties of AGB stars and supergiants with detached rings (Class\,III) or irregular extended (Class\,IV) emission at 70~$\mu$m and/or 160~$\mu$m. 
Notes and references as in Table~1.
$^*$ These objects show a combination of two types of interaction: an inner spherical ring together with an outer extended structure.
$^{**}$ Either typical Fermata bow shock or a spherical ring.}\label{tb:properties1b}
\endfirsthead
\caption{continued.}\\
\endhead
\endfoot
\\
\hline\hline
IRAS identifier		 & Name		& \multicolumn{1}{l}{Distance\tablefootmark{A}}    & \Mdot			   & $z$\tablefootmark{E}   & $n_H$\tablefootmark{D}  	& $\mu$\tablefootmark{H}      	& $v_\star$\tablefootmark{G}    & P.A.	 	&   $i$      & $v_w$\tablefootmark{F}    &\multicolumn{2}{c}{predicted $R_0$\tablefootmark{B}}	&&\multicolumn{3}{c}{observed $R_0$\tablefootmark{C}} 	& $n_H$\tablefootmark{I} & Type\tablefootmark{J} 	& Binary  \\ \cline{12-13}\cline{15-17}
	     		 &		&  (pc)	   					   & ($10^{-7} M_\odot$ yr$^{-1}$) & (pc)      		    & (cm$^{-3}$) 		&(mas yr$^{-1}$)  	& (\kms)      	    		&(\degr\ N-E)	&  (\degr)   & (\kms)       		 &(\arcmin)	& (pc)        				&&(\arcmin)& (pc)       & ($\theta$) 			&(cm$^{-3}$)	&      	& 	  \\ \hline 
\hline
{\bf III ``Rings''}\\ \cline{1-1}
 \object{00248+3518} & AQ And $^*$&  781$\pm$555$^6$,	  {\bf 825}    & 6.5$^7$			       & -360  &  0.05      &  13.4   & 54.1	& 218.0      &  17	  &		 & 1.5     & 0.37	 &   & 0.87    & 0.21	   & --        &   0.2      & C 	   &	      \\   
 \object{01159+7220} & S Cas	  &			    943        & 22$^{15}$			       & 176   &  0.34      &  19.4   & 91.0	& 190.1      & -15	  & 22.0$^{13}$  & 0.7     & 0.20	 &   & 0.83    & 0.23	   & --        &   0.25     & O - S	   &	      \\
 \object{01246-3248} & R Scl $^*$ &			    370$^5$    & (80$^{10}$), 16$^2$		       & -350  &  0.06      &  36.7   & 66.3	& 208.3      & -15	  & 17.0$^2$	 & 4.4     & 0.48	 &   & 0.23    & 0.025     & --        &   22.5     & C 	   &	      \\   
 \object{03374+6229} & U Cam      &			    430$^6$    & 10$^7$ 			       &  60   &  1.10      &  4.2    & 9.8	& 350.4      &  49	  & 20.6$^{21}$  & 5.3     & 0.66	 &   & 0.12    & 0.015     & --        &   30	    & C 	   & yes$^{1,14}$\\
 \object{05028+0106} & W Ori $^*$ &  377$\pm$135$^h$		       & 2.3$^7$			       &  -131 &  0.51      &  10.5   & 18.8	& 91.5       &  0	  & 8.6$^{16}$   & 1.4     & 0.15	 &   & 0.7     & 0.077     & --        &   2.1      & C 	   &	      \\  
 \object{10329-3918} & U Ant	  & {\bf 268$\pm$39}$^h$, 260$^6$      & (7.6$^7$),  40$^{10}$  	       &  90   &  0.82      &  19.3   & 36.5	& 285.8      &  41	  & 20.5$^2$	 & 5.3     & 0.41	 &   & 1.5     & 0.12	   & --        &   10.0     & C 	   &	      \\   
 \object{10350-1307} & U Hya $^*$ & {\bf 208$\pm$10}$^h$, 161$^2$      & (0.92$^1$, 2.0$^{14}$), 0.49	$^2$   & 143   &  0.48      &  66.8   & 71.9	& 118.5      & -26	  & 8.5$^2$	 & 0.3     & 0.02	 &   & 1.9     & 0.12	   & --        &   0.01     & C 	   &	      \\   
 \object{12427+4542} & Y CVn	  &  321$\pm$35$^h$		       & 1.5$^{14}$			       &  319  &  0.08      &  17.2   & 32.7	& 32.2       &  40	  & 7.5$^2$	 & 1.8     & 0.17	 &   & 3.2     & 0.30	   & --        &   0.03     & C 	   &	      \\
 \object{13462-2807} & W Hya	  &  104$\pm$12$^h$		       & 0.78$^2$			       &  72   &  0.98      &  43.9   & 48.2	& 207.6      &  58	  & 8.5$^2$, (6.5$^{17}$)&0.8& 0.03	 &   & 1.1/3.8 & 0.033/0.12& --        &   0.6      & O - M7e	   & yes$^1$  \\   
 \object{15094-6953} & X TrA	  &  360$\pm$57$^h$, {\bf 243}         & 1.3$^{16}$			       &  -29  &  1.49      &  14.2   & 16.4	& 68.2       & -7	  & 8.7$^{16}$   & 1.1     & 0.08	 &   & 2.5     & 0.18	   & --        &   0.3      & C 	   &	      \\
 \object{18476-0758} & S Sct	  &  386$\pm$85$^h$		       & 1.4$^x$			       &  -8   &  1.85      &  5.4    & 17.5	& 84.9       &  59	  & 8.0$^2$, (17.3$^{10}$)&0.6& 0.07	 &   &        & 	  &	       &	    & C 	   & yes$^1$  \\   
 \object{19390+3229} & TT Cyg	  &  {\bf 436}, 610$^6$ 	       & (1.1$^7$),  50$^{10}$  	       &  52   &  1.19      &  9.4    & 38.3	& 251.8      & -52	  & 12.6$^{22}$  & 2.2     & 0.28	 &   & 0.55    & 0.070     & --        &   19.7     & C 	   & yes$^1$  \\   
 \object{20141-2128} & RT Cap $^{**}$& 291$\pm$52$^h$		       & (1.0$^{16}$), 9.4$^7$  	       &  -121 &  0.60      &  5.2    & 21.6	& 184.5      & -69	  & 8.0$^{16}$   & 2.9     & 0.25	 &   &  1.5    & 0.13	   & --        &   2.2      & C 	   &	      \\ 
 \object{21358+7823} & S Cep	  &  407$\pm$107$^h$		       & 15$^{16}$)			       &  150  &  0.45      &  3.2    & 23.1	& 148.5      & -56	  & 22.0$^{16}$  & 5.4     & 0.64	 &   &  1.5    & 0.18	   & --        &   5.8	    & C 	   &	      \\ 
 \object{23438+0312} & TX Psc $^*$&  275$\pm$30$^h$		       & 0.91$^{12}$			       & -212  &  0.24      &  50.2   & 66.4	& 247.7      &  11	  & 10.5$^{12}$  & 0.3     & 0.03	 &   & 0.27    & 0.022     & --        &   1.0      & C - C5-7,2   &	      \\ 
\\  						         	   											         							  									   %
{\bf IV ``Irregular''}\\ \cline{1-1}
 \object{01556+4511} & AFGL 278   & 174$\pm$24$^h$		       &				       &  -33  &  1.44      &  27.1   & 22.3	& 56.7       &  5	  & 8.3$^{25}$   & 0.6     & 0.03	 &   &        & 	  &	       &	    & O - M7III    &	       \\
 \object{04497+1410} & $o$ Ori    & 200$\pm$28$^h$, {\bf 163}	       & $<$ 0.4$^{13}$ 		       &  -36  &  1.39      &  41.1   & 38.8	& 187.0      & -34	  &		 & 1.9     & 0.09	 &   &        & 	  &	       &	    & O - M3 S     & yes$^5$   \\
 \object{15492+4837} & ST Her	  & 293$\pm$51$^h$		       & 1.3$^{15}$			       &  238  &  0.19      &  27.2   & 38.4	& 169.8      & -7	  & (9.1$^{13}$), 8.5$^{15}$ & 1.1 & 0.09&   &        & 	  &	       &	    & O - M6-7IIIa S &         \\
 \object{20120-4433} & RZ Sgr	  & 730$^2$			       & (150$^{15}$), 5.8$^2$  	       & -384  &  0.04      &  16.2   & 63.2	& 111.6      & -28	  & 9.0$^2$, (11.4$^{13}$)&1.2& 0.26	 &   & 2.     & 0.43	  & --         &   0.02     & S - S4,4     &	       \\
 \object{20396+4757} & V Cyg	  & 420$^5$, 271$^2$, {\bf 366}        & (9.4$^1$), 4$^2$		       &  39   &  1.35      &  19.4   & 35.6	& 223.1      &  25	  & 15.0$^2$	 & 0.8     & 0.09	 &   &        & 	  &	       &	    & C - C5,3     &	       \\
 \object{22196-4612} & $\pi^1$ Gru& {\bf 163$\pm$20}$^h$, 152$^2$      & 8.5$^2$			       & -119  &  0.61      &	8.4   & 11.9	& 91.8       & -68	  & 30.0$^2$	 & 17.1    & 0.81	 &   &        & 	  &	       &	    & S - S5+G0V   & yes$^{10}$\\
 \object{23412-1533} & R Aqr	  & {\bf 250}$^{13}$, 197$^{28}$       & 0.6$^{(28)}$			       & -220  &  0.22      &  27.5   & 45.0	& 136.4      & -43	  & 16.7$^{14}$  & 0.5     & 0.05	 &   &        & 	  &	       &	    & O-M7IIIpev   & yes$^3$   \\
 \hline
\end{longtable}
\end{landscape}
}

\longtabL{3}{
\scriptsize
\begin{landscape}
\begin{longtable}{lllllllllllrrll}
\caption{AGB stars that show no obvious bow shocks are assigned to Class X pending further notice.
Notes and references as in Table~1. Note that Y\,Lyn, SW\,Vir, and AFGL\,5379 are classified X pending 
planned observations.}\label{tb:properties2}
\endfirsthead
\caption{continued.}\\
\endhead
\endfoot
\hline\hline
IRAS identifier&   Name		          & \multicolumn{1}{l}{distance\tablefootmark{A}}  & \Mdot      & $z^{23}$	& $n_H^D$ &  $\mu$         & $v_\star$ & P.A.	         & $i$     & $v_w^{13}$   & \multicolumn{2}{c}{predicted $R_0^B$}& Type\tablefootmark{I} & Binary\\  	    
     	       &  			  & (pc)	       	        & ($10^{-7} M_\odot$ yr$^{-1}$) &   (pc)	&(cm$^{-3}$)& (mas yr$^{-1}$)& (\kms)    & (\degr\ N-E)  & (\degr)& (\kms)     &(\arcmin)  & (pc)	 		    &	      		    & 	    \\ \hline			    
\hline
{\bf X ``Non-detection''}&   \\ \cline{1-1}	  			        												       
 \object{01037+1219}  &  WX Psc 		 & 740$^1$		       & 190$^2$, 77$^{24}$    &  -553 & 0.01	  &	      & $\geq$9.0  &		 &	     & 19.8$^2$        & $\leq$9.5	 &$\leq$2.0	  &  O-M9      &	   \\
 \object{01144+6658}  &  AFGL 190		 & 2790$^{14}$  	       & 638$^{14}$	       &   37  & 1.38	  &	      & $\geq$21.0 &		 &	     & 18.1$^{13}$     &		 &		  &  C         &	   \\	
 \object{01304+6211}  &  OH 127.8+0.0		 & 2100$^{23}$  	       & 500$^{23}$	       &   14  & 1.73	  &	      & $\geq$55.3 &		 &	     & 12.5$^{23}$     & $\leq$1.1	 &$\leq$0.7	  &  O         &	   \\	
 \object{02270-2619}  &  R For  		 & 690$^{13}$		       & 9.0$^{24}$	       &  -625 & 0.00	  & 6.9       & 22.6	   &   120.8	 &    -5     & 16.1$^{16}$     & $\leq$0.8	 &$\leq$0.2	  &  C         &	   \\	
 \object{03112-5730}  &  TW Hor 		 & 322$\pm$38$^h$	       &1.5$^{24}$, 0.24$^{14}$&  265  & 0.14	  & 34.3      & 52.4	   &   58.8	 &     1     & 7.5$^{15}$      &       0.5	 &	0.05	  &  C         & yes$^2$   \\	
 \object{04020-1551}  &  V Eri  		 & 439$\pm$133$^h$	       &		       &  -290 & 0.11	  & 16.7      & 39.8	   &   205.6	 &   -31     & 13$^{25}$       &       0.9	 &	0.12	  &  O M5/M6IV & yes$^2$   \\
 \object{04361-6210}  &  R Dor  		 & 55$\pm$3$^h$ 	       & 6.1$^{24}$, 1.3$^{17}$&  -20  & 1.64	  & 146.4     & 39.4	   &   217.1	 &    10     & 6.0$^{17}$      &       4.8	 &	0.08	  &  O M8IIIe  & yes$^1$   \\
 \object{04566+5606}  &  TX Cam 		 & 380$^2$		       & 65$^2$,34$^{24}$      &  72   & 0.98	  &	      & $\geq$10.8 &		 &	     & 21.2$^2$        & $\leq$10.7	 &$\leq$1.2	  &  O-M8.5    & yes$^{13}$\\
 \object{04573-1452}  &  R Lep  		 & 413$\pm$174$^h$	       & 13$^{24}$	       &  -200 & 0.27	  & 6.5       & 18.5	   &   116.5	 &    42     & 17.4$^6$        &       2.9	 &	0.35	  &  C         &	   \\	
 \object{07209-2540}  &  VY CMa 		 & 1140$^{27}$, {\bf 562}      & 2800$^2$	       &   -35 & 1.41	  & 12.4      & 42.9	   &   86.9	 &    36     & 46.5$^2$        &       31.0	 &	5.1	  &  O-M3/M4III& yes$^1$   \\	
 \object{09076+3110}  &  RS Cnc\tablefootmark{a} &{\bf 143$\pm$11}$^h$, 122$^{28}$  &27-36/ 0.55-0.74$^{13}$& 111 & 0.66  & 18.3      & 14.1	   &   154.4	 &    30     & 8.0$^6$         &       2.7	 &	0.11	  &  O-M6IIIaeS&	   \\  
 \object{10131+3049}  &  RW LMi 		 & 320$^{14}$, {\bf 440}$^1$   & 19.2, 59$^2$	       &  384  & 0.04	  &	      & $\geq$1.6  &		 &	     & 20.8$^2$        & $\leq$71.8	 &$\leq$9.2	  &  C         &	   \\	
 \object{10491-2059}  &  V Hya  		 & 600$^{14}$		       &2.8$^{24}$,83.1,610$^2$&  399  & 0.04	  & 7.2       & 27.9	   &   302.3	 &   -38     & 30$^2$	       &       7.8	 &	1.6	  &  C         & yes$^{11}$\\	
 \object{11331-1418}  &  HD100764		 & 290$\pm$83$^h$	       &		       &  218  & 0.23	  & 17.6      & 24.3	   &   126.6	 &     1     &  	       &       2.2	 &	0.2	  &  C         &	   \\	
 \object{12544+6615}  &  RY Dra 		 & 431$\pm$110$^h$	       & 2.1$^{24}$	       &  351  & 0.06	  & 23.6      & 49.2	   &   116.8	 &    -9     & 10.0$^{15}$     &       0.4	 &	0.05	  &  C         &	   \\	
 \object{14219+2555}  &  RX Boo 		 & 190$\pm$23$^h$	       & 3.6$^2$,6.2$^{24}$    &  193  & 0.29	  & 58.7      & 52.9	   &   142.9	 &     2     & 9.0$^2$         &       1.3	 &	0.07	  &  O-M7.5    &	   \\
 \object{15194-5115}  &  II Lup 		 & 590$^{14}$, {\bf 500}$^2$   & 181$^{24}$, 39$^2$    &  56   & 1.15	  &	      & $\geq$15.0 &		 &	     & 23$^2$	       & $\leq$6.5	 &$\leq$0.9	  &  C         &	   \\	
 \object{15477+3943}  &  V CrB  		 & 630$^{13}$		       & 7.1$^{24}$	       &  506  & 0.01	  & 15.5      & 114.2	   &   151.9	 &   -64     & 8.1$^{13}$      &       0.02	 &	0.02	  &  C         &	   \\	
 \object{18050-2213}  &  VX Sgr 		 & 262$\pm$187$^h$	       & 610$^2$, 140$^{24}$   &  10   & 1.80	  & 4.1       & 8.5	   &   28.1	 &    50     & 24.3$^2$        &       31.6	 &	2.4	  &  O-M4eIa   & yes$^1$   \\
 \object{18240+2326}  &  AFGL 2155		 & 920$^{14}$		       & 45.6$^{14}$	       &  265  & 0.14	  &	      & $\geq$60.0 &		 &	     & 16.1$^{25}$     &		 &		  &  C         &	   \\	
 \object{18306+3657}  &  T Lyr  		 & 719$\pm$254$^h$,{\bf 650}$^{26}$ & 9.4$^{14}$, 0.7$^{15}$&232 & 0.20   & 8.5       & 26.8	   &   225.8	 &    17     & 11.5$^{15}$     &       0.24	 &	0.05	  &  C         &	   \\	
 \object{18348-0526}  &  OH 26.5+0.6		 & 1370$^2$		       & 97$^2$, 350$^{24}$    &  30   & 1.49	  &	      & $\geq$27.4 &		 &	     & 17.0$^2$        & $\leq$1.8	 &$\leq$0.7	  &  O         &	   \\	
 \object{20077-0625}  &  IRC-10 529		 & 620$^2$		       & 45$^2$, 90$^{24}$     &  -201 & 0.27	  &	      & $\geq$18.0 &		 &	     & 16.5$^2$        & $\leq$5.6	 &$\leq$1.0	  &  O         &	   \\	
 \object{21197-6956}  &  Y Pav  		 & 403 $\pm$96$^h$	       & 1.6$^{15}$	       & -234  & 0.19	  &   2.9     & 5.8	    & 124.0	 &   -44     & 8.0$^{15}$      &       7.3	 & 0.86 	  &  C         &	   \\	
 \object{23166+1655}  &  LL Peg 		 & 1000$^{14}$, {\bf 980}$^2$  & 144$^{14}$, 310$^2$   & -620  & 0.00	  &	      & $\geq$31.0 &		 &	     & 16.0$^2$        & $\leq$3.8	 &$\leq$1.1	  &  C         & yes$^{12}$\\	
 \object{23320+4316}  &  LP And 		 & 780$^{14}$, {\bf 630}$^2$   & 46$^2$, 150$^{15}$    & -171  & 0.36	  &	      & $\geq$17.0 &		 &	     & 14.0$^2$        & $\leq$3.9	 &$\leq$0.7	  &  C         &	   \\	
     	              & \object{NML Cyg}	 & 1220$^2$		       & 870$^2$, 350$^{24}$   &  -26  & 1.54     &	       & $\geq$2.0  &		  &           & 33.0$^2$	& $\leq$71.5	  &$\leq$25.4	   &  O-M6IIIe  &	    \\   
\hline \hline																		           
 \object{07245+4605}  &  Y Lyn  		 & 253$\pm$61$^h$	       &5.6$^{13}$/0.36$^{13}$ &  123  & 0.58	  & 11.5      & 13.9	   &   32.6	 &    -2     & 5.4$^{13}$      &       4.1	 &	0.3	  &  O-M6SIb-II& yes$^{15}$\\
 \object{13114-0232}  &  SW Vir 		 & 143$\pm$17$^h$	       & 1.4$^x$, 4$^t$        &  138  & 0.50	  & 15.8      & 14.7	   &   301.5	 &   -49     & 7.8$^a$, 7.5$^t$&       1.4	 &	0.06	  & O-M7III    &	   \\
 \object{17411-3154}  &  AFGL 5379		 & 1190$^{2,h}$ 	       & 280$^2$, 350$^x$      &  -13  & 1.76	  &	      & $\geq$22.7 &		 &	     & 25.0$^2$        & $\leq$5.4	 &$\leq$1.9	  & O	       &	   \\ 
\hline
\multicolumn{15}{p{18cm}}{\tablefoot{\tablefoottext{a}{The central source is possible extended. However, no detached emission is detected.}}}\\
\end{longtable}
\end{landscape}
}

\section{Observational diagnostics of wind-ISM interaction regions}\label{sec:morphology}

\subsection{Morphological classification of the detached far-infrared emission}\label{subsec:classification}

From the observations (Figs.~\ref{fig:firstmap} to~\ref{fig:lastmap}) one can immediately recognise different overall shapes
of the detected (detached) extended emission. The two most obvious cases are {\it arcs or ``fermata''} and {\it ``rings''}.
Two additional distinct morphologies are the double opposing arcs or {\it ``eye''} and the {\it ``irregular''} emission. The
morphological classification is summarised in Table~\ref{tb:classes} and indicated for all sources in
Tables~\ref{tb:properties1} and~\ref{tb:properties1b}.

\emph{``Fermata''} (Class\,I) interaction is characterised by a relatively smooth elliptical arc spanning an azimuthal opening
angle $\geq$120\degr. This class resembles closest the wind-ISM bow shock shape predicted by theoretical models
(Sect.~\ref{sec:bowshock}). We note that there are different cases within this class reminiscent of the different shapes that
occur due to variations in the stellar and interstellar parameters. This will be discussed in detail in Sect.~\ref{sec:hydro}.
For example, X\,Pav, EP\,Aqr and X\,Her show a peculiar bullet-like shape with back flow emission in the wake of the bow shock.
Others show clear signatures of Kelvin-Helmholtz (KH) and/or Rayleigh-Taylor (RT) instabilities (see Sect.~\ref{sec:hydro}). 
Class\,II (\emph{``eyes''}) includes objects with two elliptical non-concentric arcs observed at opposing sides of the central
source, both have a covering angle of $\leq 180$\degr. In two cases, VY\,UMa and U\,Cam, the two arcs are connected 
and there is even tentative
evidence for a jet structure in the mid plane. Class\,III consists of spherical structures, \ie\ circular \emph{``rings''}.
This category includes typical detached spherical-shell objects, but could also include true rings (\ie\ not spheres). From
the seven, well-studied CO-detected detached shells around carbon stars (\eg\ U\,Ant, U\,Cam, TT\,Cyg, R\,Scl, S\,Sct,
V644\,Sco, DR\,Ser; \citealt{2010A&A...511A..37M}), only the latter two are not included in the MESS survey. All known and observed
detached CO shell objects show a spatially resolved dust ring, co-spatial with the gas emission. Larger, and thus presumably
older, rings are also detected around Y\,CVn and AQ\,And. For these objects no corresponding CO shell has been detected,
possibly due to the photo-dissociation of CO by the interstellar radiation field (\citealt{2007MNRAS.380.1161L};
\citealt{2010A&A...518L.140K}). Sources with diffuse irregular extended emission are classified \emph{``irregular''}
(Class\,IV). All other targets, for which no evidence of diffuse extended emission has been observed, are assigned to Class\,X,
\emph{``non-detection''}.

Interestingly, for several objects, such as R\,Scl, U\,Cam, TX\,Psc we detect small detached rings in addition to the
classical bow shock region further away from the central source. These objects are subsequently assigned to both Class\,I and
Class\,III (Tables~\ref{tb:properties1} and~\ref{tb:properties1b}). For these objects it may be the case that we observe a
young spherical shell (originating from a wind-wind interaction  due to a recent thermal pulse) expanding in a relatively low
density medium. In this scenario the local environment of the star has been blown out by the earlier wind/mass-loss which
swept out ISM material and created a wind-ISM bow shock. At a later stage a thermal pulse produced a density / temperature
enhancement traced by the infrared emission. Alternatively, the small inner ring represents a structure delineating the
interface between the (back flowing) termination shock and the free expansion zone. For other objects, such as CW\,Leo and
$\alpha$\,Ori, the Herschel/PACS observations reveal irregular multiple incomplete shells in the inner regions of the stellar
wind envelope (\eg\ \citealt{2011A&A...534A...1D}).

\begin{table}[t!]
\begin{center}
\caption{Morphological classification.}  
\begin{tabular}{llc}\hline\hline
Class		   & Description		       & Shape  				      \\ \hline
I    	           & Fermata\tablefootmark{a}	       & \includegraphics[height=9pt]{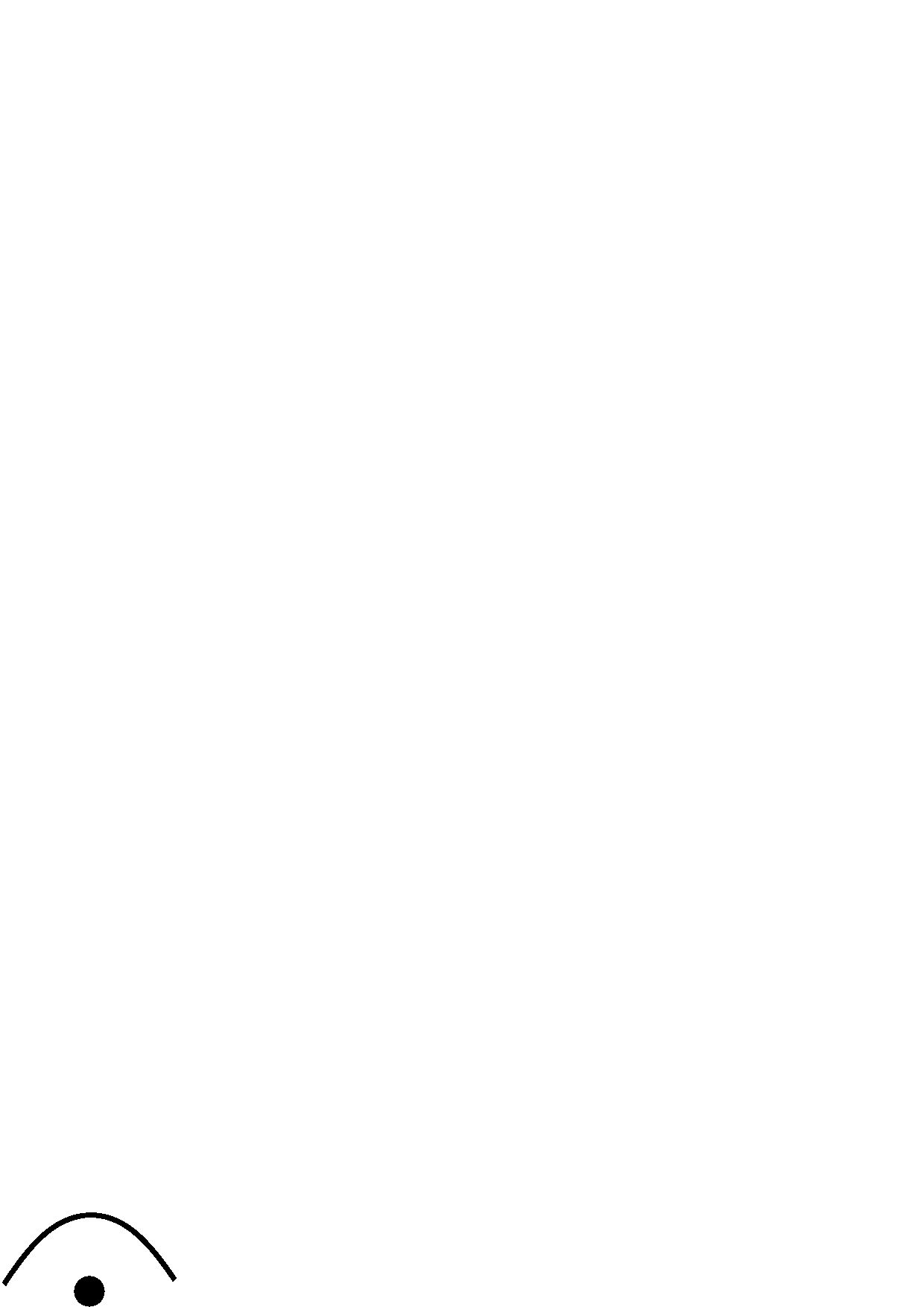} \\
II		   & Eye - double non-concentric arcs  & \includegraphics[height=9pt]{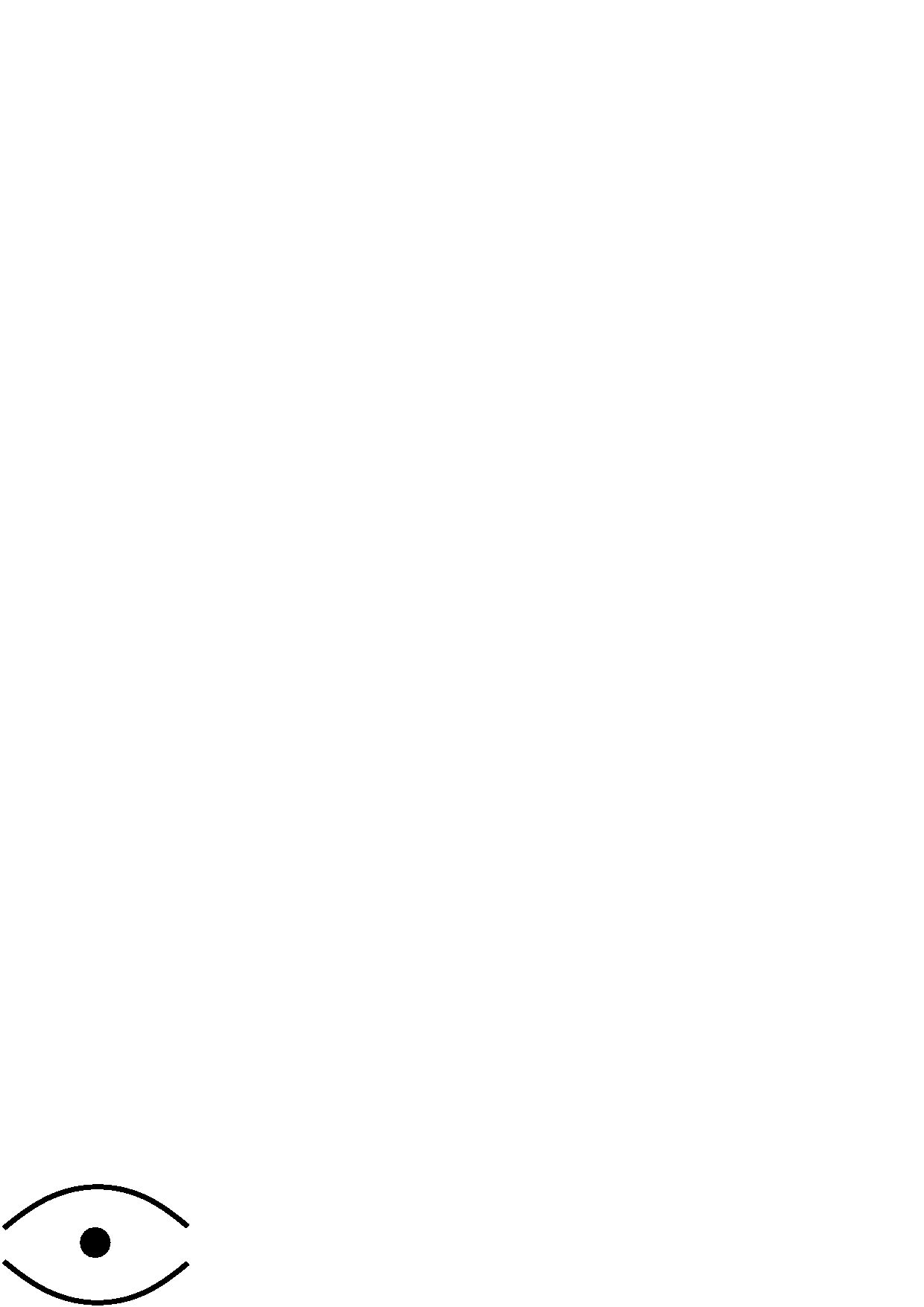} \\	  
III		   & (circular) Ring 		       & \includegraphics[height=9pt]{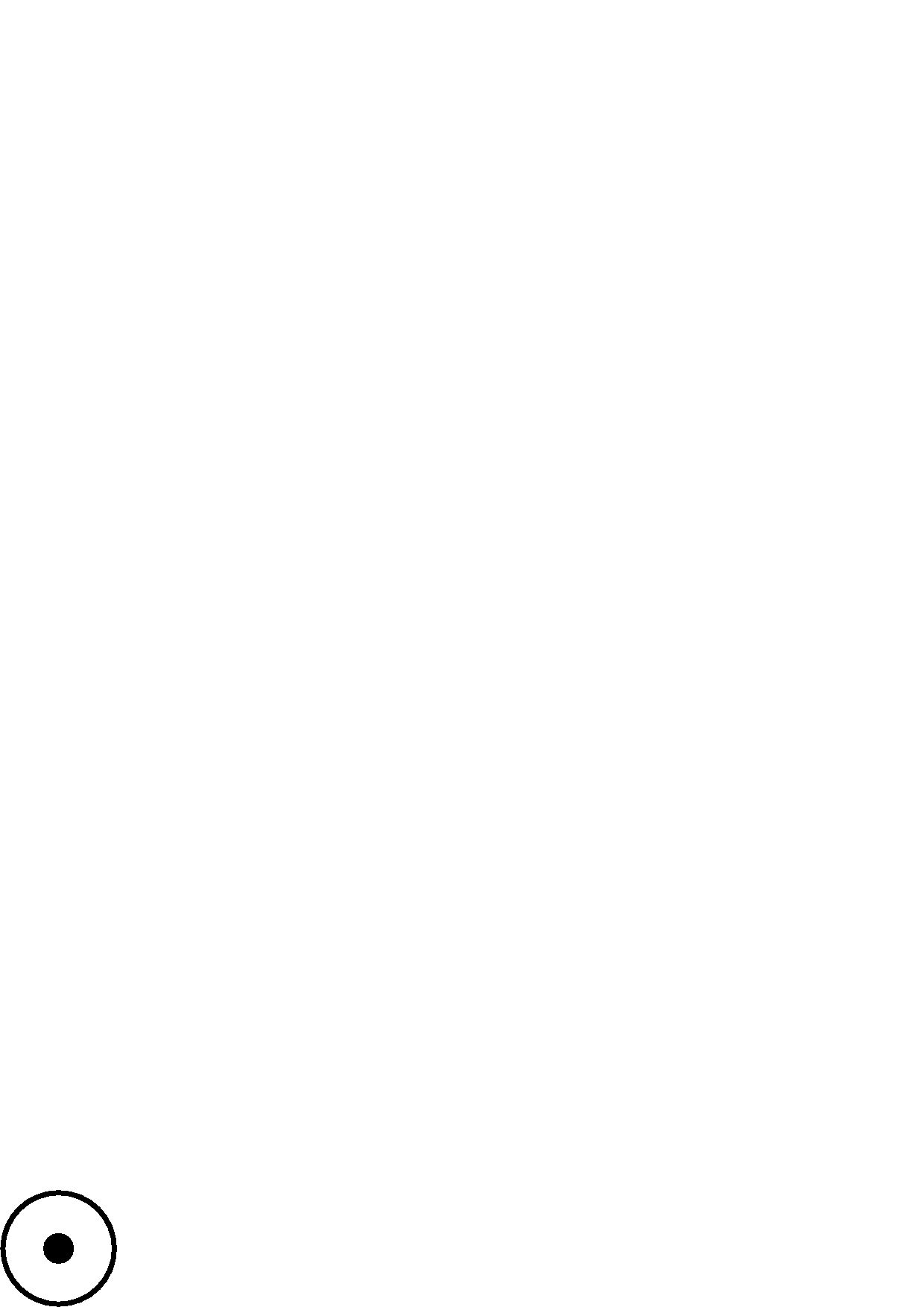} \\
IV		   & Irregular (diffuse)	       &					      \\
\hline
\end{tabular}
\label{tb:classes}
\tablefoot{\tablefoottext{a}{A ``fermata'' is a musical sign.}}
\end{center}
\end{table}

\begin{figure}[h]
\centering
\includegraphics[width=0.9\columnwidth]{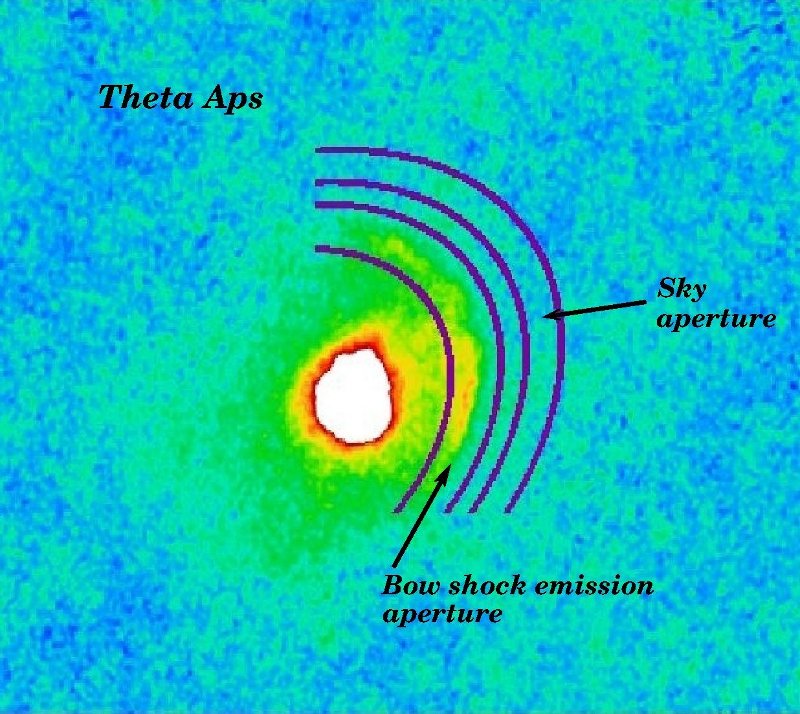}
\caption{Illustration of the extended flux measurements of the observed bow shock of $\theta$\,Aps. 
Both the elliptical bow shock aperture and the sky aperture are indicated.}
\label{fig:apertureflux}
\end{figure}

\subsection{Radial distance of arcs and detached shells}\label{subsec:radialprofile}

For some objects the extended emission is very faint with a low contrast with respect to the sky background. In order to
improve the measurement of the angular distance/size of arcs and shells we have constructed radial profiles for Class\,II and
Class\,III objects. Table~\ref{tb:extendedflux} gives the radii of the \emph{``eyes''} (II) and \emph{``rings''} (III) as
identified in the azimuthally averaged radial profiles shown next to the image panels in Figs.~\ref{fig:eyes}
and~\ref{fig:rings} (adopted azimuth angles are given at the top of each radial profile panel; for the \emph{``rings''} objects
it is alway 360\degr. 
For the \emph{``eyes''} the radii and radial profiles of both arcs are given. For a few cases with very
faint extended emission, the assignment is ambiguous. V\,Pav possibly belongs to Class\,III instead of Class\,II as the dust
emission faintly traces a full circle. RT\,Cap is tentatively assigned to Class\,III but its ring is not complete and rather
consists of two distinct arcs. However, the arcs are concentric, which is not the case for true \emph{``eye''} objects. For
S\,Cep the ring is also not complete but resembles more a full circle with gaps. The detached shell around Y\,CVn is very
faint and barely visible at a radius of $\sim$190\arcsec, but in this case the detection is supported by the radial profile
and its earlier detection with ISO/ISOPHOT by \citet{1996A&A...315L.221I}. Tentatively, a faint arc bow shock structure
(\emph{``fermata''}) extending over $\sim$100\degr\  (from position angle $\sim$30\degr\ to 130\degr) is present at 9\arcmin\
east of the central star. X\,TrA is also listed in Class\,III although also here the ring is faint, with only a brighter arc
to the east. Further observations are necessary to resolve these ambiguities.

\begin{table*}[th!]
\centering
\caption{
Aperture flux of observed bow shocks (Class\,I and~II) and detached rings (Class\,III). 
Irregular Class\,IV is excluded. An elliptical annulus was applied in order to measure flux in the non-spherical arcs. 
The annuli of the detached rings were nearly spherical.
The shape of the elliptical annulus was identified from the image with the most significant detection and then 
directly applied to the other band. 
For the fermata the flux is only integrated over a limited azimuthal angle covering the observed emission.}
\label{tb:extendedflux}
\resizebox{\textwidth}{!}{
\begin{tabular}{llllllrrllll}\hline\hline
IRAS id		&  Object	& Class  & \multicolumn{2}{c}{Radius\tablefootmark{a}} & Dust annulus	   & \multicolumn{2}{c}{Flux (Jy)}    & \multicolumn{2}{c}{$M_{\rm dust+gas}$\tablefootmark{b} ($10^{-4}$ M$_\odot$)} & $M_{\rm ISM}$\tablefootmark{c}  & $M_{\rm ISM}$\tablefootmark{d} \\
		&		&        & (\arcsec)		  & (pc)	       & (\arcsec)	   & 70~$\mu$m        & 160~$\mu$m    & \@70~$\mu$m &\@160~$\mu$m& ($10^{-4}$ M$_\odot$) & ($10^{-4}$ M$_\odot$)\\ \hline
00248+3518     	&  AQ And  	& I+III  & 52			  & 0.21	       & 40-62 (circle)    & 2.20$\pm$0.01    & 1.28$\pm$0.01 & 0.9	    & 1.1	& 1.1	      & 4.4 	\\	  
01159+7220	&  S Cas	& III    & 50\tablefootmark{e}    & 0.23	       & 32-64  	   & 2.26$\pm$0.01    & 1.11$\pm$0.02 & 1.1	    & 1.1	& 12  	      & 54  	\\
01246-3248  	&  R Scl	& I      & 54\tablefootmark{j}    & 0.10	       & 51-68 (arc)	   & 1.27$\pm$0.01    & 0.49$\pm$0.01 & 0.2	    & 0.2 	& 0.1         & 0.3 	\\
		&  		& III    & 14\tablefootmark{f}    & 0.03 	       &		   &		      & 	      & 	    &		&	      & 	\\
02168-0312	&  $o$ Cet	& I      & 82\tablefootmark{j}    & 0.04	       & 70-150 (arc)	   &47.64$\pm$0.03    & 9.75$\pm$0.05 & 2.2	    & 1.0	& 0.5 	      & 0.2 	\\
03374+6229 	&  U Cam	& III    & 7\tablefootmark{f}	  & 0.02 	       &		   &		      & 	      & 	    &		&	      & 	\\
		&		& II     & 57/62\tablefootmark{i} & 0.12/0.13	       & 80-140 	   & 2.87$\pm$0.02    & 2.44$\pm$0.05 & 0.6	    & 1.1 	& 40  	      & 1110	\\
03507+1115	&  NML Tau	& I      & 85			  & 0.10	       & 95-130 (arc)	   & 5.23$\pm$0.02    & 3.04$\pm$0.03 & 0.6	    & 0.8	& 3.5         & 62 	\\
04459+6804 	&  ST Cam	& II     & 67/84\tablefootmark{i} & 0.14/0.17	       & 84-122 	   & 1.28$\pm$0.02    & 0.61$\pm$0.03 & 0.3	    & 0.3	& 3.1	      & 2.2	\\
05028+0106 	&  W Ori	& I/III  & 92			  & 0.17	       & 70-120 	   & 1.99$\pm$0.02    & 0.62$\pm$0.03 & 0.4	    & 0.3	& 8.2	      & 32  	\\
05418-4628 	&  W Pic	& I      & 34\tablefootmark{j}    & 0.08	       & 62-90  	   & 0.92$\pm$0.01    & 0.25$\pm$0.02 & 0.2	    & 0.1	& 2.7	      & 13 	\\
05524+0723	&  $\alpha$ Ori & I      & 397\tablefootmark{j}   & 0.25	       & 510-660 (arc)	   &56.68$\pm$0.19    &22.64$\pm$0.38 & 3.8	    & 3.2	& 201 	      & 490  	\\  
06331+3829	&  UU Aur	& I      & 82\tablefootmark{j}    & 0.14	       & 100-140 (arc)     & 5.44$\pm$0.02    & 2.50$\pm$0.03 & 0.9	    & 0.9	& 14 	      & 88  	\\
09448+1139 	&  R Leo	& I      & 93\tablefootmark{j}    & 0.03	       & 94-134 	   & 9.44$\pm$0.02    & 2.91$\pm$0.03 & 0.2 	    & 0.1	& 0.1         & 2.1 	\\
09452+1330	&  CW Leo\tablefootmark{k}& I & 507\tablefootmark{j}  &	 0.29	       & 560-710 (arc)	   & 6.88$\pm$0.08    &10.13$\pm$0.11 & 0.4	    & 1.4	& 76  	      & 31  	\\  
		&		& I	 &			  & 		       &          	   &		      & 6.00$\pm$0.13 & 	    & 0.8	&	      & 	\\  
10329-3918      &  U Ant	& III    & 42			  & 0.06 	       & 30-55 (circle)    &16.32$\pm$0.01    & 4.68$\pm$0.01 & 2.2	    & 1.3	& 0.4         & 5.3 	\\				      
10350-1307      &  U Hya	& I+III  & 114  		  & 0.12	       & 100-133 (circle)  &17.44$\pm$0.03    & 9.33$\pm$0.03 & 1.8	    & 2.1	& 1.7         & 0.03	\\
10416+6740 	&  VY UMa	& II     & 38/46\tablefootmark{i} & 0.07/0.09	       & 38-88  	   & 3.60$\pm$0.01    & 2.33$\pm$0.02 & 0.7	    & 1.0	& 0.7         & 2.5 	\\
10580-1803 	&  R Crt	& II     & $\sim$140		  & 0.18	       & 165-270	   & 5.07$\pm$0.07    & 6.42$\pm$0.11 & 0.7	    & 1.8	& 21 	      & 58  	\\
12427+4542 	&  Y CVn	& III    & $\sim$190		  & 0.30	       & 150-260	   & 5.14$\pm$0.07    & 3.79$\pm$0.10 & 0.8	    & 1.3	& 7.7	      & 2.9	\\
13001+0527	&  RT Vir	& I      &			  &		       & 50-140 (circle)   & 5.15$\pm$0.03    & 3.30$\pm$0.04 & 0.4	    & 0.5	& 0.6	      &     	\\
13269-2301	&  R Hya	& I      & 96\tablefootmark{j}    & 0.05	       & 200-245 (arc)     & 4.65$\pm$0.02    & 1.93$\pm$0.03 & 0.3	    & 0.3	& 3.8         & 12 	\\
13462-2807      &  W Hya	& III    & 68,230\tablefootmark{g}& 0.03, 0.12         & 70-108 (ellipse)  &21.28$\pm$0.02    & 6.08$\pm$0.03 & 1.1 	    & 0.7	& 0.2         & 0.1 	\\
14003-7633	&  $\theta$ Aps	& I      & 76\tablefootmark{j}    & 0.04	       & 118-146 (arc)     & 2.15$\pm$0.01    & 0.76$\pm$0.02 & 0.1 	    & 0.1	& 1.3         & 0.4 	\\
15094-6953 	&  X Tra	& III    & 150\tablefootmark{h}   & 0.18	       & 60-210 	   & 9.70$\pm$0.08    & 6.89$\pm$0.12 & 1.8	    & 2.7	& 106	      & 21  	\\
16011+4722	&  X Her	& I      & 45\tablefootmark{j}    & 0.03	       & 40-90 (ellipse)   & 9.17$\pm$0.01    & 3.23$\pm$0.02 & 0.6	    & 0.5 	& 0.2 	      & 0.2 	\\
17389-5742 	&  V Pav	& II     & 97/100\tablefootmark{i}& 0.17/0.18	       & 95-140 	   & 1.30$\pm$0.03    & 1.50$\pm$0.04 & 0.2	    & 0.6	& 21 	      & 12 	\\
18476-0758	&  S Sct	& III    & 70    		  & 0.13	       & 30-90 (circle)	   &14.09$\pm$0.02    & 8.85$\pm$0.03 & 2.7	    & 3.7 	& 13 	      &    	\\  
19126+3247 	&  W Aql	& II     & 45/75\tablefootmark{i} & 0.07/0.12	       & 36-86 (ellipse)   &11.10$\pm$0.02    & 4.25$\pm$0.03 & 1.9	    & 1.6	& 5.8	      & 144	\\
19233+7627 	&  UX Dra	& II     & 76/54\tablefootmark{i} & 0.14/0.10	       & 50-110 	   & 2.89$\pm$0.01    & 1.11$\pm$0.03 & 0.6	    & 0.5	& 4.4	      & 2.5	\\
19314-1629 	&  AQ Sgr	& I      & 57			  & 0.09 	       & 50-100 	   & 3.44$\pm$0.02    & 2.53$\pm$0.02 & 0.6	    & 0.9	& 5.4         & 3.7 	\\
19390+3229      &  TT Cyg  	& III    & 33			  & 0.07 	       & 26-43 (circle)    & 1.74$\pm$0.01    & 0.86$\pm$0.01 & 0.4	    & 0.4	& 1.3         & 21 	\\
20038-2722	&  V1943 Sgr	& I      & 66\tablefootmark{j}    & 0.06	       & 50-130 (arc)	   & 5.98$\pm$0.03    & 2.30$\pm$0.03 & 0.6	    & 0.5	& 2.6	      & 0.8 	\\
20075-6005 	&  X Pav	& I      & 50\tablefootmark{j}    & 0.07 	       & 94-122 	   & 5.74$\pm$0.01    & 2.69$\pm$0.02 & 0.6	    & 0.5	& 3.1         & 19 	\\
20141-2128 	&  RT Cap	& III    & 92			  & 0.13	       & 62-118 	   & 2.21$\pm$0.02    & 0.91$\pm$0.04 & 0.3	    & 0.3	& 4.0         & 15 	\\
20248-2825	&  T Mic	& I      &			  &		       & 40-100 (ellipse)  & 5.48$\pm$0.02    & 0.58$\pm$0.03 & 0.6	    & 0.1	& 1.2	      &    	\\
21358+7823 	&  S Cep	& III    & 90			  & 0.18	       & 70-130 	   & 2.14$\pm$0.03    & 	      & 0.4	    &		& 11 	      & 142	\\
21419+5832 	&  $\mu$ Cep	& I      & 78\tablefootmark{j}    & 0.15	       & 100-150	   &28.64$\pm$0.04    & 11.5$\pm$0.12 & 5.6	    & 4.8	& 56 	      & 637	\\
21439-0226 	&  EP Aqr	& I      & 43\tablefootmark{j}    & 0.03	       & 45-68  	   & 4.65$\pm$0.01    & 1.36$\pm$0.01 & 0.2	    & 0.1	& 0.1         & 1.2 	\\
23438+0312	&  TX Psc	& III    & 16\tablefootmark{f}    & 0.02 	       &		   &		      & 	      & 	    &		&	      & 	\\  
		&  		& I      & 38\tablefootmark{j}    & 0.05	       & 12-58 (ellipse)   & 4.79$\pm$0.01    & 1.07$\pm$0.02 & 0.7	    & 0.3	& 0.2 	      & 0.7 	\\  
23558+5106	&  R Cas	& I      & 97\tablefootmark{j}    & 0.06	       & 100-160 (arc)	   & 8.35$\pm$0.01    & 3.52$\pm$0.04 & 0.5	    & 0.5	& 2.5 	      & 2.9 	\\  
\hline						 												     
\end{tabular}
}
\tablefoot{   
\tablefoottext{a}{Radii of rings and fermata derived from the azimuthally averaged radial profiles (Fig.~1)}
\tablefoottext{b}{Derived from Eq.~\ref{eq:dust-mass} using the total integrated flux at 70 and 160~$\mu$m, respectively, adopting
		  a gas-to-dust ratio of 200 and a dust temperature of 30~K.}
\tablefoottext{c}{$M_{\rm ISM}$ = $\frac{4}{3} \pi r^3\ \rho_{\rm ISM}$, with $n_H$ taken from Eq.~\ref{eq:nH2}.}
\tablefoottext{d}{$M_{\rm ISM}$, with $n_H$ taken from the local densities inferred from the measured stand-off distance and Eq.~\ref{eq:standoff}; 
		  Tables~\ref{tb:properties1}~and~\ref{tb:properties1b}.}
\tablefoottext{e}{Central source is offset by 0\arcsec\ Ra, 5\arcsec\ Dec.}
\tablefoottext{f}{Ring radius from the deconvolved image.}
\tablefoottext{g}{Inner and outer (at 160~$\mu$m) rings, respectively.}
\tablefoottext{h}{Central source is offset by 12\arcsec\ Ra, 5\arcsec\ Dec.}
\tablefoottext{i}{Radial distances are quoted for both north and south arcs (east-west for W\,Aql).}
\tablefoottext{j}{projected distance $A$ (Fig.~\ref{fig:wilkinoids}).}
\tablefoottext{k}{CW\,Leo has been observed two times at 160~$\mu$m.}
}
\end{table*}


\subsection{Far-infrared dust emission}\label{subsec:flux}

Cold dust grains emit strongly at mid- to far-infrared wavelengths. Dust emission can, in a simple approximation, be described as a  blackbody modified
by the grain emissivity; $F_\nu \propto B_\nu \cdot Q_\nu$, with $Q_\nu \approx \nu^\beta$, where $\beta$  depends on the type of dust considered. For
example, $\beta = 1.1$ for amorphous carbon in the ISM (\citealt{1991ApJ...377..526R}) as well as for carbon grains in carbon stars
(\citealt{1986ApJ...303..327J}), while astronomical silicates have $\beta = 2$ (\citealt{1988ApJ...331..435V}). 

Only for fast shocks ($v_{\rm shock} > 50$~\kms) high shock temperatures are reached that give rise to UV radiation (\eg\ observed for the bow shock of
CW\,Leo) and possibly a strong [\ion{O}{i}] 63~$\mu$m line. But, as the grains are only weakly coupled to the gas, due to the low densities,
$T_{\rm dust} \ll T_{\rm gas}$. Low velocity shocks on the other hand give primarily rise to dust emission in the infrared. Despite observational
efforts with Spitzer and Herschel, no infrared line emission has been detected in AGB star bow shocks yet (\citealt{Ueta2011_AGB}; Decin et al. 2011),
and we will assume here that the observed emission is entirely due to thermal dust emission (see also Sect.~\ref{sec:bowshock}).

In Table~\ref{tb:extendedflux} we give the aperture flux measured in both PACS bands for the elliptical bow shocks and circular rings. To determine the
appropriate aperture we fit the extended dust emission with an ellipse. The inner and outer annulus of the aperture are subsequently derived from the
azimuthally averaged radial profile. For detached rings the fitted ellipse is circular and the aperture is taken over the entire annulus. Bow shocks
have a nearly elliptical shape, and their aperture flux is determined over a limited azimuthal angular range of the annulus.  An illustrative example
of a bow shock aperture measurement is shown for $\theta$\,Aps in Fig.~\ref{fig:apertureflux}.
Table~\ref{tb:extendedflux} shows that the 70 over 160~$\mu$m flux density ($F_\nu$) ratio varies between 1.2 and 3.5, 
corresponding to a dust temperature of $\sim$30 to 40~K (for $\beta = 2$).


The observed infrared flux -- for optically thin emission -- gives a direct measure of the emitting dust mass, and thus -- given a 
gas-to-dust ratio -- the total mass. Assuming specific dust properties one can write (\citealt{2005AIPC..761..123L}):
\begin{equation}
M_{\rm dust} = \frac{d^2 F_\lambda}{\kappa(\lambda) B_\lambda(T)} = \frac{d^2 F_\nu}{\kappa(\nu) B_\nu(T)} \label{eq:dust-mass}
\end{equation}

with $F_\nu$ the observed flux (in erg s$^{-1}$ cm$^{-2}$ Hz$^{-1}$), $\lambda$ the wavelength (in cm), $d$ the distance (in
cm), $T_\mathrm{dust}$ the dust temperature (in K), $B_\nu$ the Planck black body curve in frequency units, and
and$\kappa_\nu$ the dust opacity at the observed wavelengths. 
We adopt a dust opacity of 60 and 10~cm$^2$ g$^{-1}$ at 70 and 160~$\mu$m, respectively (from \citealt{2001ApJ...554..778L} who 
give $\kappa(\lambda) = 2.92 10^5 (\lambda/\mu\mathrm{m})^{-2}$, \ie\ $\beta = 2$). 
For the dust temperature we adopt 35~K (as derived above), which means the dust is slightly heated with respect to the cirrus 
dust ($T_{\rm dust} \sim$15 to 20~K; \citealt{2001IAUS..204...47B}).
The estimated total dust and gas masses (adopting an average gas-to-dust ratio of 200) emitting in the bow shock region or detached shell are
summarised in Table~\ref{tb:extendedflux}. The total observed masses range from $\sim 1 \times 10^{-5}$ to $\sim 6 \times 10^{-4}$~M$_\odot$.
We note that the dust mass inferred from the measured infrared emission is sensitivy to the dust temperature, which is not well constraint 
by the avaialbe data, the dust emissivity law ($\kappa(\lambda)$), which is uncertain by order of magnitude and strongly dependent on 
the chemical composition of the dust, as well as the gas-to-dust ratio.


The total mass of the ambient medium that could be swept up by the stellar wind is roughly the volume set by the stand-off
distance of the bow shock or the radius of the detached ring, $r$, times the mass volume density: $M_{\rm shell,ISM} = 4/3
\pi\ r^3\ \rho_{\rm ISM}$. In Table~\ref{tb:extendedflux} we give $M_\mathrm{ISM}$ using both the ISM densities, $n_H$, derived
using eq.~\ref{eq:nH2} as well as using $n_H$ inferred from the measured stand-off distance and Eq.~\ref{eq:standoff}.
For may cases the observed gas \& dust mass is less than the potential swept-up ISM mass, but in a few cases the derived masses are similar.
The potentially swept-up ISM mass, $M_\mathrm{ISM}$ ranges from 10$^{-5}$ to 10$^{-1}$~M$_\odot$.
In comparison, assuming constant mass-loss rates, the stellar mass loss after 1000~years amounts too $1$ -- $50 \times 10^{-4}$~M$_\odot$,
again similar to the observed total mass.
This consistency is not surprising, since to first order, the total mass accumulated in the bow shock depends on its age
 (\ie\ mass of stellar material piled-up) and the ISM density (mass of swept-up ambient ISM material). However, for bow shocks 
 (spanning a limited azimuthal angle), only a fraction -- typically less than 
half -- of the stellar wind and ISM material is entrained into the bow shock region. Furthermore, both ISM and wind material flow from 
the bow shock apex along the contact discontinuity to be finally shed in the tails. For detached shells, both the ISM and stellar wind 
mass are presumably preserved, unless either formation or destruction of dust grains and molecules alters the gas-to-dust ratio and its chemical 
composition (\eg\  photo-destruction of CO by the ISRF or processing of grains in the shock).

\section{Interaction between stellar winds and the ISM}\label{sec:bowshock}

Bow shocks are common in astrophysical contexts and can occur where two material flows -- of different density, velocity, or
viscosity -- collide. Examples are bow shocks around compact \ion{H}{ii} regions due to stellar winds
(\citealt{1990ApJ...353..570V}, \citealt{1991ApJ...369..395M}; \citealt{1995MNRAS.273..422R}), but also due to winds in binary
systems (\citealt{1992ApJ...386..265S}; \citealt{2008MNRAS.388.1047P}; \citealt{2009MNRAS.396.1743P};
\citealt{2011A&A...527A...3V}), or movement of an object with a magnetic field through a medium
(\citealt{1971SPhD...15..791B}, \citealt{1993Natur.362..133C}), between slow and fast winds (\citealt{1998A&A...337..149S};
\citealt{2000A&A...357..180S}), and between stellar winds and the surrounding ISM (\citealt{1989PThPh..81..810M};
\citealt{1995MNRAS.277...53B}; \citealt{1998A&A...338..273C}).

In this section we will discuss the formation of bow shocks due to interaction of a stellar wind with a low density medium.
First the general case of a star with a stellar wind moving through the ISM is considered. We discuss also the special case of
a stationary star with a stellar wind expanding into either an older stellar wind or the interstellar medium.

\subsection{The size and shape of a bow shock due to ``wind-ISM'' interaction}\label{sec:standoff}


If a supersonic stellar wind and the ambient medium collide, a bow shock interface can be created at a distance from the
moving star where the ram pressures and momentum fluxes of the wind and the ISM balance each other. This is called the contact
discontinuity. It is a result of interaction between two fluids, but shocks are not necessary. In the
absence of shocks, there will be only collisional heating and no H$\alpha$ emission. For shocks to occur the Mach number $M$ of
the fluid velocity $v$ ($M = v / v_{\rm sound}$) should be higher than unity. I.e. the relative velocity between the stellar
wind and ISM needs to be higher than the speed of sound in the ISM. Note however that the speed of sound in the low density
isothermal warm neutral medium (WNM) is very low ($v_{\rm sound}\sim 1$~\kms) and thus even a slow AGB wind ($v_w\sim
10$~\kms) will move supersonically. Because the sound speed  (\emph{adiabatic} case; $\gamma = 5/3$) scales 
with $(P/\rho)^{1/2}$ (or T$^{1/2)}$) the sound speed will be
 lower in the diffuse cold neutral medium (CNM). In the non-adiabatic regime the ram pressure balance is given
by $\rho_w v_w^2 = \rho_{\rm ISM} v_{\rm ISM}^2$. Assuming that the layers mix and that the post-shock cooling is efficient 
(\ie\ instantaneous cooling), the thickness of the dense shell is negligible with respect to the distance from the star. This
is, for example, valid for a slow stellar wind interacting with a hot low-density medium (\citealt{1992ApJ...400..222B}). In
this approximation the stand-off distance, $R_0$, defined as the distance between the star and the apex of the contact
discontinuity or bow shock, is given by (\eg\ \citealt{1971SPhD...15..791B}; \citealt{1975Ap&SS..35..299D};
\citealt{1995MNRAS.273..422R}, \citealt{1996ApJ...459L..31W}):

\begin{equation}
R_0 = \sqrt{\frac{\dot{M}\ v_w}{4 \pi\ \rho_{\rm ISM}\ v^2_\star}} \label{eq:standoff}
\end{equation}

\noindent with \Mdot\ the rate of mass loss and $v_w$ the velocity of the isotropic stellar wind (with respect to the rest-frame of the star), $\rho_{\rm ISM}$ the {\it mass} density of the ambient ISM, $v_\star$ the relative space velocity of the star with respect to the ISM. 
The standard bow shock morphology consists of a forward shock separating the  unshocked and shocked ISM, a wind termination shock that separates the free-streaming wind from the shocked wind, and between them a contact discontinuity separating the shocked wind from the shocked ISM
(see \eg\ \citealt{1977ApJ...218..377W} and \citealt{1999isw..book.....L}).
If the cooling of the shocked stellar wind is inefficient, a thick hot, low-density gas layer will exist between the free-flowing wind and the
bow shock. The termination (reverse) shock and the bow shock, delineating the contact discontinuity, both travel away from this
contact discontinuity; the bow shock in forward direction and the termination shock in backward direction with respect to the relative motion
of the star in a stationary ISM.


As mentioned above, for efficient cooling, the physical thickness of the shocked region remains small and thus the two shock
fronts delineating the contact discontinuity are not resolved. In Sect.~\ref{sec:hydro} we present hydrodynamical simulations
of wind-ISM interaction which explore the effect of varying the parameters used in Eq.~\ref{eq:standoff} (\Mdot, $v_\star$,
$n_\mathrm{H}$) as well as other parameters such as the dust-to-gas ratio and the temperature of the ISM. These simulations also
confirm that, despite the formation of turbulent instabilities and inefficient cooling in some cases, Eq.~\ref{eq:standoff} gives in
general a rather accurate prediction for the stand-off distance.

Adopting appropriate values for the mass-loss rate, stellar wind velocity, the star's velocity, and the ISM density
(Sect.~\ref{sec:properties}), we can directly predict the stand-off distance, $R_0$ from Eq.~\ref{eq:standoff}. This predicted
value for $R_0$ can then be compared directly to the measured $R_0$ obtained by de-projecting the observed minimum distance
between the star and the bow shock outline. Vice versa, we can use the measured $R_0$ and Eq.~\ref{eq:standoff} to derive
$n_{\rm ISM}$. This is important as the local ISM density is difficult to determine observationally (see
Sect.~\ref{sec:properties}).

\begin{figure}[!t]
\includegraphics[angle=0, width=.8\columnwidth]{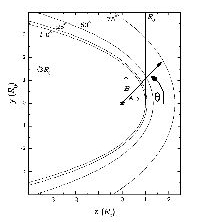}
\caption{Observable projected outlines of {\it wilkinoids} for an $R_{0}$ of 1 and star-ISM motion inclinations 
of $i=0, 25, 50,$ and $75\degr$ (with respect to the line of sight). The stellar position is indicated by a star.
 $A$ is the observed (projected) stand-off distance between the star and the bow shock apex.
 $B$ is the observed (projected) distance between the star and the bow shock for $\theta = 90$~\degr.
 $\theta$ for any point $x$ on the projected outline is the angle between the star-apex line and the line connecting the star with the point $x$.}
\label{fig:wilkinoids}
\end{figure}

For example, let us apply this to $\alpha$\,Ori. The star is at a distance, $d = 197$~pc ($z$ = -14~pc), moving at an
intermediate velocity of 28.3~\kms\ through the ISM (\citealt{2008PASJ...60S.407U}; \citealt{2011ApJ...734L..26V}). Its
mass-loss rate is $3 \times 10^{-6}$~M$_\odot$ yr$^{-1}$ and the stellar wind velocity, $v_w = 14.5$~\kms\ 
(Table~\ref{tb:properties1}). From the observed de-projected stand-off distance, $R_0$ of 5.0\arcmin\ or 0.3~pc, the local ISM
density is then found to be 4.2~cm$^{-3}$. This estimate is a factor of two higher than the density 
derived from the models of the ISM (Eq.~\ref{eq:nH2}, $n_H = 1.7$~cm$^{-3}$). 
The predicted stand-off distance is close to that obtained from a detailed simulation of $\alpha$\,Ori (see \eg\ 
\citealt{2011ApJ...734L..26V} and simulation~A in Sect.~\ref{sec:hydro}).

The 2D-shape of the contact discontinuity (and thus also the termination shock and bow shock) can be solved analytically in
the optically thin approximation (\citealt{1991ApJ...369..395M}; \citealt{1995MNRAS.273..422R}, \citealt{1996ApJ...459L..31W}):

\begin{equation}
R(\theta) = R_0 \frac{\sqrt{3 (1 - \theta \cot \theta)}}{\sin \theta} \label{eq:Rtheta}
\end{equation}

\noindent with $\theta$ the latitudinal angle from the apex of the bow shock as seen from the position of the central star,
and $R_0$, the stand-off distance defined in Eq.~\ref{eq:standoff}. Alternatively, it can be written as $y(z) = \sqrt{3} R_0 \sqrt{1 - z /
R_0}$ (\eg\ \citealt{1991ApJ...369..395M}; Fig.~\ref{fig:wilkinoids}). Eq.~\ref{eq:Rtheta} is valid only for the case of a
relative star-ISM motion in the plane of the sky (\ie\ a ``side view'' of the bow shock or a star-ISM motion inclination of $i
= 0$\degr\ with respect to the plane of the sky). In this case, $R_0$ is directly the angular separation, $A$, of the star to
the observed bow shock outline (Fig.~\ref{fig:wilkinoids}). Similarly, when there is no inclination($i$ = 0\degr), the angular
distance between the star and the surface of the bow shock parabola at an angle of $\theta$ = 90\degr\  (perpendicular to the
apex direction from the star) can be defined as $B$, where $B = R(90\degr) = y(0) = R_0 \sqrt{3}$
(\citealt{1996ApJ...459L..31W} and Fig.~\ref{fig:wilkinoids}).
In this geometry, $R_0$ can also be derived from measuring the angular distance between the star and the bow shock interface at
an angle of $\theta$ = 90\degr\ with respect to the star-ISM motion direction ($B$; as illustrated in
Fig.~\ref{fig:wilkinoids}), since $R$(90\degr) = y(0) = $R_0 \sqrt{3}$ (\citealt{1996ApJ...459L..31W}). 

\begin{figure}[t!]
\centering
\includegraphics[angle=0, width=.9\columnwidth]{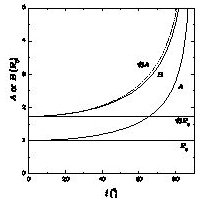}
\caption{Observable bow shock properties $A$ and $B$ as a function of inclination $i$ for a fixed $R_{0}=1$.}
\label{fig:bowshockproperties}
\end{figure}

In general the relative star-ISM motion will be inclined with respect to the plane of the sky, thus $i \neq 0$. In order to
quantify the effects of an inclined bow shock with respect to the plane of the sky, we simulated the appearance of the three
dimensional bow shock surface. This hollow paraboloid {\it wilkinoid} surface is defined by rotating the two dimensional
shape, \ie\ the R($\theta$) curve, around its axis of symmetry (which is defined by  the line connecting the star and the bow
shock apex).  We numerically construct a 3D {\it wilkinoid} surface. These points
are then projected onto the plane of the sky under a given inclination angle to compute the observable outline of the bow
shock in the optically thin approximation.  This outline thus traces the loci of tangential line of sight to the bow shock
paraboloid at a given inclination angle. Four such constructed {\it wilkinoids} with an $R_0$ of 1 and star-ISM motion
inclinations with respect to the plane of the sky of $i = 0, 25, 50$, and $70$\degr\ are shown in Fig.~\ref{fig:wilkinoids} to
illustrate the change in shape with increasing inclination angle. These simulations show indeed that for an increasingly
higher inclination $i$ (corresponding to a smaller viewing angle with respect to the line-of-sight), the projected size of the
bow shock outline becomes increasingly larger, which is qualitatively consistent with earlier bow shock simulations for
compact \ion{H}{ii} regions (\citealt{1991ApJ...369..395M}). As noted above, in Fig.~\ref{fig:wilkinoids} we also indicate the
two observable quantities, $A$ and $B$. $A$ is the projected -- minimal -- distance between the star and the bow shock outline
in the direction of relative motion, which can be measured directly from the observed image.  $B$ is the distance
perpendicular to $A$, \ie\ at $\theta = 90$\degr, from the star to the bow shock outline.   
It is important to note here that due to the inclination of the bow shock surface with respect to the line-of-sight, a ``pseudo'' apex   
of the observable outline appears where the line-of-sight becomes tangent to the rotated 3D-Wilkin paraboloid.   
The observable $A$ is in fact the distance from the star to this ``pseudo''-apex projected onto the plane of the sky
(\ie\ not to the real apex of the {\it wilkinoid}).
At high inclination, but less than 90\degr, both the ``pseudo''-apex and true apex of the observable outline
would appear -- in the model -- as two brightness enhancements. 
We refer the reader to Figs.~5 and~11 in \citet{1991ApJ...369..395M} for an excellent illustration of this effect.
However, due to superposition of the true apex and the very bright central star the latter will not be detectable for our stars. In addition, the faint ``pseudo''-apex will also be difficult to observe. For
$i=90$\degr\ there is no tangent to the paraboloid, thus only the bow shock apex would theoretically appear as a spherical
central peak brightness distribution, where it not for the fact that it will not be visible due to superposition with the very
infrared bright central star. For the $i = 0$\degr\  (side view) the simulation gives indeed $A = R_0$ (\ie\ the projected and
de-projected stand-off distance are the same) and $B = R_0 \sqrt{3}$ (also here the true apex is observed).

In Fig.~\ref{fig:bowshockproperties} the $A$ and $B$ values, in units of $R_0$, are given as a function of the inclination $i$.
Measuring both $A$ and $B$ would then, in principle, directly constrain both $R_0$ and $i$. Unfortunately, measuring both
quantities in the detected bow shocks is often not possible due to irregularities and incompleteness of the bow shock. Also
the change of outline with varying $i$ is a small effect, thus making the exact inclination angle and stand-off distance
highly degenerate. For inclination angles $i < 45$~\degr\ the change in $A$ or $B$ is less than 10\%
(Fig.~\ref{fig:bowshockproperties} and \citealt{1991ApJ...369..395M}). Only at large  inclinations (or small viewing angles)
does the observed projected distance $A$ to the bow shock outline differ significantly from $R_0$, the de-projected stand-off
distance. Consequently, using these relations and the kinematical properties of the measured stars and its local ISM,
inclination-corrected $R_0$ values can be derived by measuring $A$ and/or $B$ together with a calculated or assumed $i$. Thus,
first a plausible inclination is derived from the space motion, secondly $A$ is measured from the observed image and thirdly,
$R_0$ is obtained from the $R_0$-$A$ relation in Fig.~\ref{fig:bowshockproperties}.  Ideally the star's kinematical properties
(proper motion and radial motion) would give the heliocentric 3D space motion vector of the star, wile the observed bow
structure would -- independently -- yield the heliocentric 3D orientation of the bow, whose apex's 3D orientation represents the
heliocentric relative space motion of the star, with respect to the ISM. Thus, the difference of the two would yield the
heliocentric 3D ISM flow vector. However, in order to circumvent the high degeneracy between $i$ and $R_0$ we use the space
motion vector inclination angle $i$ to constrain $R_0$ from the observed bow shock shape, which thus neglects a potential ISM
flow (see Sect.~\ref{sec:distance} for a discussion on the local LSR ISM velocity). Table~\ref{tb:properties1} gives the
de-projected $R_0$ obtained via {\it wilkinoid} fitting as discussed above for the objects which show emission in the shape of
an arc or shell, and thus potentially trace the dust emission outline of a bow shock.

\subsection{Special cases of ``wind-ISM'' and ``wind-wind'' interaction}

In the case of a (nearly) stationary star (with respect to the local ISM and compared to the stellar wind velocity), 
spherical symmetry of the stellar wind-ISM interaction is
preserved (\eg\ \citealt{2007MNRAS.380.1161L}). Nevertheless, there is a relative velocity difference between the stellar wind
and the ambient medium. The interaction of a spherical outflow with external matter leads to the formation of a region of
compressed material within two spherical boundaries  (\citealt{1977ApJ...218..377W}). As before, this region consists of a
termination shock (where the freely expanding (supersonic) stellar outflow is abruptly slowed down by compressed material), a
contact discontinuity (separating the circumstellar and interstellar matter) and the bow shock (external boundary at which the
external medium is compressed by the expanding shell). It is thus a special case of a ``wind-ISM'' interaction scenario
described in the previous section. The total mass of the detached shell would consist of both circumstellar and interstellar
matter. The latter can be estimated from the total hydrogen mass that would have been present in a sphere of radius equal to
the radius of the detached ring (Sect.~\ref{subsec:flux}). 

However, the fact that the ``external'' material of the
stellar wind is interacting with is moving together with the source (thus keeping the spherical symmetry) may also suggest that in
fact the ``external'' matter is not genuine ISM but rather material remaining from an older mass-loss event, perhaps during the RGB phase.
The thermal pulse scenario leads to an interaction between a slow and a fast wind, \ie\ ``wind-wind'' interaction. In this case,
$v_\star$ from Eq.~\ref{eq:standoff} is the relative velocity between the slow and fast wind. In the rest frame of the star,
the second wind is computationally equal to a flowing ISM. A detached dust shell is believed to result from an intense -- relatively short --
episode of enhanced mass-loss rate and wind velocity (\eg\ initiated by a He-shell flash), whose wind has a higher outflow
velocity and will thus interact with the previous slower wind. The subsequent sharp drop in mass-loss rate and outflow
velocity for a few thousand years directly following the thermal pulse or He-flash will lead to a detached dust shell (\eg\
\citealt{1990A&A...230L..13O}; \citealt{1993ApJ...413..641V}; \citealt{2000A&A...357..180S}; \citealt{2007A&A...470..339M}).
The (relative) jump in velocity and mass-loss rate from pre-flash to flash-peak is possibly the most critical parameter
governing the formation of a detached shell (\citealt{2007A&A...470..339M}). The latter scenario seems more appropriate for
the formation of geometrically thin detached molecular shells with high expansion velocities showing little interaction with
the surrounding medium. Indeed, the detection of bow shocks together with a small detached ring gives further support to this
hypothesis. 
Another possibility could be that the ring represents the (inner) termination shock associated to the bow shock located further outwards.
Other scenarios, such as constant mass-loss in combination with non-isotropic mass-loss events and clumpy dust
formation have also been detected in the form of non-concentric spherical shells (Decin et al. 2011).

\nocite{DecinBetelgeuse}

\section{Stellar, circumstellar and interstellar properties}\label{sec:properties}

The relation (Eq.~\ref{eq:standoff}) between the stand-off distance of the bow shock region and the star's mass-loss rate, \Mdot, terminal wind velocity, $v_w$, peculiar velocity ($v_\star$), and the ISM density ($n_\mathrm{H}$) is powerful in its simplicity. For observed bow
shocks around stars with known (observed) mass-loss properties and space motion, the local ISM density -- which is often the
most difficult to determine accurately -- can be inferred directly from the measured stand-off distance. On the other hand, for
stars with known stellar mass-loss, known space motions and (assumed) ISM densities, the stand-off distance can be predicted.
In this section we review the relevant and available stellar properties of the observed AGB stars and supergiants as well as
the properties of the local ISM. The interplay between these various physical properties determines the final shapes and sizes
of the wind-ISM interaction zone, in particular the contact discontinuity as discussed in the Sect.~\ref{sec:standoff}. 
Adopting the stellar, circumstellar and interstellar properties presented in this section for each
object in our survey, both the predicted ISM densities for observed bow shocks as well as predicted stand-off distances for
all targets in the survey are given in Tables~\ref{tb:properties1} to~\ref{tb:properties2}.

\subsection{Stellar distance and (relative) space velocity}\label{sec:distance}

The parallax, proper motion, and the radial local standard of rest (LSR) velocities give a direct estimate of the distance, $d$,
and the absolute \emph{peculiar} space velocities, $v_\star$ (as per \citealt{1987AJ.....93..864J}), for the majority of the
AGB stars and red supergiants. The LSR velocities have been corrected for the solar motion $v_\odot$ [($U,V,W$) = (11.1,
12.24, 7.25) \kms] (\citealt{2010MNRAS.403.1829S})\footnote{Note that other common values adopted for the solar motion from
\eg\ \citet{1968gaas.book.....M}, \citet{1998MNRAS.298..387D}, or \citet{2009ApJ...700..137R} give results for $v_\star$ that
differ slightly, by a few \kms.}. Parallaxes from re-processed Hipparcos data (\citealt{2007A&A...474..653V}) were taken,
when available, if the uncertainties are less than 20\%. For several (distant) stars, the parallactic distances derived from
Hipparcos data have large uncertainties or have not been obtained at all. For these targets distances can be derived, for
example, from the (pulsation) period-luminosity($K$) relation, and observed apparent $K$ magnitudes. A generic Galactic
$P$-$K$ relation valid for both oxygen- and carbon-rich (Mira) variables was established by \eg\
\citet{2000MNRAS.319..728W,2008MNRAS.386..313W}: $M_K = -3.51 [ {\log}\ P - 2.38] - 7.25$. \citet{2003A&A...407..213V}
obtained VLBI parallaxes for four Miras that were more consistent with distances derived via the $P$-$K$ relation than those
obtained with Hipparcos. 
Tables~\ref{tb:properties1} to~\ref{tb:properties2} give the derived peculiar velocities and peculiar absolute proper motions 
with their associated position angles.

Radial (LSR) velocities are taken from CO line surveys where available. Primary sources are \citet{2010A&A...523A..18D} and
\citet{2006MNRAS.369..783M}, which have good agreement for sources in common to both studies (typically the differences are
less than 3~\kms).

At this point we assume that the relative peculiar velocity between the ISM and the star, $v_\star$, is determined entirely by the
star space velocity with respect to the local standard of rest (LSR) (\ie\ the stationary ISM). 
In other words, we assume that there is no flow of the ISM itself. This
could be a simplification for some cases where the ISM may have an appreciable flow velocity as matter is being blown away by
super-bubbles (see \citealt{2009ASPC..418..117U,2009ASPC..418..463U} for the case of $\alpha$~Ori).

To test this, we estimate the local LSR ISM velocity ($v_r, v_l, v_b$) from the galactic rotation
(\citealt{1928BAN.....4..269O}, \citealt{1997MNRAS.291..683F}) at the position of the stars and correct $v_\star$ to obtain
the local $v_{\star - \rm ISM}$. For the majority of the objects the $v_{\star - \rm ISM}$ is within $\pm$20\% of $v_\star$,
thus decreasing/increasing the predicted stand-off distance by the same factor. Only in a few cases this correction gives a
significant change in peculiar velocity; for U\,Cam, V\,Hya, T\,Lyr, and $\mu$\,Cep this gives  relative space velocities that
are approximately a factor of 1.5 higher, and for Y\,Pav a factor of 2 lower.

The average space velocity, $v_\star$, for our sample of stars is $\sim$36~\kms\  (Tables~\ref{tb:properties1}
to~\ref{tb:properties2}), consistent with the average velocity of $\sim$30~\kms\ for Galactic AGB stars
(\citealt{2000MNRAS.317..460F}).

\subsection{Mass-loss rate}\label{sec:mass-loss}

The mass-loss rate of evolved stars is important for the enrichment of the ISM. It also directly affects the shape and size
of the stellar wind-ISM interaction (Sects.~\ref{sec:bowshock} and~\ref{sec:hydro}). Mass-loss rates can be estimated directly
via modeling of observed CO line profiles (\eg\ \citealt{1998ApJS..117..209K}; \citealt{1998A&A...337..797G};
\citealt{2002A&A...390..511G}; \citealt{2010A&A...523A..18D}).  In general the derived (gas) mass-loss rates depend on the
line strength, terminal gas expansion velocity ($v_w^2$), and distance ($d^2$). The average gas mass-loss rate from $\sim$300
Galactic carbon stars is $1.1 \times 10^{-5}$~M$_\odot$ yr$^{-1}$ (\citealt{2002A&A...390..511G}). The latter should probably
be reduced by an order of magnitude since new Hipparcos results show that the distances adopted by \citet{2002A&A...390..511G}
are over-estimated by a factor of two to four. Using CO-derived mass-loss rates, an empirical relation has been established
between the mass-loss rate and luminosity variability period (\eg\ \citealt{2010A&A...523A..18D}): log(\Mdot) = -7.37 + 3.42
$\times$ 10$^{-3}$ P (for P $\leq$ 800 days) and log(\.M) = -4.46 (for P~$\geq$~850 days), with \Mdot\ in units of M$_\odot$
yr$^{-1}$. However, the scatter on this relation is rather large with a typical uncertainty of a factor of 10  (see
\citealt{2010A&A...523A..18D} for details). Similar results are obtained for carbon Mira variables by
\citet{1998A&A...337..797G}: log(\Mdot) = 4.08 log~$P$ - 16.54.  Adopted mass-loss rates -- based on CO observations or
luminosity period -- are included in Tables~\ref{tb:properties1} to~\ref{tb:properties2}. If neither \Mdot\ nor
period are available, we adopt a generic value of $5 \times 10^{-7}$~M$_\odot$ yr$^{-1}$.  Uncertainties in \Mdot\ arise
predominantly from inaccurate distance determinations for distant stars.

\subsection{Stellar wind velocities}

For the (terminal) wind velocity, $v_w$, we adopt the terminal velocity of the CO envelope (\eg\
\citealt{2010A&A...523A..18D}; Tables~\ref{tb:properties1} to~\ref{tb:properties2} and references therein).
\citet{2002A&A...390..511G}  find an average expansion velocity of 18.7$\pm$6.1~\kms\ for a sample of 330 carbon stars. For a
set of 24 oxygen-rich and 13 carbon-rich AGB stars, \citet{2010A&A...523A..18D} find mean $v_w$ values of 14.5 and 15.4~\kms,
respectively. These values are 4-10~\kms\ higher than those of \citet{2006A&A...454L.103R} for a set of 77 oxygen-rich and 61
carbon-rich stars. For carbon-rich stars, this average is $\sim$3~\kms\ lower than the one found by \citet{2002A&A...390..511G}.
If no CO terminal velocities are available, we adopt a generic value of $v_w = 15$~\kms, for both oxygen- and carbon-rich stars.

\subsection{Local interstellar medium density}

The ambient (uniform) mass density of the ISM is defined as $\rho_{\rm ISM} = \mu_\mathrm{H}\ m_\mathrm{H}\ n_\mathrm{H}$, with $\mu_\mathrm{H} = 1.4$, the mean
nucleus number per hydrogen  atom for the local medium, $m_H$ is the mass of the hydrogen nucleus, and $n_\mathrm{H}$ the interstellar
hydrogen nucleus density. Although the ISM is far from uniform, there is a general dependence between the interstellar matter
(space-averaged) volume density,  $n_{\rm ISM}$, and $z$, the distance from the Galactic plane\footnote{$z$ can be expressed
as $z$(pc) = $d$ sin $b$ + $z_\odot$, with $d$ the distance, $b$ the Galactic latitude, and $z_\odot$ the sun's vertical
displacement from the Galactic plane. The exact value of $z_\odot$ depends strongly on the assumed underlying Galactic model
and/or the observational data selection criteria. However, the value is converging towards $\sim$15 -- 20~pc (13$\pm$7; Brand
\& Blitz 1993; \citealt{1995AJ....110.2183H}: 20.5$\pm$3.5~pc; \citealt{2006JRASC.100..146R}: 19.6$\pm$2.1~pc;
\citealt{2007MNRAS.378..768J}: 17$\pm$3~pc). We adopt a value of 15~pc.}. 

The different phases of the ISM have different average densities, scale-heights and filling factors (see \eg\
\citealt{1995ASPC...80..233S}; \citealt{2001IAUS..204...47B}; \citealt{2001RvMP...73.1031F}; \citealt{2004come.book...33W}; \citealt{2005ARA&A..43..337C}). 
\citet{1990ARA&A..28..215D} used radio observations of \ion{H}{i} to derive the structure of the
CNM, WNM and WIM (warm ionised medium):. 

\begin{equation}
n_H (z) = 0.39\ e^{-\left(\frac{z}{127 \rm pc}\right)^2}\ +\ 0.11\ e^{-\left(\frac{z}{318 \rm pc}\right)^2} \ +\ 0.06\ e^{-\frac{|z|}{403 \rm pc}} \label{eq:nH1}
\end{equation}

\noindent Based on UV observations of \ion{H}{i} and H$_2$, \citet{1977ApJ...216..291S} and \citet{1994ApJ...427..274D} arrive at a similar result for the WNM as
\citet{1990ARA&A..28..215D}. An alternative relation is given by \citet{1993A&AS...99..291L} which is based on
\citet{1978ppim.book.....S} (density $n_\mathrm{H}$(z=0) = 2~cm$^{-3}$) and \citet{1981gask.book.....M} (scale height of 100~pc):

\begin{equation}
n_H (z) = 2.0\ e^{-\frac{|z|}{100 \rm pc}}  \label{eq:nH2} 
\end{equation}

\noindent These two relations are shown further below in Fig.~\ref{fig:z_nH} (Sect.~\ref{sec:discussion}). Thus, we obtain some first
insights into the local ISM density for our objects (Tables~\ref{tb:properties1} to~\ref{tb:properties2}). Evidently, one
should be cautious using these relations for specific cases as there is structure in the ISM on all spatial scales, with
different phases that have different filling factors.  For example, diffuse to molecular clouds ($n_H \sim 10 -
1000$~cm$^{-3}$) have an order of magnitude lower filling factor than the low density WNM and WIM surrounding it. Thus,
statistically about 10\% of the objects in our survey could be moving through a denser medium than inferred from the above
relation for the WNM. Generally, these relations do indicate that $n_\mathrm{H}$ drops as one moves away from the Galactic plane,
leading to -- on average -- lower volume densities of the ISM. We use Eq.~\ref{eq:nH2} to get a first estimate of the local ISM
density  for the objects in this survey (Tables~\ref{tb:properties1} to~\ref{tb:properties2}). Accurate measurements of the
local ISM density for all stars in our sample would be desirable, but are currently -- to the best of our knowledge -- unavailable.
In fact, we will show that for certain cases the observations of bow shocks can potentially be used, via the measurements of the
stand-off distances, to derive estimates of the local ISM density.

\subsection{Circumstellar chemistry, spectral type and binarity}

In addition to the above stellar properties that have a direct impact on the theoretical stand-off distance via
Eq.~\ref{eq:standoff}, other possibly pertinent information such as spectral types of the central star and dominant
circumstellar chemistry  (oxygen versus carbon rich), and binarity is also included in Tables~\ref{tb:properties1}
to~\ref{tb:properties2}. If and how these stellar attributes could affect the occurrence and shaping of bow shocks is
discussed further in Sect.~\ref{sec:discussion}.

\begin{table*}[tp!]
\centering
\caption{Parameters for the moving star simulations. }
\label{tb-hdsims}
\begin{tabular}{lllllllllll}\hline\hline
label & description     		&  $\dot{M}$		        & $M_{\rm dust}/M_{\rm gas}$  &$v_\star$ & $n_{\rm ISM}$ & $T_{\rm ISM}$ & physical space& basic grid	& $R_0$\tablefootmark{a}   & $R_s$\tablefootmark{b}\\
      &                			& ($\mathrm{M}_\odot$ yr$^{-1}$)& (\%)    				  &(\kms)    & (cm$^{-3}$)   & (K)	     & (pc)          &  		& (pc)  	    	   & (pc)		\\  \hline
A     &    basic model\tablefootmark{c} &  $10^{-6}$    	        & 1       				  & 25       & 2	     & 1	     &  1.5x1        & $120\times80$	&   0.26 $\pm$ 0.01 	   & 0.35	        \\
B     &    high $\dot{M}$    		&  $10^{-5}$    	        & 1       				  & 25       & 2	     & 1	     &  2x2          & $160\times160$	&   0.59 $\pm$ 0.03 	   & 0.73	        \\
C     &    low  $\dot{M}$    		&  $10^{-7}$    	        & 1       				  & 25       & 2	     & 1	     &  2x1          & $160\times80$	&  0.090 $\pm$ 0.001	   & 0.10	        \\
D     &    low dust          		&  $10^{-6}$    	        & 0.1     				  & 25       & 2	     & 1	     &  2x1          & $160\times80$	&   0.33 $\pm$ 0.03 	   & 0.33	        \\
E     &    high $v_\star$    		&  $10^{-6}$    	        & 1       				  & 75       & 2	     & 1	     &  1.5x1        & $120\times80$	&   0.11 $\pm$ 0.02 	   & 0.13	        \\
F     &    low $n_{\rm ISM}$ 		&  $10^{-6}$    	        & 1       				  & 25       & 0.2	     & 1	     &  3x2          & $240\times160$	&   0.76 $\pm$ 0.08 	   & 0.92	        \\
G     &    warm ISM          		&  $10^{-6}$    	        & 1       				  & 25       & 2	     & 8\,000	     &  2x2          & $160\times80$	&   0.31 $\pm$ 0.06 	   & 0.45	        \\
\hline
\end{tabular}
\tablefoot{
\tablefoottext{a}{Estimated. Large-scale instabilities make $R_0$ (contact discontinuity) a time-dependent property. 
In particular, for simulations E and G the mixing is very efficient, eliminating in effect the contact discontinuity altogether.}
\tablefoottext{b}{$R_s$, the location of the forward shock of the shocked gas region, where the ISM transition from unshocked to shocked gas can
be more accurately measured.}
\tablefoottext{c}{$v_w = 15$~\kms\ for all models.}
}
\end{table*}

\section{Hydrodynamical models of interaction between the slow stellar wind of a moving star and the ISM}\label{sec:hydro}

Hydrodynamical simulations offer the opportunity to explore the effect of varying physical properties of either the stars
(\eg\ mass-loss rate, wind and (relative) space velocity) and/or the ISM (density, temperature) in a coherent systematic way.
For example, \citet{2000A&A...357..180S} and \citet{2007MNRAS.380.1161L} used hydrodynamical simulations to show that a brief episode of
increased mass-loss rate could give rise to an expanding, geometrically thin shell. \citet{2006MNRAS.372L..63W} applied
numerical simulations of a two-wind model to explain the observed structure around R\,Hya. Simulations by
\citet{2007MNRAS.382.1233W} and \citet{2007ApJ...660L.129W} indicate that a higher mass-loss rate (for similar ISM density and space velocity)
will result in more pronounced Kelvin-Helmholtz instabilities (see their Fig.~1). \citet{2006MNRAS.366..387W} obtain bullet-shaped emission
structures simulating a PN (their Figs.~3 and~5). \citet{2003ApJ...585L..49V} include time-dependent
mass-loss in simulations of wind-ISM interaction, leading to time-dependent stand-off distances.

One particular strength of hydrodynamical simulations is that they allow to study the formation, growth and dissipation of
fluid instabilities -- like Rayleigh-Taylor (RT) and Kelvin-Helmholtz (KH) instabilities -- important in wind-ISM shock
interactions. RT instabilities occur when a dense, heavy fluid is accelerated by a light fluid. Normally two plane-parallel
fluid layers are meta-stable, but a small perturbation can destroy this delicate equilibrium. This manifests itself as
so-called inter-penetrating ``RT-fingers''. Such instabilities could be quenched by a restoring force such as a magnetic
field, thus preventing these instabilities to grow \citep{1961hhs..book.....C}. The KH instability results from a velocity
shear between two fluid layers. This flow of one fluid over another will induce a centrifugal force which leads to changes in
pressure which amplifies the ripple. Together with a RT instability, the KH instability will form structures in the shape of
mushroom caps on the end of the RT fingers. The KH time-dependent turbulent eddies can form a complex structure arising in a
steady flow, if they become large enough to influence that large-scale morphology of the shocked gas.  Other instabilities
that can occur are the non-linear thin shell instability (\citealt{1994ApJ...428..186V}) and the transverse acceleration 
instability (\citealt{1996ApJ...461..372D}) as shown numerically by \eg\ \citet{1998NewA....3..571B} and
\citet{1998A&A...338..273C}.

\subsection{Hydrodynamical simulations}

\begin{figure}[t!]
\centering
\includegraphics[width=.9\columnwidth]{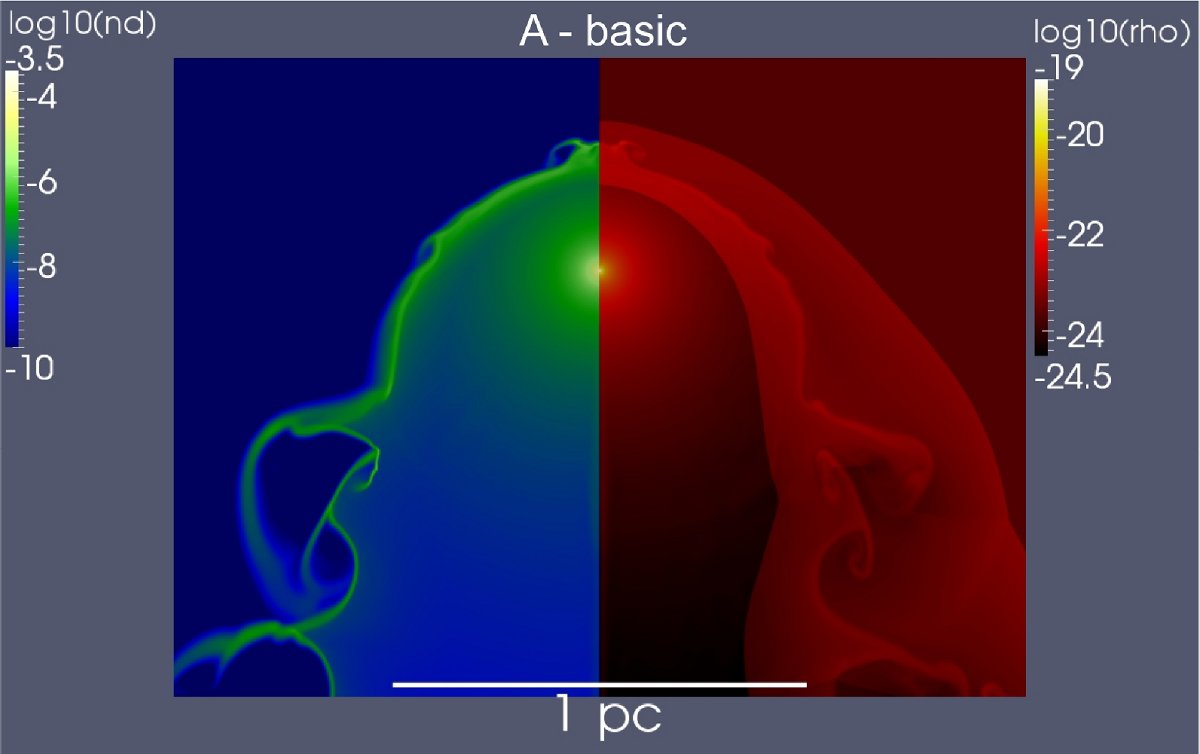}
\includegraphics[width=.9\columnwidth]{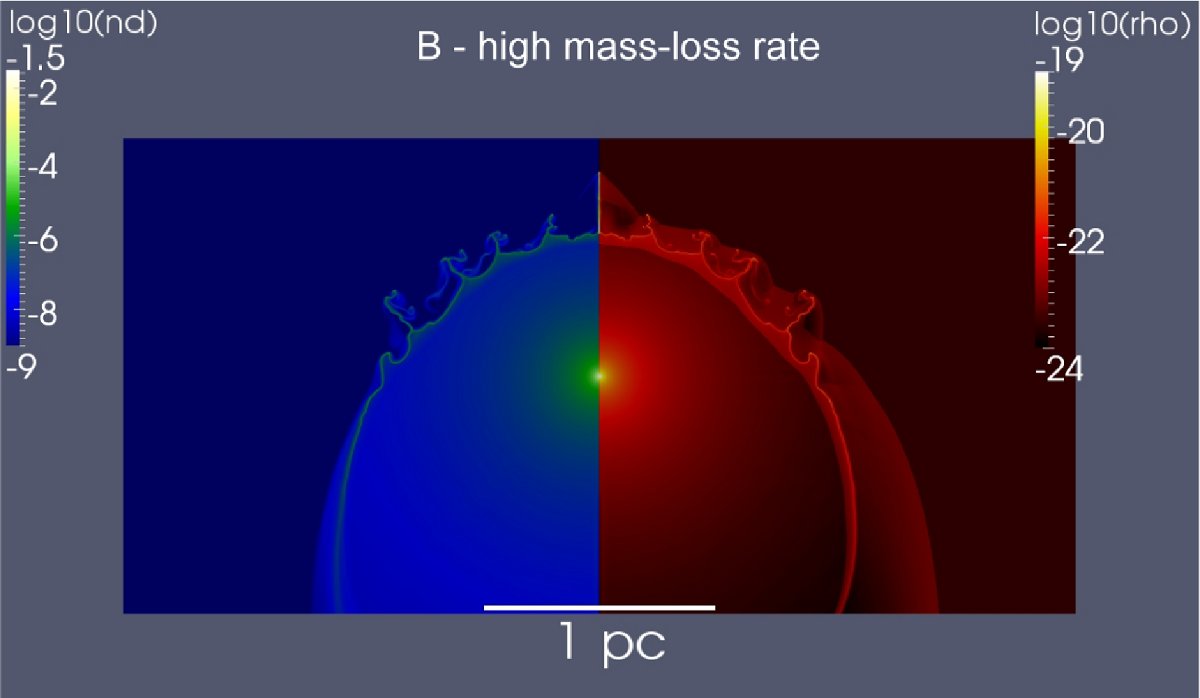}
\caption{Top panel (A): Gas density in g/cm$^3$ (right) and dust grain particle density in cm$^{-3}$ (left) 
   for simulation A after 1.37$\times10^5$\,years; basic model. 
   The bow shock is smooth, but at the contact discontinuity both Rayleigh-Taylor and Kelvin-Helmholtz instabilities are visible.
   Bottom panel (B): Similar to top panel, but for simulation~B after 5.0$\times10^4$\,years; high \Mdot. 
   The instabilities are now primarily of the Rayleigh-Taylor type. 
   Due to the stronger wind the bow shock lies further from the star.}
\label{fig:basic}
\label{fig:highdm}
\end{figure}

\begin{figure}[th!]
\centering
\includegraphics[width=.9\columnwidth]{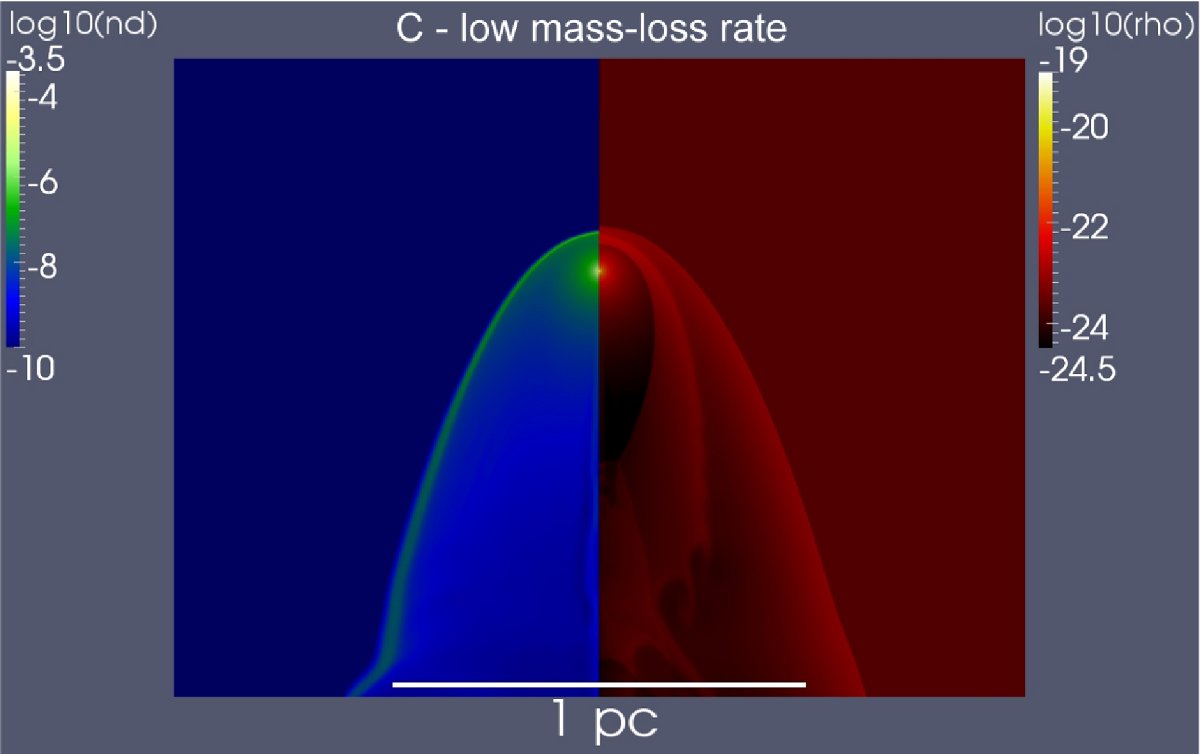}
\includegraphics[width=.9\columnwidth]{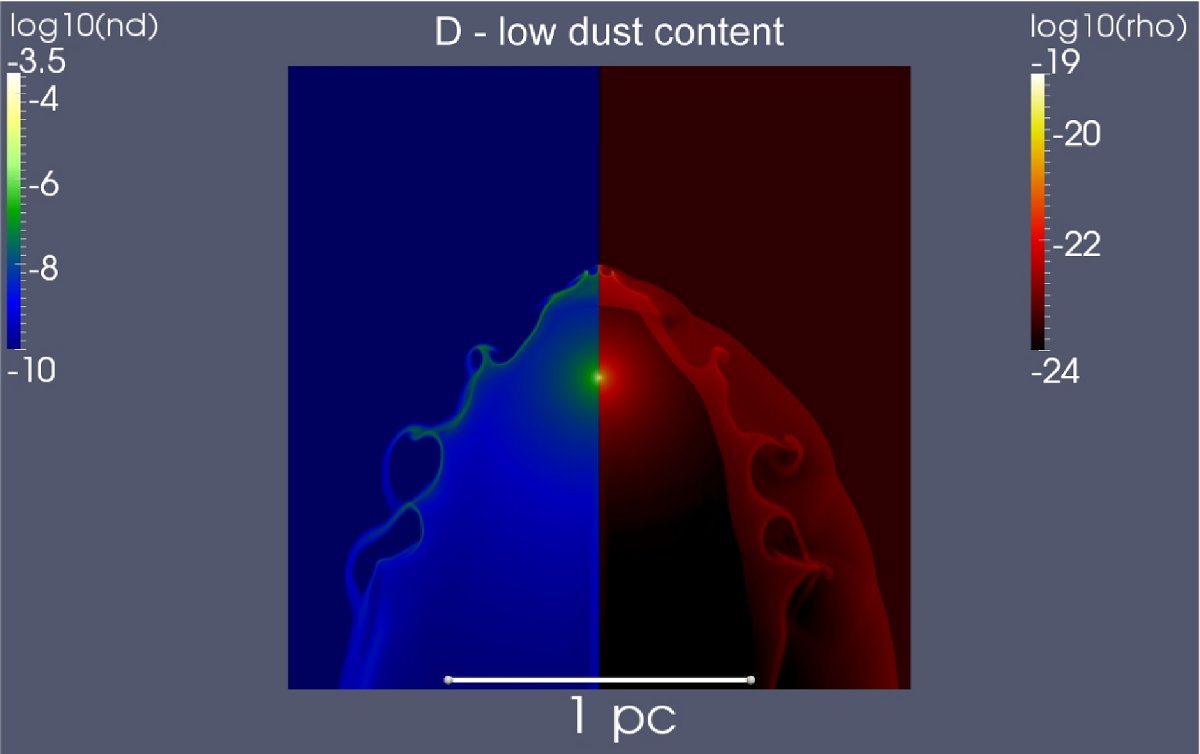}
\caption{Similar to Fig.~\ref{fig:basic}, but for simulation~C  after 1.0$\times10^5$\,years(top) and 
   simulation~D after 5.0$\times10^4$\,years (bottom).
   Top panel (C): Because of the weaker wind the bow shock is very close to the star. The bow shock morphology 
   is completely stable. The shocked gas fills up the cavity behind the star due to low ram pressure of the wind,
   and thermal pressure of the shocked gas, because the shocks are almost completely adiabatic.
   Bottom panel (D): The bow shock is more conical than for the simulations with a stronger dust component 
   and shows more instabilities. The instabilities are primarily of the Kelvin-Helmholtz type.}
\label{fig:lowdm}\label{fig:lowdust}
\end{figure}

\begin{figure}[th!]
\centering
\includegraphics[width=.9\columnwidth]{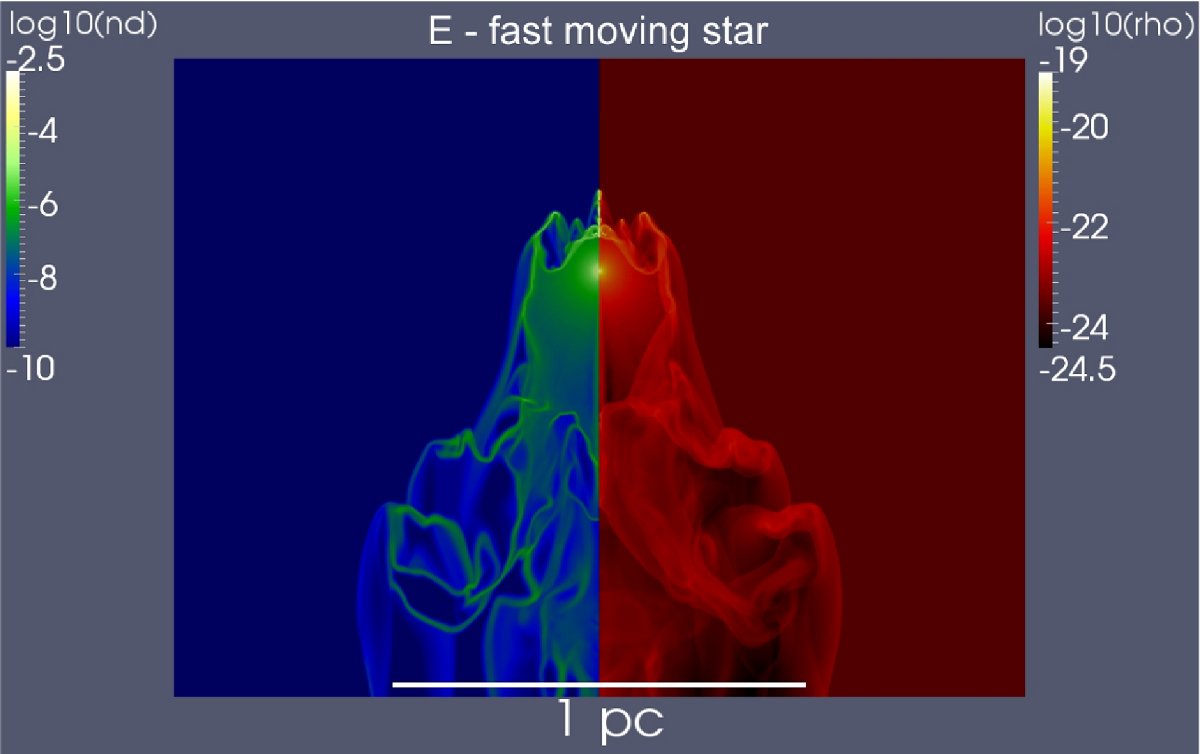}
\includegraphics[width=.9\columnwidth]{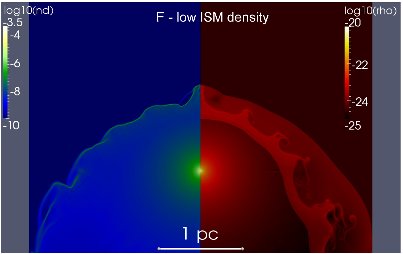}
\includegraphics[width=.9\columnwidth]{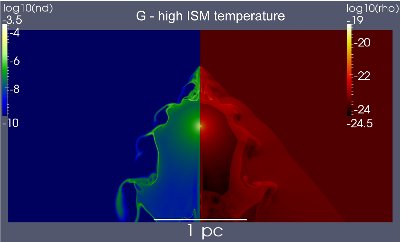}
\caption{Similar to Fig.~\ref{fig:basic}, but for simulation~E after 6.5$\times10^4$\,years (top), 
    F after 5.0$\times10^4$\,years (middle), and G after 1.0$\times10^5$\,years (bottom). 
   Top panel (E): The entire shocked gas region is completely unstable due to a combination of radiative shocks 
   and the high ram-pressure from the stellar motion in the ISM. The high ISM ram pressure brings 
   the bow shock close to the star and the region behind the star is filled up with shocked gas due 
   to the turbulent motion of the bow shock.
   Middle panel (F): The low density ISM results in a very extended bow shock region.
   Bottom panel (G): Due to the high ISM temperature, the radiative cooling in the shocked ISM is relatively ineffective. 
   Since the shocked ISM maintains a high thermal pressure, due to the enforced minimum temperature 
   of 8\,000~K, the shocked gas fills the void behind the moving star. }
      \label{fig:faststar}  \label{fig:lowism} \label{fig:highTism}
\end{figure}

Here we present a series of seven simulations of the interaction between the ISM and the circumstellar medium of moving,
evolved stars in order to find out how the morphology of the bow shock varies with the various stellar wind and ISM
parameters. The different pertinent parameters for the simulations are summarised in Table~\ref{tb-hdsims}. For our
hydrodynamical simulations we use the MPI-AMRVAC code \citep{Keppens2011}. This code solves the conservation equations of
hydrodynamics on an adaptive mesh (AMR) grid.  We use a 2-D cylindrical grid in the r-z plane. The basic resolution is set at
80 grid points per parsec, but allows four additional levels of refinement, each doubling the effective resolution. This gives
us a maximum effective resolution of 1280 grid points per parsec. Since some of the models require a larger physical space
we use grids of different sizes for those simulations.  In those cases we increase the number of grid points on the basic
level to maintain the same resolution. The expanding stellar wind is inserted by filling a small circle with wind
material.  The motion of the star is handled by giving the ambient medium around the star a velocity in the z-direction.
Therefore, we are simulating the wind interaction in the frame-of-reference of the star. Since all models are 2-D and the
simulated space lies along the direction of motion of the star, the snapshots (Figs.~\ref{fig:basic} through
\ref{fig:highTism}) show projections for $i=0$. 

To the basic conservation equations we have added the effect of optically thin radiative cooling. This is necessary since some of
the shocks are strongly radiative, which changes the morphology of the shocked gas. In the case of simulation~G, where the ISM has
a high temperature, we put a lower limit on the cooling in the ISM, so that in the ISM the temperature does not fall below
8\,000~K. More importantly, we have also added a (simplified) dust component (\citealt{2011ApJ...734L..26V}). This is done by using
a two-fluid approximation, with the dust represented as a gas without internal pressure. For simplicity we only include dust in the
wind and neglect the dust component of the ISM. This gives us a set of five partial differential equations (not counting the vector
components) for the pressure balance for the gas and dust. For the gas we have conservation of mass, momentum and energy, while for
the dust we only have conservation of mass and momentum. The appropriate equations are given in \citet{2011ApJ...734L..26V}.
Radiative losses depend on the hydrogen and electron particle densities (derived from $\rho$ assuming full ionization with
hydrogen  mass and involve a temperature dependent cooling curve $\Lambda(T)$. This cooling curve has been calculated with the {\tt
CLOUDY} code \citep{1998PASP..110..761F} and includes radiative losses through IR radiation from the dust. We also include a drag
force linking the gas and dust, which is derived from a combination of  Epstein's drag law for  the subsonic regime and Stokes'
drag law for the  supersonic regime (\citealt{1975ApJ...198..583K}, see also \citealt{2011ApJ...734L..26V}). The drag force depends
on the dust particle density and radius (0.005~$\mu$m), the velocity difference between gas and dust, as well as the thermal speed of the gas
(\citealt{1975ApJ...198..583K}). For further details we refer the reader to \citet{2011ApJ...734L..26V}.

Since our equations are purely hydrodynamical, we don't take into account magnetic fields or the effect of radiation on the
dust particles. We also neglect destruction and creation of dust particles. Radiative processes are not included and thus the
dust temperature cannot be treated appropriately (currently only collisional heating of dust could be accounted for). Also
the gas is purely heated by collisions, and cooled radiatively; photo-ionisation is not included. Introducing radiative
cooling by dust will cause the bow shock to become thinner and more unstable.

Adding destruction processes such as collisional heating or UV irradiation could, in principle, be included, but at a high
cost in the required computing time. Both mechanisms are not expected to be very efficient as grains are not easily destroyed,
because they are very effective radiators (emitting all heat immediately in the infrared) and the shocks discussed here are
not very strong, and the radiation field would not be strong enough to destroy grains. Furthermore, dust can also be created
in the high density shock regions, thus lowering the effective dust destruction rate. Further details are given in
\citet{2011ApJ...734L..26V}.

The input and grid parameters are summarised in Table \ref{tb-hdsims}. We start with a basic model (simulation~A), which has
input parameters based on the observations of $\alpha$-Orionis \citep{2008PASJ...60S.407U}.  Using this model as a starting
point, we vary individual parameters to investigate the effect on the morphology of the bow shock. Note that for a stationary
star, $v_\star = 0$~\kms, the shock between the wind and ISM will drive a spherical shell of wind material sweeping up the ISM
material, thus leading to detached shells / rings (not shown).

\subsection{The morphology of the bow shock}

As shown in Figs.~\ref{fig:basic} through~\ref{fig:highTism}, the collision between the expanding wind and the moving (in the
rest-frame of the star) ISM creates a bow shock.  The location and morphology of the shocked gas depend on the exact input
parameters. These figures show the gas density and dust particle density in the circumstellar environment after the bow shock 
has reached its (semi-) permanent equilibrium distance from the star. Typically, this takes between 50\,000 and 150\,000\,years,
depending on the parameters of the simulation. 
Our basic model (Simulation~A, Fig.~\ref{fig:basic}) shows a standard bow shock morphology: a forward shock separating the 
unshocked and shocked ISM, a wind termination shock that separates the free-streaming wind from the shocked wind, and between 
them a contact discontinuity separating the shocked wind from the shocked ISM, with the temperatures in the shocked wind typically
lower than in the shocked ISM due to the higher density of the former. 
This pattern is repeated in all subsequent models, with the exception of the fast moving star (Simulation~E, Fig.~\ref{fig:faststar}) 
and the high ISM temperature cases (simulation~G, Fig.~\ref{fig:highTism}), both of which show a chaotic bow  shock with strong 
turbulent behaviour. 

Behind the star a relatively structureless -- empty -- region remains. This is only filled by the free-streaming wind of the
star.   Note that in our simulation the wind velocity is always less than the stellar velocity. Exceptions occur if the shocked
gas region has a high thermal pressure, which pushes it into the empty region  (Simulation~C, Fig.~\ref{fig:lowdm} and
simulation~G, Fig.~\ref{fig:highTism}), or if the shocked gas is very turbulent  (Simulation~E, Fig.~\ref{fig:faststar}).

The location of the shock can be approximated by the wind and ISM conditions as described in Eq.~\ref{eq:standoff}. In
Table~\ref{tb-hdsims} we give the location of the contact discontinuity ($R_0$) as well as the location of the  forward shock
of the shocked-gas region ($R_s$), where the ISM transition from unshocked to shocked gas is located. In some cases $R_0$ has to be
approximated since instabilities can make it a time-dependent value. For simulation A (the basic model), the analytical
approximation puts the contact discontinuity at 0.42~pc.  In our simulation the position of the contact discontinuity 
deviates from this prediction, lying at about 0.26~pc (while the forward shock lies further ahead at $\sim$0.35~pc).  
The difference can be explained as a result of radiative energy loss in the
shock.  The scaling, as described by Eq.~\ref{eq:standoff}, appears to be quite accurate.  For example, for an order of
magnitude decrease in ISM density (simulation~F, Fig.~\ref{fig:lowism}) or increase/decrease of mass-loss rate (simulation~B,
Fig.~\ref{fig:highdm}), the location of the bow shock moves away/closer with a factor of about $\sqrt{10}$. 

The dust grains, which start with the same velocity as the wind, are initially carried along.  When the wind reaches the
termination shock, the gas slows down abruptly.  Because they are not subject to the shock, the dust grains keep moving, but are
now subject to drag force due to the  difference in velocity with the gas. This causes them to slow down over time.  As a
result they tend to pile up at the contact discontinuity (\citealt{2011ApJ...734L..26V}).  Once the grains reach the contact
discontinuity, they tend to follow the local instabilities and are eventually carried downstream.  Those grains, which crossed
the contact discontinuity, will usually go downstream faster, since in most of our simulations the shocked ISM has a  higher
velocity than the shocked wind. The influence of the grain size on the gas-dust interaction is investigated in
\citet{2011ApJ...734L..26V}.

\subsection{Instabilities in the bow shock}

In most of our simulations, instabilities occur along the contact discontinuity. These instabilities consist of a combination
of RT and KH effects (see above). In our case, the (relatively) low-density-shocked ISM exerts a force on the denser shocked
wind, which leads to the formation of RT fingers. As a result, the wind material starts to flow into the shocked ISM. Here the
wind material is subject to a sheer-force due to the relative motion of the star with respect to the ISM. These
instabilities are enhanced by the dust (albeit this is a secondary effect), which tends to continue in a straight line in an attempt
to cross from the shocked wind into the shocked ISM, dragging the gas along with it. This is best observed in the case of
simulation~B (Fig.~\ref{fig:highdm}). The velocity difference between the shocked gas layers causes KH instabilities to
develop wherever the RT instabilities cause a local displacement of gas. These lead to the characteristic ``cyclonic''
features (see Figs.~\ref{fig:basic}, \ref{fig:lowdust} and \ref{fig:highTism}). The presence of dust slows down the formation
of KH instabilities, since the circular motion of the KH instabilities has difficulty overcoming the inertia of the dust
grains. Therefore, they are most visible in simulation~D (Fig.~\ref{fig:lowdust}). Also, the KH instabilities can form more
easily if the low density region (the shocked ISM) is extended, so that the instabilities do not hit the forward shock
(simulation~G, Fig.~\ref{fig:highTism}).

In the case of the star with a low mass-loss rate (simulation C, Fig.~\ref{fig:lowdm}), such instabilities are absent. The low
mass-loss rate leads to a reduced density in the shocked wind. As a result, the density difference between the shocked wind and
the shocked ISM is small, which reduces the RT effect. Also, due to the low density, the radiative cooling, which scales with
the density squared, is less effective \citep{vanMarle201144}. This is important because radiative cooling tends to favor the
formation of small, high density clumps, which in turn can  serve as a start of other instabilities.  Because the shocks are
nearly adiabatic, the shocked-gas region is wide and its thermal pressure pushes material into the area behind the star, rather
than leaving it empty. 

The fast moving star (simulation~E, Fig.~\ref{fig:faststar}) has a far more irregularly shaped bow shock. In this case the
shocks on both sides are highly radiative, leading to a very compressed shocked gas region. This, combined with the strong
ram-pressure from the ISM, causes the entire shell to become unstable, leading  to a ragged form which will change
considerable over time as local instabilities grow to a size where they dominate the entire structure of the shell. This
turbulent motion also causes the gas (and the dust grains) to end up behind the star.

\section{Discussion}\label{sec:discussion}

\subsection{Comparison between observed and hydrodynamical bow shock morphology}

The simulations presented in Sect.~\ref{sec:hydro} only cover part of the parameter space spanned by all stellar,
circumstellar and interstellar properties. In particular, the simulations focus on the effects of increasing the star's
peculiar velocity with respect to the ISM, changing mass-loss rate, increasing the ISM temperature and lowering the ISM
density.

One evident issue revealed by the simulation is that the shocked gas region can have considerable spatial extent
(this occurs when the shocks are adiabatic - as in most cases presented here - as opposed to radiative). In the
approximation  of efficient cooling, the shock region will be thin and unresolved; however, the simulations show that this is
not necessarily valid for all cases and can lead -- depending also on the exact cooling law -- to extended interaction zones.
Depending on the exact morphology and the temperatures in the shocked gas region, one could -- theoretically -- be observing
either the forward shock, reverse shock or contact discontinuity. However, the current far-infrared observations are most
sensitive to cold dust grains. Due to the limited spatial resolution, it is not clear which region is actually represented
by the far-infrared emission. It may represent the contact discontinuity (\ie\ the entire unresolved shocked region), the
bow shock, or the termination shock (\citealt{2007MNRAS.380.1161L}). Our simulations in Sect.~\ref{sec:hydro} show that most
dust grains pile-up at the contact discontinuity where they tend to follow the local instabilities which carry the grains
downstream. Thus, the simulations suggest that  the observed far-infrared dust emission primarily traces the contact
discontinuity. On the other hand, if grains are destroyed in the shock they will be found in the unshocked stellar wind region
and thus far-infrared dust emission would delineate the termination shock (and the contact discontinuity if the shock region
is physically thin enough). This might prove useful in explaining the spatial offset between observed UV emission and
far-infrared emission in the bow shock of CW\,Leo (\citealt{2010A&A...518L.141L}). Further work on the destruction and formation 
of dust grains (c.q. alteration of the dust size distribution) in shocks is warranted. As stipulated above, the simulations do not
include radiative transfer and can thus not provide appropriate dust temperatures needed to simulate the infrared emission.

\begin{table}[t!]
\centering
\caption{Comparison between observed stand-off distances derived from Herschel/PACS maps and previous studies with IRAS, 
Spitzer \& AKARI. The predicted values (this work) are given as a function of $n_H$ (cm$^{-3}$). The local interstellar ISM density, $n_H$,
is estimated from the predicted $R_0/\sqrt{n_H}$ and the observed $R_0$ and compared to $n_H$ given by Eq.~\ref{eq:nH2}.}
\label{tb:compareR0}
\resizebox{\columnwidth}{!}{
\begin{tabular}{lllllll}\hline\hline
Object		& \multicolumn{3}{c}{De-projected $R_0$}			      &     Ref     & \multicolumn{2}{c}{$n_H$}	\\ \cline{2-4} 
		& \multicolumn{2}{c}{this work}	   		& literature	      &  	    & \multicolumn{2}{c}{(cm$^{-3}$)} \\ \cline{2-3}\cline{4-4} \cline{6-7}
		& observed (\arcmin)&  predicted (\arcmin)	& (\arcmin) 	      &  	    &   Eq.~\ref{eq:standoff}     &   Eq.~\ref{eq:nH2}	\\ \hline
$o$\,Cet\tablefootmark{a}& 1.2  &  0.7/$\sqrt{n_H}$          	& $\sim$3      	      &     (1)	    &	 0.4 	  &    1.1 	\\
$\alpha$ Ori	& 5.0           &  10.6/$\sqrt{n_H}$           	& $\sim$4             &     (2)	    &	 4.6	  &    1.9	\\
CW Leo		& 6.6           &  3.9/$\sqrt{n_H}$           	& 5.9        	      &     (3)	    &	 0.3	  &    0.7	\\
R Hya		& 1.6           &  2.4/$\sqrt{n_H}$         	& 1.6 $\pm$ 0.1       &     (4)	    &	 2.5	  &    0.8	\\ 
R Cas		& 1.5       	&  2.2/$\sqrt{n_H}$           	& 1.4 $\pm$ 0.1       &     (5)	    &	 2.1	  &    1.8	\\
\hline
\end{tabular}
}
\tablebib{References: (1) \mbox{\citealt{2008ApJ...685L.141R}}; (2) \mbox{\citealt{2008PASJ...60S.407U}}, Decin et al. in preparation; 
(3) \mbox{\citealt{2010A&A...518L.141L}}, \mbox{\citealt{2010ApJ...711L..53S}}; (4) \citet{2006ApJ...648L..39U}; (5) \mbox{\citealt{2010A&A...514A..16U}}}
\tablefoot{
\tablefoottext{a}{Note that for $o$\,Cet the observed features delineate most likely the termination shock and not contact discontinuity, this measure is thus a lower limit to
$R_0$, which will be larger than 1.2\arcmin (\eg\ more like 2\arcmin\ as observed in the UV) leading to a lower density of $n_H = 0.2$~cm$^{-3}$.}
}
\end{table}

\subsubsection*{RT and KH instabilities}

The presented PACS infrared observations reveal RT and KH instabilities in astrophysical bow shocks. RT fingers can be seen in
the shock region of R\,Scl. KH ``wiggles'' or density knots are more common, and can be seen in the bow shock regions of, for
example, UU\,Aur, R\,Hya, X\,Pav, EP\,Aqr, $\mu$\,Cep, R\,Leo, RT\,Vir, X\,Her, and V1943\,Sgr.

Almost spherical bow shocks spanning a large azimuthal angular range correspond closest to simulations A, B and F,  the nominal, high
\Mdot\ and low $n_\mathrm{H}$ cases, respectively. For low $n_\mathrm{H}$ (simulation~F) the KH instabilities are more pronounced, while
the RT instabilities are more evident in the high \Mdot\ case. The smooth, spherical bow shocks of $\alpha$\,Ori,  UU\,Aur, and X\,Pav are
similar to the morphology of the  simulation~A (nominal case; Fig.~\ref{fig:basic}). UU\,Aur and X\,Pav show back flow emission also seen
in the simulation. The ``wiggles'' in the bow shocks of R\,Hya and R\,Leo correspond to those seen in simulation~F (Fig.~\ref{fig:lowism})
which contradicts the high ISM density derived for R\,Leo (see Sect.~\ref{subsec:comparison}). Both X\,Her and V1943\,Sgr have high space velocities leading to morphologies
much like that seen in simulation~E (Fig.~\ref{fig:faststar}).

Further comparisons are hampered on the one hand by the (unknown) projection effects in the observations and on the other hand
by the, currently, limited parameter space covered with the few simulations at hand. A more extensive, multi-parameter grid of
simulations and/or a detailed modeling of individual cases will be required to match observations with simulations to confirm
whether or not the latter accurately predict the former.

\subsection{Proper motion and inclination}

In most cases the proper motion vector (if known) points roughly (within $\sim$20\degr-30\degr)  in the direction of the
observed (Class\,I) bow shock apex. Exceptions are $\mu$\,Cep and $o$\,Cet. 
One possible explanation is that, like $o$\,Cet, $\mu$\,Cep is also a long-period (tens of years) binary, for which
the present-day proper motion is not pointing in the same direction as it was when the bow-shock matter was expelled.
This conjecture needs to be further tested by further observation and modeling of these systems and their observed bow shocks.

For the majority of the Class\,II objects, the proper motion direction is aligned roughly along the symmetry axis
(perpendicular to the arc-star-arc line). Only for V\,Pav the proper motion points to the north-west arc. For Class\,IV the
proper motion direction coincides with the location of extended irregular emission. The proper motion for Class\,III objects
is not consistently aligned with, for example, shell features. In a few cases (TT\, Cyg, U\,Ant) there is some stronger shell
emission in the direction of the proper motion, but this is not observed for the other detected rings.

There is no clear indication for a dependency on the inclination angle, $i$. All five classes include sources with a range of
inclinations between almost 0\degr\ up to 60-70\degr. Except for VY\,UMa ($i$=-1\degr), there are no objects with small
inclination angles among the \emph{``eyes''} (all have $i \geq 28$\degr), whereas the \emph{``fermata''}, \emph{``rings''} and
\emph{``irregular''} all have a more uniform distribution over inclination angles. However, we stress that at this point the
sample is too small to derive any further conclusions.

For about one third of the \emph{``non-detection''} class, the proper motions are not known and thus their space velocities
are, at best, lower limits based on radial velocities. This is in part a distance bias. The objects with non-detections are on
average further away and consequently their proper motions have been too small to be measured. This means also that the
predicted stand-off distances are upper limits.

\subsection{Binary interaction, circumstellar chemistry, and magnetic fields?}

\subsubsection*{Binary interaction}

There is little evidence for binary interaction in our sample. Even though for several stars in our sample, there is solid
evidence for  close companions (see Tables~\ref{tb:properties1} to~\ref{tb:properties2}), none show suspicious features in
their bow shock interfaces. However, many of these appear in the category  \emph{``irregular''}, which could indicate that
such wind-wind-ISM interaction is only visible on small spatial scales.  It is difficult to arrive at any firm conclusion as
the number of binaries in the survey is low. Excluding the optical ``visual'' binaries, as these are not yet confirmed, 
out of the 25 objects in the \emph{``fermata''} class only one is a known binary (Mira) which indeed reveals a peculiar shape  of the interaction zone
(Mayer et al. 2011). Three others, W\,Aql, EP\,Aqr and $\theta$\,Aps are solid candidates for binarity. 
None of  the \emph{``rings''} are binaries.  Noteworthy, 3 out of 6 of the \emph{``irregulars''} are
confirmed binaries, and 2 of the 6 \emph{``eyes''} are potentially binaries. Finally, also 6 out of 30 non-detections are
binary systems. Including visual binaries adds 3 binaries to each of the \emph{``fermata''}, \emph{``eyes''}, and \emph{``non-detection''} 
classes, and 4 binaries to the \emph{``rings''}.
In this case, 5 out of 7 \emph{``eyes''} show evidence for binarity, whether this means there is a connection between binarity
and the \emph{``eyes''} morphology can not be confirmed nor excluded at this point.

\subsubsection*{Oxygen-rich versus carbon-rich chemistry}

One key property of the AGB stars and red supergiants in our sample is their circumstellar envelope chemistry.
This can be either carbon-rich (the dust is predominantly made up of amorphous carbon, graphite and silicon
carbide) or oxygen-rich (the dust is composed of silicates and oxides). Although we do not find a strong
dependence on the presence versus absence of bow shocks and rings with respect to chemistry, there does appear to
be a distinction between the shape of the extended emission found around carbon and oxygen-rich objects. The
\emph{``fermata''} and \emph{``irregular''} classes include a high fraction, 22 out of 31, of O-rich stars. The
\emph{``eyes''} and \emph{``rings''}, on the other hand, include only few, 3 out of 22, O-rich stars.
It seems that at least the \emph{``ring''} structure is typical of C-rich chemistry, as already pointed out by
\cite{2000A&A...353..583O}. 
A further argument favoring such a link between \emph{``ring''} structure  and C-richness is that \emph{all}
\emph{``fermata''} stars also displaying  a \emph{``rings''} structure are C-rich stars (they are noted with an 
asterisk in Table~\ref{tb:properties1}, and are also listed in Table~\ref{tb:properties2}).

Could the circumstellar envelope structure be related to chemistry through differences in mass loss rates?
It has been shown (\citealt{1985ApJ...293..273K}) that oxygen-rich stars  have a wider range of
mass-loss rates than C-rich stars for which the \Mdot\ is globally more homogeneous. On average, mass loss rates 
of C stars are higher than those of MS and S stars (\citealt{2010A&A...513A...4G}).  Apart from the fact that the 
mass-loss rate standard deviation of O-rich stars  is indeed found to be somewhat larger than that of C-rich stars, no
clear dichotomy emerges between the mass loss rates of O-rich stars and C-rich stars  in the present sample.

Furthermore, the higher emissivity (\citealt{1984ApJ...285...89D,2001ApJ...554..778L}) and lower gas-to-dust
ratio of 160 for oxygen-rich dust compared to the higher ratio of 400 for carbon-rich dust 
(see \eg\ \citealt{1985ApJ...293..273K, 2005A&A...439..171H}) will, for O-rich dust, give rise to brighter infrared emission for a given
total dust and gas mass. One could then argue that the O-rich stars mass loss is more easily  detected even with
a patchy structure. This could perhaps explain why O-rich stars are detected preferentially among both \emph{``irregulars''}
and \emph{``fermata''} classes.

An alternative explanation is that the \emph{``Ring''} (and possible the \emph{``Eyes'')} morphology, are
explained through (i) the interaction between a fast wind sweeping out matter from a previous slower
wind, or (ii) a phase of drastically enhanced mass loss, \eg\ caused by a thermal pulse. 
If this explanation holds, and if one assumes that most thermal pulses events are followed by a third dredge-up
(TDUP) episode, then the discriminant character would not be O-rich or C-rich,  but instead pre-TDUP or post-TDUP
(\ie, thermally-pulsing (TP) AGB) stars). All intrinsic S-type stars, C-type stars and technetium-rich M stars in
Tables~\ref{tb:properties1} and~\ref{tb:properties2} are TP-AGB stars. In fact, all \emph{``Ring''} stars are TP-AGB stars.

\subsubsection*{Stellar and interstellar magnetic fields}

Some wind-ISM models include the effects of an interstellar magnetic field on the shaping of the bow shock region.
\citet{1980MNRAS.191..761H} predicted that the ISM magnetic pressure could deform the outer shells of very large spherical
halos of PNe into \emph{``lemon''} shapes, with the axis of symmetry inclined to the ordered ISM magnetic field direction.
\citet{1997ApJ...484..277S} and \citet{1998RMxAC...7..149D} elaborated on this scenario predicting different bow shock shapes
depending on the relative magnitudes of the  star's space velocity, $v_\star$, the wind velocity, $v_w$, and the Alv\'en
speed, $v_A$. The double arc objects (Class\,II) such as VY\,UMa and AQ\,Sgr could perhaps provide evidence for this scenario,
although both are at relatively high latitudes and their symmetry axes are not aligned with the Galactic plane (assuming that
at these locations the magnetic field is also parallel to the Galactic plane).  It is also possible that the star has its
own magnetic field and thus may exhibit axis-symmetric mass loss. Preliminary simulations (private communication A.-J. Van Marle) 
indicate that double arcs as well as jets (tentatively identified for VY\,UMa and W\,Hya) could be formed by a (rotating) 
star with a magnetic field.

\subsection{Detached spherical shells} 

The observed detached shell objects (Class\,III) show quite a range of expansion ages calculated out of their measured or assumed
expansion velocities, distances and angular sizes. Ranging from the youngest shells like U\,Cam or R\,Scl with ages of the order
of 1000 years and much older ones like AQ\,And, Y\,CVn or UX\,Dra with ages of a few 10\,000 years we see also a significant trend
in the detectability in fossil mm-CO. Whereas all objects with shell ages below some 1000 years are prominent in CO, no detached
shell older than 10\,000 years was successfully detected yet. This could be understood in terms of photo-dissociation by the
interstellar radiation field in the older extended shells with sizes of 0.2~pc or more (\citealt{2010A&A...518L.140K}). 
\citet{2011ApJ...729L..19A} find that the detached shell of U\,Ant detected with PACS is not the outermost one found with AKARI.
Thus, the shell detected with PACS is either a density enhancement due to a two-wind interaction or it could represent the
termination shock, similar to the scenario proposed for Y\,CVn by \citet{2007MNRAS.380.1161L}.

Another interesting finding with respect to Class\,III objects are the objects that have both a detached shell and a bow shock
interaction region, with R\,Scl being the showcase object. Seeing the bow shock far out off the detached shell more or less excludes 
one of the explanations of detached shells, namely their interpretation as being the wind-ISM interface
(\citealt{2007MNRAS.380.1161L}). 

\begin{figure}[htp!]
\centering
\includegraphics[angle=-90,width=\columnwidth]{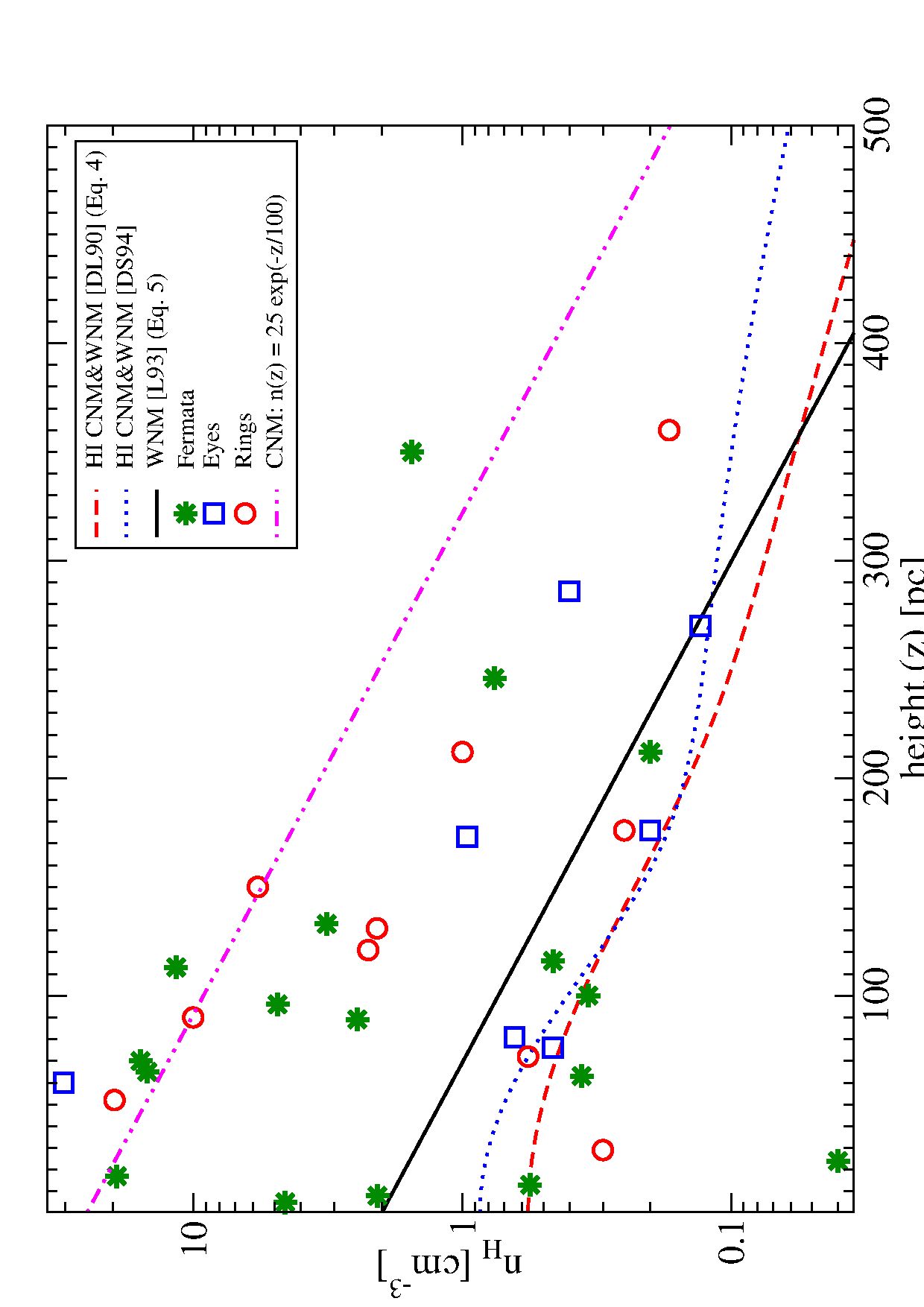}
\caption{ISM density derived from the observed stand-off distance $R_0$ as a function of the height above the Galactic plane, $z$.
Different density scaling height models are also shown (\eg\ Eqs.~\ref{eq:nH1} and~\ref{eq:nH2}).
References: [DL90]: \citet{1990ARA&A..28..215D}; [DS94]: \citet{1994ApJ...427..274D}; [L93]: \citet{1993A&AS...99..291L}.}
\label{fig:z_nH}
\end{figure}

\subsection{Comparison between predicted and observed $R_0$: Implications for the ISM density}\label{subsec:comparison}

For a number of objects in our sample, previous studies have reported on the detection of wind-ISM bow shocks with IRAS, Spitzer
or AKARI.  The measured deprojected angular stand-off distance obtained with PACS agree very well with earlier results
(Table~\ref{tb:compareR0}). The predicted values for $R_0$ (as  function of $\sqrt{n_H}$) give ISM densities that are somewhat
higher than those computed from Eq.~\ref{eq:nH2}. For $\alpha$\,Ori, \citet{2008PASJ...60S.407U} predict $n_\mathrm{H}$ = 1.5 --
1.9~cm$^{-3}$ which is indeed in line with $n_\mathrm{H}$ = 1.9~cm$^{-3}$ derived from Eq.~\ref{eq:nH2}, but
higher than that inferred from the stand-off distance, $n_\mathrm{H}$ = 4.6~cm$^{-3}$ (Table~\ref{tb:compareR0}).
\citet{2010A&A...518L.141L} derive $n_H \geq 2$~cm$^{-3}$ for CW~Leo, which is a factor of 2 higher than derived from both 
our predicted stand-off distance as well as from Eq.~\ref{eq:nH2}
($n_\mathrm{H}$ = 0.5 -- 1.0~cm$^{-3}$). 

Adopting the values for the different parameters given in Tables~\ref{tb:properties1} to~\ref{tb:properties2} yields $R_0$ less
than 8\arcmin\ for all objects with detected bow shock interaction. For all detected wind-ISM interaction objects, the
predicted stand-off distances range from about 0.02 to 0.9~pc, in line with the observed de-projected values. This gives some
credibility in the predicted stand-off distances (from Eq.~\ref{eq:standoff}) for the objects in Class\,X (\emph{``non-detections''}). 
These results suggest that additional bow shocks should have been detectable for other stars within 300~pc, in particular for R\,Dor,
Y\,Lyn, RS\,Cnc, and possibly HD\,100764 and RX\,Boo.
We note that higher sensitivity observations could possibly still reveal the expected extended emission.

\begin{figure}[t!]
 \centering
 \includegraphics[angle=-90,width=\columnwidth]{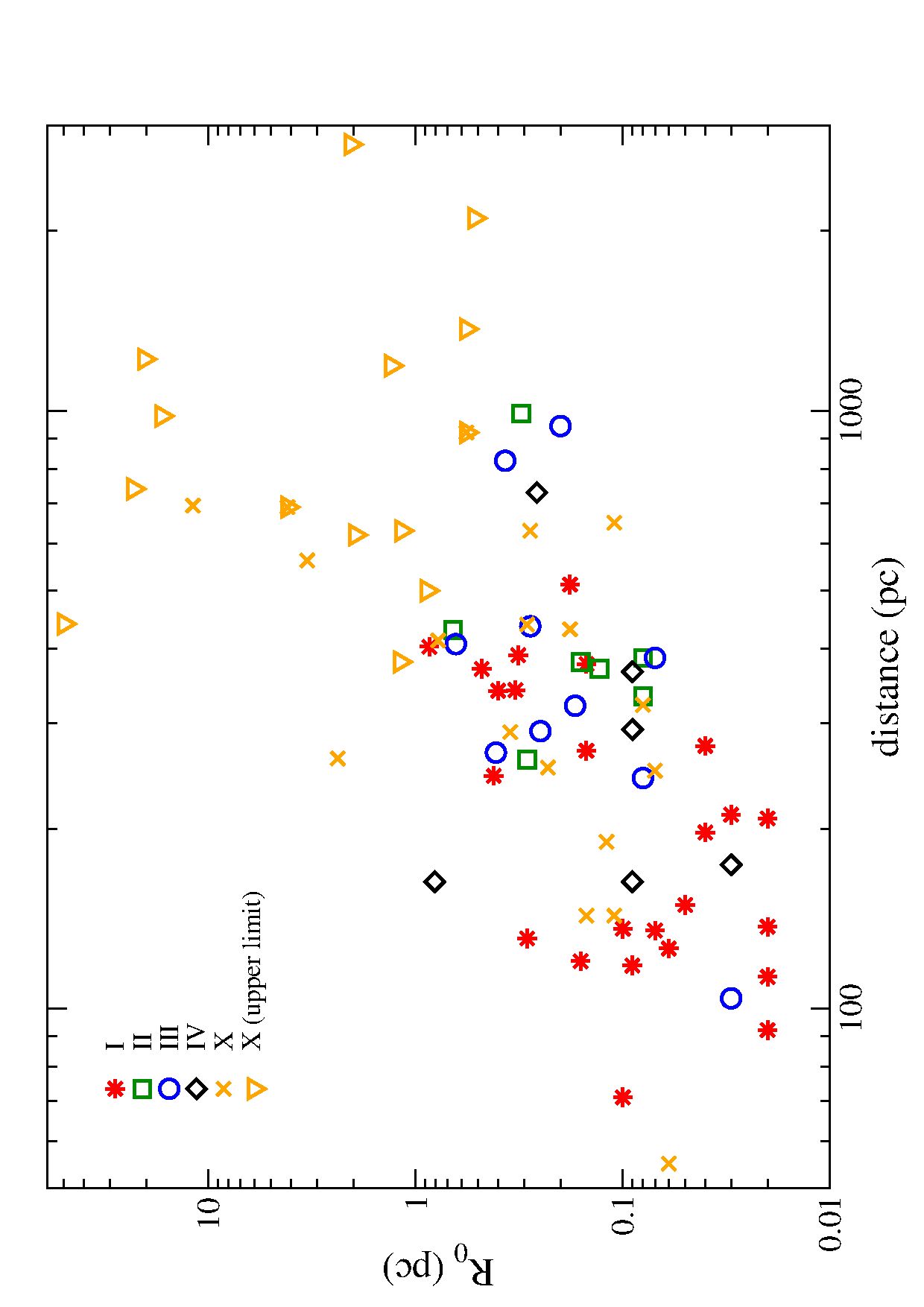}
 \includegraphics[angle=-90,width=\columnwidth]{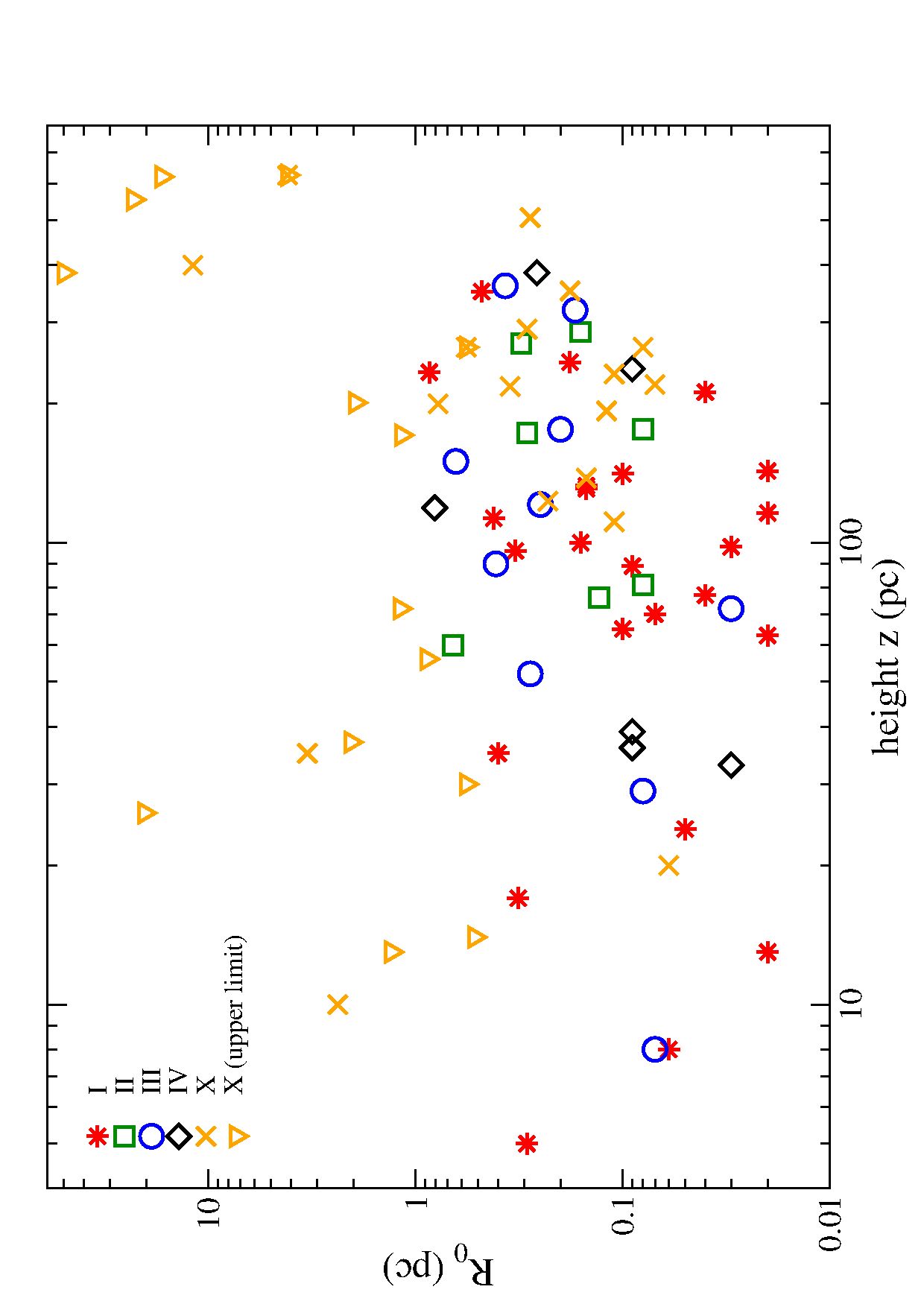}
 \caption{Predicted stand-off distance as a function of distance (top) and as a function of height above
  the Galactic plane, $z$ (bottom). All in units of parsec.
  Classes: I (stars), II (squares), III (circles), IV (diamonds), and X (crosses \& triangles).}
 \label{fig:standoff-distance}
\end{figure}

The uncertainty of the predicted stand-off distance depends on the different stellar and ISM parameters. The space velocity,
$v_\star$ (for nearby stars), as well as the  wind velocity, $v_w$ are well determined observationally with small ($\sim$10\%)
uncertainties. The uncertainties in the mass-loss rates can be an order of magnitude, in particular for the targets with
inaccurate distance estimates (Sect.~\ref{sec:properties})  which introduces errors of a factor of three in $R_0$. Finally, the
local ISM density is difficult to derive observationally. Without knowledge of the exact phase of the ISM the star is traversing,
estimates of the density could be off by several orders of magnitude (see the discussion on densities, scale heights, and
filling factors in Sect.~\ref{sec:properties}).  Table~\ref{tb:properties1} lists $n_\mathrm{H}$ derived from equating the
observed de-projected $R_0$ with the theoretical $R_0$, taking $n_\mathrm{H}$ as the unknown. The comparison of
$n_\mathrm{H}$ derived from Eq.~\ref{eq:nH2} with that derived from the observed and predicted $R_0$ shows that these generally
agree within a factor of three (Fig.~\ref{fig:z_nH}). In a few cases, the derived ISM density is an order of magnitude higher ($n_H
\approx 5$ to 35~cm$^{-3}$) than given by Eqs.~\ref{eq:nH1} and \ref{eq:nH2}. Possibly, for these cases, the star is moving through
a diffuse cold medium (CNM) with typical densities $n_H \sim 10-100$~cm$^{-3}$, an order of magnitude higher than those of the WNM
(Sect.~\ref{sec:properties}). Alternatively, this apparent discrepancy could also be resolved if the ISM itself has a peculiar
flow velocity of $v_\star$-$v_{\rm ISM}$  with respect to the local standard of rest. If the peculiar velocity between the two
media is higher, the density required to arrive at the same stand-off distance is lower, since $v_\star \propto 1/\sqrt{n_H}$. On
the other hand, the derived ISM density also scales linearly with the adopted \Mdot, thus an order of magnitude over-estimate of
the mass-loss rate leads to the same order of magnitude over-estimates of $n_\mathrm{H}$. This may explain the high $n_\mathrm{H}$
for some objects with high adopted \Mdot, such as W\,Aql, but not easily for others, such as R\,Leo, which  have already low
values for \Mdot. For R\,Leo, the turbulent features in the bow shock indicate a higher space velocity which is at odds  with the
observed $v_\star = 15$~\kms. If indeed the relative star-ISM space velocity for R\,Leo is higher, the inferred ISM density will
be correspondingly lower (now $n_H \propto v_\star^{-2}$).  For example, for a much higher peculiar velocity of 45~\kms\  (more in
line with the bow shock morphology), $n_H = 1.6$~cm$^{-3}$ which is  close to $n_H = 1.1$~cm$^{-3}$ given by Eq.~\ref{eq:nH2}. In
Sect.~\ref{sec:properties} we roughly estimated the LSR velocity of the local ISM and found that the corrections to $v_\star$ are
small ($\leq$~20\%)  for most objects, with a few exceptions. Only for two cases do these higher relative velocities lead to much
smaller values for the ISM density;  $n_\mathrm{H}$ = 10.1 and 13.5~cm$^{-3}$ for $\mu$\,Cep and U\,Cam, respectively. The ISM
velocity correction does not alter the high densities obtained for \eg\ R\,Leo, W\,Aql and EP\,Aqr.
Another explaination could be that this `denser' medium is a remnant of an earlier, slower stellar wind.
How and if the (tentative) binarity of both W\,Aql and EP\,Aqr plays a role is as of yet not clear.

\subsection{Presence versus absence of wind-ISM bow shocks}

In the previous sections, we have established different morphological classes of bow shocks and examined the basic physics
giving rise to a bow shock and discussed the various parameters affecting their size and shape. In this section we
explore different properties of the stellar objects as well as the ISM in order to understand which conditions are required to
be able to detect bow shocks and/or detached shells around AGB stars and red supergiants.

For 50 out of 78 (63\%) AGB stars and red supergiant, we detect detached extended far-infrared emission suggestive of a bow
shock, detached ring or irregular extended emission. Limiting the sample to nearby objects ($d \leq 500$~pc) eliminates many
uncertainties and the ``detection-rate'' improves to 43 out of 56 objects, or 78\%. Restricting the distance further to, for
example, $d \leq 300$~pc, only marginally improves the detection rate to 28 out of 34 (80\%).

\begin{figure}[t!]
 \centering
 \includegraphics[width=0.8\columnwidth,clip]{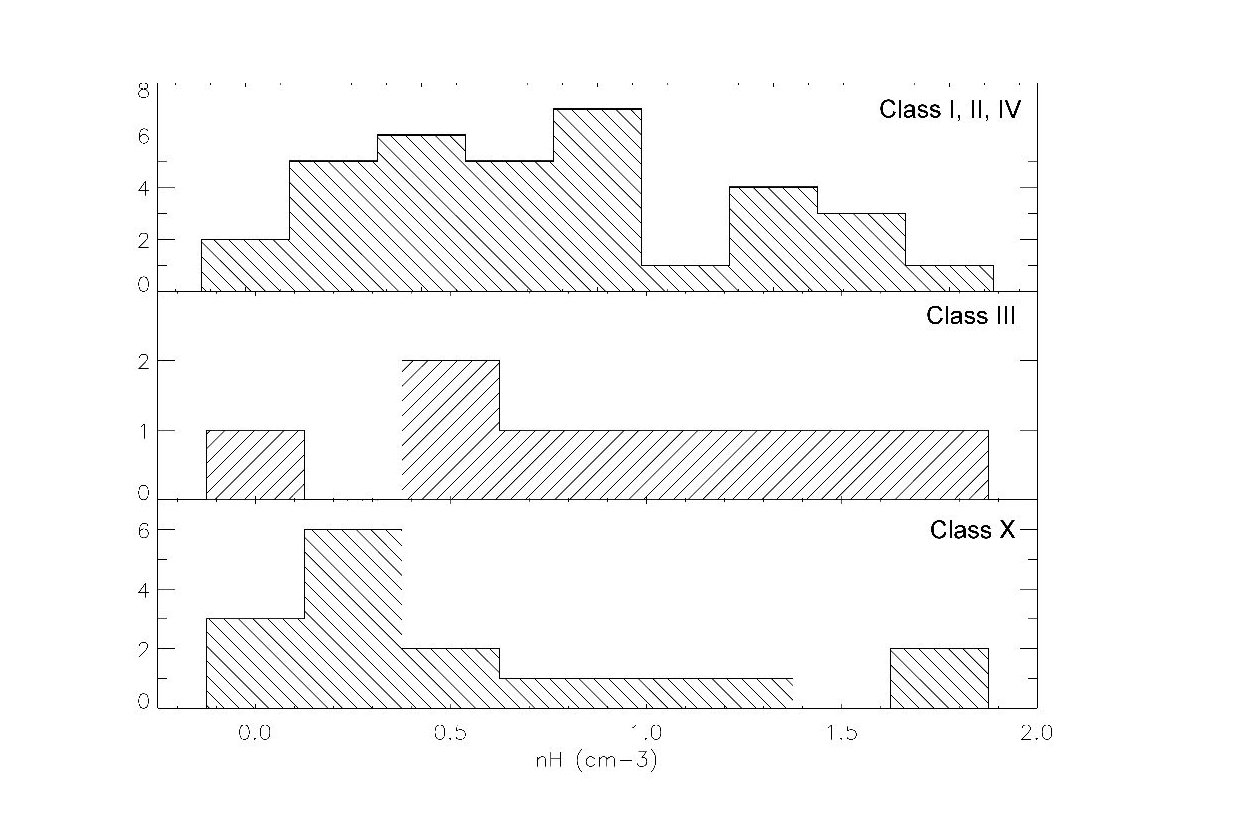}
 \caption{Histogram of the ``predicted'' ISM density, $n_\mathrm{H}$, for the \emph{``detected''} Class\,I, II, IV (top panel)
  and Class\,III (middle panel) \emph{``non-detection''} Class\,X (bottom panel) objects.}
  \label{fig:histo_nH}
\end{figure}

\begin{figure}[t!]
 \centering
 \includegraphics[width=0.8\columnwidth,clip]{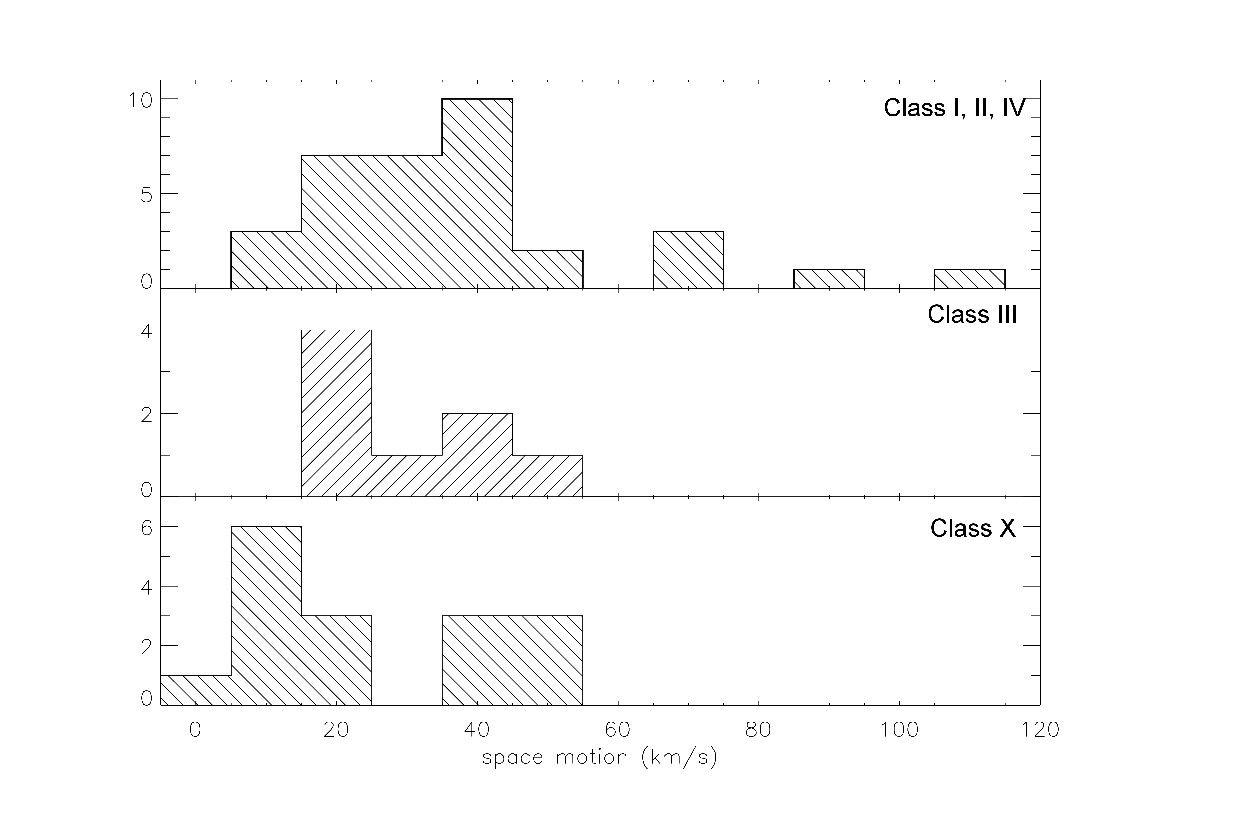}
 \caption{Histogram of the space velocity, $v_\star$, for the \emph{``detected''} Class\,I, II, IV (top panel)
  and Class\,III (middle panel) \emph{``non-detection''} Class\,X (bottom panel) objects.}
  \label{fig:histo_vstar}
\end{figure}

\begin{figure}[t!]
 \centering
 \includegraphics[width=0.8\columnwidth,clip]{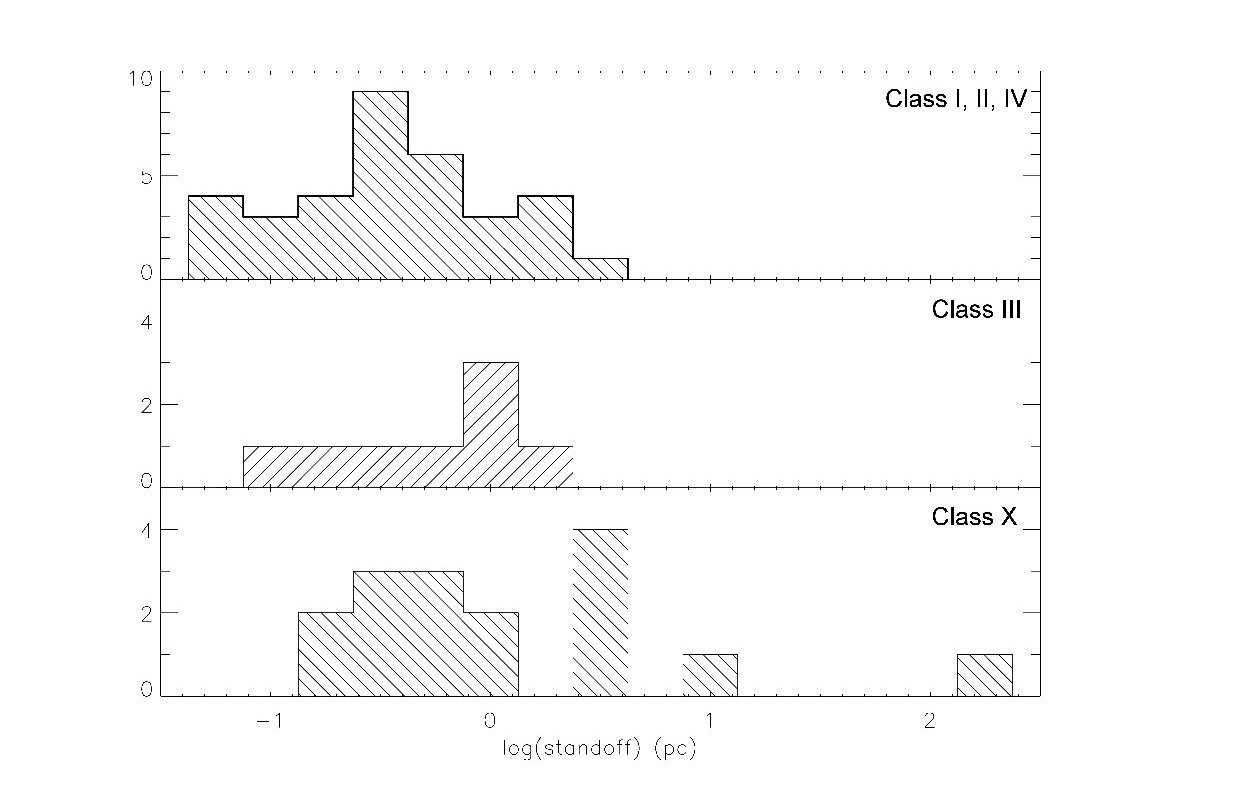}
 \caption{Histogram of the standoff distance, $R_0$ (pc), for the \emph{``detected''} Class\,I, II, IV (top panel)
  and Class\,III (middle panel) \emph{``non-detection''} Class\,X (bottom panel) objects.}
  \label{fig:histo_standoff}
\end{figure}

To test whether the ``detection'' and ``non-detection'' samples have similar or statistically different distributions as a
function  of $d$, $z$ or $n_\mathrm{H}$, $v_\star$, $i$, \Mdot, and $R_0$, we use the two-sample Kolmogorov-Smirnov ($K$-$S$) test.
The results are given in Table~\ref{tb:kolmogorov}. In order to eliminate the effect of large uncertainties in the distances, we
perform the $K$-$S$ test on a distance limited ($d \leq 500$~pc) sample.   The $K$-$S$ probabilities of ``detections'' versus
``non-detections'' are particularly low (and thus indicative that the two samples have different distributions) for 
$n_\mathrm{H}$/$z$, $v_\star$, and $R_0$.  
The corresponding distribution histograms for the most distinctive parameters, 
$n_\mathrm{H}$, $v_\star$, and $R_0$, are shown for the different classes (detection/non-detection) in 
Figs.~\ref{fig:histo_nH} to~\ref{fig:histo_standoff}.
This points towards the scenario that the presence of bow shocks is strongly dependent
on the stellar velocity (relative to the local medium), the local ISM density and  the resulting standoff distance.  The latter
sets the size of the bow shock region and is apparently determined predominantly by $v_\star$ and $n_\mathrm{H}$ and not so much
by the star's mass-loss properties (cq. evolutionary phase).

All objects with observed wind-ISM interaction zones (Class\,I to IV) have $R_0$~$\leq$~1~pc, while many class X sources have
$R_0$~$\geq$~1~pc. The distribution of the observed de-projected $R_0$ values is very similar to the distribution of
predicted $R_0$  for sources in Class\,I to~IV. However, a quantitative comparison between the predicted and observed
stand-off distances does not reveal a strong correlation. 

Irrespective of the absolute distance (and thus apparent angular stand-off distance) it appears that a necessary, but not sufficient, 
condition for the detection of bow shocks and rings is that their physical size has to be smaller than about 1~pc.
This points towards a physical effect  (\eg\ reduced surface brightness) instead of an
observational limit, although it does not explain the absence of bow shock emission around nine of the observed nearby AGB stars 
(with known $v_\star$ and $R_0 < 1$pc) included in Class\,X.

Concluding, we find that bow shocks are detected and predicted to occur for most nearby objects ($<$~500~pc) as well as for objects whose 
stellar and local ISM properties yield relatively small stand-off distances ($<$~1~pc). For more distant objects the detection is (likely)
hampered by lower sensitivity and lower spatial resolution, though we could not find any reasons to suggest that these objects
would not have bow shock interaction. Indeed, several distant objects such as AQ\,And ($\sim$800~pc), S\,Cas ($\sim$940~pc),
RZ\,Sgr ($\sim$730~pc) reveal evidence for wind-ISM interaction, but these detections might have been fortuitous.

\begin{table}[tp!]
\centering
\caption{Two-sample Kolmogorov-Smirnov test applied to different stellar, circumstellar and interstellar parameters
possibly relevant to ``detection'' and ``non-detection'' samples (distance $d \leq 500$~pc).}\label{tb:kolmogorov}
\resizebox{\columnwidth}{!}{
\begin{tabular}{lllll}\hline\hline
          			&\multicolumn{4}{c}{Kolmogorov probability $p$~\tablefootmark{a}} 	\\ \cline{2-5}
				& (I,II,IV)	     & (I,II,IV)	& (III)            & (I-IV)	\\         
				& vs. (III)          & vs. (X)          & vs. (X)          & vs. (X)    \\         
				& n=(34,8)	     & n=(34,16)	& n=(8,16)         & n=(42,16)  \\ \hline
critical value ($\alpha=0.01$)	& 0.70	             & 0.56		& 0.76	           &		\\ \hline
$d$ (pc)			& 0.25  	     & 0.18		& 0.98             & 0.28	\\
$z$ (pc) / $n_H$ (cm$^{-3}$)	& 0.99  	     & 0.03		& 0.19             & 0.02	\\
$v_\star$ (\kms)		& 0.43  	     & 0.02 		& 0.09             & 0.02	\\
inclination $i$ (\degr)		& 0.01  	     & 0.52		& 0.19	           & 0.68	\\
\Mdot (M$_\odot$ yr$^{-1}$)	& 0.55  	     & 0.48		& 0.59             & 0.57	\\
$v_w$ (\kms)			& 0 78  	     & 0.74		& 0.84             & 0.78	\\
$R_0$ predicted (pc)		& 0.40  	     & 0.07		& 0.84             & 0.09	\\
$R_0$ predicted (arcmin)	& 0.94  	     & 0.02		& 0.19             & 0.02	\\
\hline
\end{tabular}
}
\tablefoot{\tablefoottext{a}{A value close to unity indicates that the two samples (detection and non-detection) have a 
high probability to be from the same parent distribution. A low probability suggests their distributions differ significantly,
thus indicating the parameter is decisive in observing bow shock interaction in our survey. The second to fifth column
give the $K$-$S$ results between different sub-sets of classes (see Table~4 for the classification). The number of data-points
in each (sub)set are given in the fourth row.}
}
\end{table}

\section{Conclusions}\label{sec:conclusion}

This paper presents a morphological inventory of ``wind-ISM'' bow shocks and ``wind-wind'' interactions detected in the
far-infrared with Herschel/PACS. Five main classes are identified:

\begin{itemize}

\item[I] \emph{``Fermata''}. These objects are characterised by a large arc or shell-like structure spanning an angle of at
least 90\degr. Several objects in this class show the presence of turbulence, \ie\ Rayleigh-Taylor or Kelvin-Helmholtz
instabilities. Some objects, such as X\,Pav, EP\,Aqr, and X\,Her, reveal the presence of back flowing material in the wake (behind) the star, 
leading to a distinct bullet-shape.

\item[II] \emph{``Eyes''}. These are characterised by two arcs on opposing sides of the central object. These arcs are
elliptical and non-concentric. In two cases the arcs connect and form a \emph{``lemon''} shape morphology.

\item[III] \emph{``Rings''}. This class includes the well known detached shell objects, typical of wind-wind interaction and
dust pile-up. Younger, smaller rings have a co-spatial counterpart molecular gas ring, while the larger dust shells do not.

\item[IV] \emph{``Irregular''}. This class includes all sources which show extended (non-detached), irregular emission at 70
or 160~$\mu$m.

\item[X] \emph{``Non-detection''}. All objects for which no large extended far-infrared emission has been observed (c.q.
resolved) in the form of a bow shock or ring with the PACS instrument.

\end{itemize}

Oxygen rich stars give rise predominantly to ``fermata'' or ``irregular'' morphologies.
There is tentative evidence that all ``ring'' stars are thermally-pulsing AGB stars.
We identify a few cases for which both a detached (inner) shell is found within the (outer) bow shock, suggesting that
the detached shell is not due to the wind-ISM interaction, but suggest rather a wind-wind scenario.

The presence or absence of bow shocks is determined by the stand-off distance which depends on relative values of the stellar
parameters  ($v_\star$, $v_w$ and \Mdot) as well as those of the ISM ($n_\mathrm{H}$) as given by the theoretical approximation as
well as the hydrodynamical simulations.  The distribution of stand-off distances for the entire survey sample shows a clear
separation between objects with and without detected extended emission. Indeed most of the objects (44) assigned to Classes\,I
to~IV have $R_0 <$ 1.0~pc. No extended bow shock or detached shell emission has been detected for the 15 objects with predicted
$R_0 > 1$~pc and/or $d > 1000$~pc. Limiting the sample to $d < 500$~pc, we derive a $R_0 < 1.0$~pc, and thus predict
the presence of bow shocks, for 8 stars in the \emph{``non-detection''} class: TW\,Hor, V\,Eri, R\,Dor, R\,Lep, RS\,Cnc, HD\,100764, RY\,Dra, and
RX\,Boo. Many of these show extended shells at about 1~MJy sr$^{-1}$ sensitivity in Spitzer and AKARI maps
(private communication: T. Ueta for Spitzer and H. Izumiura for AKARI).

The angular size of the predicted stand-off distances for 6 other objects in Class\,X are larger than the obtained PACS image
maps. In particular, all-sky survey missions, such as AKARI (\citealt{2007PASJ...59S.369M}) and WISE
(\citealt{2010AJ....140.1868W}) could help in observing extended  bow shocks and shells around these objects.  Observations of bow
shocks and detached shells for distant ($>$1~kpc) AGB stars and red supergiants require thus higher sensitivities to detect
extended emission, but not necessarily higher spatial resolution as most have predicted stand-off distances larger than 1\arcmin.

The observed infrared emission indicates the presence of moderate amounts of dust ($2 \times 10^{-7}$ to $2.5 \times
10^{-4}$~M$_\odot$), implying only up to a few tenths of solar masses of dust and gas in the bow shock interaction region. 
The observed dust and gas masses are similar to the potentially swept-up ISM material.

This survey represents only a first step in fully characterising and understanding the formation and shaping of bow shocks and
detached shells around AGB stars and red supergiants. It is clear that additional simulations trying to represent as closely
as possible the different observed morphologies are required in particular to understand the formation of instabilities.
Detailed observations of the bow shock spectral energy distribution (dust) and line emission (gas) is needed to quantify the
physical conditions and composition of the material in these shocked regions.  Potentially, detailed observations of bow
shocks around larger samples of AGB stars can be used to independently probe the local ISM density  (and possibly the magnetic
field) distribution in the Galaxy.

\begin{acknowledgements}

This research has made extensive use of the Simbad and Vizier, operated at CDS, Strasbourg, France, 
as well as of NASA's Astrophysics Data System.
We thank Wang Ye at the Department of Physics \& Astronomy, University of Kentucky, for providing us with the radiative cooling curve.
This work was supported in part by the Belgian Federal Science Policy Office via the PRODEX Programme of ESA (no. C90371).
F.K.\ acknowledges funding by the Austrian Science Fund FWF under project 
number P23586-N16, R.O.\ under project number I163-N16.
A.J.v.M.\ acknowledges support from FWO, grant G.0277.08 and K.U.Leuven GOA/09/009.
PACS has been developed by a consortium of institutes led by MPE (Germany) and including UVIE (Austria); 
KUL, CSL, IMEC (Belgium); CEA, OAMP (France); MPIA (Germany); IFSI, OAP/AOT, OAA/CAISMI, LENS, SISSA
(Italy); IAC (Spain). This development has been supported by the funding agencies BMVIT (Austria), 
ESA-PRODEX (Belgium), CEA/CNES (France), DLR (Germany), ASI (Italy), and CICT/MCT (Spain).

\end{acknowledgements}

\bibliographystyle{aa}  
\bibliography{/lhome/nick/Desktop/ReadingMaterial/Astronomy/Bibtex/bibtex} 

\end{document}